\newcommand{ \logg} {$\log{g}$}
\newcommand{ \rsun }{$R_{\sun}$}

\newcommand{ \rstar }{$R_{\star}$}

\newcommand{ \tstar}{$T_{\star}$}
\newcommand{ \rp}{$R_{p}$}

\newcommand{ \Splanet}{$S_{p}$}

\newcommand{ \re}{$R_{\earth}$}
\newcommand{ \Rearth}{$R_{\earth}$}

\newcommand{ \se}{$S_{\earth}$}
\newcommand{ \kepler}{\emph{Kepler}}

\newcommand{ \Ktwo}{\emph{K2}}
\newcommand{ \injtce}{injTCE}
\newcommand{ \scrtce}{scrTCE}
\newcommand{ \invtce}{invTCE}
\newcommand{ \opstce}{obsTCE}
\newcommand{ \injtces}{injTCEs}
\newcommand{ \scrtces}{scrTCEs}
\newcommand{ \invtces}{invTCEs}
\newcommand{ \opstces}{obsTCEs}

\newcommand{ \ntces}{34,032}        
\newcommand{ \ntcesnorogue}{32,534}        
\newcommand{ \npredrtwentyfivekois}{8,826}  
\newcommand{ \nkebs}{2,605}         
\newcommand{ \nephemmatch}{1,859}   
\newcommand{ \nonlyephemmatch}{106} 
\newcommand{ \nscrtces}{13782} 
\newcommand{ \ninvtces}{14953} 
\newcommand{ \ninjecttargs}{146,294} 
\newcommand{ \nkoisdr}{8054} 

\newcommand{\nasa} {NASA Ames Research Center, Moffett Field, CA 94035, USA}
\newcommand{\seti}{SETI Institute, 189 Bernardo Ave, Suite 200, Mountain View, CA 94043, USA}
\newcommand{\stsci}{Space Telescope Science Institute, 3700 San Martin Drive, Baltimore, MD 21218}
\newcommand{\nexsci}{IPAC-NExScI, Mail Code 100-22, Caltech, 1200 E. California Blvd. Pasadena, CA 91125}
\newcommand{\mitadd}{MIT Kavli Institute for Astrophysics and Space Research, 77 Massachusetts Avenue, 37-241, Cambridge, MA 02139}
\newcommand{\orbital}{Orbital Insight, 100 W Evelyn Ave \#110, Mountain View, CA 94041}
\newcommand{\bishop}{Dept. of Physics and Astronomy, Bishop's University, 2600 College St., Sherbrooke, QC, J1M 1Z7, Canada}
\newcommand{\bari}{Bay Area Environmental Research Institute, 625 2nd St., Ste 209, Petaluma, CA 94952, USA}
\newcommand{\nsffellow}{NSF Astronomy \& Astrophysics Postdoctoral Fellow}
\newcommand{\cfa}{Harvard-Smithsonian Center for Astrophysics, 60 Garden Street, Cambridge MA 02138, USA}
\newcommand{\aarhus}{Stellar Astrophysics Centre, Dept. of Physics and Astronomy, Aarhus University, Ny Munkegade 120, 8000 Aarhus C, Denmark}
\newcommand{\nasagoddard}{NASA Goddard Space Flight Center, 8800 Greenbelt Road, Greenbelt, MD 20771}
\newcommand{\wyle}{KRBwyle, 2400 Nasa Parkway, Houston, TX 77058 USA}
\newcommand{\pennstate}{Dept. of Astronomy \& Astrophysics, 525 Davey Laboratory, The Pennsylvania State University, University Park, PA, 16802, USA} 
\newcommand{\habworlds}{Center for Exoplanets and Habitable Worlds, 525 Davey Laboratory, The Pennsylvania State University, University Park, PA, 16802, USA} 

\lccode`1=`1
\lccode`2=`2
\lccode`3=`3

\documentclass[twocolumn,tighten]{aastex61}
\let\underscore\_
\renewcommand{\_}{\discretionary{\underscore}{}{\underscore}}  

\usepackage{hyperref}
\usepackage{amsmath}
\usepackage{url}
\usepackage{color}
\usepackage{soul}
\usepackage{marginnote}
\usepackage[T1]{fontenc}  
\usepackage{makecell}  

\newcommand\Kepler{\textit{Kepler}}

\begin{document}

\title{Planetary Candidates Observed by \Kepler. VIII.\\
A Fully Automated Catalog With Measured Completeness and Reliability\\Based on Data Release 25 }

\author[0000-0001-7106-4683]{Susan E. Thompson}
\altaffiliation{a.k.a. Susan E. Mullally, email: smullally@stsci.edu}
\affiliation{\seti}
\affiliation{\nasa}
\affiliation{\stsci}

\author[0000-0003-1634-9672]{Jeffrey L. Coughlin}
\affiliation{\nasa}
\affiliation{\seti}

\author{Kelsey Hoffman}
\affiliation{\seti}

\author{Fergal Mullally}
\affiliation{\seti}
\affiliation{\nasa}
\affiliation{\orbital}

\author{Jessie L. Christiansen}
\affiliation{\nexsci}

\author{Christopher J. Burke}
\affiliation{\nasa}
\affiliation{\seti}
\affiliation{\mitadd}

\author{Steve Bryson}
\affiliation{\nasa}

\author{Natalie Batalha}
\affiliation{\nasa}

\author{Michael R. Haas}
\altaffiliation{NASA Ames Associate}
\affiliation{\nasa}

\author{Joseph Catanzarite}
\affiliation{\seti}
\affiliation{\nasa}

\author[0000-0002-5904-1865]{Jason F. Rowe}
\affiliation{\bishop}

\author{Geert Barentsen}
\affiliation{\bari}

\author{Douglas A. Caldwell}
\affiliation{\seti}
\affiliation{\nasa}


\author{Bruce D. Clarke}
\affiliation{\seti}
\affiliation{\nasa}

\author[0000-0002-4715-9460]{Jon M. Jenkins}
\affiliation{\nasa}

\author{Jie Li}
\affiliation{\seti}

\author[0000-0001-9911-7388]{David W. Latham}
\affiliation{\cfa}

\author{Jack J. Lissauer}
\affiliation{\nasa}

\author{Savita Mathur}
\affiliation{Space Science Institute, 4750 Walnut Street, Suite 205, Boulder, CO 80301, USA}

\author{Robert L. Morris}
\affiliation{\seti}
\affiliation{\nasa}

\author{Shawn E. Seader}
\affiliation{Rincon Research Corporation,101 N Wilmot Rd, Tucson, AZ 85711}

\author{Jeffrey C. Smith}
\affiliation{\seti}
\affiliation{\nasa}

\author{Todd C. Klaus}
\affiliation{\nasa}

\author{Joseph D. Twicken}
\affiliation{\seti}
\affiliation{\nasa}

\author{Jeffrey E. Van Cleve}
\affiliation{\seti}

\author{Bill Wohler}
\affiliation{\seti}
\affiliation{\nasa}

\author[0000-0001-9674-1564]{Rachel Akeson}
\affiliation{\nexsci}

\author[0000-0002-5741-3047]{David R. Ciardi}
\affiliation{\nexsci}

\author[0000-0001-9662-3496]{William D. Cochran}
\affiliation{McDonald Observatory and Department of Astronomy, University of Texas at Austin, Austin, TX 78712}

\author{Christopher E. Henze}
\affiliation{\nasa}

\author{Steve B. Howell}
\affiliation{\nasa}

\author{Daniel Huber}
\affiliation{Institute for Astronomy, University of Hawai`i, 2680 Woodlawn Drive, Honolulu, HI 96822, USA}
\affiliation{Sydney Institute for Astronomy (SIfA), School of Physics, University of Sydney, NSW 2006, Australia}
\affiliation{\seti}
\affiliation{\aarhus}

\author[0000-0002-1913-0281]{Andrej Pr\v sa}
\affiliation{Villanova University, Dept.~of Astrophysics and Planetary Science, 800 Lancaster Ave, Villanova PA 19085}

\author{Solange V. Ram\'{\i}rez}
\affiliation{\nexsci}

\author{Timothy D. Morton}
\affiliation{Department of Astrophysical Sciences, Princeton University, 4 Ivy Lane, Princeton, NJ 08544, USA}

\author[0000-0001-7139-2724]{Thomas Barclay}
\affiliation{\nasagoddard}

\author{Jennifer R. Campbell}
\affiliation{\nasa}
\affiliation{\wyle}

\author{William J. Chaplin}
\affiliation{School of Physics and Astronomy, University of Birmingham, Edgbaston, Birmingham B15 2TT, UK}
\affiliation{\aarhus}

\author[0000-0002-9003-484X]{David Charbonneau}
\affiliation{\cfa}

\author{J{\o}rgen Christensen-Dalsgaard}
\affiliation{\aarhus}

\author{Jessie L. Dotson}
\affiliation{\nasa}

\author{Laurance  Doyle}
\affiliation{Institute for the Metaphysics of Physics, Principia College, One Maybeck Place, Elsah, Illinois 62028}
\affiliation{\seti}

\author{Edward W. Dunham}
\affiliation{Lowell Observatory, 1400 W Mars Hill Rd, Flagstaff, AZ 86001}

\author[0000-0002-8985-8489]{Andrea K. Dupree}
\affiliation{\cfa}

\author[0000-0001-6545-639X]{Eric B. Ford}
\affiliation{\pennstate} 
\affiliation{\habworlds} 
\affiliation{Center for Astrostatistics, 525 Davey Laboratory, The Pennsylvania State University, University Park, PA, 16802, USA} 
\affiliation{Institute for CyberScience, The Pennsylvania State University}

\author{John C. Geary}
\affiliation{\cfa}

\author{Forrest R. Girouard}
\affiliation{Orbital Sciences Corporation, 2401 East El Segundo Boulevard, Suite 200, El Segundo, CA 90245, USA}
\affiliation{\nasa}

\author[0000-0002-0531-1073]{Howard Isaacson}
\affiliation{Dept. of Astronomy, UC Berkeley, Berkeley, CA 94720, USA}

\author{Hans Kjeldsen}
\affiliation{\aarhus}


\author{Elisa V. Quintana}
\affiliation{\nasagoddard}

\author{Darin Ragozzine}
\affiliation{Brigham Young University, Department of Physics and Astronomy, N283 ESC, Provo, UT 84602, USA}

\author[0000-0003-1179-3125]{Megan Shabram}
\altaffiliation{NASA Postdoctoral Program Fellow, Administered by Universities Space Research Association}
\affiliation{\nasa}

\author{Avi Shporer}
\affiliation{Division of Geological and Planetary Sciences, California Institute of Technology, Pasadena, CA 91125, USA}

\author{Victor Silva Aguirre}
\affiliation{\aarhus}

\author{Jason H. Steffen}
\affiliation{University of Nevada, Las Vegas, 4505 S Maryland Pkwy, Las Vegas, NV 89154}

\author{Martin Still}
\affiliation{\bari}

\author{Peter Tenenbaum}
\affiliation{\seti}
\affiliation{\nasa}

\author{William F. Welsh}
\affiliation{Department of Astronomy, San Diego State University, 5500 Campanile Drive, San Diego, CA 92182-1221}

\author[0000-0003-2862-6278]{Angie Wolfgang}
\altaffiliation{\nsffellow}
\affiliation{\pennstate} 
\affiliation{\habworlds}

\author{Khadeejah A Zamudio}
\affiliation{\nasa}
\affiliation{\wyle}

\author{David G. Koch}
\altaffiliation{deceased}
\affiliation{\nasa}

\author{William J. Borucki}
\altaffiliation{NASA Ames Associate}
\affiliation{\nasa}

\begin{abstract}
We present the Kepler Object of Interest (KOI) catalog of transiting exoplanets based on searching four years of \Kepler{} time series photometry (Data Release 25, Q1--Q17). The catalog contains \nkoisdr{} KOIs of which 4034 are planet candidates with periods between 0.25~and 632~days. Of these candidates, 219 are new and include two in multi-planet systems (KOI-82.06 and KOI-2926.05), and ten high-reliability, terrestrial-size, habitable zone candidates. This catalog was created using a tool called the Robovetter which automatically vets the DR25 Threshold Crossing Events \citep[TCEs,][]{Twicken2016}. The Robovetter also vetted simulated data sets and measured how well it was able to separate TCEs caused by noise from those caused by low signal-to-noise transits. We discusses the Robovetter and the metrics it uses to sort TCEs. For orbital periods less than 100\,days the Robovetter completeness (the fraction of simulated transits that are determined to be planet candidates) across all observed stars is greater than 85\%. For the same period range, the catalog reliability (the fraction of candidates that are not due to instrumental or stellar noise) is greater than 98\%.  However, for low signal-to-noise candidates between 200 and 500 days around FGK dwarf stars, the Robovetter is 76.7\% complete and the catalog is 50.5\% reliable. The KOI catalog, the transit fits and all of the simulated data used to characterize this catalog are available at the NASA Exoplanet Archive.

\end{abstract}


\section{Introduction}

\Kepler{'s} mission to measure the frequency of Earth-size planets in the Galaxy is an important step towards understanding the Earth's place in the Universe.  Launched in 2009, the \Kepler{} Mission \citep{Koch2010,Borucki2016} stared almost continuously at a single field for four years (or 17, $\approx$90~day quarters), recording the brightness of $\approx$200,000 stars ($\approx$160,000 stars at a time) at a cadence of 29.4 minutes over the course of the mission. \Kepler{} detected transiting planets by observing the periodic decrease in the observed brightness of a star when an orbiting planet crossed the line of sight from the telescope to the star. \Kepler{'s} prime-mission observations concluded in 2013 when it lost a second of four reaction wheels, three of which were required to maintain the stable pointing.  From the ashes of \Kepler{} rose the \Ktwo{} mission which continues to find exoplanets in addition to a whole host of astrophysics enabled by its observations of fields in the ecliptic \citep{Howell2014,VanCleve2016K2}. 
While not the first to obtain high-precision, long-baseline photometry to look for transiting exoplanets \citep[see e.g.,][]{Barge2008,ODonovan2006}, \Kepler{} and its plethora of planet candidates revolutionized exoplanet science. The large number of Kepler planet detections from the same telescope opened the door for occurrence rate studies and has enabled some of the first measurements of the frequency of planets similar to the Earth in our Galaxy.  To further enable those types of studies, we present here the planet catalog that resulted from the final search of the Data Release 25 (DR25) \Kepler{} mission data along with the tools provided to understand the biases inherent in the search and vetting done to create that catalog.

First, we put this work in context by reviewing some of the scientific achievements accomplished using \Kepler{} data.  Prior to \Kepler{,} most exoplanets were discovered by radial velocity methods \citep[e.g.][]{Mayor1995}, which largely resulted in the detection of Neptune- to Jupiter-mass planets in orbital periods of days to months. The high precision photometry and the four-year baseline of the \Kepler{} data extended the landscape of known exoplanets.  To highlight a few examples, \citet{Barclay2013} found evidence for a moon-size terrestrial planet in a 13.3 day period orbit, \citet{Quintana2014} found evidence of an Earth-size exoplanet in the habitable zone of the M dwarf Kepler-186, and \citet{Jenkins2015} statistically validated a super-Earth in the habitable zone of a G-dwarf star. Additionally, for several massive planets \Kepler{} data has enabled measurements of planetary mass and atmospheric properties by using the photometric variability along the entire orbit \citep{Shporer2011,Mazeh2012,Shporer2017}. \Kepler\ data has also revealed hundreds of compact, co-planar, multi-planet systems, e.g., the six planets around Kepler-11 \citep{Lissauer2011}, which collectively have told us a great deal about the architecture of planetary systems \citep{Lissauer2011b,Fabrycky2014}.   Exoplanets have even been found orbiting binary stars, e.g., Kepler-16~(AB)~b \citep{Doyle2011}.

Other authors have taken advantage of the long time series, near-continuous data set of 206,150\footnote{This tally only includes the targeted stars and not those observed by ``accident'' in the larger apertures.} stars to advance our understanding of stellar physics through the use of asteroseismology. Of particular interest to this catalog is the improvement in the determination of stellar radius \citep[e.g.,][]{Huber2014a,Mathur2017ApJS} which can be one of the most important sources of error when calculating planetary radii. \Kepler{} data was also used to track the evoluation of star-spots created from magnetic activity and thus enabled the measurement of stellar rotation rates \citep[e.g.][]{Aigrain2015,Garcia2014,McQuillan2014,Zimmerman2017}. Studying stars in clusters enabled \citet{Meibom2011} to map out the evolution of stellar rotation as stars age. \Kepler{} also produced light curves of 2876\footnote{This represents the number reported in the Kepler Binary Catalog, \url{http://keplerebs.villanova.edu}, in August 2017.} eclipsing binary stars \citep{Prsa2011,Kirk2016} including unusual binary systems, such as the eccentric, tidally-distorted, Heartbeat stars \citep{Welsh2011,Thompson2012,Shporer2016hb} that have opened the doors to understanding the impact of tidal forces on stellar pulsations and evolution \citep[e.g.,][]{Hambleton2017,Fuller2017}.

The wealth of astrophysics, and the size of the \Kepler{} community, is in part due to the rapid release of \Kepler{} data to the NASA Archives: the Exoplanet Archive \citep[][]{Akeson2013} and the MAST (Mikulski Archives for Space Telescopes).  The \Kepler{} mission released data from every step of the processing \citep{Thompson2016KAM,Stumpe2014,Bryson2010b}, including its planet searches. The results of both the original searches for periodic signals (known as the TCEs or Threshold Crossing Events) and the well-vetted KOIs (Kepler Objects of Interest) were made available for the community. The combined list of \Kepler{'s} planet candidates found from all searches can be found in the cumulative KOI table\footnote{\url{https://exoplanetarchive.ipac.caltech.edu/cgi-bin/ TblView/nph-tblView?app=ExoTbls\&config=cumulative}}.  The KOI table we present here is from a single search of the DR25 light curves\footnote{doi:10.17909/T9488N}. While the search does not include new observations, it was performed using an improved version of the \Kepler{} Pipeline \citep[version 9.3,][]{Jenkins2017}. For a high-level summary of the changes to the \Kepler{} Pipeline, see the DR25 data release notes \citep{DRN25,KDCH}. The \Kepler{} Pipeline has undergone successive improvements since launch as the data characteristics have become better understood.

The photometric noise at time scales of the transit is what limits \Kepler{} from finding small terrestrial-size planets. Investigations of the noise properties of \Kepler{} exoplanet hosts by \citet{Howell2016} showed that those exoplanets \added{around dwarf FGK-type stars} with the radii $\leq$1.2\re{} are only found around the brightest, most photometrically quiet stars. As a result, the search for the truly Earth-size planets are limited to a small subset of \Kepler{'s} stellar sample.  Analyses by \citet{Gilliland2011,Gilliland2015} show that the primary source of the observed noise was indeed inherent to the stars, with a smaller contributions coming from imperfections in the instruments and software. Unfortunately, the typical noise level for 12$^{th}$ magnitude solar-type stars is closer to 30\,ppm \citep{Gilliland2015} than the 20\,ppm expected prior to launch \citep{Jenkins2002a}, causing \Kepler{} to need a longer baseline to find a significant number of Earth-like planets around Sun-like stars.   Ultimately, this higher noise level impacts \Kepler{'s} planet yield. And, because different stars have different levels of noise, the transit depth to which the search is sensitive varies across the sample of stars. This bias must be accounted for when calculating occurrence rates, and is explored in-depth for this run of the \Kepler{} Pipeline by the transit injection and recovery studies of \citet[][]{Burke2017b,Burke2017a} and \citet{Christiansen2017}.

To confirm the validity and further characterize identified planet candidates, the \Kepler{} mission benefited from an active, funded, follow-up observing program. This program used ground-based radial velocity measurements to determine the mass of exoplanets \citep[e.g.,][]{Marcy2014} when possible and also ruled out other astrophysical phenomena, like background eclipsing binaries, that can mimic a transit signal.  Both funded \added{and unfunded high-resolution imaging studies have covered $\approx$90\% of known KOIs \citep[see e.g.,][]{Furlan2017,Law2014,Barnec2016,Ziegler2017} }to identify close companions (bound or unbound) that would be included in \Kepler{'s} rather large 3.98\,$\arcsec$ pixels.  The extra light from these companions must be accounted for when determining the depth of the transit and the radii of the exoplanet.  While the \Kepler{} Pipeline accounts for the stray light from stars in the Kepler Input Catalog (\citealt{Brown2011}; and see flux fraction in \S2.3.1.2 of the Kepler Archive Manual; \citealt{Thompson2016KAM}), the sources identified by these high-resolution imaging studies were not included. \added{The resulting DR25 planet catalog also does not include the results of these studies because high resolution imaging is only available for stars with KOIs, and if included, could incorrectly bias occurrence rate measurements.} Based on the analysis by \citet{Ciardi2015}, where they considered the effects of multiplicity, planet radii are underestimated by a factor averaging\,$\simeq$1.5 for G dwarfs prior to vetting, or averaging $\simeq$1.2 for KOIs that have been vetted with high-resolution imaging and Doppler spectroscopy.  The effect of unrecognized dilution decreases for planets orbiting the K and M dwarfs, because they have a smaller range of possible stellar companions

Even with rigorous vetting and follow-up observations, most planet candidates in the KOI catalogs cannot be directly confirmed as planetary. The stars are too dim and the planets are too small to be able to measure a radial velocity signature for the planet.   Statistical methods study the likelihood that the observed transit could be caused by other astrophysical scenarios and have succeeded in validating thousands of \Kepler{} planets \citep[e.g.][]{Morton2016,Torres2015,Rowe2014,Lissauer2014}.  

The Q1--Q16 KOI catalog \citep{Mullally2015cat} was the first with a long enough baseline to be significantly impacted by another source of false positives, the long-period false positives created by the instrument itself.  In that catalog (and again in this one), the majority of long-period, low SNR TCEs are ascribed to instrumental effects incompletely removed from the data before the TCE search. \Kepler\ has a variety of short timescale (on the order of a day or less), non-Gaussian noise sources including focus changes due to thermal variations, signals imprinted on the data by the detector electronics, noise caused by solar flares, and the pixel sensitivity changing after the impact of a high energy particle (known as a sudden pixel sensitivity drop-out, or SPSD). Because the large number of TCEs associated with these types of errors, and because the catalog was generated to be intentionally inclusive (i.e. high completeness), many of the long-period candidates in the Q1--Q16 KOI catalog are expected to simply be noise.  We were faced with a similar problem for the DR25 catalog and spent considerable effort writing software to identify these types of false positives, and for the first time we include an estimate for how often these signals contaminate the catalog.

The planet candidates found in \Kepler{} data have been used extensively to understand the frequency of different types of planets in the Galaxy. Many studies have shown that small planets ($<4$\re) in short period orbits are common, with occurrence rates steadily increasing with decreasing radii \citep{Burke2016,Howard2012,Petigura2013b,Youdin2011}.  \citet{Dressing2013,Dressing2015}, using their own search, confined their analysis to M dwarfs and orbital periods less than 50\,d and determined that multi-planet systems are common around these low mass stars.  Therefore planets are more common than stars in the Galaxy (due, in part, to the fact that low mass stars are the most common stellar type). 
\citet{Fulton2017}, using improved measurements of the stellar properties \citep{Petigura2017}, looked at small planets with periods of less than 100\,d and showed that there is a valley in the occurrence of planets near 1.75\re{}. This result improved upon the results of \citet{Howard2012} and \citet{Lundkvist2016} and further verified the evaporation valley predicted by \citet{Owen2013} and \citet{Lopez2013} for close-in planets.

Less is known about the occurrence of planets in longer period orbits. Using planet candidates discovered with \Kepler{}, several papers have measured the frequency of small planets in the habitable zone of sun-like stars \citep[see e.g.][]{Burke2015,ForemanMackey16,Petigura2013b} using various methods. \citet{Burke2015} used the Q1--Q16  KOI catalog \citep{Mullally2015cat} and looked at G and K stars and concluded that 10\% (with an allowed range of 1--200\%) of solar-type stars host planets with radii and orbital periods within 20\% of that of the Earth. \citet{Burke2015} considered various systematic effects and showed that they dominate the uncertainties and concluded that improved measurements of the stellar properties, the detection efficiency of the search, and the reliability of the catalog will have the most impact in narrowing the uncertainties in such studies.

\subsection{Design Philosophy of the DR25 catalog}

The DR25 KOI catalog is designed to support rigorous occurrence rate studies. To do that well, it was critical that we not only identify the exoplanet transit signals in the data but also measure the catalog reliability (the fraction of transiting candidates that are not caused by noise), and the completeness of the catalog (the fraction of true transiting planets detected).

The measurement of the catalog completeness has been split into two parts: the completeness of the TCE list (the transit search performed by the \Kepler\ Pipeline) and the completeness of the KOI catalog (the vetting of the TCEs). The completeness of the \Kepler\ Pipeline and its search for transits has been studied by injecting transit signals into the pixels and examining what fraction are found by the \Kepler{} Pipeline \citep{Christiansen2017, Christiansen2015b,Christiansen2013a}. \citet{Burke2015} applied the appropriate detection efficiency contours \citep{Christiansen2015} to the 50--300\,d period planet candidates in the Q1--Q16 KOI catalog \citep{Mullally2015cat} in order to measure the occurrence rates of small planets. However, that study was not able to account for those transit signals correctly identified by the \Kepler{} Pipeline but thrown-out by the vetting process. Along with the DR25 KOI catalog, we provide a measure of the completeness of the DR25 vetting process. 

\Kepler{} light curves contain variability that is not due to planet transits or eclipsing binaries. While the reliability of \Kepler\ catalogs against astrophysical false positives is mostly understood \citep[see e.g.][]{Morton2016}, the reliability against false alarms (a term used in this paper to indicate TCEs caused by intrinsic stellar variability, over-contact binaries, or instrumental noise, i.e., anything that does not look transit-like) has not previously been measured. Instrumental noise, statistical fluctuations, poor detrending, and/or stellar variability can conspire to produce a signal that looks similar to a planet transit. When examining the smallest exoplanets in the longest orbital periods, \citet{Burke2015} demonstrated the importance of understanding the reliability of the catalog, showing that the occurrence of small, earth-like-period planets around G dwarf stars changed by a factor of $\approx$10 depending on the reliability of a few planet candidates.  In this catalog we measure the reliability of the reported planet candidates against this instrumental and stellar noise.  

The completeness of the vetting process is measured by vetting thousands of injected transits found by the \Kepler{} Pipeline. Catalog reliability is measured by vetting signals found in scrambled and inverted \Kepler{} light curves and counting the fraction of simulated false alarms that are dispositioned as planet candidates. This desire to vet both the real and simulated TCEs in a reproducible and consistent manner demands an entirely automated method for vetting the TCEs.  

Automated vetting was introduced in the Q1--Q16 KOI catalog \citep{Mullally2015cat} with the Centroid Robovetter and was then extended to all aspects of the vetting process for the DR24 KOI catalog \citep{Coughlin2016}. Because of this automation, the DR24 catalog was the first with a measure of completeness that extended to all parts of the search, from pixels to planet candidates.  Now, with the DR25 KOI catalog and simulated false alarms, we also provide a measure of how effective the vetting techniques are at identifying noise signals and translate that into a measure of the catalog reliability. As a result, the DR25 KOI catalog is the first to explicitly balance the gains in completeness against the loss of reliability, instead of always erring on the side of high completeness. 

\subsection{Terms and Acronyms}
\label{abbrev}
We try to avoid unnecessary acronyms and abbreviations, but a few are required to efficiently discuss this catalog.  Here we itemize those terms and abbreviations that are specific to this paper and are used repeatedly. The list is short enough that we choose to group them by meaning instead of alphabetically. 

\begin{itemize}

\item[] \textbf{TCE}: Threshold Crossing Event. Periodic signals identified by the transiting planet search (TPS) module of the \Kepler{} Pipeline \citep{JenkinsKDPH}.
\item[] \textbf{\opstce}: Observed TCEs. TCEs found by searching the observed DR25 \Kepler\ data and reported in \citet{Twicken2016}. \added{See \S\ref{s:tces}.}
\item[] \textbf{\injtce}: Injected TCEs. TCEs found that match a known, injected transit signal \citep{Christiansen2017}. See \S\ref{s:injecttce}.
\item[] \textbf{\invtce}: Inverted TCEs. TCEs found when searching the inverted data set in order to simulate instrumental false alarms \citep{Coughlin2017a}. \added{See \S\ref{s:tcefalsealarms}.}
\item[] \textbf{\scrtce}: Scrambled TCEs. TCEs found when searching the scrambled data set in order to simulate instrumental false alarms \citep{Coughlin2017a}. \added{\S\ref{s:tcefalsealarms}.}
\item[] \textbf{TPS}: Transiting Planet Search module. This module of the \Kepler{} Pipeline performs the search for planet candidates. Significant, periodic events are identified by TPS and turned into TCEs.
\item[] \textbf{DV}: Data Validation. Named after the module of the \Kepler\ Pipeline \citep{JenkinsKDPH} which characterizes the transits and outputs one of the detrended light curves used by the Robovetter metrics.  DV also created two sets of transit fits: original and supplemental (\S\ref{s:fits}).
\item[] \textbf{ALT}: Alternative. As an alternative to the DV detrending, the \Kepler\ Pipeline implements a detrending method that uses the methods of \citet{Garcia2010} and the out-of-transit points in the pre-search data conditioned (PDC) light curves to detrend the data. The \Kepler{} Pipeline performs a trapezoidal fit to the folded transit on the ALT detrended light curves (\S\ref{s:fits}).
\item[] \textbf{MES}: Multiple Event Statistic. A statistic that measures the combined significance of all of the observed transits in the detrended, whitened light curve assuming a linear ephemeris \citep[][]{Jenkins2002b}.
\item[] \textbf{KOI}: Kepler Object of Interest. Periodic, transit-like events that are significant enough to warrant further review. A KOI is identified with a KOI number and can be dispositioned as a planet candidate or a false positive. The DR25 KOIs are a subset of the DR25 \opstce{s}. \added{See \S\ref{s:assemble}.}
\item[] \textbf{PC}: Planet Candidate. A TCE or KOI that passes all of the Robovetter false positive identification tests. Planet candidates should not be confused with confirmed planets where further analysis has shown that the transiting planet model is overwhelmingly the most likely astrophysical cause for the periodic dips in the \Kepler{} light curve.\added{See \S\ref{s:robovetter}.}
\item[] \textbf{FP}: False Positive. A TCE or KOI that fails one or more of the Robovetter tests. Notice that the term includes all types of signals found in the TCE lists that are not caused by a transiting exoplanet, including eclipsing binaries and false alarms.\added{See \S\ref{s:robovetter}.}
\item[] \textbf{MCMC}: Markov chain Monte Carlo. This refers to transit fits which employ a MCMC algorithm in order to provide robust errors for fitted model parameters for all KOIs
 \citep{Hoffman2017}. \added{See \S\ref{s:mcmc}.}

\end{itemize}

\subsection{Summary and Outline of the Paper}

The DR25 KOI catalog is a uniformly-vetted list of planet candidates and false positives found by searching the DR25 \Kepler\ light curves and includes a measure of the catalog completeness and reliability. In the brief outline that follows we highlight how the catalog was assembled, how we measure the completeness and reliability, and discuss those aspects of the process that are different from the DR24 KOI catalog \citep{Coughlin2016}.

In \S\ref{s:tces} we describe the observed TCEs (\opstce s) which are the periodic signals found in the actual \Kepler\ light curves. For reference, we also compare them to the DR24 TCEs. To create the simulated data sets necessary to measure the vetting completeness and the catalog reliability, we ran the \Kepler\ Pipeline on light curves that either contained injected transits, were inverted, or were scrambled. This creates \injtce s, \invtce s, and \scrtce s, respectively (see \S\ref{s:simulated}).  

We then created and tuned a Robovetter to vet all the different sets of TCEs. \S\ref{s:robovetter} describes the metrics and the logic used to disposition TCEs into PCs and FPs.  Because the DR25 \opstce\ population was significantly different than the DR24 \opstce{s}, we developed new metrics to separate the PCs from the FPs (see Appendix \ref{s:metrics} for the details on how each metric operates.) Several new metrics examine the individual transits for evidence of instrumental noise (see \S\ref{s:indivtrans}.) As in the DR24~KOI catalog, we group FPs into four categories (\S\ref{s:majorflags}) and provide minor false positive flags (Appendix \ref{s:minorflags}) to indicate why the Robovetter decided to pass or fail a TCE.  New to this catalog is the addition of a disposition score (\S\ref{s:scores}) that gives users a measure of the Robovetter's confidence in each disposition.

Unlike previous catalogs, for the DR25 KOI catalog the choice of planet candidate versus false positive is no longer based on the philosophy of ``innocent until proven guilty''. We accept certain amounts of collateral damage (i.e., exoplanets dispositioned as FP) in order to achieve a catalog that is uniformly vetted and has acceptable levels of both completeness and reliability, especially for the long period and low signal-to-noise PCs. In \S\ref{s:optimize} we discuss how we tuned the Robovetter using the simulated TCEs as populations of true planet candidates and true false alarms. We provide the Robovetter source code and all the Robovetter metrics for all of the sets of TCEs (\opstces, \injtces, \invtces, and \scrtces) to enable users to create a catalog tuned for other regions of parameter space if their scientific goals require it. 

We assemble the catalog (\S\ref{s:assemble}) by federating to previously known KOIs before creating new KOIs. Then to provide planet parameters, each KOI is fit with a transit model which uses a Markov Chain Monte Carlo (MCMC) algorithm to provide error estimates for each fitted parameter (\S\ref{s:mcmc}). In \S\ref{s:summary} we summarize the catalog and discuss the performance of the vetting using the \injtce, \invtce, and \scrtce\ sets. We show that both decrease significantly with decreasing number of transits and decreasing signal-to-noise.  We then discuss how one may use the disposition scores to identify the highest quality candidates, especially at long periods (\S\ref{s:crscores}.)  We conclude that not all declared planet candidates in our catalog are actually astrophysical transits, but we can measure what fraction are caused by stellar and instrumental noise. Because of the interest in terrestrial, temperate planets, we examine the high quality, small candidates in the habitable zone in \S\ref{s:hz}. Finally, in \S\ref{s:occurates} we give an overview of what must be considered when using this catalog to measure accurate exoplanet occurrence rates, including what information is available in other \Kepler{} products to do this work.

\section{The Q1--Q17 DR25 TCEs}
\label{tcesec}
\subsection{Observed TCEs}

\label{s:tces}
As with the previous three Kepler KOI catalogs \citep{Coughlin2016,Mullally2015cat,Rowe2015cat}, the population of events that were used to create KOIs and planet candidates are known as \opstce s. These are periodic reductions of flux in the light curve that were found by the TPS module and evaluated by the DV module of the \Kepler{} Pipeline \citep{JenkinsKDPH}\footnote{The source code of the entire Pipeline is available at \url{https://github.com/nasa/kepler-pipeline}}. The DR25 \opstce{s}  were created by running the SOC 9.3 version of the \Kepler\ Pipeline on the DR25, Q1--Q17 \Kepler\ time-series.  For a thorough discussion of the DR25 TCEs and on the pipeline's search see \citet{Twicken2016}. 

The DR25 \opstce s, their ephemerides, and the metrics calculated by the pipeline are available at the NASA Exoplanet Archive \citep{Akeson2013}.  In this paper we endeavor to disposition these signals into planet candidates and false positives.  Because the \opstce s act as the input to our catalog, we first describe some of their properties as a whole and reflect on how they are different from the \opstce{} populations found with previous searches.

\begin{figure*}[htb]
 \begin{center}
  \includegraphics[width=\linewidth]{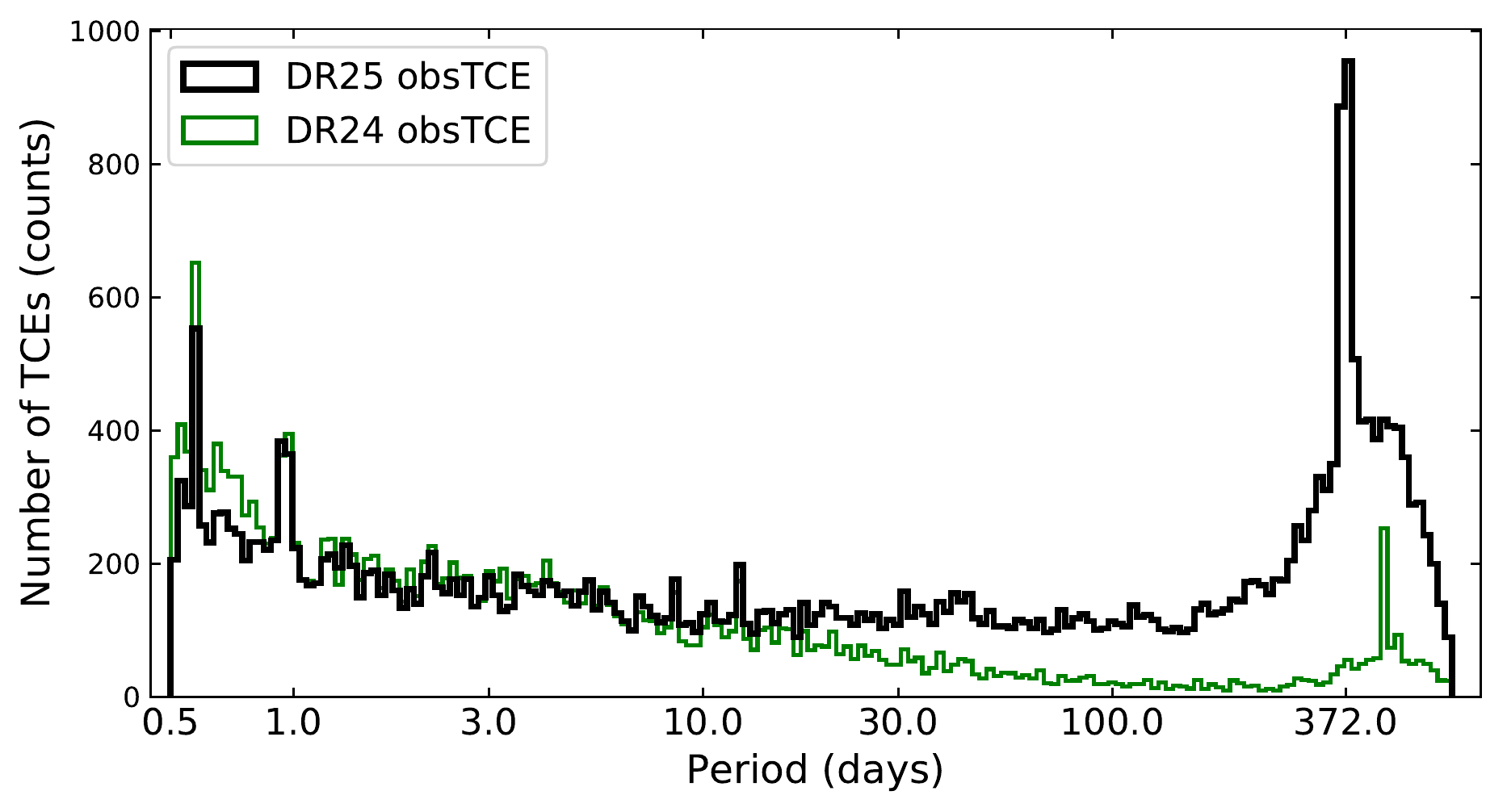}
  \caption{Histogram of the period in days of the DR25 \opstce{s} (black) using uniform bin space in the base ten logarithm of the period. The DR24 catalog obsTCEs \citep{Seader2015} are shown in green for comparison. The number of long-period TCEs is much larger for DR25 and includes a large spike in the number of TCEs at the orbital period of the spacecraft (372 days). The long and short period spikes for both distributions are discussed in \S\ref{s:tces}.}
  \label{f:obstces} 
 \end{center}
 \end{figure*}

We have plotted the distribution of the \ntcesnorogue\ \opstce s in terms of period in Figure~\ref{f:obstces}. Notice that there is an excessive number of short and long period \opstce{s} compared to the number of expected transiting planets. Not shown, but worth noting is that the number of \opstce{s} increases with decreasing MES.

As with previous catalogs, the short period ($<10$\,d) excess is dominated by true variability of stars due to both intrinsic stellar variability (e.g., spots or pulsations) and contact/near-contact eclipsing binaries. The long period excess is dominated by instrumental noise. For example, a decrease in flux following a cosmic ray hit (known as an SPSD; \citealt{KDCH}), can match up with other decrements in flux to produce a TCE. Also, image artifacts known as rolling-bands are very strong on some channels \citep[see \S6.7 of][]{KIH}  and since the spacecraft rolls approximately every 90\,d, causing a star to move on/off a \Kepler\ detector with significant rolling band noise, these variations can easily line up to produce TCEs at \Kepler{'s} heliocentric orbital period ($\approx$372 days, 2.57 in log-space). This is the reason for the largest spike in the \opstce\ population seen in Figure~\ref{f:obstces}. The narrow spike at 459 days (2.66 in log-space) in the DR24 \opstce{} distribution is caused by edge-effects near three equally spaced data gaps in the DR24 data processing.  The short period spikes in the distribution of both the DR25 and DR24 \opstce{s} is caused by contamination by bright variable stars (see \S\ref{s:ephemmatch} and \citealt{Coughlin2014a}).

Generally, the excess of long period TCEs is significantly larger than it was in the DR24 TCE catalog \citep{Seader2015}, also seen in Figure~\ref{f:obstces}. Most likely, this is because DR24 implemented an aggressive veto known as the bootstrap metric \citep{Seader2015}.  For DR25 this metric was calculated, but was not used as a veto. Also, other vetoes were made less strict causing more TCEs across all periods to be created. 

To summarize, for DR25 the number of false signals among the \opstce{s} is dramatically larger than in any previous catalog. This was done on purpose in order to increase the Pipeline completeness by allow more transiting exoplanets to be made into \opstce{s}. 

\subsection{Rogue TCEs}
The DR25 TCE table at the NASA Exoplanet Archive contains \ntcesnorogue\ \opstce{s} and 1498 rogue TCEs\footnote{See the tce\_rogue\_flag column in the DR25 TCE table at the exoplanet archive.} for a total of \ntces. The rogue TCEs \added{are three-transit TCEs that were only created because of a bug in the TPS module of the Kepler Pipeline. This bug prevented certain three-transit events from being vetoed and as a result they were returned as TCEs.}  This bug was not in place when characterizing the Pipeline using flux-level transit injection \citep[see][]{Burke2017a,Burke2017b} and because the primary purpose of this catalog is to be able to accurately calculate occurrence rates, we do not use the rogue TCEs in the creation and analysis of the DR25 KOI catalog. Also note that all of the TCE populations (observed, injection, inversion, and scrambling, see the next section) had rogue TCEs that were removed prior to analysis. The creation and analysis of this KOI catalog only rely on the non-rogue TCEs. Although they are not analyzed in this study we encourage the community to examine the designated rogue TCEs as the list does contain some of the longest period events detected by \Kepler.

\subsection{Simulated TCEs}
\label{s:simulated}
In order to measure the performance of the Robovetter and the \Kepler\ Pipeline, we created simulated transits, simulated false positives, and simulated false alarms.  The simulated transits are created by injecting transit signals into the pixels of the original data. The simulated false positives were created by injecting eclipsing binary signals and positionally off-target transit signals into the pixels of the original data (see \citealt{Coughlin2017a} and \citealt{Christiansen2017} for more information). The simulated false alarms were created in two separate ways: by inverting the light curves, and by scrambling the sequence of cadences in the time series. The TCEs that resulted from these simulated data are available at the Exoplanet Archive on the Kepler simulated data page.\footnote{\label{kepsimpagefn}\url{https://exoplanetarchive.ipac.caltech.edu/docs/KeplerSimulated.html}}


\subsubsection{True Transits -- Injection}
\label{s:injecttce}

We empirically measure the completeness of the \Kepler\ Pipeline and the subsequent vetting by injecting a suite of simulated transiting planet signals into the calibrated pixel data and observing their recovery, as was done for previous versions of the \Kepler{} Pipeline \citep{Christiansen2013a,Christiansen2015,Christiansen2016}. The full analysis of the DR25 injections are described in detail in \citet{Christiansen2017}. In order to understand the completeness of the Robovetter, we use the on-target injections \citep[Group 1 in][]{Christiansen2017}; we briefly describe their properties here. For each of the \ninjecttargs{} targets, we generate a model transit signal using the \citet{Mandel2002} formulation, with parameters drawn from the following uniform distributions: orbital periods from 0.5--500 days (0.5--100 days for M dwarf targets), planet radii from 0.25--7 \re{} (0.25--4 \re{} for M dwarf targets), and impact parameters from 0--1. After some re-distribution in planet radius to ensure sufficient coverage where the \Kepler{} Pipeline is fully incomplete (0\% recovery) to fully complete (100\% recovery), 50\% of the injections have planet radii below 2\,\re{} and 90\% below 40\,\re{}. The signals are injected into the calibrated pixels, and then processed through the remaining components of the \Kepler{} Pipeline in an identical fashion to the original data. Any detected signals are subjected to the same scrutiny by the Pipeline and the Robovetter as the original data. By measuring the fraction of injections that were successfully recovered by the Pipeline and called a PC by the Robovetter with any given set of parameters (e.g., orbital period and planet radius), we can then correct the number of candidates found with those parameters to the number that are truly present in the data. While the observed population of true transiting planets is heavily concentrated towards short periods, we chose the 0.5--500 day uniform period distribution of injections because more long-period, low signal-to-noise transits are both not recovered and not vetted correctly --- injecting more of these hard-to-find, long-period planets ensures that we can measure the Pipeline and Robovetter completeness. In this paper we use the set of on-target, injected planets that were recovered by the \Kepler{} Pipeline (the \injtce{s}, whose period distribution is shown in Figure~\ref{f:simtces}) to measure the performance of the Robovetter. Accurate measurement of the Robovetter performance is limited to those types of transits injected and recovered.

It is worth noting that the injections do not completely emulate all astrophysical variations produced by a planet transiting a star.  For instance, the injected model includes limb-darkening, but not the occultation of stellar pulsations or granulation, which has been shown to cause a small, but non-negligible, error source on measured transit depth \citep{Chiavassa2017} for high signal-to-noise transits.

\subsubsection{False Alarms -- Inverted and Scrambled} 
\label{s:tcefalsealarms}
\begin{figure*}[hp]
 \begin{center}
  \includegraphics[width=0.975\linewidth]{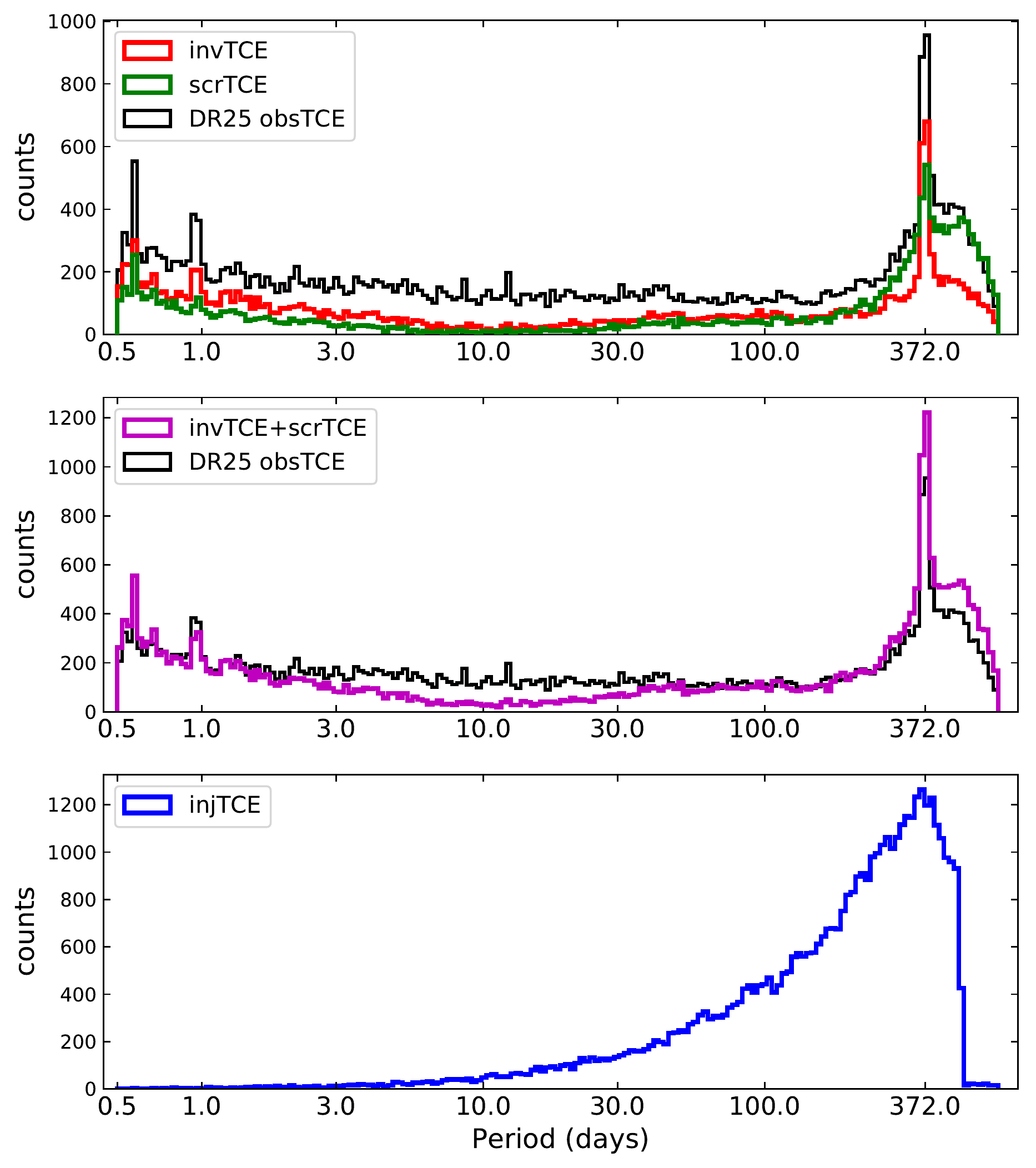}
  \caption{Histogram of the period in days of the cleaned \invtce{s} (top, red), the cleaned \scrtce{s} (top, green), and \injtce{s} (bottom, blue) in uniform, base-ten logarithmic spacing. The middle plot shows the union of the \invtces{} and the \scrtces{} in magenta.  The DR25 \opstce{s} are shown for comparison on the top two figures in black.  At shorter periods ($< 30$\,days) in the top figure, the difference between the simulated false alarm sets and the observed data represents the number of transit-like KOIs; at longer periods we primarily expect false alarms. Notice that the \invtce{s} do a better job of reproducing the one-year spike, but the \scrtce{s} better reproduce the long-period hump. Because the \injtce{s} are dominated by long-period events (significantly more long-period events were injected), we are better able to measure the Robovetter completeness for long-period planets than short-period planets.}
  \label{f:simtces} 
 \end{center}
 \end{figure*}

To create realistic false alarms that have noise properties similar to our \opstce{s}, we inverted the light curves (i.e., multiplied the normalized, zero-mean flux values by negative one) before searching for transit signals. Because the pipeline is only looking for transit-like (negative) dips in the light curve, the true exoplanet transits should no longer be found. However, quasi-sinusoidal signals due to instrumental noise, contact and near-contact binaries, and stellar variability can still create detections. In order for inversion to exactly reproduce the false alarm population, the false alarms would need to be perfectly symmetric (in shape and frequency) under flux inversion, which is not true. For example, stellar oscillations and star-spots are not sine waves and SPSDs will not appear the same under inversion. However, the rolling band noise that is significant on many of \Kepler's channels is mostly symmetric.  The period distribution of these \invtce{s} is shown in Figure~\ref{f:simtces}. The distribution qualitatively emulates those seen in the \opstce{s}; however there are only $\sim$60\% as many.  This is because the population does not include the exoplanets nor the eclipsing binaries, but it is also because many of the sources of false alarms are not symmetric under inversion.  The one-year spike is clearly seen, but is not as large as we might expect, likely because the broad long-period hump present in the DR25 \opstce{} distribution is mostly missing from the \invtce{} distribution. We explore the similarity of the \invtce{s} to \opstce{s} in more detail in \S\ref{s:simularity}.

Another method to create false alarms is to scramble the order of the data. The requirement is to scramble the data enough to lose the coherency of the binary stars and exoplanet transits, but to keep the coherency of the instrumental and stellar noise that plagues the \Kepler\ data set. Our approach was to scramble the data in coherent chunks of one year. The fourth year of data (Q13--Q16) was moved to the start of the light curve, followed by the third year (Q9--Q12), then the second (Q5--Q8), and finally the first (Q1--Q4). Q17 remained at the end. Within each year, the order of the data did not change. Notice that in this configuration each quarter remains in the correct \Kepler\ season preserving the yearly artifacts produced by the spacecraft. 

Two additional scrambling runs of the data, with different scrambling orders than described above, were performed and run through the \Kepler{} pipeline and Robovetter, but are not discussed in this paper, as they were produced after the analysis for this paper was complete. These runs could be very useful in improving the reliability measurements of the DR25 catalog --- see \citealt{Coughlin2017a} for more information.

\vspace{2em}
\subsubsection{Cleaning Inversion and Scrambling}
\label{s:clean}
As will be described in \S\ref{s:relcalc}, we want to use the \invtce\ and \scrtce\ sets to measure the reliability of the DR25 catalog against instrumental and stellar noise. In order to do that well, we need to remove signals found in these sets that are not typical of those in our \opstce\ set. For inversion, there are astrophysical events that look similar to an inverted eclipse, for example the self-lensing binary star, KOI~3278.01 \citep{Kruse2014}, and Heartbeat binaries \citep{Thompson2012}. With the assistance of published systems and early runs of the Robovetter, we identified any \invtce\ that could be one of these types of astrophysical events; 54 systems were identified in total. Also, the shoulders of inverted eclipsing binary stars and high signal-to-noise KOIs are found by the Pipeline, but are not the type of false alarm we were trying to reproduce, since they have no corresponding false alarm in the original, un-inverted light curves. We remove any \invtce{s} that were found on stars that had 1) one of the identified astrophysical events, 2) a detached eclipsing binary listed in \citet{Kirk2016} with morphology values larger than 0.6, or 3) a known KOI.  After cleaning, we are left with \ninvtces\ \invtce{s}; their distribution is plotted in the top of Figure~\ref{f:simtces}.

For the scrambled data, we do not have to worry about the astrophysical events that emulate inverted transits, but we do have to worry about triggering on true transits that have been rearranged to line up with noise. For this reason we remove from the \scrtce\ population all that were found on a star with a known eclipsing binary \citep{Kirk2016}, or on an identified KOI.  The result is \nscrtces\ \scrtce s; their distribution is plotted in the middle panel of Figure~\ref{f:simtces}. 
This will not remove all possible sources of astrophysical transits. Systems with only two transits (which would not be made into KOIs), or systems with single transits from several orbiting bodies would not be identified in this way. For example, KIC~3542116 was identified by \citet{Rappaport2017} as a star with possible exocomets, and it is a \scrtce\ dispositioned as an FP.  We expect the effect of not removing these unusual events to be negligible on our reliability measurements relative to other systematic differences between the \opstce{s} and the \scrtce{s}.


After cleaning the \invtce{s} and \scrtce{s}, the number of \scrtce{s} at periods longer than 200\,d closely matches the size and shape of the \opstce{} distribution, except for the one-year spike. The one-year spike is well represented by the \invtce{s}.  The distribution of the combined \invtce{} and \scrtce{} data sets, as shown in the middle plot of Figure~\ref{f:simtces}, qualitatively matches the relative frequency of false alarms present in the DR25 \opstce\ population. Tables\,\ref{t:invclean} and \ref{t:scrclean} lists those \invtce{s} and \scrtce{s} that we used when calculating the false alarm effectiveness and false alarm reliability of the PCs.

\begin{deluxetable}{lrrc}
\tablecolumns{4}
\tablewidth{\linewidth}
\tabletypesize{\scriptsize}
\tablecaption{\invtce{s} used in the analysis of catalog reliability \label{t:invclean}}

\tablehead{
\colhead{TCE-ID} & \colhead{Period} & \colhead{MES} & \colhead{Disposition}\\
\colhead{(KIC-PN)} & \colhead{days} &  &\colhead{PC/FP}\\
}
\startdata
000892667-01 &    2.261809 &       7.911006 &   FP \\
000892667-02 &  155.733356 &      10.087069 &   FP \\
000892667-03 &  114.542735 &       9.612742 &   FP \\
000892667-04 &  144.397127 &       8.998353 &   FP \\
000892667-05 &   84.142047 &       7.590044 &   FP \\
000893209-01 &  424.745158 &       9.106225 &   FP \\
001026133-01 &    1.346275 &      10.224972 &   FP \\
001026294-01 &    0.779676 &       8.503883 &   FP \\
001160891-01 &    0.940485 &      12.176910 &   FP \\
001160891-02 &    0.940446 &      13.552523 &   FP \\
001162150-01 &    1.130533 &      11.090898 &   FP \\
001162150-02 &    0.833482 &       8.282225 &   FP \\
001162150-03 &    8.114960 &      11.956621 &   FP \\
001162150-04 &    7.074370 &      14.518677 &   FP \\
001162150-05 &    5.966962 &      16.252800 &   FP \\
     \nodata &     \nodata &      \nodata & \nodata \\
\enddata
\tablecomments{The first column is the TCE-ID and is formed using the KIC Identification number and the TCE planet number (PN). This table is published in its entirety in the machine-readable format. A portion is shown here for guidance regarding its form and content.}
\end{deluxetable}

\begin{deluxetable}{lrrc}
\tablecolumns{4}
\tablewidth{\linewidth}
\tabletypesize{\scriptsize}
\tablecaption{\scrtce{s} used in the analysis of catalog reliability \label{t:scrclean}}
\tablehead{
\colhead{TCE-ID} & \colhead{Period} & \colhead{MES} & \colhead{Disposition}\\
\colhead{(KIC-PN)} & \colhead{days} &  &\colhead{PC/FP}\\
}
\startdata
000757099-01 &    0.725365 &     8.832907 &   FP \\
000892376-01 &  317.579997 &    11.805184 &   FP \\
000892376-02 &    1.532301 &    11.532692 &   FP \\
000892376-03 &  193.684366 &    14.835271 &   FP \\
000892376-04 &  432.870540 &    11.373951 &   FP \\
000892376-05 &  267.093312 &    10.308785 &   FP \\
000892376-06 &    1.531632 &    10.454597 &   FP \\
000893004-01 &  399.722285 &     7.240176 &   FP \\
000893507-02 &  504.629640 &    15.434824 &   FP \\
000893507-03 &  308.546946 &    12.190248 &   FP \\
000893507-04 &  549.804329 &    12.712417 &   FP \\
000893507-05 &  207.349237 &    11.017911 &   FP \\
000893647-01 &  527.190559 &    13.424537 &   FP \\
000893647-02 &  558.164884 &    13.531707 &   FP \\
000893647-03 &  360.260977 &     9.600089 &   FP \\
     \nodata &     \nodata &      \nodata & \nodata\\
\enddata
\tablecomments{The first column is the TCE-ID and is formed using the KIC Identification number and the TCE planet number (PN). This table is published in its entirety in the machine-readable format. A portion is shown here for guidance regarding its form and content.}
\end{deluxetable}

\vspace{3em}
\subsection{TCE Transit Fits}
\label{s:fits}
The creation of this KOI catalog depends on four different transit fits: 1) the original DV transit fits, 2) the trapezoidal fits performed on the ALT \citet{Garcia2010} detrended light curves, 3) the supplemental DV transit fits, and 4) the MCMC fits (see \S\ref{s:mcmc}).  The \Kepler\ Pipeline fits each TCE with a \citet{Mandel2002} transit model using \citet{Claret2000} limb darkening parameters. After the transit searches were performed for the observed, injected, scrambled, and inverted TCEs, we discovered that the transit fit portion of DV was seeded with a high impact parameter model that caused the final fits to be biased towards large values, causing the planet radii to be systematically too large (for further information see \citealt{Christiansen2017} and \citealt{Coughlin2017a}). Since a consistent set of reliable transit fits are required for all TCEs, we refit the transits.  The same DV transit fitting code was corrected for the bug and seeded with the \Kepler{} identification number, period, epoch, and MES of the original detection. These ``supplemental'' DV fits do not have the same impact parameter bias as the original.  Sometimes the DV fitter fails to converge and in these cases we were not able to obtain a supplemental DV transit fit, causing us to fall back on the original DV fit. Also, at times the epoch wanders too far from the original detection; in these cases we do not consider it to be a successful fit and again fall back on the original fit.

Because the bug in the transit fits was only discovered after all of the metrics for the Robovetter were run, the original DV and trapezoidal fits were used to disposition all of the sets of TCEs. These are the same fits that are available for the \opstce{s} in the DR25 TCE table at the NASA Exoplanet Archive. \replaced{Most}{Nearly all of the} Robovetter metrics are agnostic to the parameters of the fit, and so the supplemental DV fits would only change a few of the Robovetter decisions, \added{namely the V-shape Metric, as it relies on the radius ratio and impact parameter (see \S\ref{s:shapemetric}), and the Model-Shift metrics since that test utilizes the transit model fit (see \ref{s:ms}). The impact on all the Model-Shift Tests is negligible since they only utilize the general shape of the transit fit, and not the fitted parameters themselves. Note, if the supplemental fits were used for either test, we would have chosen different thresholds for the metrics so as to obtain a very similar catalog. } While the Robovetter itself runs in a few minutes, several of the metrics used by the Robovetter (see Appendix \ref{s:metrics}) require weeks to compute, so we chose not to update the metrics in order to achieve \added{a small improvement in the consistency of our products}. For all sets of TCEs, the original DV fits are listed in the Robovetter input files\footnote{Robovetter input files have the format kplr\_dr25\_obs\_robovetter\_input.tar.gz and can be found in the Robovetter github repository, \url{https://github.com/nasa/kepler-robovetter}}. The supplemental fits are used to understand the completeness and reliability of the catalog as a function of fitted parameters (such as planet radii or insolation flux). For all sets of TCEs, the supplemental DV fits are available as part of the Robovetter results tables linked from the TCE documentation page\footnote{The Robovetter results files are linked under the Q1-Q17 DR25 Information on the page \url{https://exoplanetarchive.ipac.caltech.edu/docs/Kepler\_TCE\_docs.html}} for the \opstce{s} and from the simulated data page\footnote{\url{https://exoplanetarchive.ipac.caltech.edu/docs/KeplerSimulated.html}} \citep[see][]{Christiansen2017,Coughlin2017a} for the injected, inverted, and scrambled TCEs.

\begin{figure}[htb]
\centering
\begin{tabular}{c}
\includegraphics[width=\linewidth]{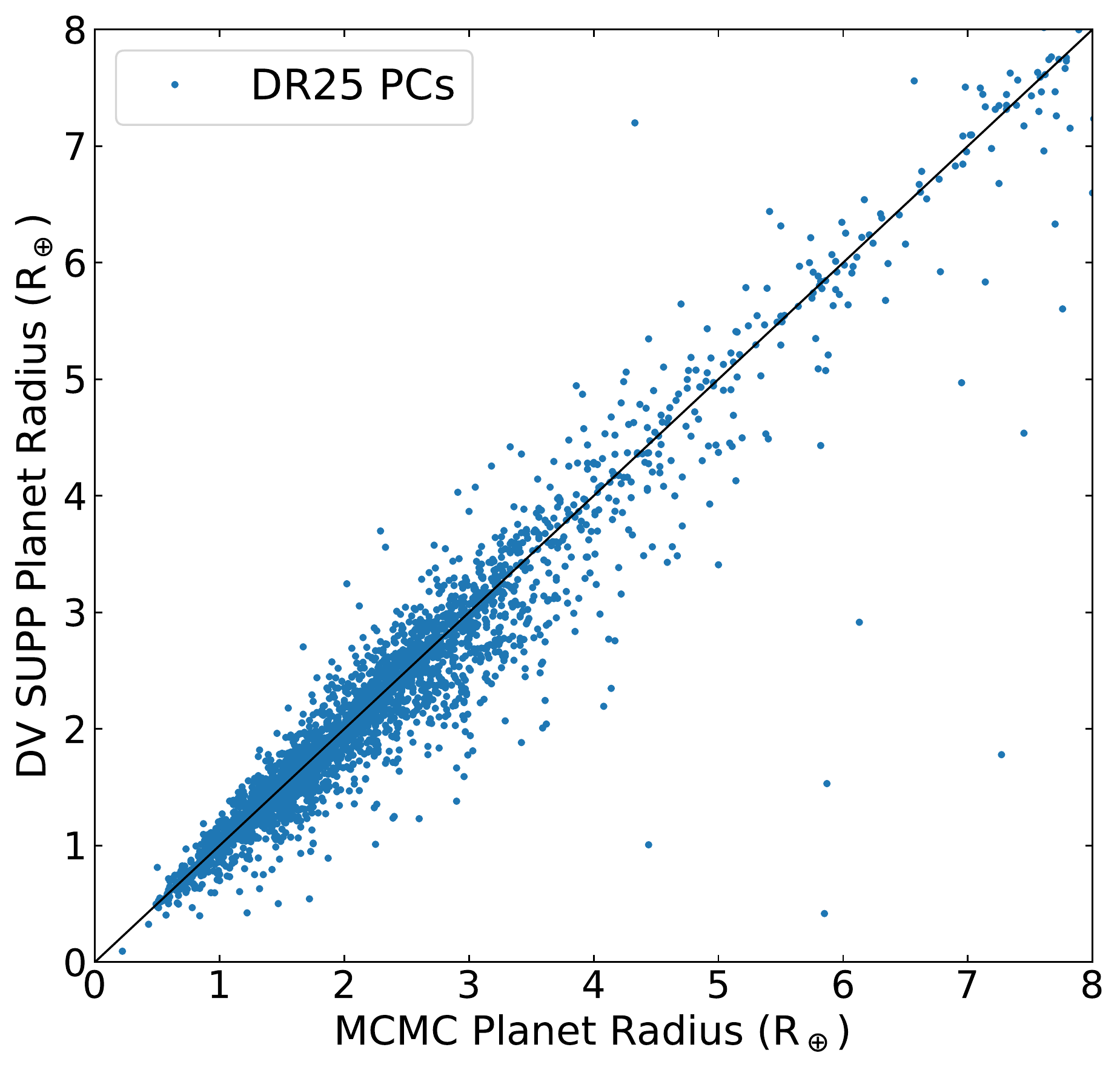} \\[2em]
\includegraphics[width=\linewidth]{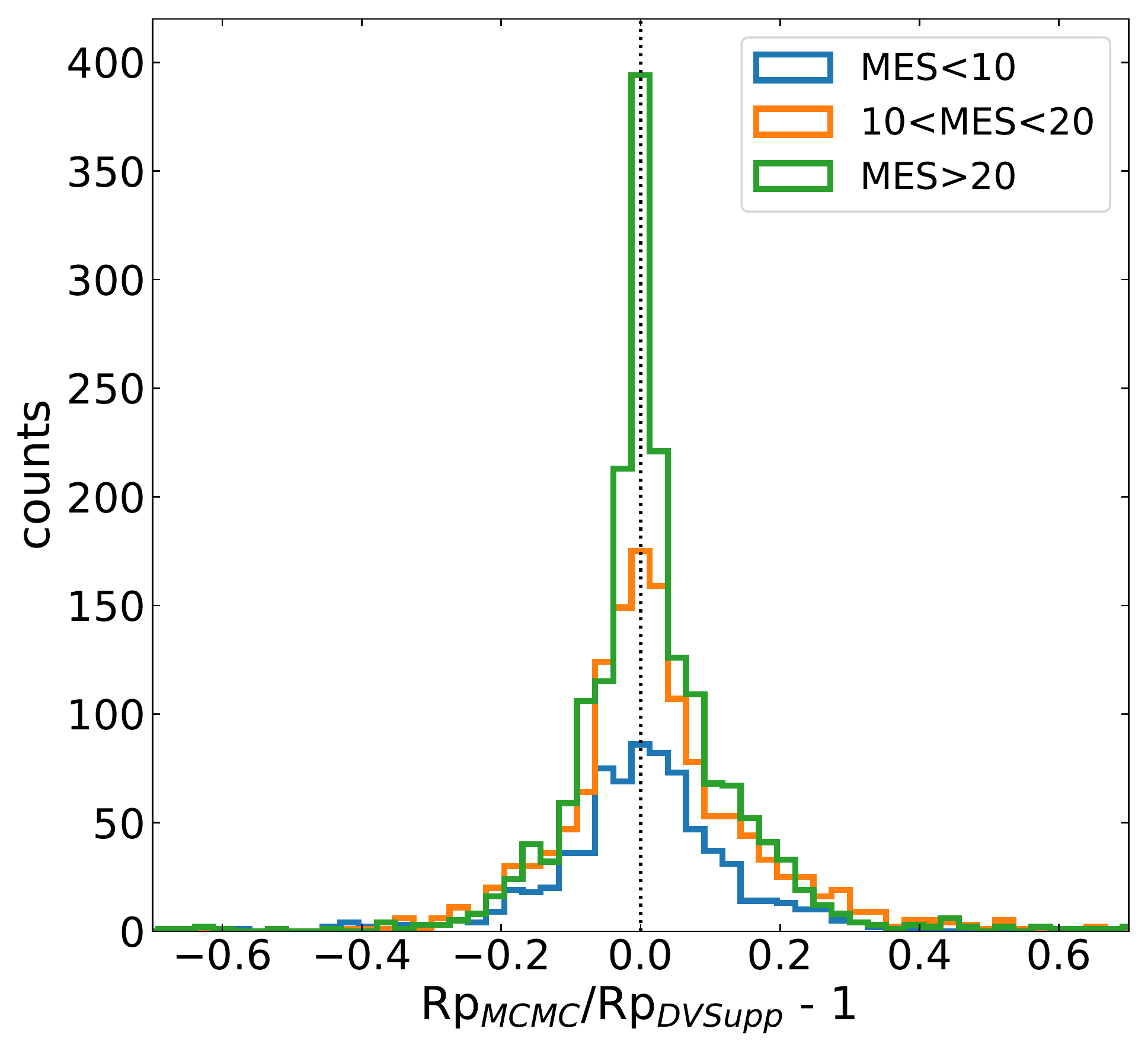}
\end{tabular}
\caption{Top: Comparison of the DR25 PCs fitted planet radii measured by the MCMC fits and the DV supplemental fits. The 1:1 line is drawn in black. Bottom: Histogram of the difference between the MCMC fits and the DV fits for the planet candidates in different MES bins. While individual objects have different fitted values, as a group the planet radii from the two fits agree. }
\label{f:mcmcsupp}
\end{figure}

The MCMC fits are only provided for the KOI population and are available in the DR25 KOI table\footnote{\url{https://exoplanetarchive.ipac.caltech.edu/cgi-bin/TblView/nph-tblView?app=ExoTbls\&config=q1\_q17\_dr25\_koi}} at the NASA Exoplanet Archive. The MCMC fits have no consistent offset from the supplemental DV fits.  To show this, we plot the planet radii derived from the two types of fits for the planet candidates in DR25 and show the distribution of fractional change in planet radii; see Figure~\ref{f:mcmcsupp}. The median value of the fractional change is 0.7\% with a standard deviation of 18\%. While individual systems disagree, there is no offset in planet radii between the two populations. The supplemental DV fitted radii and MCMC fitted radii agree within 1-sigma of the combined error bar (i.e., the square-root of the sum of the squared errors) for 78\% of the KOIs and 93.4\% of PCs (only 1.8\% of PC's radii differ by more than 3-sigma).  The differences are caused by discrepancies in the detrending and because the MCMC fits include a non-linear ephemeris in its model when appropriate (i.e., to account for transit-timing variations).

\subsection{Stellar Catalog}
\label{s:stars}
Some of the derived parameters from transit fits (e.g., planetary radius and insolation flux) of the TCEs and KOIs rely critically on the accuracy of the stellar properties (e.g., radii, mass, and temperature). For all transit fits used to create this catalog we use the DR25 Q1--Q17 stellar table provided by \citet{Mathur2017ApJS}, which is based on conditioning published atmospheric parameters on a grid of Dartmouth isochrones \citep{Dotter2008}. The best-available observational data for each star is used to determine the stellar parameters; e.g., asteroseismic or high-resolution spectroscopic data, when available, is used instead of broad-band photometric measurements. Typical uncertainties in this stellar catalog are $\approx$27\% in radius, $\approx$17\% in mass, and $\approx$51\% in density, which is somewhat smaller than previous catalogs.

After completion of the DR25 catalog an error was discovered: the metallicities of 780 KOIs were assigned a fixed erroneous value ([Fe/H] = 0.15 dex). These targets can be identified by selecting those that have the metallicity provenance column set to "SPE90". Since radii are fairly insensitive to metallicity and the average metallicity of \Kepler{} stars is close to solar, the impact of this error on stellar radii is typically less than 10\% and does not significantly change the conclusions in this paper. Corrected stellar properties for these stars will be provided in an upcoming erratum to \citet{Mathur2017ApJS}. The KOI catalog vetting and fits rely exclusively on the original DR25 stellar catalog information. Because the stellar parameters will continue to be updated (with data from missions such as \emph{Gaia}, \citealt{gaia1,gaia2}) we perform our vetting and analysis independent of stellar properties where possible and provide the fitted information relative to the stellar properties in the KOI table.  A notable exception is the limb darkening values; precise transit models require limb darkening coefficients that depends on the stellar temperature and gravity. However, limb-darkening coefficients are fairly insensitive to the most uncertain stellar parameters in the stellar properties catalog (e.g., surface gravity; \citealt{Claret2000}).

\section{The Robovetter: Vetting Methods and Metrics}
\label{s:robovetter}
The dispositioning of the TCEs as PC and FP is entirely automated and is performed by the Robovetter\footnote{\label{rvgithub}\url{https://github.com/nasa/kepler-robovetter}}. This code uses a variety of metrics to evaluate and disposition the TCEs.  


Because the TCE population changed significantly between DR24 and DR25 (see Figure~\ref{f:obstces}), the Robovetter had to be improved in order to obtain acceptable performance.  Also, because we now have simulated false alarms (\invtces\ and \scrtces) and true transits (\injtces), the Robovetter could be tuned to keep the most \injtce{s} and remove the most \invtce{s} and \scrtce{s}. This is a significant change from previous KOI catalogs that prioritized completeness above all else.  In order to sufficiently remove the long period excess of false alarms, this Robovetter introduces new metrics that evaluate individual transits (in addition to the phase-folded transits), expanding the work that the code Marshall \citep{Mullally2016} performed for the DR24 KOI catalog.

Because most of the Robovetter tests and metrics changed between DR24 and DR25, we fully describe all of the metrics.  In this section we summarize the important aspects of the Robovetter logic and only provide a list of each test's purpose. The details of these metrics, and more details on the Robovetter logic, can be found in Appendix~\ref{s:metrics}. We close this section by explaining the creation of the ``disposition score'', a number which conveys the confidence in the Robovetter's disposition.

\subsection{Summary of the Robovetter}

\begin{figure*}[htb]
\centering
\includegraphics[width=\linewidth]{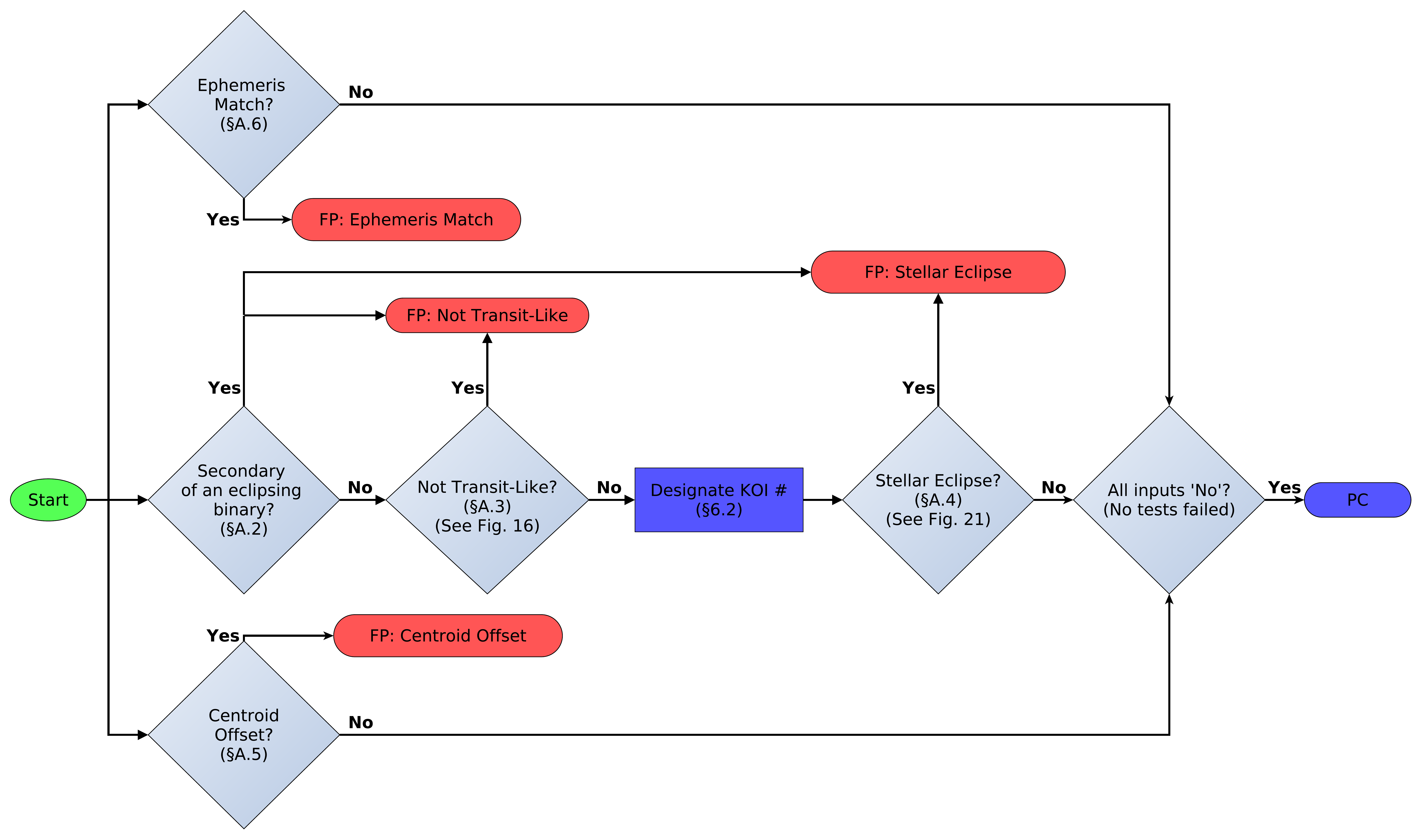}
\caption{Overview flowchart of the Robovetter. Diamonds represent ``yes'' or ``no'' decisions that are made with quantitative metrics. A TCE is dispositioned as an FP if it fails any test (a ``yes'' decision) and is placed in one or more of the FP categories. (A TCE that is identified as being the secondary eclipse of a system is placed in both the Not Transit-Like and Stellar Eclipse categories.) If a TCE passes all tests (a ``no'' decision for all tests) it is dispositioned as a PC. The section numbers on each component correspond to the sections in this paper where these tests are discussed. More in-depth flowcharts are provided for the not transit-like and stellar eclipse modules in Figures~\ref{robovetter-transitlike-fig} and \ref{robovetter-sigsec-fig}.}
\label{robovetter-overview-fig}
\end{figure*}

In Figure~\ref{robovetter-overview-fig} we present a flowchart that outlines our robotic vetting procedure. Each TCE is subjected to a series of ``yes'' or ``no'' questions (represented by diamonds) that either disposition it into one or more of the four FP categories, or else disposition it as a PC. Behind each question is a series of more specific questions, each answered by quantitative tests.

First, if the TCE under investigation is not the first in the system, the Robovetter checks if the TCE corresponds to a secondary eclipse associated with an already examined TCE in that system. If not, the Robovetter then checks if the TCE is transit-like. If it is transit-like, the Robovetter then looks for the presence of a secondary eclipse. In parallel, the Robovetter looks for evidence of a centroid offset, as well as an ephemeris match to other TCEs and variable stars in the \kepler{} field. 

\label{s:majorflags}
Similar to previous KOI catalogs \citep{Coughlin2016, Mullally2015cat, Rowe2015cat}, the Robovetter assigns FP TCEs to one or more of the following false positive categories:

\begin{itemize}
  \item Not Transit-Like (NT): a TCE whose light curve is not consistent with that of a transiting planet or eclipsing binary. These TCEs are usually caused by instrumental artifacts or non-eclipsing variable stars. If the Robovetter worked perfectly, all false alarms, as we have defined them in this paper, would be marked as FPs with only this Not Transit-Like flag set. 
  \item  Stellar Eclipse (SS): a TCE that is observed to have a significant secondary event, v-shaped transit profile, or out-of-eclipse variability that indicates the transit-like event is very likely caused by an eclipsing binary. Self-luminous, hot Jupiters with a visible secondary eclipse are also in this category, but are still given a disposition of PC. In previous KOI catalogs this flag was known as Significant Secondary.
  \item Centroid Offset (CO): a TCE whose signal is observed to originate from a source other than the target star, based on examination of the pixel-level data.
  \item Ephemeris Match Indicates Contamination (EC): a TCE that has the same period and epoch as another object, and is not the true source of the signal given the relative magnitudes, locations, and signal amplitudes of the two objects. See \citet{Coughlin2014}.
\end{itemize}

The specific tests that caused the TCE to fail are specified by minor flags. These flags are described in Appendix~\ref{s:minorflags} and are available for all FPs.  Table~\ref{t:metrics} gives a summary of the specific tests run by the Robovetter when evaluating a TCE.  The table lists the false positive category (NT, SS, CO or EC) of the test and also which minor flags are set by that test.  Note that there are several informative minor flags, which are listed in Appendix~\ref{s:minorflags}, but are not listed in Table~\ref{t:metrics} because they do not change the disposition of a TCE. Also, Appendix~\ref{s:minorflags} tabulates how often each minor flag was set to help understand the frequency of each type of FP. 

New to this Robovetter are several tests that look at individual transits. The tests are named after the code that calculates the relevant metric and are called: Rubble, Marshall, Chases, Skye, Zuma, and Tracker.  Each metric only identifies which transits can be considered "bad", or not sufficiently transit-like.  The Robovetter only fails the TCE if the number of remaining good transits is less than three, or if the recalculated MES, using only the good transits, drops below 7.1.

Another noteworthy update to the Robovetter in DR25 is the introduction of the v-shape metric, originally introduced in \citet{Batalha2013}.  The intent is to remove likely eclipsing binaries which do not show significant secondary eclipses by looking at the shape and depth of the transit (see \S\ref{s:shapemetric}).

\begin{deluxetable*}{llcll}
\tablecolumns{11}
\tabletypesize{\scriptsize}
\tablewidth{\linewidth}
\tablecaption{Summary of the DR25 Robovetter tests}
\tablehead{
Test Name & Section & Major & Minor & Brief\\[-3pt]
                    &                   & Flags   & Flags & Description
        }
\startdata
Is Secondary    & \ref{s:issecond}  & \makecell{\\NT \\SS}  & IS\_SEC\_TCE                      & The TCE is a secondary eclipse.\\
\tableline\\[-4pt]
LPP Metric      & \ref{s:lpp}       & NT                & \makecell[l]{LPP\_DV \\ LPP\_ALT} & The TCE is not transit-shaped.\\[3pt]
\tableline\\[-4pt]
SWEET NTL       & \ref{s:sweetntl}  & NT                & SWEET\_NTL          & The TCE is sinusoidal.\\
\tableline\\[-4pt]
TCE Chases      & \ref{s:tcechases} & NT                & ALL\_TRANS\_CHASES  & The individual TCE events are not uniquely significant.\\[3pt]
\tableline\\[-4pt]
MS$_1$       & \ref{s:ms}           &  NT               & \makecell[l]{MOD\_NONUNIQ\_DV \\ MOD\_NONUNIQ\_ALT}   & The TCE is not significant compared to red noise.\\[2pt]
\tableline\\[-4pt]
MS$_2$      & \ref{s:ms}            & NT                & \makecell[l]{MOD\_TER\_DV\  \\ MOD\_TER\_ALT}     & Negative event in phased flux as significant as TCE.\\[3pt]
\tableline\\[-4pt]
MS$_3$      & \ref{s:ms}            & NT                & \makecell[l]{MOD\_POS\_DV \\MOD\_POS\_ALT}       & Positive event in phased flux as significant as TCE.\\[3pt]
\tableline\\[-4pt]
Max SES to MES                & \ref{s:sesmes}                    & NT                  & INCONSISTENT\_TRANS & The TCE is dominated by a single transit event.\\[3pt]
\tableline\\[-4pt]
Same Period                   & \ref{s:sameperiod}                & NT                  & SAME\_NTL\_PERIOD   & Has same period as a previous not transit-like TCE.\\[3pt]
\tableline\\[-4pt]
Individual Transits           & \ref{s:indivtrans}                & NT                  & INDIV\_TRANS\_      & Has $<$~3 good transits and recalculated MES $<$~7.1.\\[2pt]
\hspace{2em}Rubble            & \ref{s:rubble}                    & \nodata             & INDIV\_TRANS\_RUBBLE   & Transit does not contain enough cadences.\\
\hspace{2em}Marshall          & \ref{s:marshall}                  & \nodata             & INDIV\_TRANS\_MARSHALL & Transit shape more closely matches a known artifact. \\
\hspace{2em}Chases            & \ref{s:chases}                    & \nodata             & INDIV\_TRANS\_CHASES   & Transit event is not significant.\\[2pt]
\hspace{2em}Skye              & \ref{s:skye}                      & \nodata             & INDIV\_TRANS\_SKYE     & Transit time is correlated with other TCE transits. \\
\hspace{2em}Zuma              & \ref{s:zuma}                      & \nodata             & INDIV\_TRANS\_ZUMA     & Transit is consistent with an increase in flux.\\[2pt]
\hspace{2em}Tracker           & \ref{s:tracker}                   & \nodata             & INDIV\_TRANS\_TRACKER  & No match between fitted and discovery transit time.\\
\tableline\\[-4pt]
Gapped Transits               & \ref{s:rocky}                     & NT                  & TRANS\_GAPPED          &  The fraction of transits identified as bad is large.\\[3pt]
\tableline\\[-4pt]
MS Secondary & \ref{s:second}  & SS & \makecell[l]{MOD\_SEC\_DV \\  MOD\_SEC\_ALT} & A significant secondary event is detected.\\[2pt]
\tableline\\[-4pt]
Secondary TCE                 & \ref{s:secondTce}                 & SS & HAS\_SEC\_TCE  & A subsequent TCE on this star is the secondary.\\[3pt]
\tableline\\[-4pt]
Odd Even    & \ref{s:oddeven} & SS                      & \makecell[l]{DEPTH\_ODDEVEN\_DV \\ DEPTH\_ODDEVEN\_ALT \\MOD\_ODDEVEN\_DV \\MOD\_ODDEVEN\_ALT}  &  The depths of odd and even transits are different.\\[2pt]
\tableline\\[-4pt]
SWEET EB                      & \ref{s:sweeteb}                   & SS                  & SWEET\_EB           & Out-of-phase tidal deformation is detected.\\[3pt]
\tableline\\[-4pt]
V Shape Metric                & \ref{s:shapemetric}               & SS                  & DEEP\_V\_SHAPE      & The transit is deep and v-shaped.\\[3pt]
\tableline\\[-4pt]
Planet Occultation\tablenotemark{\tiny{PC}} & \ref{s:sscand} & SS & \makecell[l]{PLANET\_OCCULT\_DV \\PLANET\_OCCULT\_ALT}  & Significant secondary could be planet occultation.\\
\tableline\\[-4pt]
Planet Half Period\tablenotemark{\tiny{PC}} & \ref{s:sscand} & \nodata & \makecell[l]{PLANET\_PERIOD\_IS\_HALF\_DV \\ PLANET\_PERIOD\_IS\_HALF\_ALT}  &Planet scenario possible at half the DV period.\\[2pt]
\tableline\\[-4pt]
Resolved Offset               & \ref{s:centroidrv}                & CO                  & CENT\_RESOLVED\_OFFSET & The transit occurs on a spatially resolved target.\\[3pt]
\tableline\\[-4pt]
Unresolved Offset             & \ref{s:centroidrv}                & CO                  & CENT\_UNRESOLVED\_OFFSET & A shift in the centroid position occurs during transit.\\[3pt]
\tableline\\[-4pt]
Ghost Diagnostic              & \ref{s:ghost}                     & CO                  & HALO\_GHOST           & The transit strength in the halo pixels is too large.\\[3pt]
\tableline\\[-4pt]
Ephemeris Match               & \ref{s:ephemmatch}                & EC                  & EPHEM\_MATCH          & The ephemeris matches that of another source.\\
\enddata
\tablecomments{More details about all of these tests and how they are used by the Robovetter can be found in the sections listed in the second column.}
\label{t:metrics}
\tablenotetext{PC}{\hspace{1em}These tests override previous tests and will cause the TCE to become a planet candidate.}
\end{deluxetable*}

\subsection{Disposition Scores}
\label{s:scores}
We introduce a new feature to this catalog called the Disposition Score. Essentially the disposition score is a value between 0 and 1 that indicates the confidence in a disposition provided by the Robovetter. A higher value indicates more confidence that a TCE is a PC, regardless of the disposition it was given. This feature allows one to select the highest quality PCs by ranking KOIs via the disposition score, for both use in selecting samples for occurrence rate calculations and prioritizing individual objects for follow-up. \emph{We stress that the disposition score does not map directly to a probability that the signal is a planet.} However, in \S\ref{s:crscores} we discuss how the disposition score can be used to adjust the reliability of a sample.

The disposition score was calculated by wrapping the Robovetter in a Monte Carlo routine. In each Monte Carlo iteration, for each TCE, new values are chosen for most of the Robovetter input metrics by drawing from an asymmetric Gaussian distribution \footnote{The asymmetric Gaussian distribution is created so that either side of the central value follows a Gaussian, each with a different standard deviation.} centered on the nominal value. The Robovetter then dispositions each TCE given the new values for its metrics. The disposition score is simply the fraction of Monte Carlo iterations that result in a disposition of PC. (We used 10,000 iterations for the results in this catalog.) For example, if a TCE that is initially dispositioned as a PC has several metrics that are just barely on the passing side of their Robovetter thresholds, in many iterations at least one will be perturbed across the threshold. As a result, many of the iterations will produce a false positive and the TCE will be dispositioned as a PC with a low score.  Similarly, if a TCE only fails due to a single metric that was barely on the failing side of a threshold, the score may be near 0.5, indicating that it was deemed a PC in half of the iterations.  Since a TCE is deemed a FP even if only one metric fails, nearly all FPs have scores less than 0.5, with most very close to 0.0.  PCs have a wider distribution of scores from 0.0 to 1.0 depending on how many of their metrics fall near to the various Robovetter thresholds.

To compute the asymmetric Gaussian distribution for each metric, we examined the metric distributions for the injected on-target planet population on FGK dwarf targets. In a 20 by 20 grid in linear period space (ranging from 0.5 to 500\,d) and logarithmic MES space (ranging from 7.1 to 100), we calculated the median absolute deviation (MAD) for those values greater than the median value and separately for those values less than the median value.  We chose to use MAD because it is robust to outliers. MES and period were chosen as they are fundamental properties of a TCE that well characterize each metric's variation. The MAD values were then multiplied by a conversion factor of 1.4826 to put the variability on the same scale as a Gaussian standard deviation \citep{Hampel1974,Ruppert2010}. A two-dimensional power-law was then fitted to the 20 by 20 grid of standard deviation values, separately for the greater-than-median and less-than-median directions. With this analytical approximation for a given metric, an asymmetric Gaussian distribution can be generated for each metric for any TCE given its MES and period.

\begin{figure*}[hp]
\centering
\begin{tabular}{cc}
\includegraphics[width=0.485\linewidth]{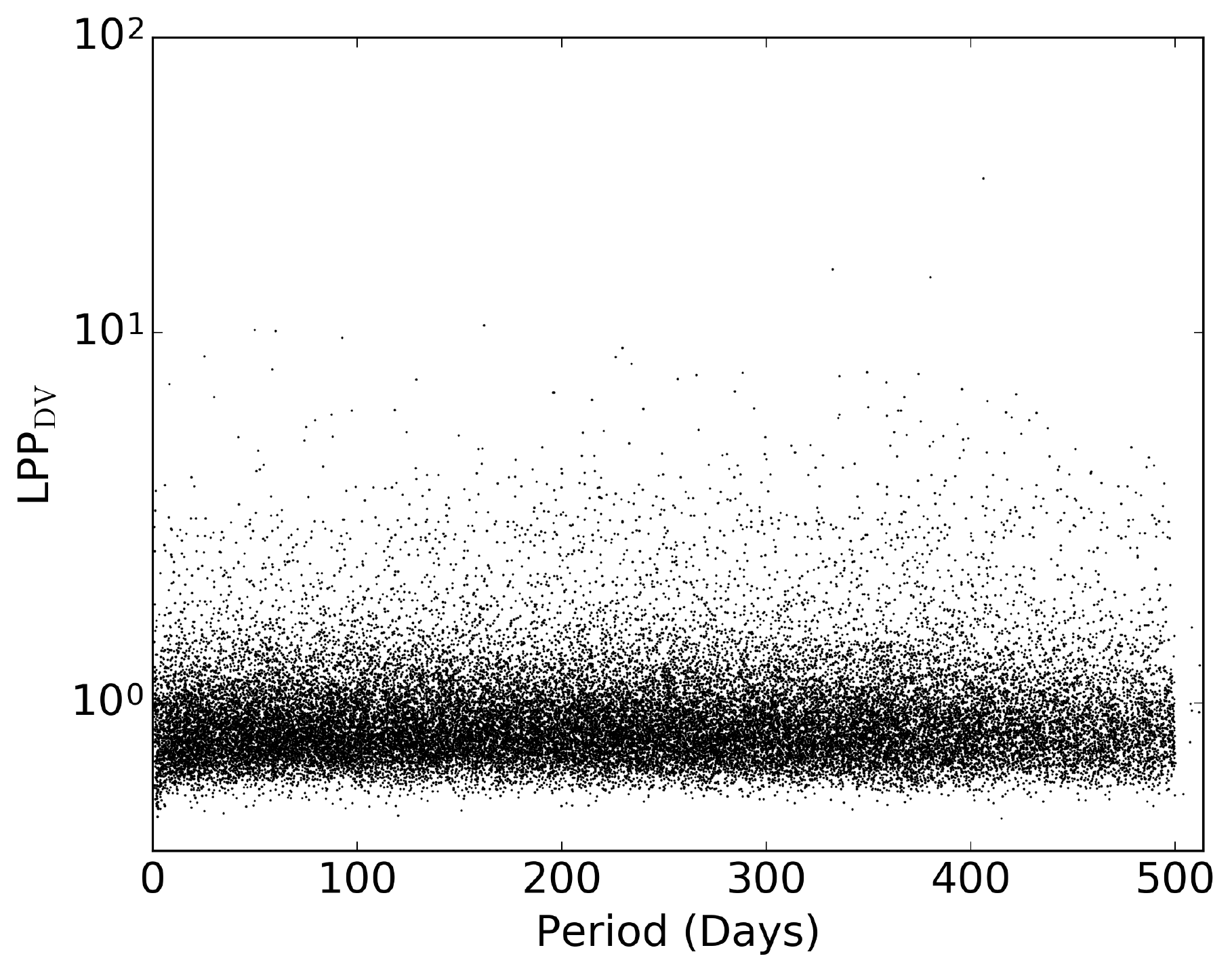} &
\includegraphics[width=0.485\linewidth]{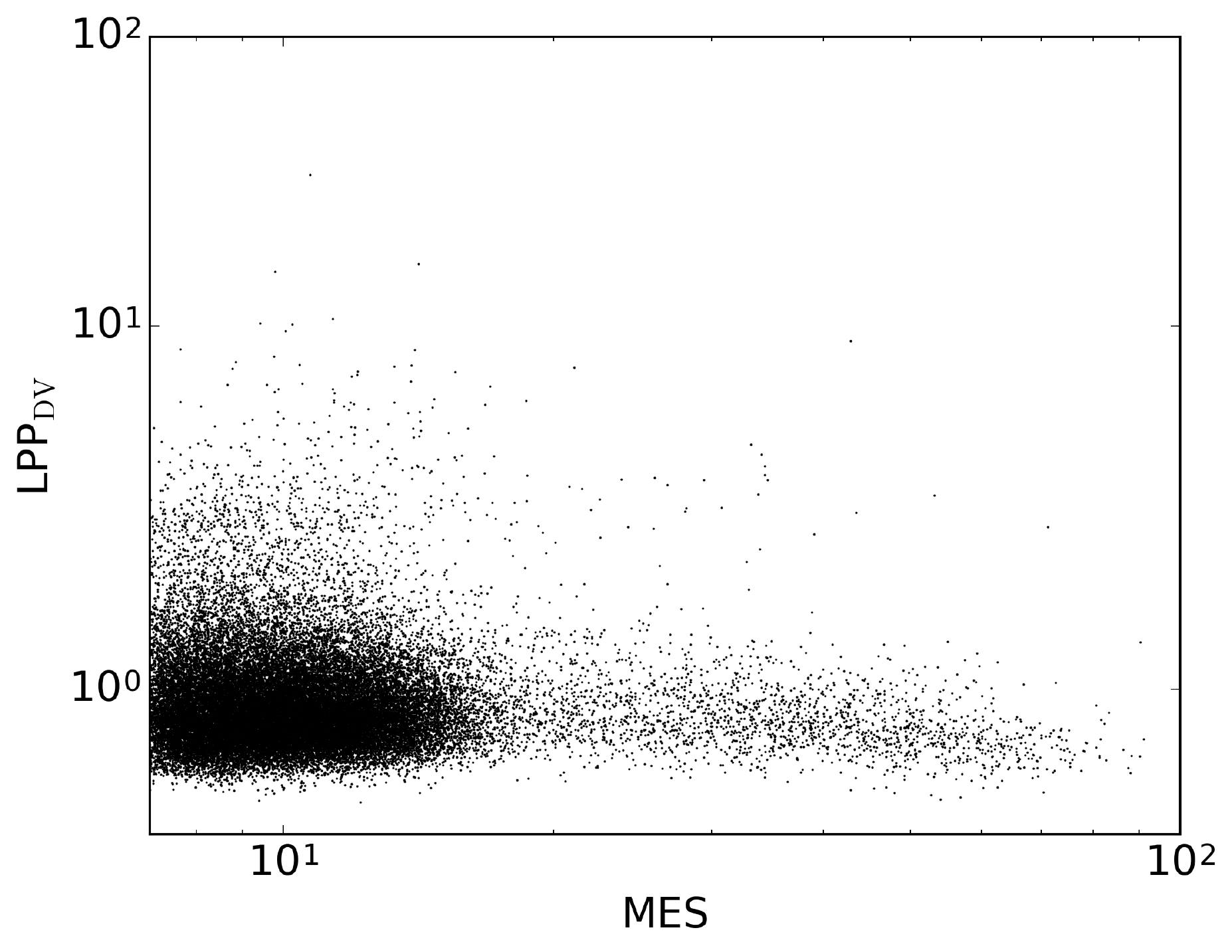} \\
\includegraphics[width=0.485\linewidth]{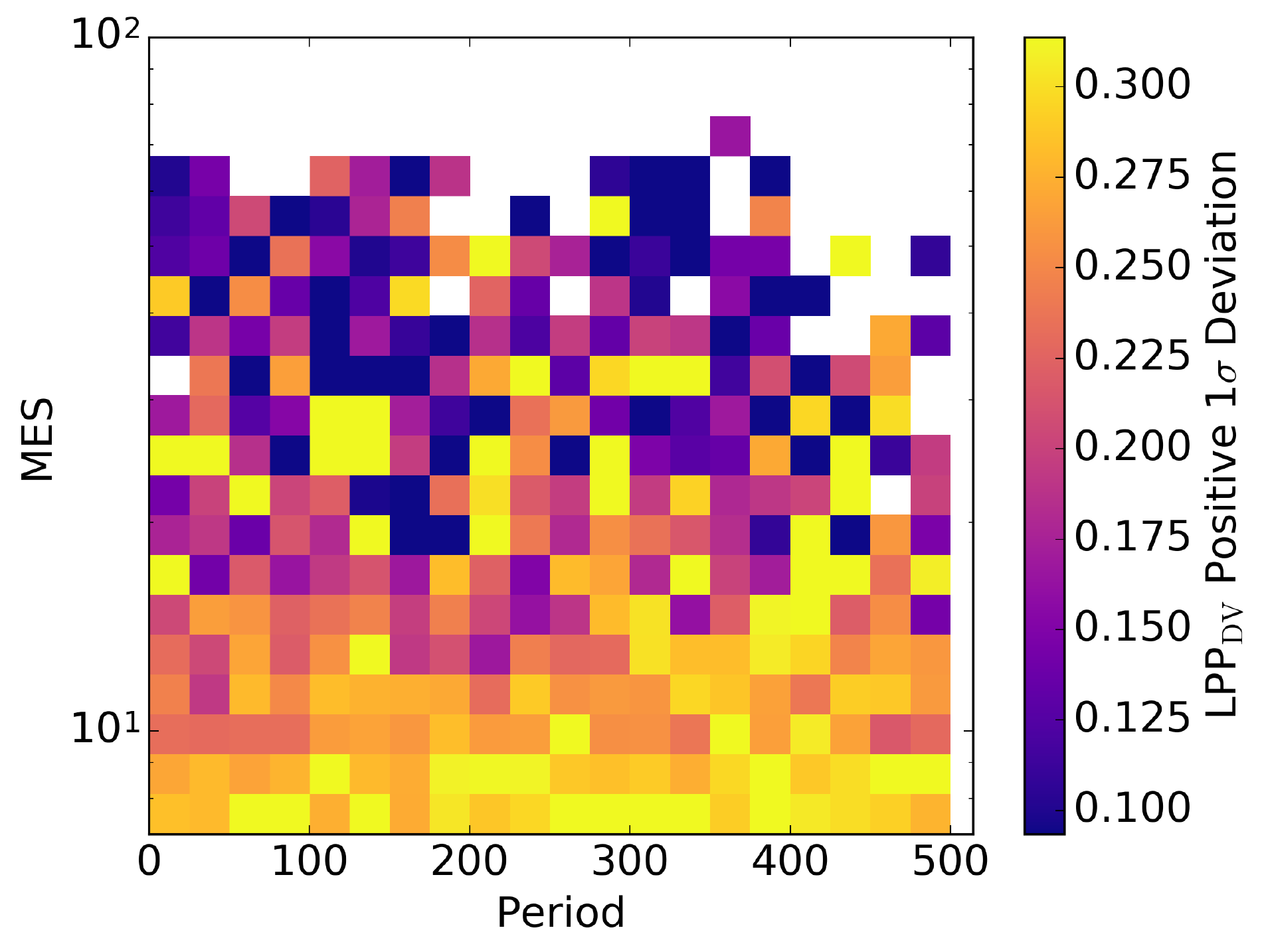} &
\includegraphics[width=0.485\linewidth]{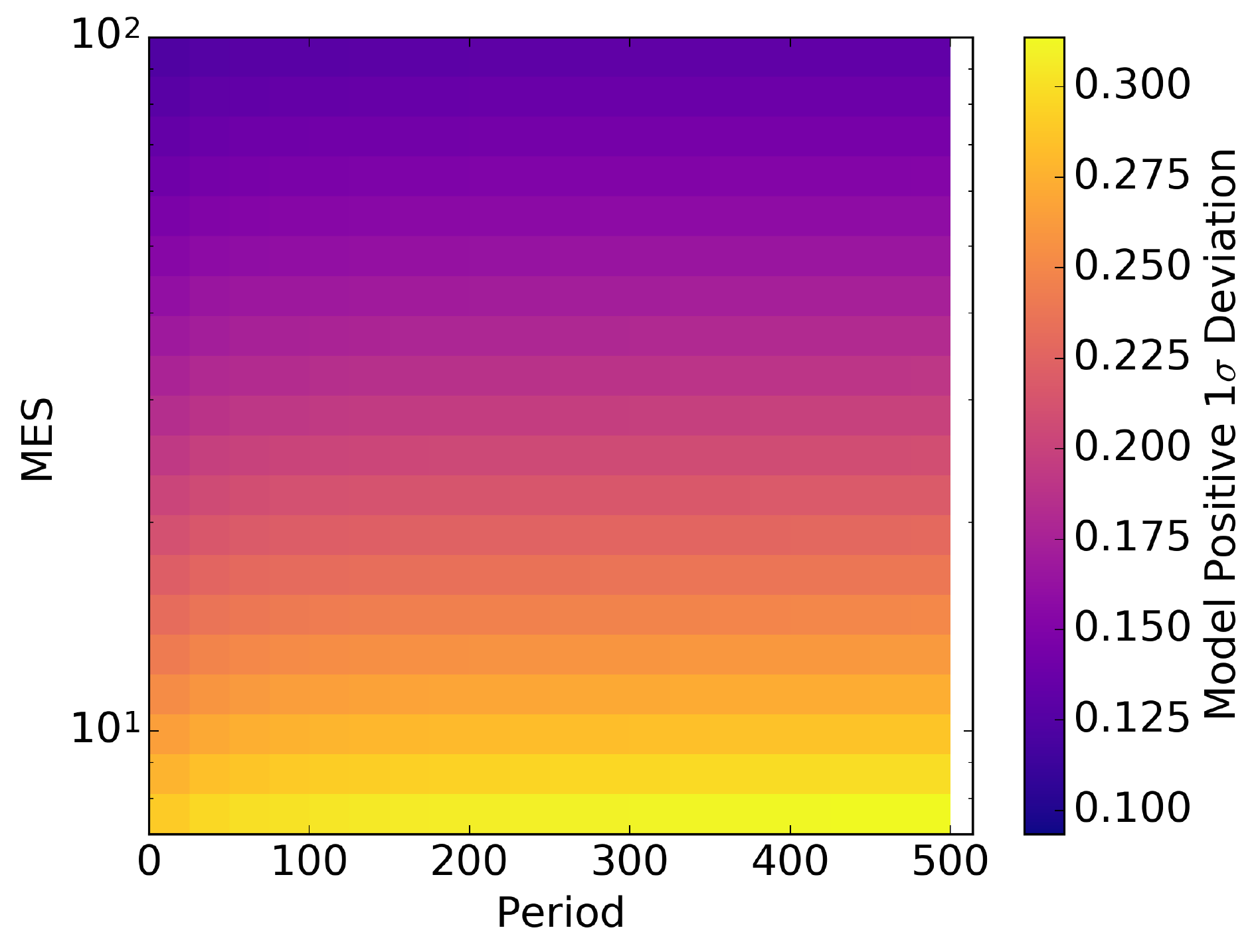} \\
\includegraphics[width=0.485\linewidth]{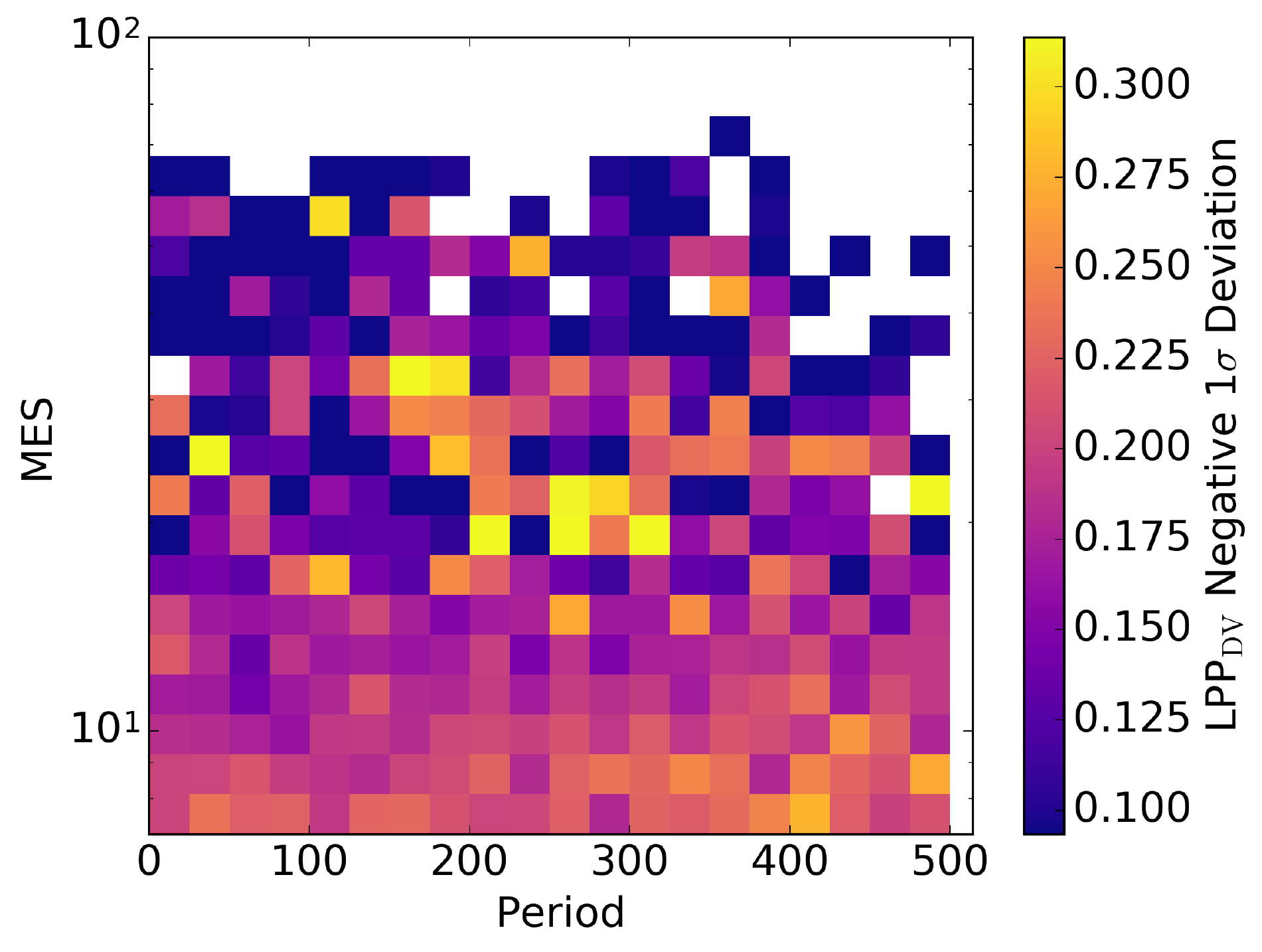} &
\includegraphics[width=0.485\linewidth]{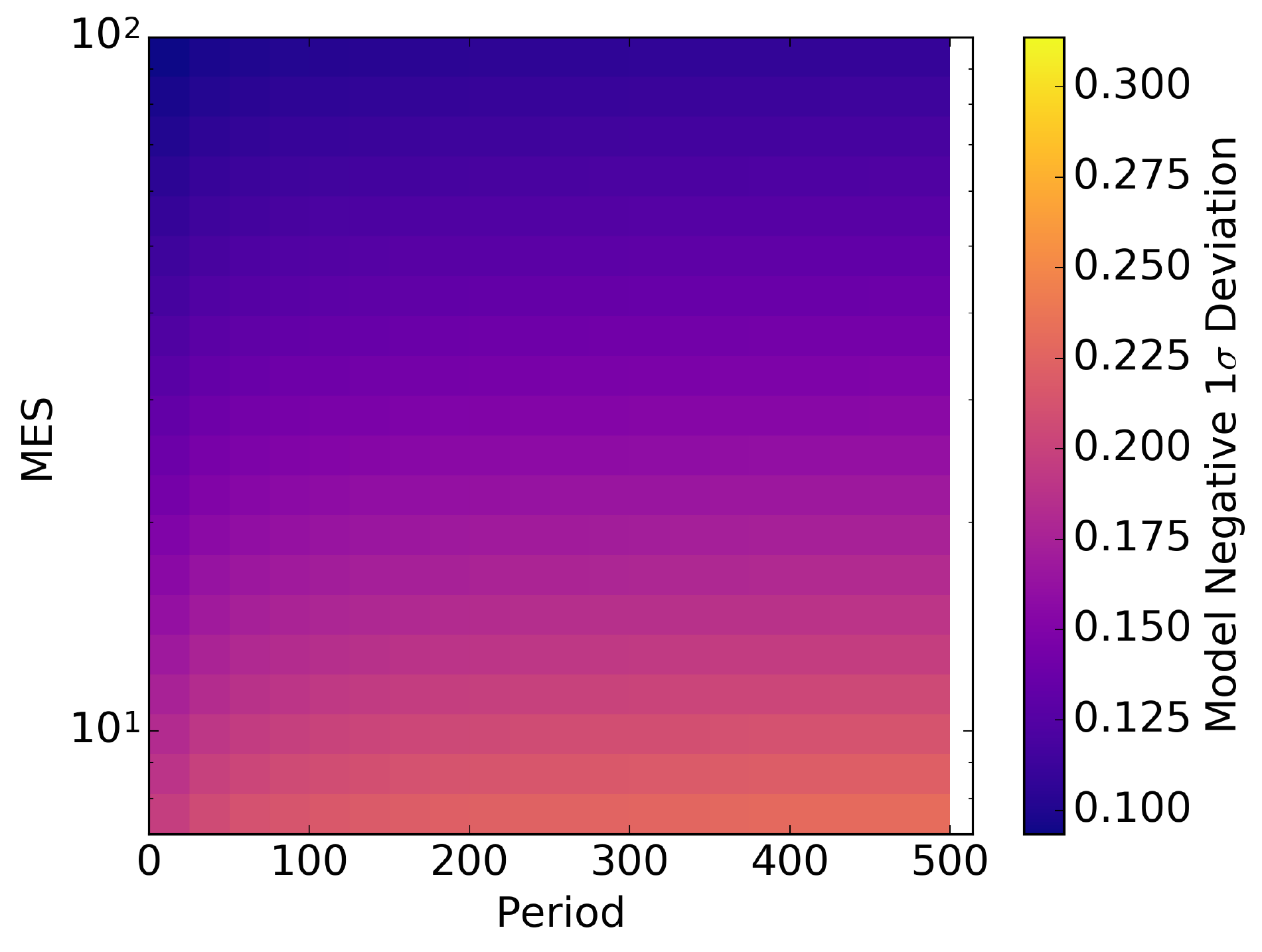}
\end{tabular}
\caption{The top-left plot shows the LPP$_{DV}$ values of all on-target injected planets on FGK dwarf targets as a function of period, and the top-right shows them as a function of MES. The middle-left plots shows the measured positive 1$\sigma$ deviation (in the same units as LPP$_{DV}$) as a function of MES and period, and the middle-right plot shows the resulting best-fit model. The bottom plots show the same thing, but for the negative 1$\sigma$ deviation (again in the same units as LPP$_{DV}$). These resulting model distributions are used when computing the Robovetter disposition score.}
\label{score-fig-1}
\end{figure*}

An example is shown in Figure~\ref{score-fig-1} for the LPP metric (Locality Preserving Projections, see \S\ref{s:lpp}) using the DV detrending. The top-left plot shows the LPP values of all on-target injected planets on FGK dwarf targets as a function of period, and the top-right shows them as a function of MES. The middle-left plot shows the measured positive 1$\sigma$ deviation (in the same units as the LPP metric) as a function of MES and period, and the middle-right plot shows the resulting best-fit model. The bottom plots show the same thing but for the negative 1$\sigma$ deviation. As can be seen, the scatter in the LPP metric has a weak period dependence, but a strong MES dependence, due to the fact it is easier to measure the overall shape of the light curve (LPP's goal) with higher MES (signal-to-noise). 

Most, but not all, of the Robovetter metrics were amenable to this approach. Specifically, the list of metrics that were perturbed in the manner above to generate the score values were: LPP (both DV and ALT), all the Model-shift metrics (MS$_1$, MS$_2$, MS$_3$, and MS Secondary, both DV and ALT), TCE Chases, max-SES-to-MES, the two odd/even metrics (both DV and ALT), Ghost Diagnostic, and the recomputed MES using only good transits left after the individual transit metrics.

\section{Calculating Completeness and Reliability}
%

\newcommand{\opsfp}{N$_{\rm FP_{\rm obs}}$}
\newcommand{\opspc}{N$_{\rm PC_{\rm obs}}$}
\newcommand{\opsN}{N$_{\rm obs}$}
\newcommand{\trueopspc}{T$_{\rm PC_{\rm obs}}$}
\newcommand{\missedfp}{T$_{\rm FP_{\rm obs}}$ - N$_{\rm FP_{\rm obs}}$}
\newcommand{\invfp}{N$_{\rm FP_{\rm inv}}$}
\newcommand{\invpc}{N$_{\rm PC_{\rm inv}}$}
\newcommand{\invN}{N$_{\rm inv}$}
\newcommand{\simfp}{N$_{\rm FP_{\rm sim}}$}
\newcommand{\simpc}{N$_{\rm PC_{\rm sim}}$}
\newcommand{\simN}{N$_{\rm sim}$}
\newcommand{\sfatce}{sfaTCE}

We use the \injtce, \scrtce, and \invtce\ data sets to determine the performance of the Robovetter and to measure the completeness and the reliability of the catalog. As a reminder, the reliability we are attempting to measure is only the reliability of the catalog against false alarms and does not address the astrophysical reliability (see \S\ref{s:occurates}). As discussed in \S\ref{s:tces}, the long-period \opstces\ are dominated by false alarms and so this measurement is crucial to understand the reliability of some of the most interesting candidates in our catalog.

Robovetter completeness, $C$, is the fraction of injected transits detected by the \Kepler\ Pipeline that are passed by the Robovetter as PCs.  As long as the \injtce{s} are representative of the observed PCs, completeness tells us what fraction of the true planets are missing from the final catalog.  Completeness is calculated by dividing the number of on-target \injtce{s} that are dispositioned as PCs (N$_{\rm PC_{\rm inj}}$) by the total number of \injtce{s} (N$_{\rm inj}$).

\begin{equation}
C \approx \frac{N_{\rm PC_{\rm inj}}}{N_{\rm inj}}
\label{e:comp}
\end{equation}

\noindent If the parameter space under consideration becomes too large and there are gradients in the actual completeness, differences between the \injtce and the \opstce{} populations will prevent the completeness measured with Equation~\ref{e:comp} from matching the actual Robovetter completeness. For example, there are more long-period \injtce{s} than short-period ones, which is not representative of the observed PCs, the true fraction of candidates correctly dispositioned by the Robovetter is not accurately represented by binning over all periods. With this caveat in mind, we use $C$ in this paper to indicate the value we can measure, as shown in Equation~\ref{e:comp}.


The candidate catalog reliability, $R$, is defined as the ratio of the number of PCs which are truly exoplanets (\trueopspc) to the total number of PCs (\opspc) from the \opstce{} data set. 
\begin{equation}
\label{eq:rel}
R = \frac{T_{\rm PC_{\rm obs}}}{N_{\rm PC_{\rm obs}}}
\end{equation}

Calculating the reliability for a portion of the candidate catalog is not straight forward because we do not know which PCs are the true transiting exoplanets and cannot directly determine $T_{\rm PC_{\rm obs}}$. Instead, we use the simulated false alarm data sets to understand how often false alarms sneak past the Robovetter and contaminate our final catalog.

\subsection{Reliability Derivation}
\label{s:relcalc}
To assess the catalog reliability against false alarms, we will assume that the \scrtce{s} and \invtce{s} are similar (in frequency and type) to the \opstces.  One way to calculate the reliability of the catalog from our false alarm sets is to first calculate how often the Robovetter correctly identifies false alarms as FPs, a value we call effectiveness ($E$).  Then, given the number of FPs we identify in the \opstce\ set, we determine the reliability of the catalog against the type of false alarms present in the simulated sets (\invtces\ and \scrtces). This method assumes the relative frequency of the different types of false alarms is well emulated by the simulated data sets, but does not require the total number of false alarms to be well emulated.

Robovetter effectiveness ($E$) is defined as the fraction of FPs correctly identified as FPs in the \opstce\ data set,

\begin{equation}
\label{effect1}
E \equiv \frac{N_{\rm FP_{\rm obs}}}{T_{\rm FP_{\rm obs}}}
\end{equation}

\noindent where $T_{\rm FP_{\rm obs}}$ is the number of identified FPs which are truly FPs and $N_{\rm FP_{\rm obs}}$ is the total number of measured FPs. Notice we are using $N$ to indicate the measured number, and T to indicate the ``True'' number. 

If the simulated (sim) false alarm TCEs accurately reflect the \opstce\ false alarms, $E$ can be estimated as the number of simulated false alarm TCEs dispositioned as FPs (\simfp) divided by the number of simulated TCEs (\simN), 
\begin{equation}
\label{effect2}
E \approx  \frac{N_{\rm FP_{\rm sim}}}{N_{\rm sim}}
\end{equation}

\noindent For our analysis of the DR25 catalog, we primarily use the union of the \invtces\ and the \scrtces{} as the population of simulated false alarms when calculating $E$, see \S\ref{s:candr}.


Recall that the Robovetter makes a binary decision, and TCEs are either PCs or FPs. For this derivation we do not take into consideration the reason the Robovetter calls a TCE an FP (i.e., some false alarms fail because the Robovetter indicates there is a stellar eclipse or centroid offset). For most of parameter space, an overwhelming fraction of FPs are false alarms in the \opstce{} data set. Future studies will look into separating out the effectiveness for different types of FPs using the set of injected astrophysical FPs (see \S\ref{s:simulated}).

At this point we drop the \textit{obs} and \textit{sim} designations in subsequent equations, as the simulated false alarm quantities are all used to calculate $E$. The $N$ values shown below refer entirely to the number of PCs or FPs in the \opstce\ set so that $N$ = $N_{\rm PC}$ + $N_{\rm FP}$ = $T_{\rm PC}$ + $T_{\rm FP}$. We rewrite the definition for reliability (Eq.~\ref{eq:rel}) in terms of the number of true false alarms in \opstce, $T_{\rm FP}$,

\begin{equation}
\label{effect3}
R \equiv \frac{T_{\rm PC}}{N_{\rm PC}} =  1 + \frac{T_{\rm PC}-N_{\rm PC}}{N_{\rm PC}} 
= 1 + \frac{N - T_{\rm FP} - N_{\rm PC}}{N_{\rm PC}}
\end{equation}

\noindent When we substitute $N_{{\rm FP}}=N-N_{{\rm PC}}$ in Equation~\ref{effect3} we get another useful way to think about reliability, as one minus the number of unidentified FPs relative to the number of candidates,

\begin{equation}
\label{eq:rel2}
R = 1 - \frac{T_{\rm FP}-N_{\rm FP}}{N_{\rm PC}}
\end{equation}

\noindent However, the true number of false alarms in the \opstce{} data set, $T_{\rm FP}$, is not known. Using the effectiveness value (Equation \ref{effect2}) and combining it with our definition for effectiveness (Equation \ref{effect1}) we get,
\begin{equation}
T_{FP} = \frac{N_{\rm FP}}{E} 
\end{equation}

\noindent and substituting into equation \ref{eq:rel2} we get,
\textbf{
\begin{equation}
R= 1 - \frac{N_{\rm FP}}{N_{\rm PC}}\left(\frac{1-E}{E}\right)
\end{equation}
}

\noindent which relies on the approximation of $E$ from Equation~\ref{effect2} and is thus a measure of the catalog reliability using all measurable quantities.




This method to calculate reliability depends sensitively on the measured effectiveness which relies on how well the set of known false alarms match the false alarms in the \opstce{} data set. For example, a negative reliability can result if the measured effectiveness is lower than the true value. In these cases, it implies that there should be more PCs than exist, i.e., the number of unidentified false alarms is smaller than the number of remaining PCs to draw from.  


\subsection{The Similarity of the Simulated False Alarms}
\label{s:simularity}
In order to use the \scrtce\ and \invtce\ sets to determine the reliability of our catalog we must assume that the properties of these simulated false alarms are similar to those of the false alarms in the \opstce\ set.  Specifically, these simulated data should \added{mimic the not transit-like \opstces, e.g., FPs created by instrumental noise and stellar spots.} For instance, our assumptions break down if all of the simulated false alarms were long-duration rolling-band FPs, but only a small fraction of the observed FPs were caused by this mechanism.  Stated another way, the method we use to measure reliability, hinges on the assumption that for a certain parameter space the fraction of a particular type of FP TCEs is the same between the simulated and observed data sets.  This is the reason we removed the TCEs caused by KOIs and eclipsing binaries in the simulated data sets (see \S\ref{s:clean}). Inverted eclipsing binaries and transits are not the type of FP found in the \opstce\ data set.  Since the Robovetter is very good at eliminating inverted transits, if they were included, we would have an inflated value for the effectiveness, and thus incorrectly measure a higher reliability. 

Figure\,\ref{f:simtces} demonstrates that the number of TCEs from inversion and scrambling individually is smaller than the number of \opstces. At periods less than $\approx$100 days this difference is dominated by the lack of planets and eclipsing binaries in the simulated false alarm data sets.  At longer periods, where the TCEs appear to be dominated by false alarms, this difference is dominated by the cleaning (\S\ref{s:clean}). Effectively, we search a significantly smaller number of stars for instances of false alarms. The deficit is also caused by the fact that all types of false alarms are not accounted for in these simulations. For instance, the \invtce\ set will not reproduce false alarms caused by sudden dropouts in pixel sensitivity caused by cosmic rays (i.e., SPSDs). The \scrtce\ set will not reproduce the image artifacts from rolling band because the artifacts are not as likely to line-up at exactly one \Kepler -year.  However, despite these complications, the period distribution of false alarms in these simulated data sets basically resembles the same period distribution as the \opstce\ FP population once the two simulated data sets are combined. And since reliability is calculated using the fraction of false alarms that are identified (effectiveness), the overabundance that results from combining the sets is not a problem.

\begin{figure*}[hp]
    \centering
    \includegraphics[width=0.85\linewidth]{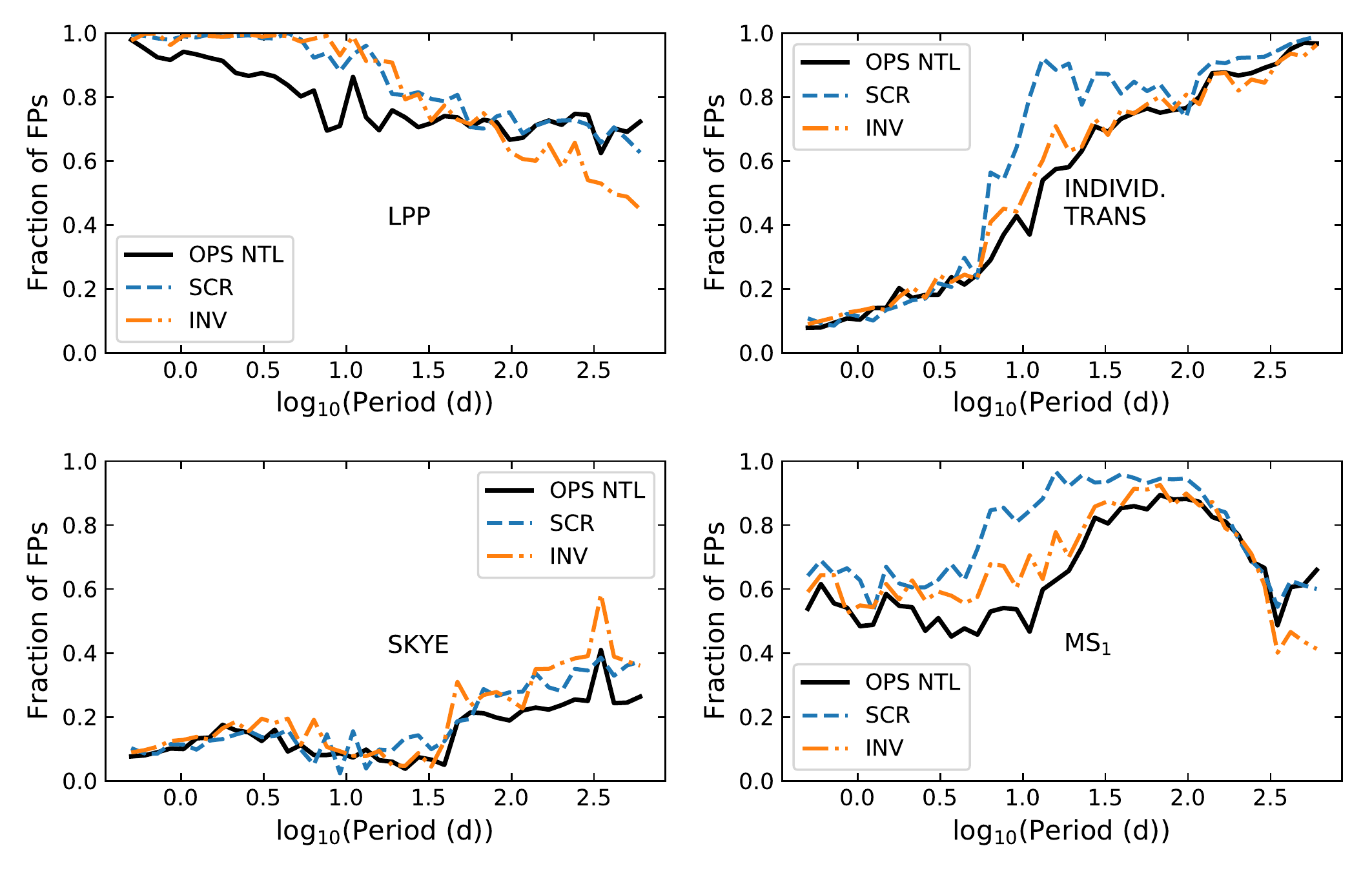}
    \includegraphics[width=0.85\linewidth]{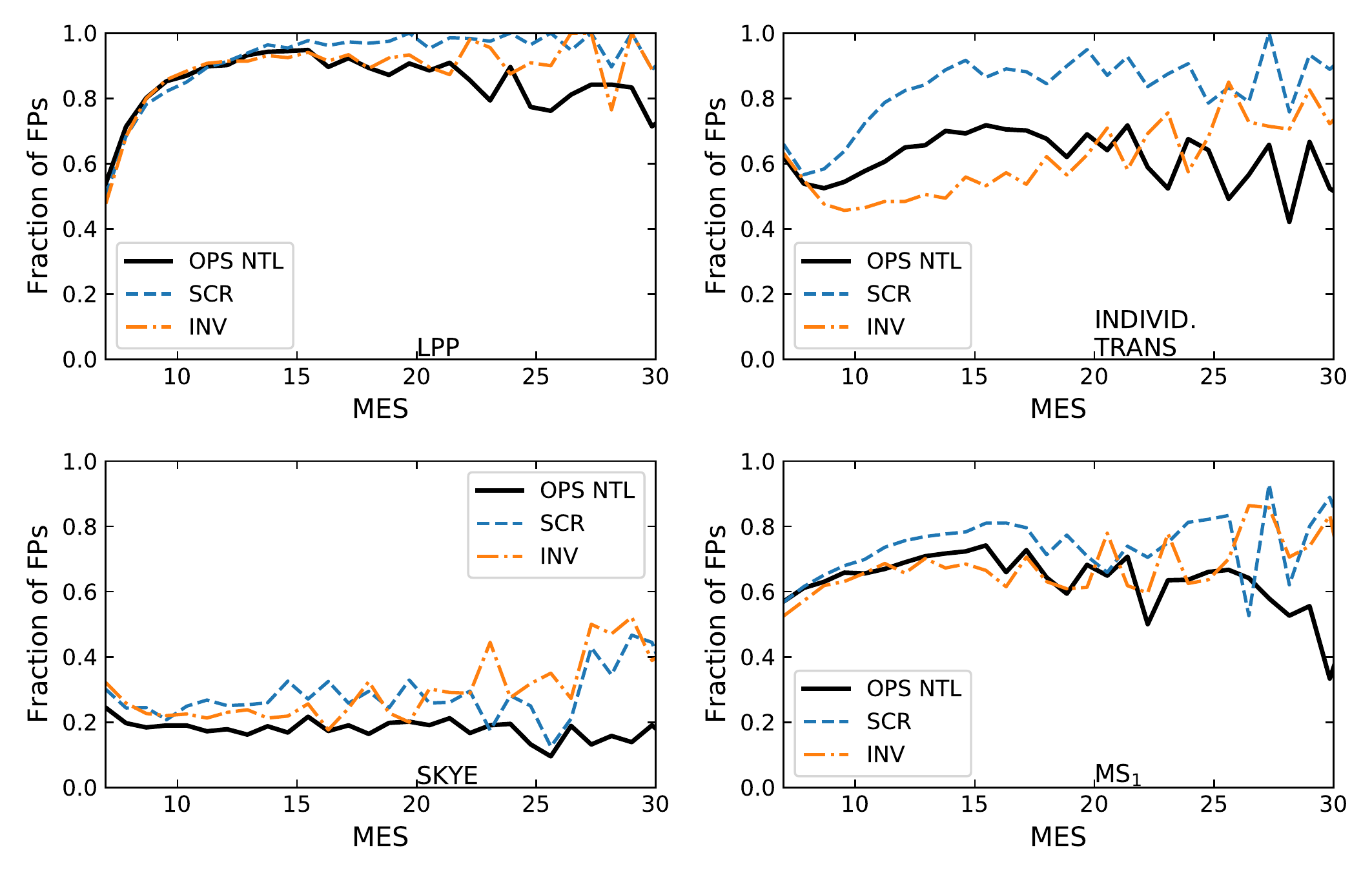}
    \caption{The fraction of not-transit-like FPs failed by a particular Robovetter metric plotted against the logarithm of the period (top two rows) or linear MES (bottom two rows).  The fraction is plotted for the \opstce\ set in black, the \scrtce\ set in blue (dashed), and the \invtce\ set in orange (dot-dashed). The metric under consideration is listed on each plot.  For each metric we include fails from either detrending (DV or ALT). Upper left: LPP metric failures. Upper Right: TCEs that fail after removing a single transit due to any of the individual transit metrics.  Lower left: TCEs that fail after removing a single transit due to the Skye metric. Lower right: Model Shift 1 metric failures. Notice that the trends are similar in all three data sets for most metrics, especially at long periods and low MES.}
    \label{f:fractionFailMetric}
\end{figure*}

Another way to judge how well the simulated data sets match the type of FP in the \opstce{s} is to look at some of the Robovetter metrics.  Each metric measures some aspect of the TCEs. For example, the LPP Metric measures whether the folded and binned light curves are transit shaped, and Skye measures whether the individual transits are likely due to rolling band noise.  If the simulated TCEs can be used to measure reliability in the way described above, then the fraction of false alarms in any period bin caused by any particular metric should match between the two sets.  In Figure~\ref{f:fractionFailMetric} we show that this is basically true for both \invtce s and \scrtce s, especially for periods longer than 100 days or MES less than 15.  Keep in mind that more than one metric can fail any particular TCE, so the sum of the fractions across all metrics will be greater than one.  The deviations between TCE sets is as large as 40\% for certain period ranges and such differences may cause systematic errors in our measurements of reliability.  But, since the types of FPs overlap, it is not clear how to propagate this information into a formal systematic error bar on the reliability.  

For our discussion of the reliability estimate, we are cautiously satisfied with this basic agreement. Given that neither of the two sets perform better across all regions of parameter space, and since having more simulated false alarms improves the precision on effectiveness,  we have calculated the catalog reliability using a union of the \scrtce\ and \invtce\ sets after they have been cleaned as described in \S\ref{s:clean}.


\section{Tuning the Robovetter for High Completeness and Reliability}
\label{s:optimize}
As described in the previous section, the Robovetter makes decisions regarding which TCEs are FPs and PCs based on a collection of metrics and thresholds.  For each metric we apply a threshold and if the TCE's metrics' values lies above (or below, depending on the metric) the threshold then the TCE is called a FP.  Ideally the Robovetter thresholds would be tuned so that no true PCs are lost and all of the known FPs are removed; however, this is not a realistic goal.  Instead we sacrifice a few \injtce{s} in order to improve our measured reliability.

How to set these thresholds is not obvious and the best value can vary depending on which population of planets you are studying.  We used automated methods to search for those thresholds that passed the most \injtce{s} and failed the most \invtce{s} and \scrtce{s}. However, we only used the thresholds found from this automated optimization to inform how to choose the final set of thresholds. This is because the simulated TCEs do not entirely emulate the observed data and many of the metrics have a period and MES dependence.  For example, the injections were heavily weighted towards long periods and low MES so our automated method sacrificed many of the short period candidates in order to keep more of the long period \injtce{s}. Others may wish to explore similar methods to optimize the thresholds and so we explain our efforts to do this below.






\vspace{2em}
\subsection{Setting Metric Thresholds Through Optimization}
\label{s:full_optimize}
For the first step in Robovetter tuning, we perform an optimization that finds the metric thresholds that \added{achieve a balance between maximizing the fraction of TCEs from the \injtce\ set that are classified as PCs (i.e., completeness) and maximizing the fraction of TCEs from the \scrtce\ and \invtce\ sets identified as FPs (i.e. effectiveness.)} Optimization varies the thresholds of the subset of the Robovetter metrics described in \S\ref{s:robovetter}. The set of metrics chosen for the joint optimization, called ``optimized metrics'' are: LPP (\S\ref{s:lpp}), the Model-shift uniqueness test metrics ($MS_{1}$, $MS_{2}$, and $MS_{3}$; \S\ref{s:ms}), Max SES to MES (\S\ref{s:sesmes}), and TCE Chases (\S\ref{s:tcechases}). Both the DV and ALT versions of these metrics, when applicable, were used in the optimization.

Metrics not used in the joint optimization are incorporated by classifying TCEs as PCs or FPs using fixed {\it a priori} thresholds prior to optimizing the other metrics.  After optimization, a TCE is classified as a PC only if it passes both the non-optimized metrics and the optimized metrics.  Prior to optimization the fixed thresholds for these non-optimized metrics pass about 80\% of the \injtce\ set, so the final optimized set can have at most 80\% completeness. Note, the non-optimized metric thresholds for the DR25 catalog changed after doing these optimizations. The overall effect was that the final measured completeness of the catalog increased (see \S\ref{s:summary}), especially for the low MES TCEs.  If the optimization were redone with these new thresholds, then it would find that the non-optimized metrics pass 90\% of the \injtce{s}.  We decided this change was not sufficient reason to rerun the optimization since it was only being used to inform and not set the final thresholds.

Optimization is performed by varying the selected thresholds, determining which TCEs are classified as PCs by both the optimized and non-optimized metrics using the new optimized thresholds, and computing $C$ and $1-E$.  Our optimization seeks thresholds that minimize the objective function $\sqrt{{(\rm 1-E})^2 + ({\rm C} - {\rm C}_0)^2}$, where ${\rm C}_0$ is the target completeness, so the optimization tries to get as close as possible to ${\rm 1-E} = 0$ and ${\rm C} = {\rm C}_0$.  We varied ${\rm C}_0$ in an effort to \replaced{reduce the ineffectiveness}{increase the effectiveness}. The thresholds are varied from random starting seed values, using the Nelder-Mead simplex algorithm via the \textsc{MATLAB} {\it fminsearch} function.  This \textsc{MATLAB} function varies the thresholds until the objective function is minimized.  There are many local minima, so the optimal thresholds depend sensitively on the random starting threshold values.  The optimal thresholds we report are the smallest of 2000 iterations with different random seed values.

Our final optimal threshold used a target of ${\rm C}_0 = 0.8$, which resulted in thresholds that yielded \replaced{${\rm 1-E} = 0.0044$}{${\rm E} = 0.9956$} and ${\rm C} = 0.799$.  We experimented with smaller values of ${\rm C}_0$, but these did not significantly \replaced{lower ineffectiveness}{increase effectiveness}.   We also performed an optimization that maximized reliability defined in \S\ref{s:relcalc} rather than \replaced{minimizing ineffectiveness}{maximizing effectiveness}. This yielded similar results. We also explored the dependence of the optimal thresholds on the range of TCE MES and period.  We found that the thresholds have a moderate dependence, while the \replaced{ineffectiveness}{effectiveness} and completeness have significant dependence on MES and period range. Exploration of this dependence of Robovetter threshold on MES and period range is a topic for future study.


\subsection{Picking the Final Robovetter Metric Thresholds}

The results of this algorithmic optimization were used as a starting point for the final thresholds chosen for the DR25 catalog. We used the Confirmed Planet table and the Certified False Positive Table at the Exoplanet Archive, as well as the results of some prominent KOIs, to manually adjust the thresholds.  Because most of the \injtce{s}, \invtce{s}, and \scrtce{s} are at long periods and low MES the automated tuning optimized the completeness and effectiveness for this part of the catalog. However, many of \Kepler{'s} PCs have short periods and high SNR. The final catalog thresholds balanced the needs of the different parts of the catalog and endeavoured to keep the completeness of the long period candidates above 70\%.

For those interested in a certain part of the KOI catalog, it may be better to re-tune the thresholds to optimize for higher reliability or to more aggressively remove certain types of false alarms.   The Robovetter code\footnote{\url{https://github.com/nasa/kepler-robovetter}} (and the Robovetter input files) are provided with the tunable thresholds listed at the top of the code.  As an example, we include Table~\ref{t:thresholds} as a list of the easily tunable thresholds for the metrics that determine whether an object is not transit-like.  The table lists the thresholds we settled on for the DR25 catalog here, but it also provides the metrics for a higher reliability (lower completeness) catalog and a higher completeness (lower reliability) catalog. (These two different sets of thresholds are also included as commented-out lines in the Robovetter code after the set of thresholds used to create the DR25 catalog.) Each metric has its own range of possible values and some are more sensitive to small adjustments than others.  Users should use caution when changing the thresholds and should endeavour to understand the different metrics, described in \S\ref{s:robovetter} and Appendix \ref{s:metrics}, before doing so.

\begin{deluxetable*}{llrrr}
\tablecolumns{5}
\tablewidth{\linewidth}
\tablecaption{Robovetter Metric Thresholds \label{t:thresholds}}
\tablehead{
\colhead{Test Name} &
\colhead{Variable Name} &
\colhead{DR25} &
\colhead{High C} &
\colhead{High R} 
}
\startdata
SWEET & SWEET\_THRESH & 50 & 50 & 50\\[2pt]
Max SES to MES & SES\_TO\_MES\_THRESH & 0.8 & 0.9 & 0.75\\[2pt]
TCE CHASES & ALL\_TRAN\_CHASES\_THRESH & 0.8 & 1.0 & 0.55 \\ [2pt]
MS$_{1}$ DV & MOD\_VAL1\_DV\_THRESH & 1.0 & 2.4 & -1.0\\[2pt]
MS$_{2}$ DV & MOD\_VAL2\_DV\_THRESH & 2.0 & 5.0 & -0.7\\[2pt]
MS$_{3}$ DV & MOD\_VAL2\_DV\_THRESH & 4.0 & 7.5 & -1.6\\[2pt]
MS$_{1}$ ALT & MOD\_VAL1\_ALT\_THRESH & -3.0 & -2.5 & -4.3 \\[2pt]
MS$_{2}$ ALT & MOD\_VAL2\_ALT\_THRESH & 1.0 & -0.5 & 2.5 \\[2pt]
MS$_{3}$ ALT & MOD\_VAL3\_ALT\_THRESH & 1.0 & 0.5 & 0.2 \\[2pt]
LPP DV & LPP\_DV\_THRESH & 2.2 & 2.8 & 2.7\\[2pt]
LPP ALT & LPP\_ALT\_THRESH & 3.2 & 3.2 & 3.2\\[0pt]
\enddata
\end{deluxetable*}

\section{Assembling the DR25 KOI Catalog}
\label{s:assemble}
The KOI catalog contains all the \opstce s that the Robovetter found to have some chance of being transit-shaped, i.e., astrophysically transiting or eclipsing systems. All of the DR25 KOIs are fit with a transit model and uncertainties for each model parameter are calculated with a MCMC algorithm.  We describe here how we decide which \opstce{s} become KOIs, how we match the \opstce{s} with previously known KOIs, and how the transit fits are performed. The KOI catalog is available in its entirety at the NASA Exoplanet Archive as the Q1--Q17 DR25 KOI Table\footnote{\url{https://exoplanetarchive.ipac.caltech.edu/cgi-bin/TblView/nph-tblView?app=ExoTbls\&config=q1\_q17\_dr25\_koi}}.
\label{koisec}

\subsection{Creating KOIs}
The Robovetter gives every \opstce\ a disposition, a reason for the disposition, and a disposition score.  However, only those that are transit-like, i.e., have some possibility of being a transiting exoplanet or eclipsing binary system, are intended to be placed in the KOI catalog. For scheduling reasons, we created the majority of KOIs before we completed the Robovetter, so some not transit-like KOIs have been included in the KOI catalog. Using the final set of Robovetter dispositions, we made sure to include the following \opstce s in the KOI table: 1) those that are ``transit-like'', i.e., are not marked with the NT-flag, and 2) KOIs that are not transit-like FPs which have a score value larger than 0.1.  This last group were included to ensure that \opstce s that marginally failed one Robovetter metric were easily accessible via the KOI catalog and given full transit fits with MCMC error bars. As in previous catalogs, all DR25 \opstce{s} that federate (\S\ref{s:federation}) to a previously identified KOI are included in the DR25 KOI table even if the Robovetter set the disposition to a not transit-like FP. All previous KOIs that were not found by the DR25 \Kepler{} Pipeline (i.e., did not trigger a DR25 \opstce{}) are not included in the DR25 KOI table at the Exoplanet Archive.

\subsection{Federating to known KOIs}
\label{s:federation}
All \opstce s that were included in the KOI catalog were either federated to known KOIs or given a new KOI number. Since KOIs have been identified before, federating the known KOIs to the TCE list is a necessary step to ensure that we do not create new KOIs out of events previously identified by the \Kepler\ pipeline.  The process has not changed substantially from the DR24 KOI catalog \citep[see \S4.2 of][]{Coughlin2016}, so we simply summarize \added{the criteria for a match} here. \added{If the TCE orbital period matches within 0.2\% of the KOI period then the two are considered federated if at least 25\% of the transit events overlap.  If the relative period mismatch was more than 0.2\%, then 70\% of the transit events need to overlap. Also, those that are found at double or half the period are also considered matches (244 KOIs in total).  In the cases where the period is double or half, then at least 95\% of the expected (every-other) transit cadences need to overlap in order to federate the two. }

In some cases our automated tools want to create a new KOI in a system where one of the other previously known KOIs in the system did not federate to a DR25 TCE.  In these cases we inspect the new system by hand and ensure that a new KOI number is truly warranted. If it is, we create a new KOI. If not, we ban the event from the KOI list.  For instance, events that are caused by video cross-talk \citep{KIH} can cause short-period transit events to appear in only one quarter each year. As a result, the \Kepler\ Pipeline finds several one-year period events for an astrophysical event that is truly closer to a few days.  In these cases we federate the one found that most closely matches the known KOI and we ban any other \opstce{s} from creating a new KOI around this star. In Table~\ref{t:banned} we report the entire list of \opstce{s} that were not made into KOIs despite being dispositioned as transit-like (or not transit-like with a disposition score $\ge$~0.1) and the automated federation telling us that one was appropriate. To identify the TCEs we specify the Kepler Input Catalog number and the planet number given by the \Kepler{} Pipeline \citep{Twicken2016}.

\begin{deluxetable}{c}
\tablecolumns{1}
\tablewidth{\linewidth}
\tablecaption{\opstce{s} Banned from Becoming KOIs (\S\ref{s:federation}) \label{t:banned}}
\tablehead{
\colhead{TCE-ID}\\
\colhead{(KIC-PN)}
}
\startdata
003340070-04\\
003958301-01\\
005114623-01\\
005125196-01\\
005125196-02\\
005125196-04\\
005446285-03\\
006677841-04\\
006964043-01\\
006964043-05\\
007024511-01\\
008009496-01\\
008956706-01\\
008956706-06\\
009032900-01\\
009301564-01\\
010223616-01\\
012459725-01\\
012644769-03\\
\enddata

\end{deluxetable}


It is worth pointing out that the \added{removal of the banned TCEs is} the one pseudo-manual step that is not repeated for all the simulated TCEs.  These banned TCEs effectively disappear when doing statistics on the catalog (i.e., these TCEs do not count as either a PC or an FP.) They are not present in the simulated data sets, nor are we likely to remove good PCs from our sample this way. Most banned TCEs are caused by either a short-period binary whose flux is contaminating our target star (at varying depths through mechanisms such as video cross-talk or reflection), or are systems with strong TTVs (transit timing variations, see \S\ref{s:mcmc}). In both cases, the Pipeline finds several TCEs at various periods, but only one astrophysical system causes the signal.  By banning these \opstces, we are removing duplicates from the KOI catalog and improving the completeness and reliability statistics reported in \S\ref{s:candr}.

\subsection{KOI Transit Model Fits}
\label{s:mcmc}

Each KOI, whether from a previous catalog or new to the DR25 catalog, was fit with a transit model in a consistent manner. The model fitting was performed in a similar to that described in \S5 of \citet{Rowe2015cat}. The model fits start by detrending the DR25 Q1--Q17 PDC photometry from MAST\footnote{\url{https://archive.stsci.edu/kepler/}} using a polynomial filter as described in \S4 of \citet{Rowe2014}. A transit model based on \citet{Mandel2002} is fit to the photometry using a Levenberg-Marquardt routine \citep{More1980} assuming circular orbits and using fixed quadratic limb darkening coefficients \citep{Claret2011} calculated using the DR25 stellar parameters \citep{Mathur2017ApJS}. TTVs are included in the model fit when necessary; the calculation of TTVs follows the procedure described in \S4.2 of \citet{Rowe2014}. The 296 KOIs with TTVs and the TTV measurements of each transit are listed in Table~\ref{t:ttv}. The uncertainties for the fitted parameters were calculated using a Markov-chain Monte Carlo (MCMC) method \citep{Ford2005AJ} with a single chain with a length of 2$\times 10^{5}$ calculated for each fit. In order to calculate the posterior distribution the first 20\% of each chain was discarded. The transit model fit parameters were combined with the DR25 stellar parameters and associated errors \citep{Mathur2017ApJS} in order to produce the reported planetary parameters and associated errors. The MCMC chains are all available at the Exoplanet Archive and are documented in \citet{Hoffman2017}. 

The listed planet parameters come from the least-squares (LS) model fits and the associated errors from the MCMC calculations. Note that not all KOIs could be successfully modelled, resulting in three different fit types: LS+MCMC, LS, and none. In the case of LS+MCMC the KOIs were fully modelled with both a least-squares model fit and the MCMC calculations were completed to provide associated errors. In the cases where the MCMC calculations did not converge, but there is a model fit, the least-square parameters are available without uncertainties (LS fit type). In the final case, where a KOI could not be modelled (e.g., cases where the transit event was not found in the detrending used by the MCMC fits) only the period, epoch, and duration of the federated TCE are reported and the fit type is listed as none.

\begin{deluxetable}{crrr}
\tabletypesize{\scriptsize}
\tablecaption{TTV Measurements of KOIs}
\tablewidth{0pt}
\tablehead{
\colhead{$n$}  & \colhead{$t_n$} & \colhead{$TTV_n$}  & \colhead{$TTV_{n\sigma}$} \\
\colhead{}     & \colhead{BJD-2454900.0}      & \colhead{days}     & \colhead{days}
}
\startdata
KOI-6.01 & & & \\
1 & 54.6961006 & 0.0774247 &  0.0147653 \\ 
2 & 56.0302021 & -0.0029102 &  0.0187065 \\ 
3 & 57.3643036 & -0.0734907 &  0.0190672 \\ 
4 & 58.6984051 & 0.0119630 &  0.0176231 \\ 
\nodata & \nodata & \nodata & \nodata\\
KOI-8.01 & & & \\
1 & 54.7046603 & -0.0001052 &  0.0101507 \\ 
2 & 55.8648130 & -0.0103412 &  0.0084821 \\ 
3 & 57.0249656 & 0.0047752 &  0.0071993 \\ 
\nodata & \nodata & \nodata & \nodata\\
KOI-8151.01 & & & \\
1 & 324.6953389 & 0.1093384 &  0.0025765 \\ 
2 & 756.2139285 & -0.3478332 &  0.0015206 \\ 
3 & 1187.7325181 & 0.0110542 &  0.0016480 \\
\nodata & \nodata & \nodata & \nodata\\
\enddata
\tablecomments{Column 1, $n$, is the transit number. Column 2, $t_n$, is the transit time in Barycentric Julian Date minus the offset 2454900.0. Column 3, $TTV_n$, is the observed - calculated (O-C) transit time. Column 4, $TTV_{n\sigma}$,  is the $1\sigma$ uncertainty in the O-C transit time.
Table \ref{t:ttv} is published in its entirety in the electronic edition of the {\it Astrophysical Journal Supplement}.  A portion is shown here for guidance regarding its form and content.}
\label{t:ttv}
\end{deluxetable}
%

\section{Analysis of the DR25 KOI Catalog}

\def \nkois {8054}
\def \ncand {4034}
\def \newkois {738}
\def \newcand {219}
\def \completeness {85.2}
\def \reliability {97.1}
\def \effectiveness {99.6}
\label{s:summary}
\vspace{-0.1em}
\subsection{Summary of the KOI Catalog}
The final DR25 KOI catalog, available at the NASA Exoplanet Archive, contains all TCEs that pass the not transit-like tests (\S\ref{nottransitlikesec}) and those that fail as not transit-like with a disposition score $\ge$ 0.1. Some overall statistics of the DR25 KOI catalog are as follows:
\vspace{-0.9em}
\begin{itemize}
\itemsep0em
    \item \nkois{}~KOIs
    \item \ncand{}~PCs
    \item \newkois{}~new KOIs
    \item \newcand{}~new PCs
    \item \completeness{}\% of \injtce{s} are PCs
    \item \effectiveness{}\% of \invtce{s} and \scrtce{s} are FPs
\end{itemize}

\begin{figure*}[htb]
    \centering
    \includegraphics[width=\linewidth]{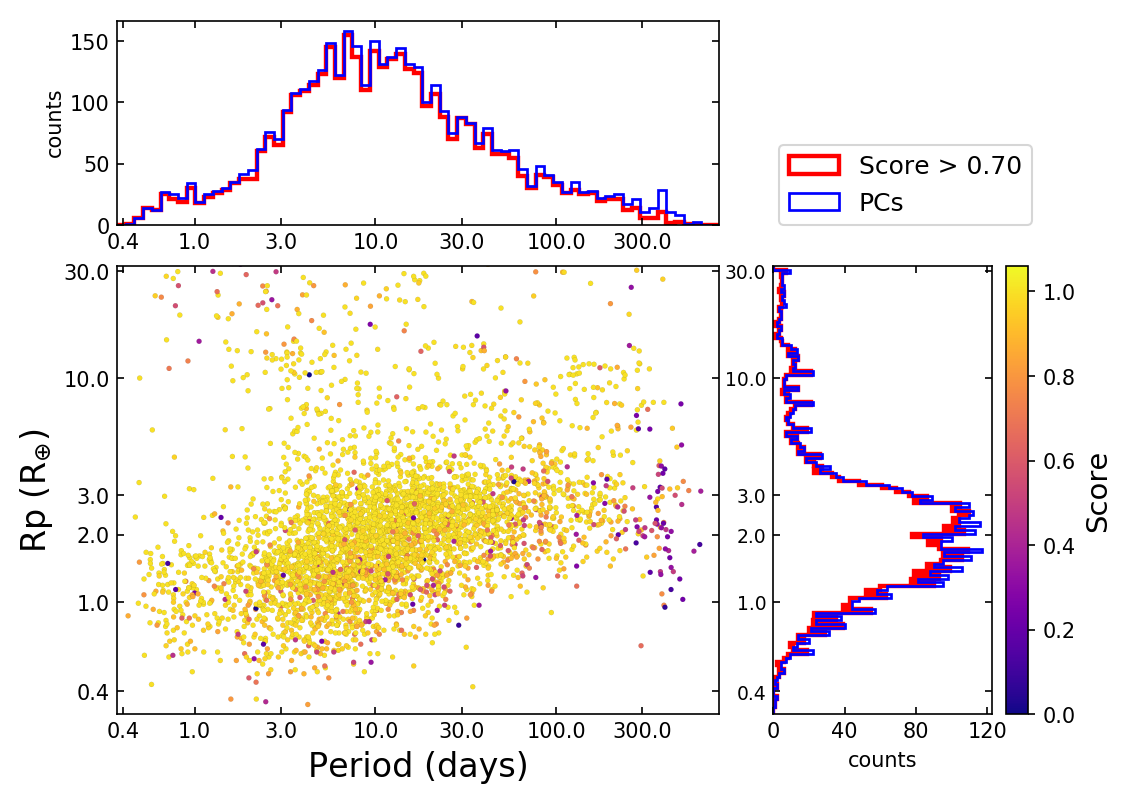}
    \caption{DR25 PCs plotted as planet radius versus period with the color representing the disposition score. The period and planet radii distributions are plotted on the top and on the left, respectively, in blue. The red line shows the distributions of those PCs with a disposition score greater than 0.7. The excess of PCs at long-periods disappears when cutting the population on disposition score. }
    \label{f:catalogPlot}
\end{figure*}

A plot of the planetary periods and radii is shown in Figure~\ref{f:catalogPlot}, with the color indicating the disposition score. The distribution of the periods and planetary radii of the planet candidates in this catalog is shown along the x- and y-axis. A clear excess of candidates exists with periods near 370\,d;  this excess disappears if we only consider those with a disposition score $>$~0.7. While the disposition score provides an easy way to make an additional cut on the PC population at long periods, when discussing the catalog PCs below we are using the pure dispositions of the Robovetter unless otherwise stated. The slight deficit of planets with radii just below 2.0\,R$_{\oplus}$ is consistent with the study of \citet{Fulton2017} where they report a natural gap in the abundance of planets between super-Earths and mini-Neptunes by applying precise stellar parameters to a subset of the \kepler\ transiting candidates \citep{CKS2,CKS1}. The new KOIs with a disposition of PC are found at all periods, but only ten have MES~$\geq$~10.

\subsection{Comparison of Dispositions to Other Catalogs}
We compare the DR25 KOI catalog to two sets of \Kepler\ exoplanets: the confirmed exoplanets and the certified FPs.  In both of these cases, additional observations and careful vetting are used to verify the signal as either a confirmed exoplanet or a certified FP \citep{Bryson2017c}. It is worth comparing the Robovetter to these catalogs as a sanity check.  

We use the confirmed exoplanet list from the Exoplanet Archive\footnote{\url{https://exoplanetarchive.ipac.caltech.edu/cgi-bin/TblView/nph-tblView?app=ExoTbls\&config=planets}} on 2017-05-24.  2279 confirmed planets are in the DR25 KOI catalog.  The DR25 Robovetter fails 44 of these confirmed planets, or less than 2\%. Half of these FPs are not transit-like fails, 16 are stellar eclipse fails, six are centroid offsets, and one is an ephemeris match. Twelve fail due to the LPP metric; all of these twelve have periods less than 50 days.  The LPP metric threshold was set to improve the reliability of the long period KOIs, an act which sacrificed some of the short period KOIs.  The reason the Robovetter failed each confirmed planet is given in the ``koi\_comment'' column at the Exoplanet Archive (see \S\ref{s:minorflags}). 

For the vast majority of the Robovetter FPs on the confirmed planet list, careful inspection reveals that there is no doubt that the Robovetter disposition is incorrect. As an example, Kepler-10b \citep[][]{Batalha2011Kepler10,Fogtmann2014Kepler10}, a rocky planet in a 0.84\,d orbit, was failed due to the LPP metric. This occurred because the detrending algorithm (the harmonic identification and removal algorithm in TPS, see \citealt{JenkinsKDPH}) used by the \Kepler\ Pipeline significantly distorts the shape of the transit, a known problem for strong, short period signals \citep{Christiansen2015}.  The LPP metric, which compares the shape of the folded light curve to known transits, then fails the TCE.

In some cases the Robovetter may be correctly failing the confirmed planet.  Many of the confirmed planets are only statistically validated \citep{Morton2016,Rowe2014}. In these cases no additional data exists proving the physical existence of a planet outside of the transits observed by \Kepler. It is possible that the DR25 light curves and metrics have now revealed evidence that the periodic events are caused by noise or a binary star. For example, Kepler-367c \citep{Rowe2014}, Kepler-1507b \citep{Morton2016}, and Kepler-1561b \citep{Morton2016} (KOIs 2173.02, 3465.01, and 4169.01, respectively) were all confirmed by validation and have now failed the Robovetter because of the new ghost metric (see \S\ref{s:ghost}), indicating that the events are caused by a contaminating source not localized to the target star.  These validations should be revisited in the light of these new results.

It is also worth noting that none of the confirmed circumbinary planets \citep[e.g.,][]{Doyle2011,Orosz2012} are in the DR25~KOI catalog. However, the eclipsing binary stars that they orbit are listed as FPs.  The timing and shape of the circumbinary planet transits vary in a complicated manner, making them unsuitable for detection by the search algorithm used by the \Kepler{} Pipeline to generate the DR25 \opstce{} list.  As a result, this catalog cannot be used for occurrence rate estimates of circumbinary planets, and their absence in the KOI catalog should not cast doubt on their veracity.

We use the Certified False Positive table\footnote{\url{https://exoplanetarchive.ipac.caltech.edu/cgi-bin/TblView/nph-tblView?app=ExoTbls\&config=fpwg}} downloaded from the Exoplanet Archive on 2017-07-11 to evaluate the performance of the Robovetter at removing known FPs. This table contains objects known to be FPs based on all available data, including ground-based follow-up information.  The Robovetter passes 106 of the 2713 certified FPs known at the time, only 3.9 per cent.  Most of those called PCs by the Robovetter are high signal-to-noise and more than half have periods less than 5 days.  The most common reason they are certified FPs is that there is evidence they are eclipsing binaries. In some cases, external information, like radial velocity measurements, provide a mass which determines that the KOI is actually a binary system. The other main reason for the discrepancy between the tables is that the certified FPs often show evidence of significant centroid offsets. In crowded fields the Centroid Robovetter (\S\ref{s:centroidrv}) will not fail observed offsets because of the potential for confusion. For the Certified False Positive table, individual cases are examined by a team of scientists who determines when there is sufficient proof that the signal is indeed caused by a background eclipsing binary.

\subsection{Catalog Completeness, Effectiveness, and Reliability}
\label{s:candr}

\begin{figure}[hp]
\begin{center}
\begin{tabular}{c}
\includegraphics[width=0.92\linewidth]{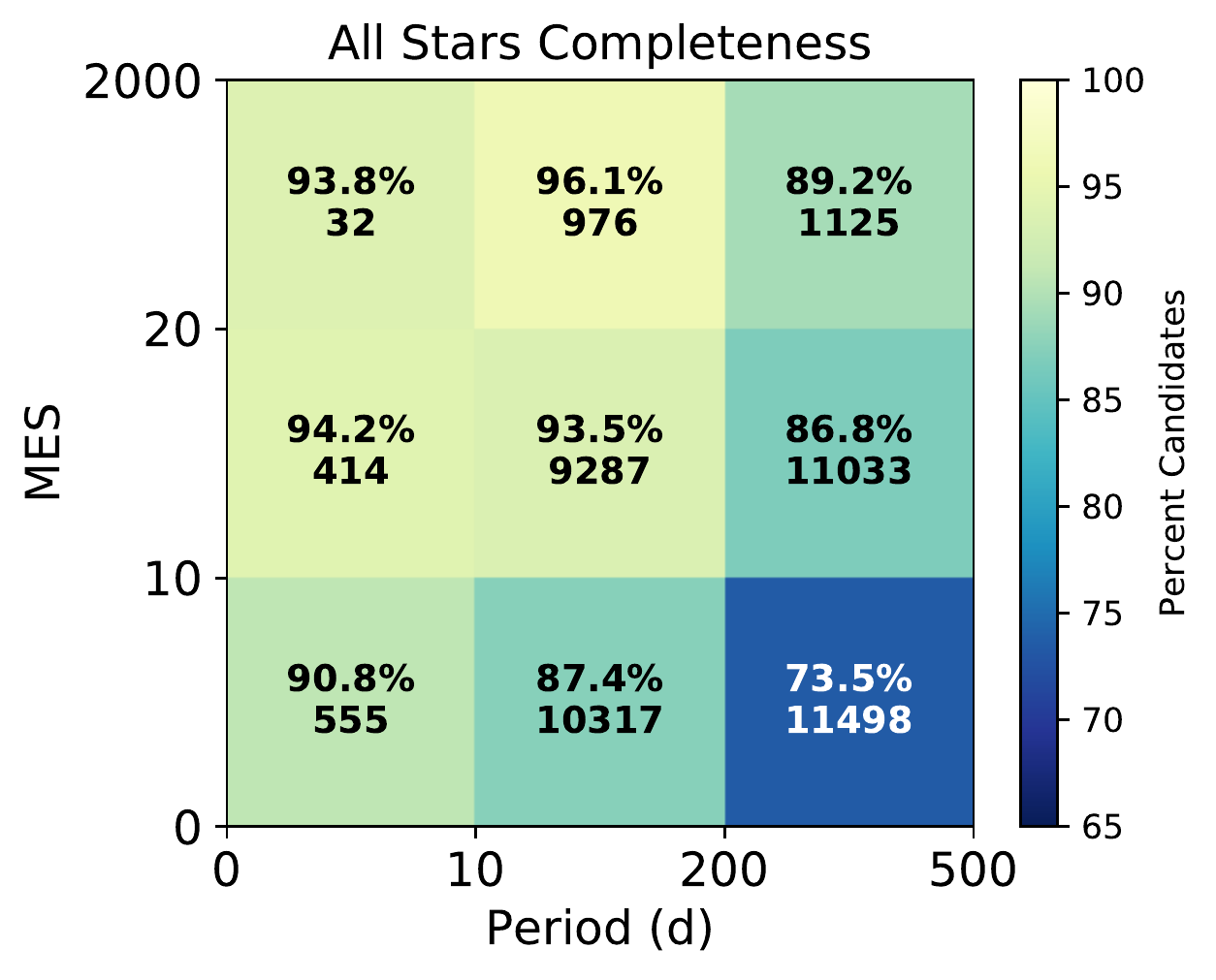} \\
\includegraphics[width=0.92\linewidth]{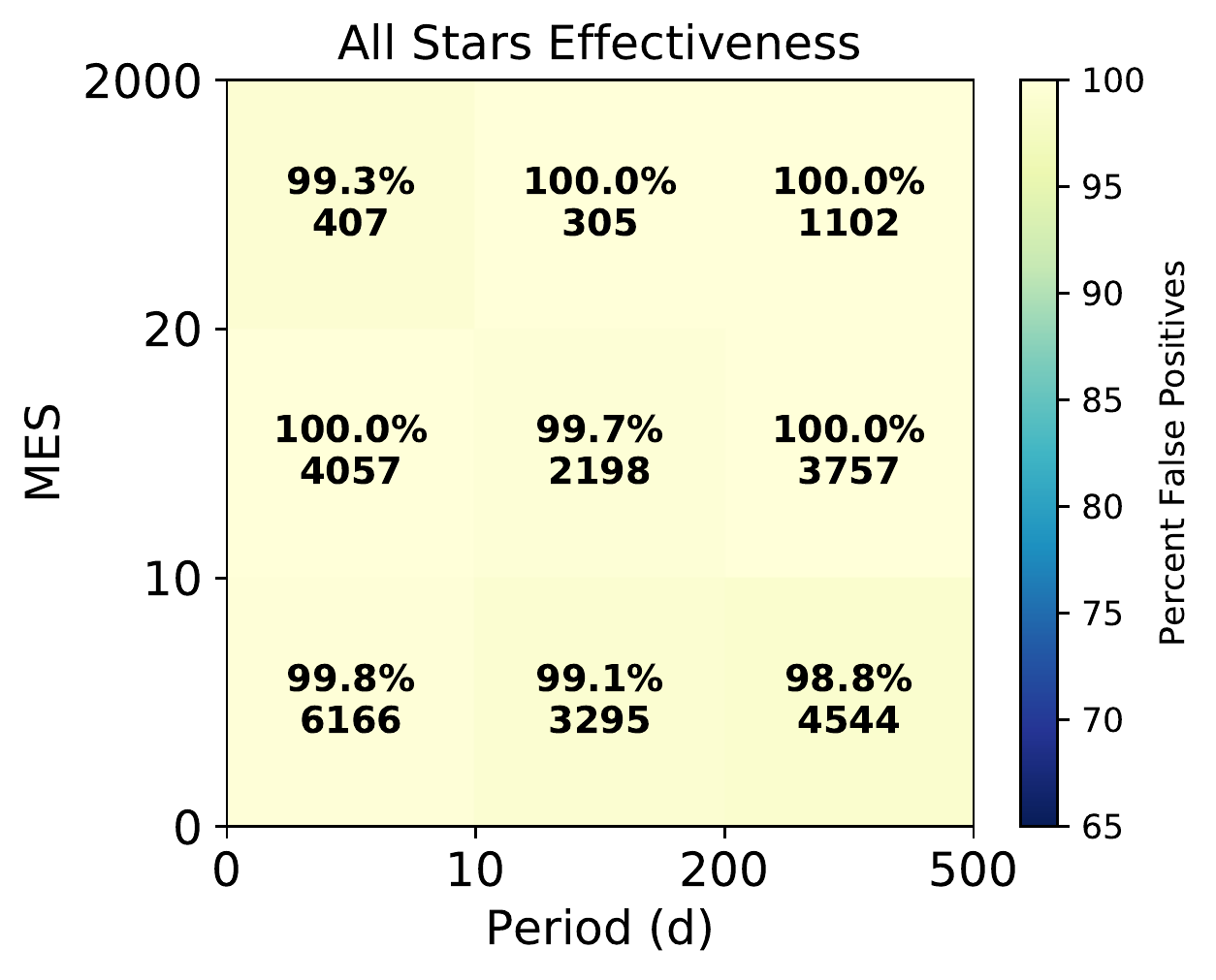} \\
\includegraphics[width=0.92\linewidth]{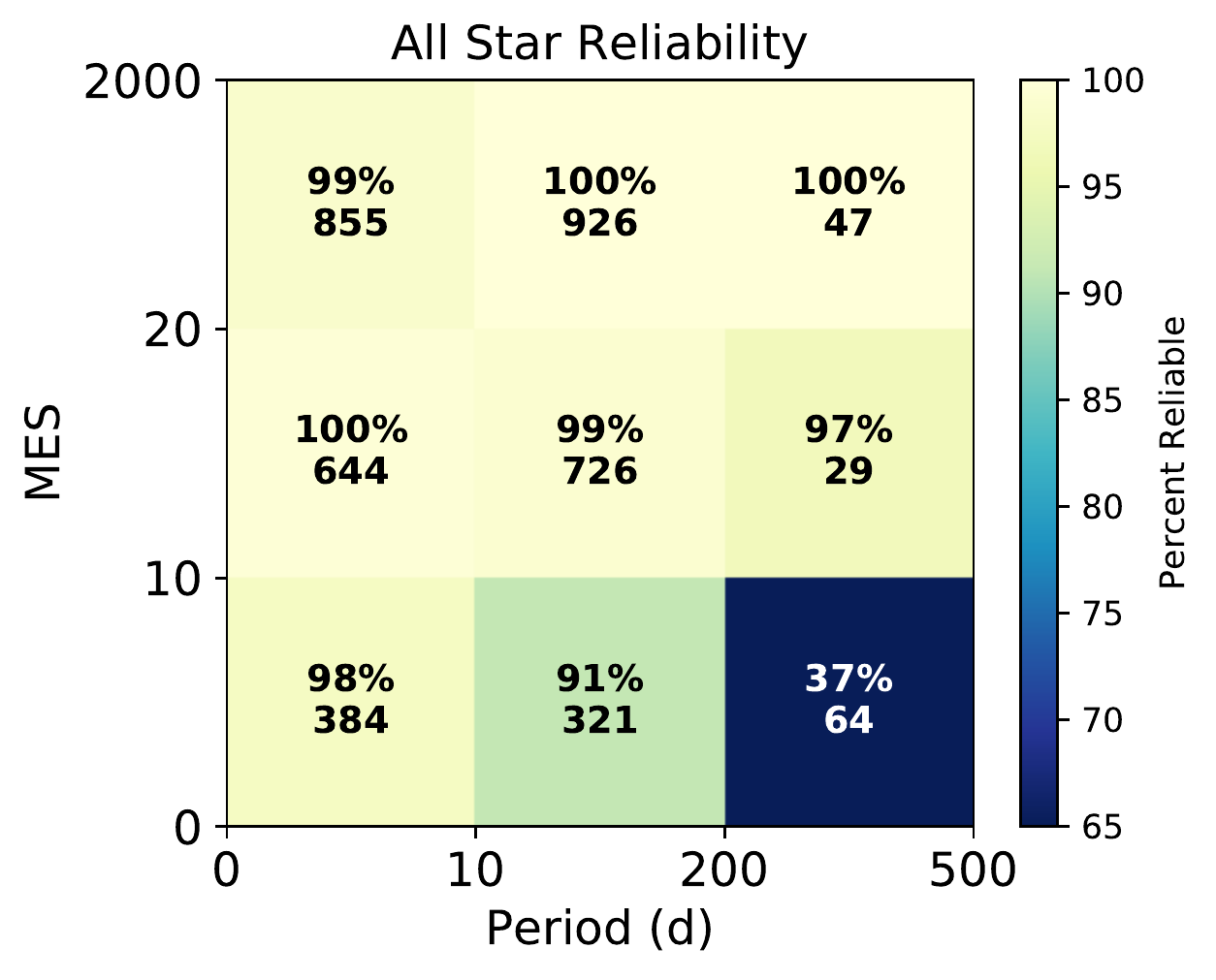}
\end{tabular}
\caption{A coarse binning of the completeness, effectiveness, and reliability for different period and MES bins (shown from top to bottom, respectively). \added{(The number of TCEs in the box is shown below the percentage for completeness and effectiveness. The number of PCs is given below the reliability.)}The effectiveness and reliability are based on the combined \invtce{} and \scrtce{} data sets. Notice that the Robovetter effectiveness at removing these false alarms is incredibly high, but for long periods and low MES the resulting reliability is lower because of the large number of false alarms and small number of true planets. For FGK dwarf stars only, the reliability is 50.3\%,\added{ the effectiveness is 98.9\%, and }the completeness is 76.7\% for planets in the longest period, lowest MES box. }
\label{f:scoregrid}
\end{center}
\end{figure}

To evaluate the performance of the Robovetter and to measure the catalog completeness and reliability, we run the Robovetter on the \injtce{s}, \invtce{s}, and \scrtce{s}. As a high level summary, Figure~\ref{f:scoregrid} provides the completeness, effectiveness ($E$), and reliability for a 3 by 3 grid across period and MES. If the same figure is made for only the FGK dwarf type stars (\logg~$\ge$~4.0 and 4000~K~$\ge$~\tstar~$<$~7000~K), the long period, low MES bin improves substantially. Giant stars are inherently noisy on time scales of planet transits \citep[see Figure~9 of][]{Christiansen2012} causing more FPs and also causing more real transits to be distorted by the noise. For FGK dwarf stars and only considering candidates with periods between 200~d and 500~d and MES~<~10, $C=76.7\%$, $1-E=1.1\%$, and $R=50.3\%$, which is a 13.1 percentage point improvement in reliability and 3 percentage point improvement in completeness compared to all stars in the same period and MES range.

\subsubsection{Completeness}
\label{s:comp}
\begin{figure*}[hp]
 \begin{center}
  \includegraphics[width=0.875\linewidth]{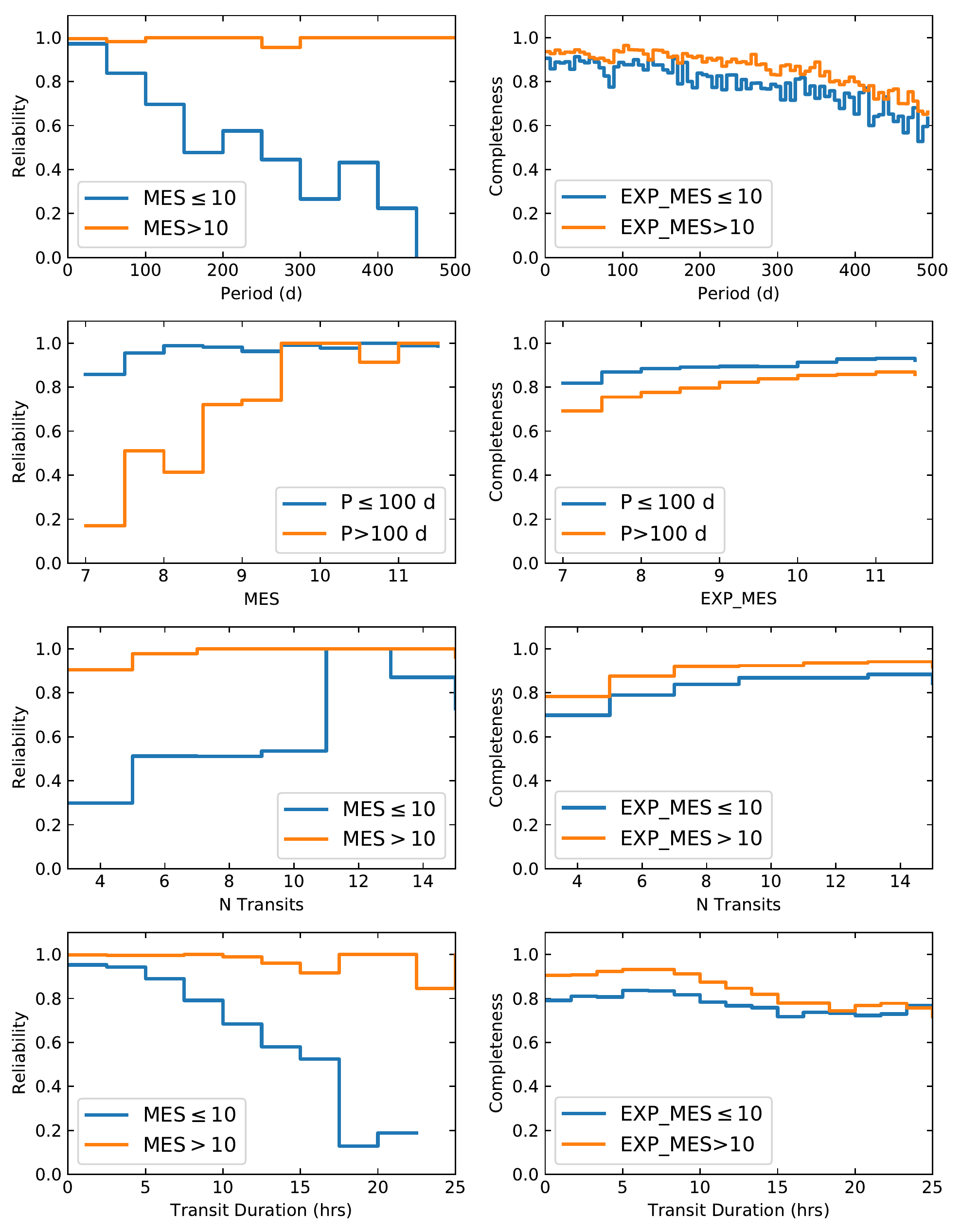}
  \caption{The reliability (left) and completeness (right)  of the DR25 catalog plotted as a function of period, MES, number of transits, and transit duration. In each case the blue line is for those with MES~$\leq$~10 or periods~$\leq$~100\,d. The orange line shows the completeness or reliability for the rest of the population (see the legend for each plot). EXP\_MES is the expected MES (see \citealt{Christiansen2017} and \S\ref{s:comp}).}
  \label{f:1dcompare}
 \end{center}
 \end{figure*}

The completeness of the vetting is measured as the fraction of \injtce{s} that are dispositioned as PCs. We discuss here the detection efficiency of the Robovetter, not the Kepler Pipeline (see \S\ref{s:occurates} for a discussion of the Pipeline completeness). Across the entire set of recovered \injtce{s} which have periods ranging from 0.5--500\,d, the Robovetter dispositioned \completeness{}\% as PC. As expected, the vetting completeness is higher for transits at shorter periods and higher MES, and lower for longer periods and lower MES. The right hand column of Figure~\ref{f:1dcompare} shows how the completeness varies with period, expected MES, number of transits, and transit duration. Note that expected MES is the average MES at which the injected transit signal would be measured in the target light curve, given the average photometric noise of that light curve and the depth of the injected transit signal --- see \citealt{Christiansen2017} for more details. The small drop in completeness just short of 90\,days is likely caused by the odd-even metric (\S\ref{s:oddeven}), which only operates out to 90\,days, confusing true transits for binary eclipses.

Because most planet occurrence rate calculations are performed using period and radius \citep[e.g.,][]{Burke2015}, we show the measured completeness binned by period and radius in Figure~\ref{f:prCompleteness}. The plot is linear in period and radius in order to emphasize the long period planets. Planetary radius is not a natural unit to understand the performance of the Robovetter since it combines the depth of the transit, the noise level of the light curve, and the stellar radius.  At the longest periods the Robovetter \replaced{more often fails the largest planets}{is more likely to disposition larger injected planets as FPs than the smaller counterparts}, though the trend is reduced when only considering the FGK dwarf stars. \added{The reason for this is that} the largest radii planets in the \injtce{} population are entirely around giant stars; large planets were not injected onto the dwarf stars \citep{Christiansen2017}. The giant stars are notoriously noisy. As a result the largest radii planets in the \injtce{s} are more likely to be dispositioned incorrectly. Also, even when only considering the dwarf stars, a larger fraction of the big planets will be around larger, more massive stars (in comparison to the small planets which will mostly be found around smaller stars). This results in a population of planets that produce longer transit durations. The Robovetter performs less well for long transit durations (see Figure~\ref{f:1dcompare}). For more figures showing the Robovetter effectiveness across different parameters, see \citet{Coughlin2017a}.

\subsubsection{Effectiveness}
The effectiveness of the Robovetter at identifying instrumental and stellar noise is calculated using the union of the \invtce{s} and \scrtce{s} (see \S\ref{s:relcalc}), after removing the TCEs specified in \S\ref{s:clean}. Across the entire set, the Robovetter dispositions 99.6\% of these simulated false alarms as FPs.  Only 119 of the 28,735 simulated false alarms are dispositioned as a PC by the Robovetter.  Most of these invPCs and scrPCs are at long periods and low MES. However, using the 4544 \invtce s and \scrtce s that have periods between 200\,d and 500\,d and MES less than 10, the Robovetter's effectiveness is 98.8\% (see Figure~\ref{f:scoregrid}).  Unfortunately, because there are so few candidates at these long periods, this translates to a relatively low reliability.  For detailed plots showing how effectiveness varies with different parameters see \citet{Coughlin2017a}.

\subsubsection{Reliability}
\label{s:reliability}
The reliability is measured according to the method described in \S\ref{s:relcalc} using the effectiveness measured from the combined \scrtce{} and \invtce{} data sets and the number of observed PCs. If one bins over the entire data set, the overall reliability of the catalog is 97\%. However, as Figure~\ref{f:1dcompare} demonstrates, the reliability for long period, and especially low MES planets, is significantly smaller.  For periods longer than 200\,d and MES less than 10, the reliability of the catalog is approximately 37\%, i.e., 6 out of 10 PCs are caused by noise. As with completeness, we plot the reliability as a function of period and planet radius in Figure~\ref{f:prReliability}. The least reliable planets are at long periods and have radii less than 4\re.

\begin{figure*}[ht]
\begin{center}
\begin{tabular}{cc}
\includegraphics[width=0.5\linewidth]{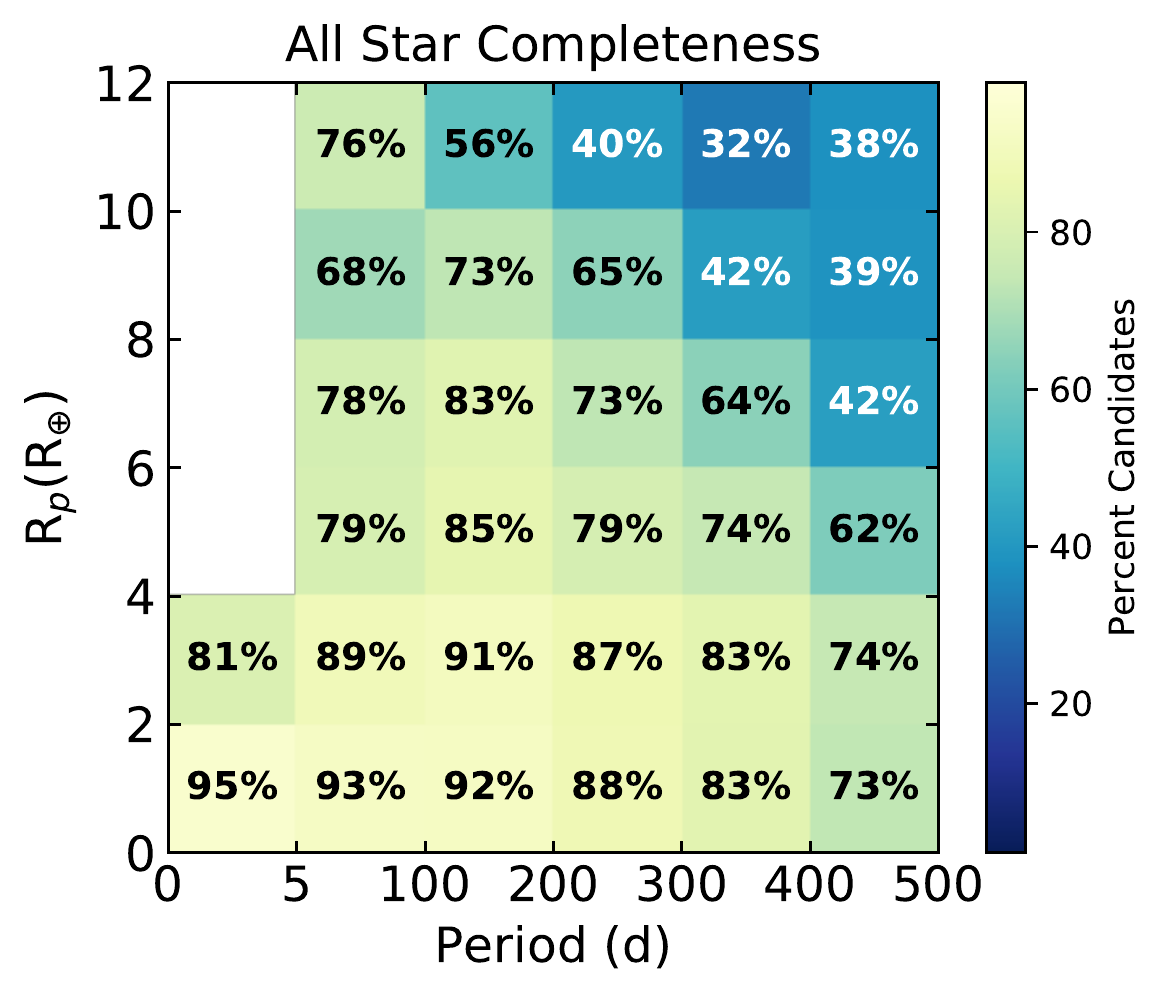} &
\includegraphics[width=0.5\linewidth]{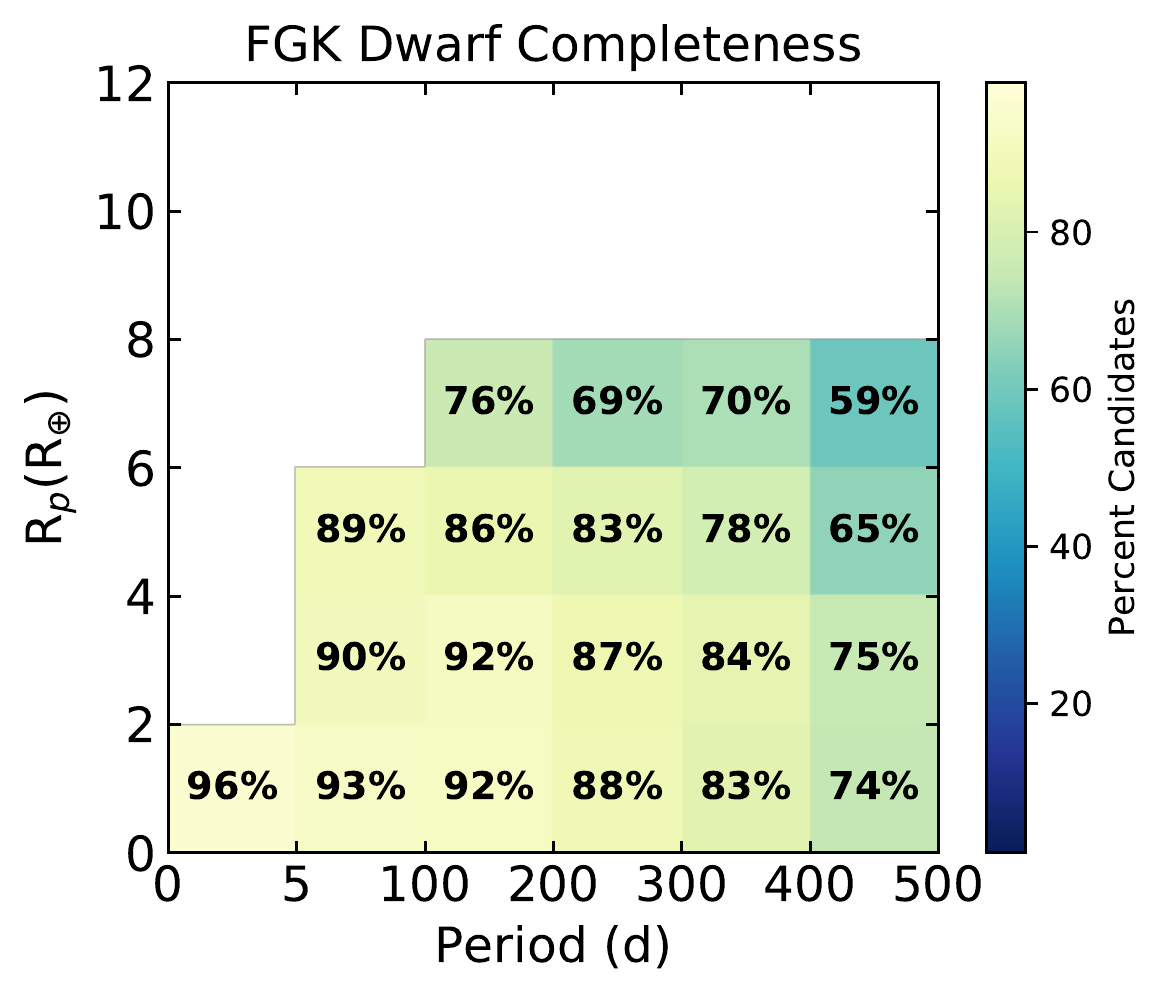}
\end{tabular}
\caption{The Robovetter completeness binned by period and planet radius for all stars (left) and for only FGK dwarf stars (right). Bins with fewer than 10 \injtce{s} are not plotted.}
\label{f:prCompleteness}
\end{center}
\end{figure*}

\begin{figure*}[htb]
\begin{center}
\begin{tabular}{cc}
\includegraphics[width=0.5\linewidth]{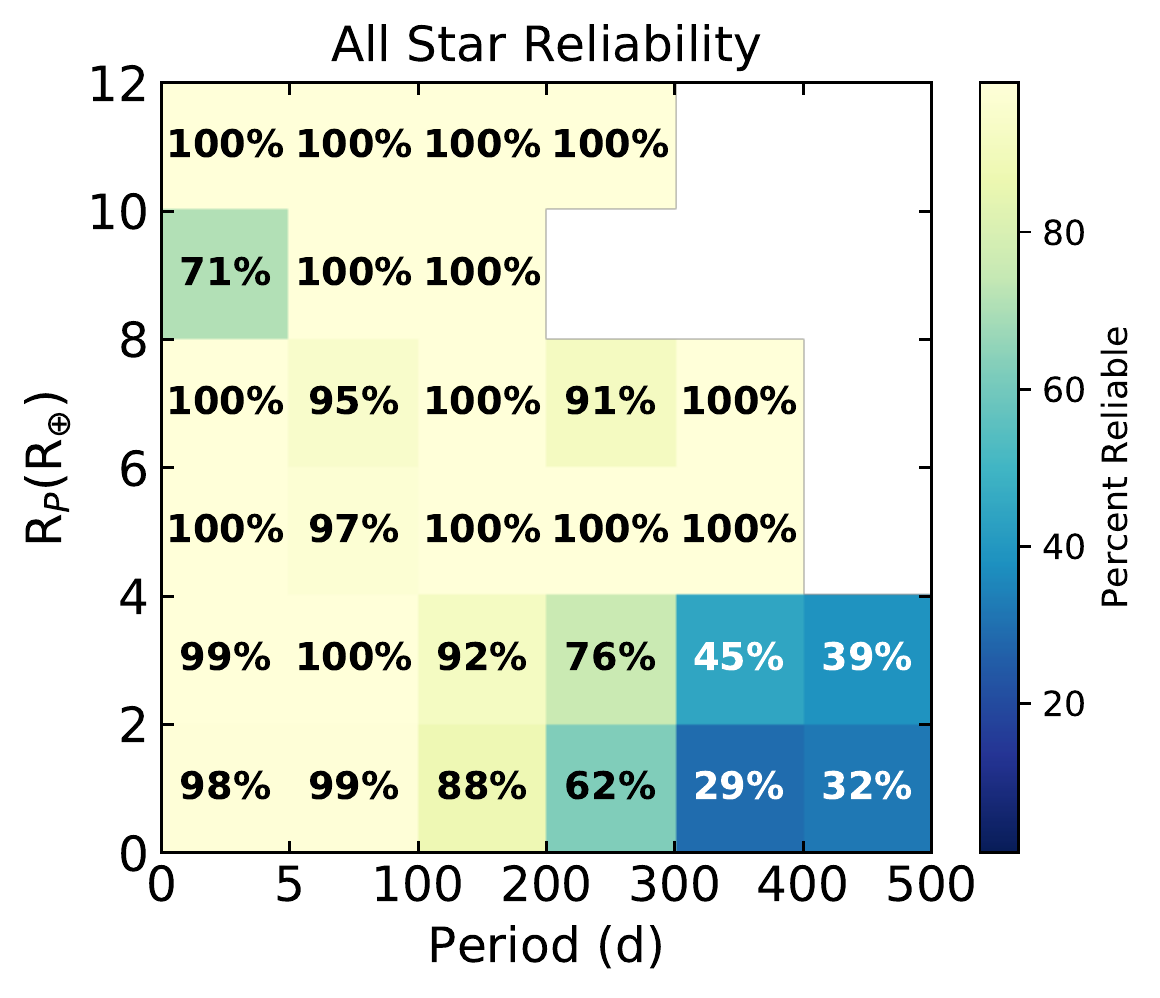} &
\includegraphics[width=0.5\linewidth]{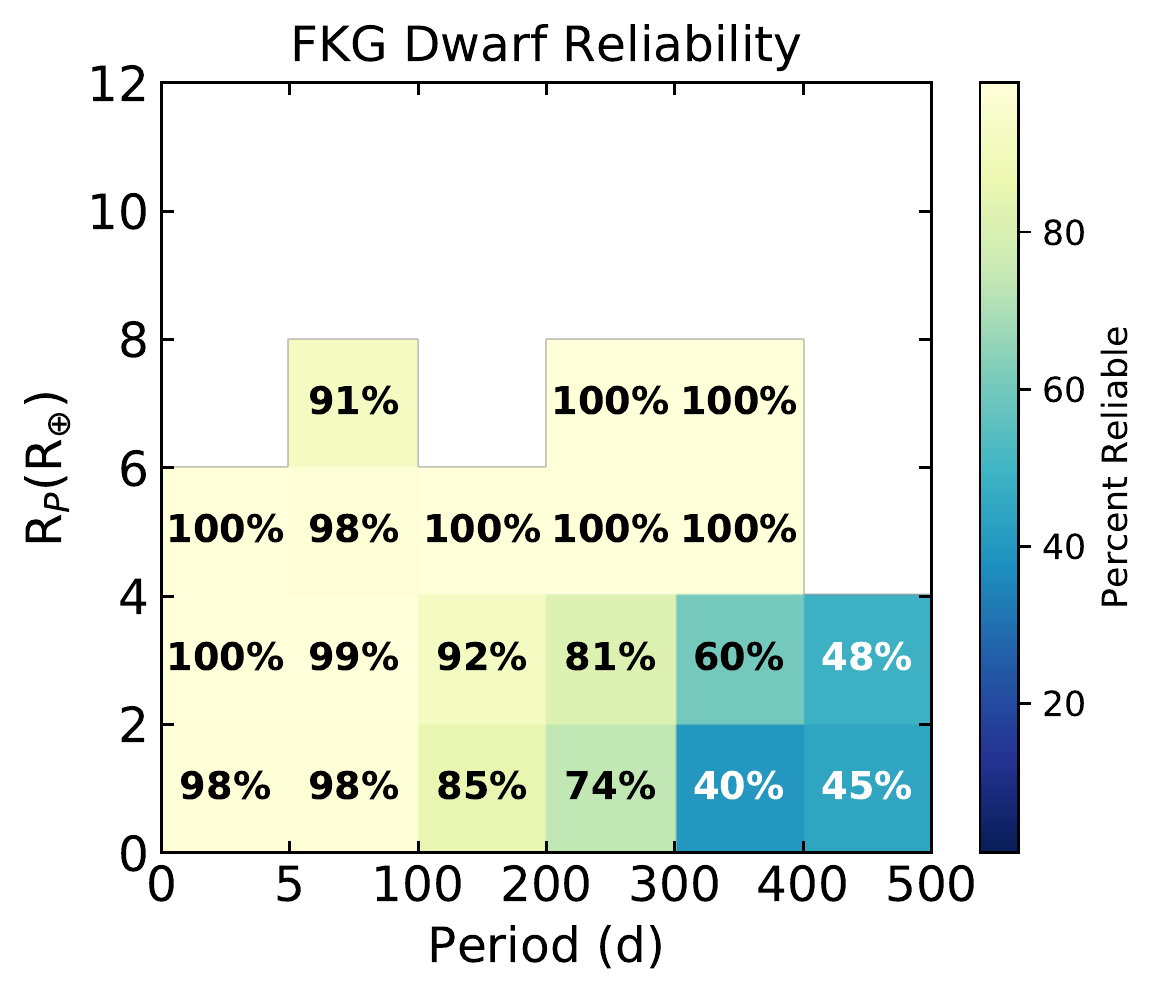}
\end{tabular}
\caption{A 2D binning of the candidate catalog reliability for period and planet radius for all stars (left) and for the FGK dwarf stars (right). Bins with fewer than 3 candidates or fewer than 20 simulated false alarms (from \invtce{} and \scrtce{}) are not plotted.}
\label{f:prReliability}
\end{center}
\end{figure*}

The uncertainty in the reliability is likely dominated by how well the false alarms in the \scrtce{} and \invtce{} sets match the false alarms in the \opstce{} data set (see \S\ref{s:simularity} for further discussion on their similarity).  One way to get a handle on the uncertainty on reliability is to calculate the reliability in three different ways for the long period (200--500\,d), low MES ($<10$) \opstce{s}.  First, we use only the \invtce{s} to measure the effectiveness at removing false alarms. This results in a lower reliability, namely $R$~=~24\% with $E$~=~98.5\%. Second, we use only the \scrtce{s} to measure the effectiveness. This results in a higher reliability, $R$~=~51\% with $E$~=~99.1\%. Third, we select, at random, half of the combined population of false alarms (\scrtce{} and \invtce{}) and calculate the reliability. After doing this random selection 100 times, we obtained $R$~=~38\% with a standard deviation of 8\%, and the distribution appears symmetric and basically Gaussian in shape.  

The Robovetter is less effective at removing the false alarms produced by inversion than those by scrambling the data. Inversion finds false alarms with periods near 372\,d, which are frequently caused by image artifacts.  Scrambling under-populates these types of false alarms, and since they are difficult to eliminate, it is not surprising that the reliability measured by inversion is worse than scrambling.  The truth likely lies somewhere in between. We encourage users of these data sets to consider ways to optimize the reliability measurement, and the error budget associated with them, when doing occurrence rate calculations. 

\added{We remind the reader that this analysis only concerns the reliability against the false alarms that are present in the \scrtce{s} and \invtce{s}. Previous studies \citep[e.g.][]{Santerne2012} discuss the reliability of previous KOI catalogs against short-period eclipsing binaries. However, since the Robovetter logic has changed considerably for this catalog (specifically the v-shaped metric was introduced and tuned to account for these false positives), the eclipsing binary false positive rate should be re-evaluated for this DR25 KOI catalog.}

\subsubsection{High Reliability Using the Disposition Score}
\label{s:crscores}

\begin{figure*}[htb]
\centering
\begin{tabular}{cc}
\includegraphics[width=0.5\linewidth]{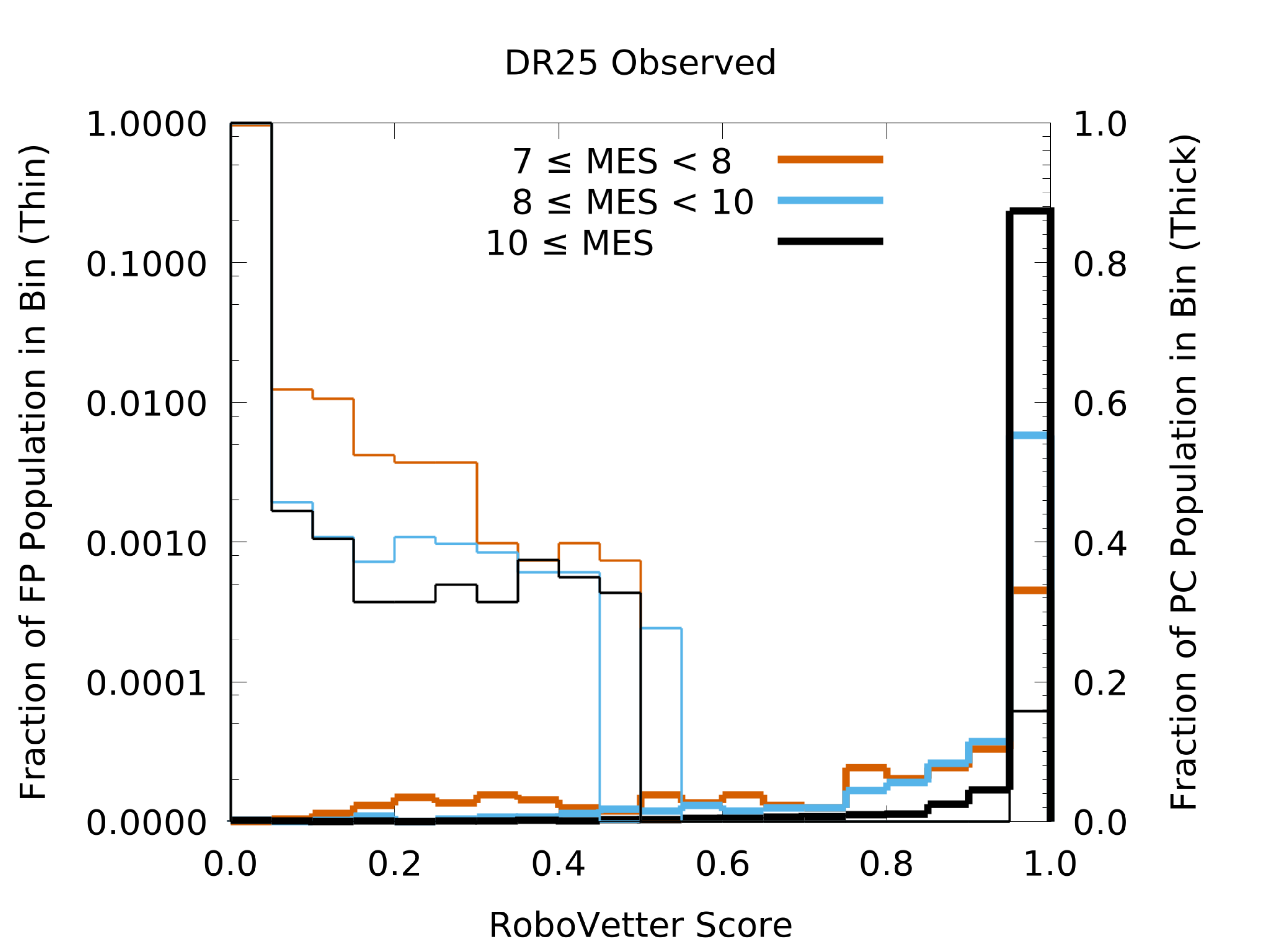} &
\includegraphics[width=0.5\linewidth]{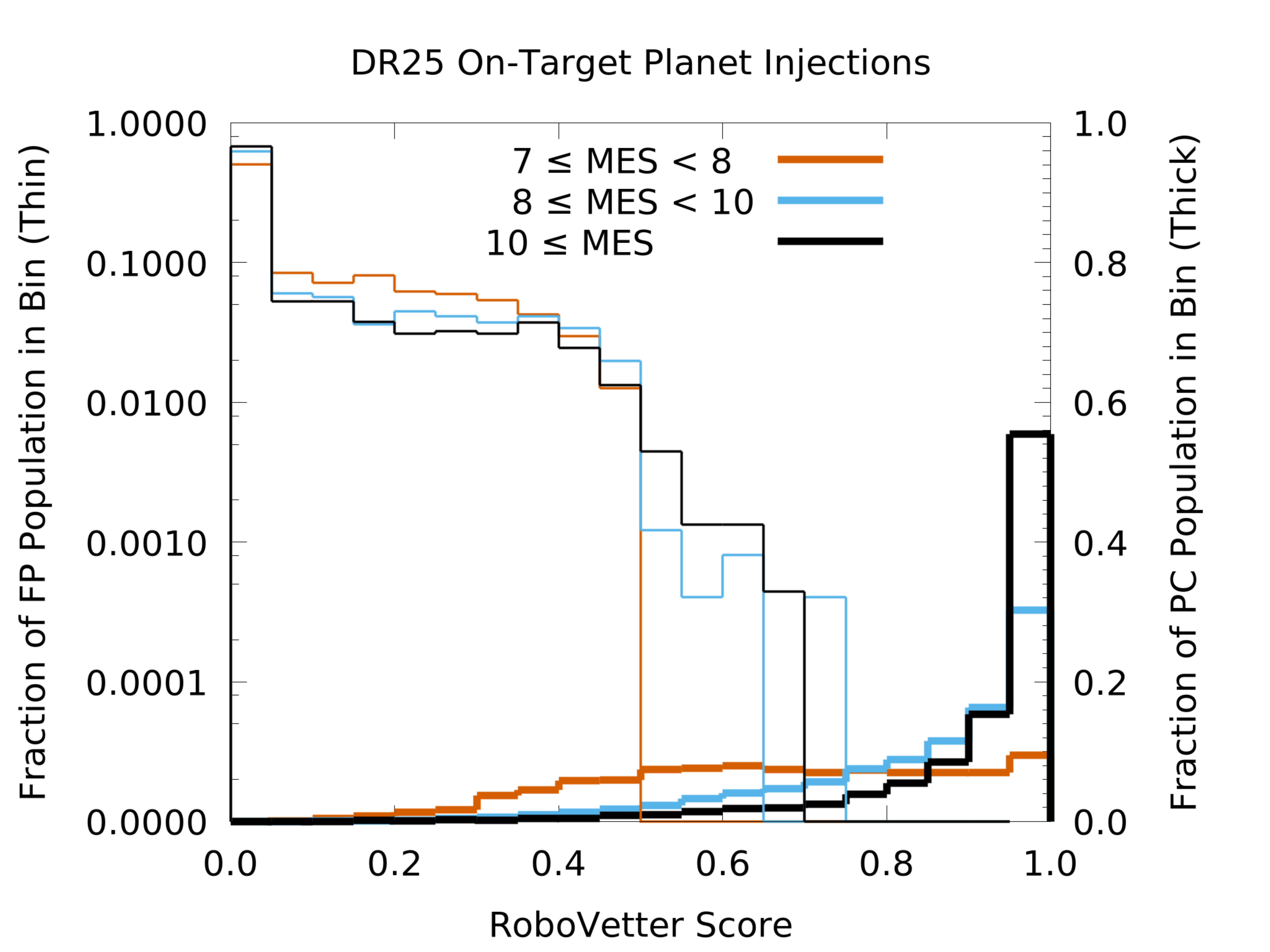} \\
\includegraphics[width=0.5\linewidth]{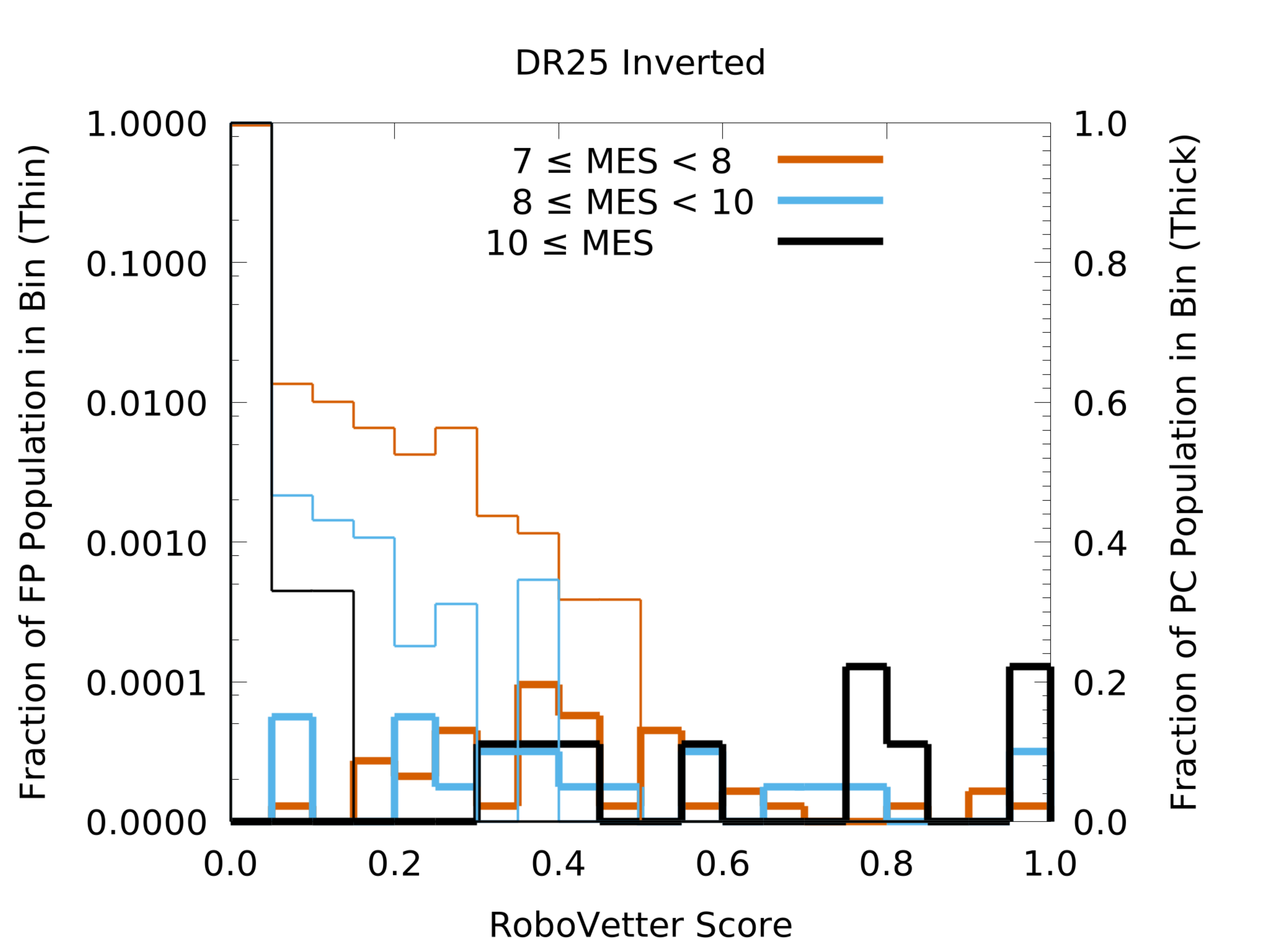} &
\includegraphics[width=0.5\linewidth]{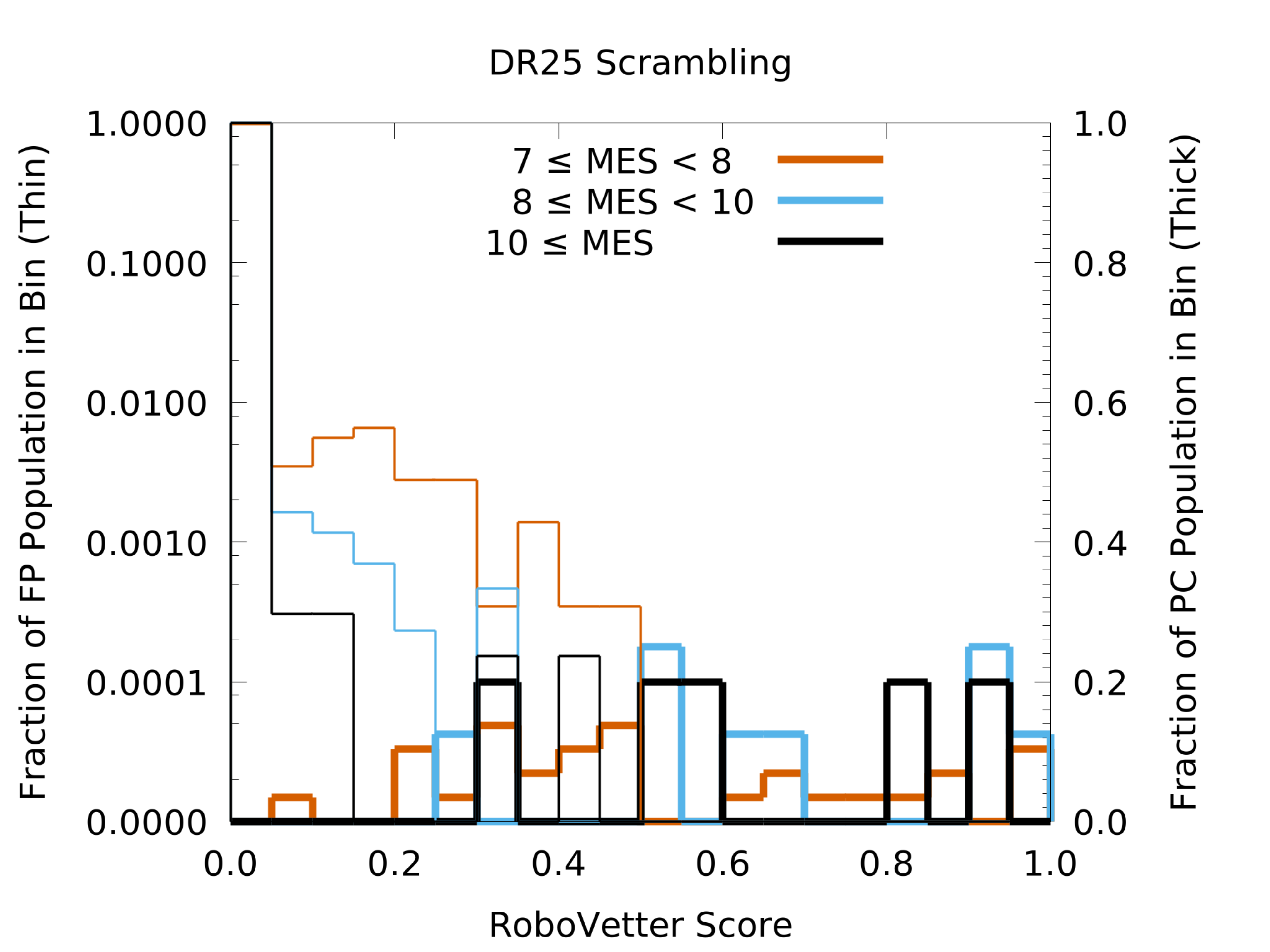} \\
\end{tabular}
\caption{Plots of the score distribution of PCs (thick lines, right y-axis) and FPs (thin lines, left y-axis, logarithmic scaling) for the observed (top-left), on-target planet injections (top-right), inverted (bottom-left), and scrambled (bottom-right) TCEs.}
\label{score-fig-2}
\end{figure*}

The disposition scores discussed in \S\ref{s:scores} can be used to select a more reliable, though more incomplete, sample of planet candidates. In Figure~\ref{score-fig-2} we show the distribution of disposition scores for the PCs and FPs from the observed, inverted, scrambled, and on-target planet injection populations. (Note, the inverted and scrambled populations have been cleaned as discussed in \S\ref{s:clean}). For all populations, the PC distribution tends to cluster near a score of 1.0 with a tail that extends towards lower score values. Lower MES values tend to have a greater proportion of lower score values. Similarly, the vast majority of FPs have a score of 0.0, with only a small fraction extending towards higher score values (note the y-axis for the FPs is logarithmic, while the y-axis for PCs is linear). Comparing the populations, the on-target planet injections have a greater concentration of score values towards 0.5 for both the PCs and FPs than other populations. Both the inverted and scrambled populations have very few PCs near high score values. We can exploit the relative distribution of PC and FP score values for the different populations to select a higher reliability catalog.

\begin{figure}[htb]
 \begin{center}
 \begin{tabular}{c}
   \includegraphics[width=.98\linewidth]{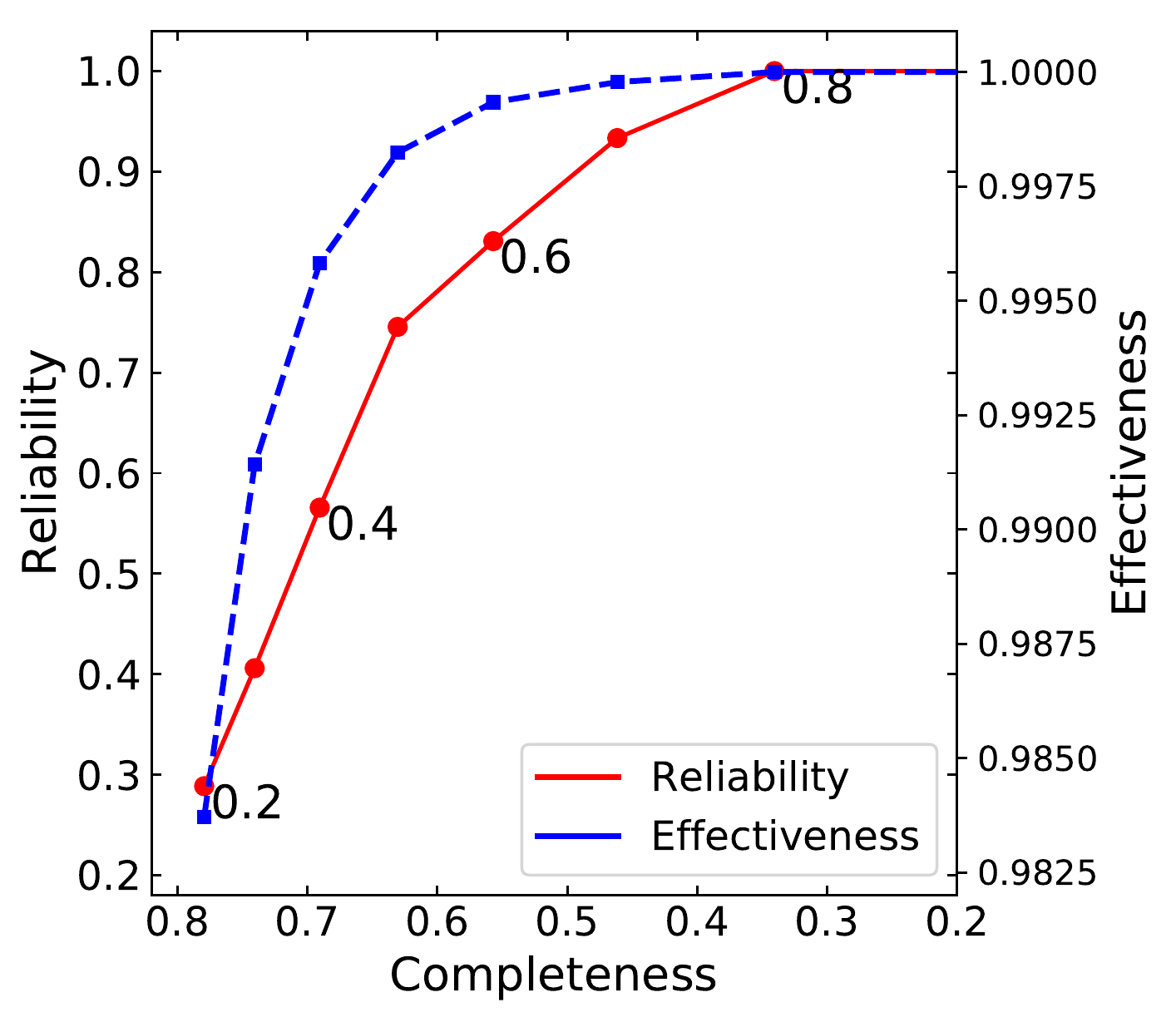} \\
  \includegraphics[width=.98\linewidth]{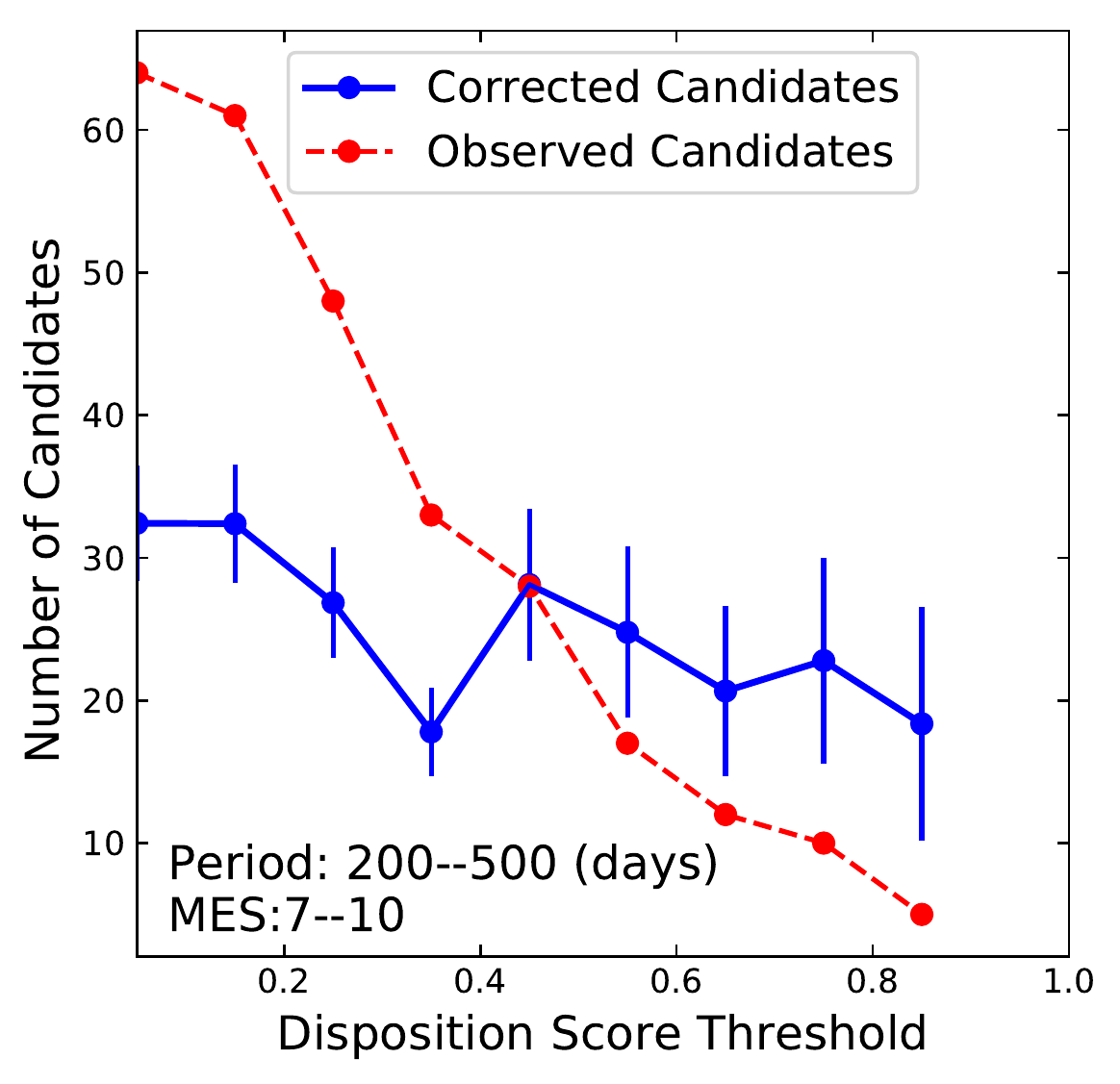}
  \end{tabular}
  \caption{\label{f:adjscore}[Top] The reliability (red) and effectiveness (blue) of the DR25 catalog as a function of Completeness for MES~$\leq$~10 and periods between 200 and 500\,d PCs that result when using different disposition score thresholds (shown as black numbers) to select the PCs. Higher disposition score thresholds result in higher reliability but lower completeness. Note, the completeness axis increases to the left.  [Bottom] The number of PCs (in red) in the same period and MES space when making a cut on different disposition scores.  The blue line corrects the number of candidates for the completeness and reliability. The error bars only reflect a Poisson error based on the number of observed planet candidates shown in red.}
 \end{center}
 \end{figure}

At the top of Figure~\ref{f:adjscore} we show how the completeness and reliability of the catalog vary for different cuts on the disposition score for MES$<$10 and periods between 200 and 500 days. The effectiveness of the Robovetter increases as the score threshold is increased. The reliability values also depend on the number of observed PCs that remain, which is why reliability does not change in step with the effectiveness. Selecting the PC sample by choosing those with a disposition score above 0.6 (see the point labeled 0.6 on the top of Figure~\ref{f:adjscore}) yields an 85\% reliability and a completeness that is still above 50\%. Doing a score cut in this way not only removes those dispositioned as a PC from the sample, but also causes a few \opstce{s} which are formally dispositioned as FPs to now be included in the sample. An FP with a high score occurs when a TCE marginally fails a single metric.  

It is interesting to note that the number of inferred candidates, i.e., the number of candidates after accounting for the Robovetter completeness and catalog reliability, does not change significantly with the score cut. In the lower plot of Figure~\ref{f:adjscore} we plot both the observed number of PCs and the corrected number of PCs that have periods between 200 and 500 days and MES less than 10.  The correction is done by taking the number of PCs and multiplying by the reliability and dividing by the completeness.  The error bars only include the Poisson counting error in the number of observed PCs and do not include errors in the measured completeness or reliability. The corrected number of PCs only varies by approximately 1$\sigma$ regardless of the score cut used.   

\subsection{Multiple-planet systems}
\citet{Lissauer2014} argues that almost all multi-planet, transit systems are real. Forty-seven, or 21\%, of the new DR25 PCs are associated with targets with multiple PCs. One of the new PCs, KOI~82.06, is part of a six candidate system around the star \Kepler-102. Five candidates have previously been confirmed \citep{Marcy2014,Rowe2014} in this system. The new candidate is the largest radius confirmed planet in the system. It also lies a bit outside the 4:3 resonance; possibly adding to the excess of planets found just wide of such first-order resonances \citep{Lissauer2011}. If verified, this would be only the 3rd system with six or more planets found by Kepler. 
The other new candidate within a high multiplicity system is KOI~2926.05. The other four candidates in this system around \Kepler-1388 have been validated by \citet{Morton2016}. This new candidate also orbits just exterior to a first-order mean motion resonance with one of the four previously known planets.



\subsection{Potentially Rocky Planets in the Habitable Zone}
\label{s:hz}
Kepler is NASA's first mission capable of detecting Earth-size planets around Sun-like stars in one-year orbits.  One of its primary science goals is to determine the occurrence rate of potentially habitable, terrestrial-size planets --- a value often referred to as eta-Earth.  Here we use the concept of a habitable zone to select a sample of planet candidates that are the right distance from their host stars and small enough to possibly have a rocky surface. A point that bears repeating is that no claims can be made regarding planetary habitability based on size and orbital distance alone.   This sample is, however, of great value to the occurrence rate studies that enable planet yield estimates for various designs of future life-detection missions \citep{stark2015}. This eta-Earth sample is provided in Table~\ref{t:hz} and shown in Figure~\ref{f:hzPlot}.

\subsubsection{Selecting the Eta-Earth Sample}

Before applying thresholds on planet properties, we first select a sample based on disposition score (see \S\ref{s:scores}) in order to produce a sample of higher reliability planets orbiting G-type stars. At long orbital period and small radius, we are vulnerable to instrumental false alarms despite the significant improvements afforded us by the latest versions of metrics like Marshall, Skye, Rubble, Chases, and Model-shift. This is evident in the FGK dwarf sample of Figures~\ref{f:prCompleteness} and~\ref{f:prReliability} by comparing the relatively low reliability (45\%--74\%) and completeness (74\% to 88\%) measurements in the bottom right boxes to others at shorter period and larger radius.  Removing candidates with score~$<$~0.5 results in a significant improvement in the sample reliability with a small degradation in the sample completeness (Figure~\ref{f:adjscore}).  The candidates reported in Table~\ref{t:hz} are $\approx$80\% reliable for the G-type stars and even higher for the K- and M-type stars. Note, there is only one late F-type star in the sample.  \replaced{Kepler was not designed to reach the habitable zones of F-type stars, nor did the target list include many such stars.}{Kepler was not designed to find small planets in the habitable zones of F-type stars and those in the DR25 catalog are of low reliability and have disposition scores less than 0.5.}

\begin{figure}[tb]
    \centering
    \includegraphics[width=\linewidth]{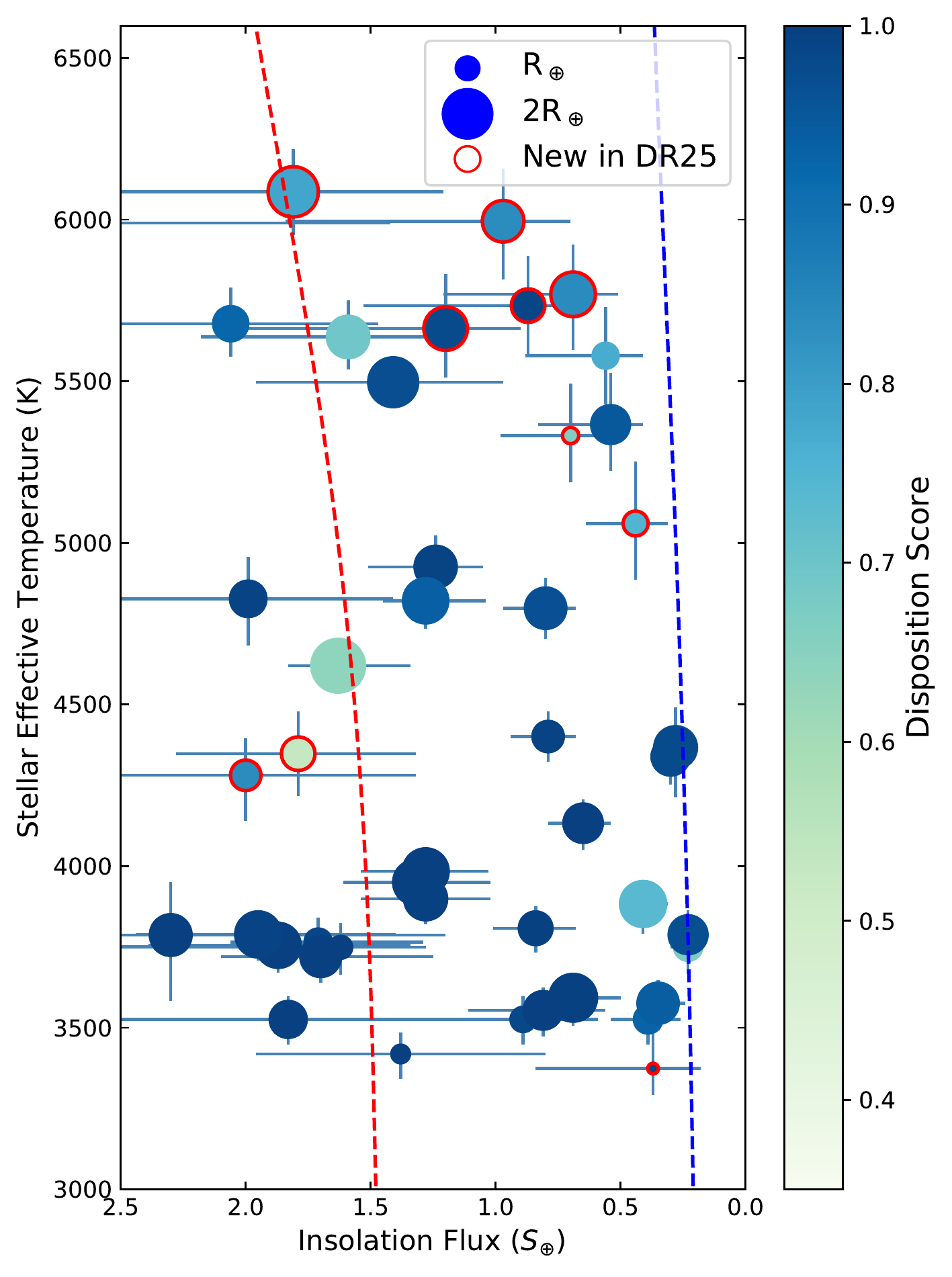}
    \caption{DR25, eta-Earth sample of PCs plotted as stellar effective temperature against insolation flux using the values reported in the DR25 KOI catalog (which uses stellar properties from the DR25 stellar catalog \citep{Mathur2017ApJS}. The size of the exoplanet is indicated by the size of the circle.  The color indicates the disposition score. Only those with disposition score greater than 0.5 are plotted.  Only objects whose error bars indicate that they could be in the habitable zone and have a radius less than 1.8~\re\ are shown. Those with a red ring are new to the DR25 catalog. }
    \label{f:hzPlot}
\end{figure}

The DR25 catalog uses the transit depth and period, along with the DR25 stellar table of \citet{Mathur2017ApJS}, to derive the planet radius and the semi-major axis of the planet's orbit.  From these we calculate the insolation flux in units of the Earth's insolation flux,
\vspace{-.3em}
\begin{equation}
S_{p} = \frac{R_{\star}^{2} \cdot (T_{\star}/5777)^{4}}{a^{2}} ,
\end{equation}

\noindent where $a$ is the semi-major axis of the planet's orbit in AU, \tstar{} is the host star temperature in Kelvin, 5777~K is the effective temperature of the Sun, R$_{\star}$ is the radius of the star in units of \rsun, and thus \Splanet{} is in units relative to the Earth's insolation flux. The errors for both insolation flux and radii include the errors from the DR25 stellar catalog. The habitable zone represents a range of orbits where the flux received by the host star allows for the possibility of surface liquid water on an Earth-size planet.  While the insolation limits for the habitable zone depends on the stellar temperature, it roughly falls from 0.2--1.7 $S_{\earth}$ (see Figure~\ref{f:hzPlot}). We use the empirical (recent Venus/early Mars) habitable zone of \citet{Kopparapu2013}.  To err on the side of inclusiveness, we include candidates whose one sigma error bars on the insolation flux overlap this empirical habitable zone.

\begin{deluxetable*}{lrlrrrrrrr}
\tablecolumns{10}
\tabletypesize{\scriptsize}
\tablewidth{\linewidth}
\tablecaption{Habitable Zone Terrestrial-Sized Planet Candidates\label{t:hz}}
\tablehead{
\colhead{KOI} &
\colhead{KIC} &
\colhead{Kepler} &
\colhead{Period} &
\colhead{\rp} &
\colhead{\Splanet} &
\colhead{\tstar} &
\colhead{\rstar} &
\colhead{MES} &
\colhead{Disp.} \\[-6pt]
\colhead{} &
\colhead{} &
\colhead{} &
\colhead{[days]} &
\colhead{[\re]} &
\colhead{[\se]} &
\colhead{[K]} &
\colhead{[\rsun]} &
&
\colhead{Score}
}
\startdata
\,~172.02 & 8692861 & Kepler-69 c & 242.46130 & 1.73$^{+0.21}_{-0.22}$ & 1.59$^{+0.59}_{-0.45}$ & 5637$^{+113}_{-101}$ & 0.94$^{+0.12}_{-0.12}$ & 18.0 & 0.693 \\ 
\,~238.03 & 7219825 & \nodata & 362.97828 & 1.96$^{+0.33}_{-0.29}$ & 1.81$^{+0.87}_{-0.60}$ & 6086$^{+133}_{-133}$ & 1.22$^{+0.20}_{-0.18}$ & 11.9 & 0.784 \\ 
\,~438.02 & 12302530 & Kepler-155 c & 52.66153 & 1.87$^{+0.11}_{-0.12}$ & 1.28$^{+0.26}_{-0.25}$ & 3984$^{+71}_{-86}$ & 0.54$^{+0.03}_{-0.04}$ & 30.6 & 1.000 \\ 
\,~463.01\tablenotemark{c} & 8845205 & Kepler-560 b & 18.47763 & 1.55$^{+0.32}_{-0.29}$ & 1.21$^{+0.72}_{-0.47}$ & 3395$^{+74}_{-67}$ & 0.28$^{+0.06}_{-0.05}$ & 78.0 & 0.001 \\ 
\,~494.01 & 3966801 & Kepler-577 b & 25.69581 & 1.70$^{+0.21}_{-0.33}$ & 2.30$^{+1.17}_{-1.10}$ & 3787$^{+163}_{-204}$ & 0.48$^{+0.06}_{-0.09}$ & 35.9 & 1.000 \\ 
\,~571.05\tablenotemark{a} & 8120608 & Kepler-186 f & 129.94410 & 1.18$^{+0.11}_{-0.14}$ & 0.23$^{+0.07}_{-0.06}$ & 3751$^{+75}_{-84}$ & 0.44$^{+0.04}_{-0.05}$ & 7.7 & 0.677 \\ 
\,~701.03 & 9002278 & Kepler-62 e & 122.38740 & 1.72$^{+0.10}_{-0.07}$ & 1.24$^{+0.27}_{-0.19}$ & 4926$^{+98}_{-98}$ & 0.66$^{+0.04}_{-0.03}$ & 35.9 & 0.994 \\ 
\,~701.04\tablenotemark{d} & 9002278 & Kepler-62 f & 267.29100 & 1.43$^{+0.08}_{-0.06}$ & 0.44$^{+0.09}_{-0.07}$ & 4926$^{+98}_{-98}$ & 0.66$^{+0.04}_{-0.03}$ & 14.3 & 0.000 \\ 
\,~812.03 & 4139816 & Kepler-235 e & 46.18420 & 1.83$^{+0.12}_{-0.15}$ & 1.32$^{+0.29}_{-0.30}$ & 3950$^{+70}_{-86}$ & 0.49$^{+0.03}_{-0.04}$ & 18.0 & 1.000 \\ 
\,~854.01 & 6435936 & Kepler-705 b & 56.05608 & 1.94$^{+0.12}_{-0.22}$ & 0.69$^{+0.15}_{-0.19}$ & 3593$^{+71}_{-86}$ & 0.49$^{+0.03}_{-0.06}$ & 19.3 & 0.996 \\ 
\,~947.01 & 9710326 & Kepler-737 b & 28.59914 & 1.83$^{+0.16}_{-0.21}$ & 1.87$^{+0.52}_{-0.53}$ & 3755$^{+75}_{-84}$ & 0.46$^{+0.04}_{-0.05}$ & 45.7 & 1.000 \\ 
1078.03 & 10166274 & Kepler-267 d & 28.46465 & 1.87$^{+0.14}_{-0.22}$ & 1.95$^{+0.49}_{-0.55}$ & 3789$^{+75}_{-82}$ & 0.46$^{+0.04}_{-0.05}$ & 22.2 & 0.992 \\ 
1298.02\tablenotemark{d} & 10604335 & Kepler-283 c & 92.74958 & 1.87$^{+0.08}_{-0.10}$ & 0.78$^{+0.15}_{-0.14}$ & 4141$^{+83}_{-91}$ & 0.58$^{+0.03}_{-0.03}$ & 10.7 & 0.000 \\ 
1404.02 & 8874090 & \nodata & 18.90609 & 0.87$^{+0.16}_{-0.21}$ & 3.03$^{+2.29}_{-1.67}$ & 3751$^{+219}_{-219}$ & 0.45$^{+0.08}_{-0.11}$ & 10.1 & 0.955 \\ 
1422.02\tablenotemark{b} & 11497958 & Kepler-296 d & 19.85029 & 1.52$^{+0.19}_{-0.23}$ & 1.83$^{+0.68}_{-0.62}$ & 3526$^{+71}_{-78}$ & 0.38$^{+0.05}_{-0.06}$ & 25.1 & 1.000 \\ 
1422.04 & 11497958 & Kepler-296 f & 63.33627 & 1.18$^{+0.15}_{-0.18}$ & 0.39$^{+0.15}_{-0.13}$ & 3526$^{+71}_{-78}$ & 0.38$^{+0.05}_{-0.06}$ & 9.1 & 0.927 \\ 
1422.05 & 11497958 & Kepler-296 e & 34.14211 & 1.06$^{+0.13}_{-0.16}$ & 0.89$^{+0.33}_{-0.30}$ & 3526$^{+71}_{-78}$ & 0.38$^{+0.05}_{-0.06}$ & 10.5 & 0.984 \\ 
1596.02 & 10027323 & Kepler-309 c & 105.35823 & 1.87$^{+0.13}_{-0.17}$ & 0.41$^{+0.09}_{-0.10}$ & 3883$^{+69}_{-93}$ & 0.50$^{+0.04}_{-0.04}$ & 16.5 & 0.738 \\ 
2162.02 & 9205938 & \nodata & 199.66876 & 1.45$^{+0.18}_{-0.18}$ & 2.06$^{+0.76}_{-0.59}$ & 5678$^{+113}_{-102}$ & 0.92$^{+0.12}_{-0.12}$ & 11.1 & 0.920 \\ 
2184.02\tablenotemark{e} & 12885212 & \nodata & 95.90640 & 2.17$^{+0.07}_{-0.12}$ & 1.63$^{+0.20}_{-0.29}$ & 4620$^{+73}_{-82}$ & 0.74$^{+0.02}_{-0.04}$ & 8.92 & 0.638 \\  
2418.01 & 10027247 & Kepler-1229 b & 86.82952 & 1.68$^{+0.12}_{-0.21}$ & 0.35$^{+0.08}_{-0.11}$ & 3576$^{+71}_{-85}$ & 0.46$^{+0.03}_{-0.06}$ & 11.7 & 0.937 \\ 
2626.01 & 11768142 & \nodata & 38.09707 & 1.58$^{+0.20}_{-0.21}$ & 0.81$^{+0.30}_{-0.25}$ & 3554$^{+71}_{-80}$ & 0.40$^{+0.05}_{-0.05}$ & 14.6 & 0.999 \\ 
2650.01 & 8890150 & Kepler-395 c & 34.98978 & 1.14$^{+0.07}_{-0.10}$ & 1.71$^{+0.35}_{-0.42}$ & 3765$^{+75}_{-83}$ & 0.52$^{+0.03}_{-0.05}$ & 10.1 & 0.985 \\ 
2719.02 & 5184911 & \nodata & 106.25976 & 1.50$^{+0.10}_{-0.16}$ & 1.99$^{+0.53}_{-0.58}$ & 4827$^{+129}_{-144}$ & 0.82$^{+0.06}_{-0.09}$ & 10.0 & 0.990 \\ 
3010.01 & 3642335 & Kepler-1410 b & 60.86610 & 1.39$^{+0.07}_{-0.10}$ & 0.84$^{+0.17}_{-0.16}$ & 3808$^{+69}_{-76}$ & 0.52$^{+0.03}_{-0.04}$ & 12.7 & 0.996 \\ 
3034.01 & 2973386 & \nodata & 31.02092 & 1.66$^{+0.12}_{-0.17}$ & 1.70$^{+0.40}_{-0.45}$ & 3720$^{+73}_{-81}$ & 0.48$^{+0.03}_{-0.05}$ & 11.9 & 1.000 \\ 
3138.01\tablenotemark{b} & 6444896 & Kepler-1649 b & 8.68909 & 0.49$^{+0.00}_{-0.00}$ & 0.47$^{+0.00}_{-0.00}$ & 2703$^{+0}_{-0}$ & 0.12$^{+0.00}_{-0.00}$ & 12.0 & 1.000 \\ 
3282.01 & 12066569 & Kepler-1455 b & 49.27684 & 1.75$^{+0.09}_{-0.13}$ & 1.28$^{+0.26}_{-0.26}$ & 3899$^{+78}_{-78}$ & 0.53$^{+0.03}_{-0.04}$ & 14.7 & 0.996 \\ 
3284.01 & 6497146 & Kepler-438 b & 35.23319 & 0.97$^{+0.06}_{-0.07}$ & 1.62$^{+0.37}_{-0.34}$ & 3749$^{+75}_{-84}$ & 0.52$^{+0.03}_{-0.04}$ & 11.9 & 1.000 \\ 
3497.01 & 8424002 & Kepler-1512 b & 20.35972 & 0.80$^{+0.12}_{-0.16}$ & 1.38$^{+0.58}_{-0.58}$ & 3419$^{+67}_{-76}$ & 0.34$^{+0.05}_{-0.07}$ & 19.6 & 1.000 \\ 
4005.01\tablenotemark{a} & 8142787 & Kepler-439 b & 178.13960 & 2.25$^{+0.22}_{-0.16}$ & 1.70$^{+0.47}_{-0.31}$ & 5431$^{+81}_{-81}$ & 0.88$^{+0.09}_{-0.06}$ & 17.8 & 0.997 \\ 
4036.01 & 11415243 & Kepler-1544 b & 168.81133 & 1.69$^{+0.10}_{-0.06}$ & 0.80$^{+0.17}_{-0.12}$ & 4798$^{+95}_{-95}$ & 0.71$^{+0.04}_{-0.03}$ & 14.8 & 0.965 \\ 
4087.01 & 6106282 & Kepler-440 b & 101.11141 & 1.61$^{+0.10}_{-0.08}$ & 0.65$^{+0.14}_{-0.11}$ & 4133$^{+74}_{-82}$ & 0.56$^{+0.03}_{-0.03}$ & 15.7 & 1.000 \\ 
4356.01\tablenotemark{a} & 8459663 & Kepler-1593 b & 174.51028 & 1.74$^{+0.14}_{-0.20}$ & 0.28$^{+0.09}_{-0.09}$ & 4367$^{+124}_{-155}$ & 0.45$^{+0.04}_{-0.05}$ & 11.0 & 0.976 \\ 
4427.01 & 4172805 & \nodata & 147.66173 & 1.59$^{+0.12}_{-0.14}$ & 0.23$^{+0.06}_{-0.05}$ & 3788$^{+76}_{-84}$ & 0.49$^{+0.04}_{-0.04}$ & 10.8 & 0.969 \\ 
4460.01 & 9947389 & \nodata & 284.72721 & 2.02$^{+0.30}_{-0.29}$ & 1.41$^{+0.55}_{-0.44}$ & 5497$^{+82}_{-74}$ & 1.08$^{+0.16}_{-0.16}$ & 10.7 & 0.972 \\ 
4550.01 & 5977470 & \nodata & 140.25194 & 1.84$^{+0.05}_{-0.12}$ & 1.28$^{+0.17}_{-0.24}$ & 4821$^{+76}_{-86}$ & 0.79$^{+0.02}_{-0.05}$ & 9.6 & 0.934 \\ 
4622.01 & 11284772 & Kepler-441 b & 207.24820 & 1.56$^{+0.09}_{-0.06}$ & 0.30$^{+0.06}_{-0.05}$ & 4339$^{+78}_{-87}$ & 0.55$^{+0.03}_{-0.02}$ & 9.7 & 0.975 \\ 
4742.01 & 4138008 & Kepler-442 b & 112.30530 & 1.30$^{+0.07}_{-0.05}$ & 0.79$^{+0.15}_{-0.11}$ & 4401$^{+78}_{-78}$ & 0.59$^{+0.03}_{-0.02}$ & 12.9 & 0.993 \\ 
7016.01 & 8311864 & Kepler-452 b & 384.84300 & 1.09$^{+0.20}_{-0.10}$ & 0.56$^{+0.32}_{-0.15}$ & 5579$^{+150}_{-150}$ & 0.80$^{+0.15}_{-0.07}$ & 7.6 & 0.771 \\ 
7223.01 & 9674320 & \nodata & 317.06242 & 1.59$^{+0.27}_{-0.12}$ & 0.54$^{+0.29}_{-0.13}$ & 5366$^{+160}_{-144}$ & 0.71$^{+0.12}_{-0.05}$ & 10.3 & 0.947 \\ 
7706.01 & 4762283 & \nodata & 42.04952 & 1.19$^{+0.08}_{-0.16}$ & 2.00$^{+0.55}_{-0.68}$ & 4281$^{+115}_{-140}$ & 0.48$^{+0.03}_{-0.06}$ & 7.5 & 0.837 \\ 
7711.01 & 4940203 & \nodata & 302.77982 & 1.31$^{+0.34}_{-0.12}$ & 0.87$^{+0.66}_{-0.22}$ & 5734$^{+154}_{-154}$ & 0.80$^{+0.21}_{-0.07}$ & 8.5 & 0.987 \\ 
7882.01 & 8364232 & \nodata & 65.41518 & 1.31$^{+0.08}_{-0.12}$ & 1.79$^{+0.49}_{-0.47}$ & 4348$^{+130}_{-130}$ & 0.65$^{+0.04}_{-0.06}$ & 7.2 & 0.529 \\ 
7894.01 & 8555967 & \nodata & 347.97611 & 1.62$^{+0.49}_{-0.15}$ & 0.97$^{+0.87}_{-0.27}$ & 5995$^{+163}_{-181}$ & 0.88$^{+0.27}_{-0.08}$ & 8.5 & 0.837 \\ 
7923.01 & 9084569 & \nodata & 395.13138 & 0.97$^{+0.12}_{-0.10}$ & 0.44$^{+0.20}_{-0.13}$ & 5060$^{+192}_{-174}$ & 0.87$^{+0.10}_{-0.09}$ & 10.0 & 0.750 \\ 
7954.01 & 9650762 & \nodata & 372.15035 & 1.74$^{+0.46}_{-0.14}$ & 0.69$^{+0.52}_{-0.18}$ & 5769$^{+155}_{-172}$ & 0.81$^{+0.21}_{-0.07}$ & 8.9 & 0.839 \\ 
8000.01 & 10331279 & \nodata & 225.48805 & 1.70$^{+0.43}_{-0.14}$ & 1.20$^{+0.90}_{-0.30}$ & 5663$^{+169}_{-152}$ & 0.78$^{+0.19}_{-0.07}$ & 8.7 & 0.975 \\ 
8012.01 & 10452252 & \nodata & 34.57372 & 0.42$^{+0.17}_{-0.12}$ & 0.37$^{+0.47}_{-0.19}$ & 3374$^{+112}_{-82}$ & 0.22$^{+0.09}_{-0.06}$ & 7.7 & 0.989 \\ 
8174.01 & 8873873 & \nodata & 295.06066 & 0.64$^{+0.07}_{-0.07}$ & 0.70$^{+0.28}_{-0.21}$ & 5332$^{+160}_{-144}$ & 0.76$^{+0.09}_{-0.09}$ & 7.4 & 0.665 \\
\enddata
\label{hzearthstab}
\vspace{-2pt}
\tablenotetext{a}{Confirmed planet properties from NASA Exoplanet Archive on May 31, 2017 place object within HZ.}
\vspace{-.5em}
\tablenotetext{b}{Confirmed planet properties from NASA Exoplanet Archive on May 31, 2017 place object exterior to the HZ.}
\vspace{-.5em}
\tablenotetext{c}{Confirmed planet with vetting score less than 0.5.}
\vspace{-.5em}
\tablenotetext{d}{Confirmed planet dispositioned as False Positive in DR25.}
\vspace{-.5em}
\tablenotetext{e}{The erratum to \citet{Mathur2017ApJS} reduces planet size, now placing the object in the eta-Earth sample. }
\end{deluxetable*}

Finally, we include only those candidates that satisfy the size constraint \rp~$- \sigma_{{R_p}\rm,low} < 1.8$~\re.  The purpose of the size constraint is to identify candidates likely to have a bulk composition similar to terrestrial planets in the solar system.  The 1.8~\re\ upper limit is taken from \citet{Fulton2017} who report a distinct gap in the radius distribution of exoplanets for planets in orbital periods of less than 100\,d.  The authors argue that the gap is the result of two (possibly overlapping) population distributions: the rocky terrestrials and the mini-Neptune size planets characterized by their volatile-rich envelopes.  Within this framework, the center of the gap marks a probabilistic boundary between having a higher likelihood of a terrestrial composition versus a higher likelihood of a volatile-rich envelope.  However, this boundary was identified using planets in orbital periods of less than 100\,days and it may not exist for planets in longer period orbits. Also, it is not entirely clear that planets on the small side of this gap are all terrestrial. \citet{Rogers2015} examined small planets with density measurements with periods less than $\approx$50\,d and showed that less than half of planets with a radii of 1.62\,\re\ have densities consistent with a body primarily composed of iron and silicates.  For our purposes of highlighting the smallest planets in this catalog, we chose to be inclusive and set the threshold at 1.8\,\re.

To summarize, Table~\ref{t:hz} lists those candidates with scores greater than 0.5 and whose error bars indicate that they could be smaller than 1.8 \re{} and lie in the habitable zone. The table also includes KOI~2184.02 because the erratum to \citet[][see \S\ref{s:stars} of this paper]{Mathur2017ApJS} reduces the stellar and planet radii so that the PC now lies in our sample. Note, the same erratum also reduces the planet radii of KOI~4460.01 and KOI~4550.01 to 2.0\,\re{} and 1.65\,\re{} respectively. The values reported in Table~\ref{t:hz} are identical to those in the KOI table at the NASA Exoplanet Archive and do not include the values reported in the erratum to \citet{Mathur2017ApJS}. Also, in order to make Table~\ref{t:hz} complete we include any \Kepler{} terrestrial-size confirmed planet that falls in the habitable zone of its star according to the confirmed planet table at the Exoplanet Archive (downloaded on 2017-05-15). The objects are included, and denoted with footnotes, even if the DR25 catalog dispositions them as FPs, or if the DR25 planetary parameters place them outside the habitable zone. However, note that statistical inferences like occurrence rates should be based on a uniform sample drawn exclusively from the DR25 catalog and its self-consistent completeness and reliability measurements (see \S\ref{s:occurates}).  


We plot the eta-Earth sample candidates in Figure~\ref{f:hzPlot}, using only the information in the DR25 KOI catalog. Notice that this final search of the \Kepler{} data not only identified previously discovered candidates around the M dwarf stars, it also yielded a handful of highly reliable candidates around the GK dwarf stars. These GK dwarf candidates have fewer transits and shallower depths, making them much more difficult to find.  Despite their lower signal-to-noise, because we provide a measure of the reliability against false alarms (along with the completeness), these candidates are available to further study the occurrence rates of small planets in the habitable zone of GK dwarf stars.

\subsubsection{Notes on the Eta-Earth Sample}
Forty-seven candidates have a score greater than 0.5 and fall in this eta-Earth sample; 10 of these are new to this catalog (KOI numbers greater than 7621.01 and KOI~238.03).  A manual review of the 10 new high-score candidates indicates that they are all low signal-to-noise with very few transits, and show no obvious reason to be called false positives. However, our reliability measurements indicate that $\approx$20\% of these targets are not caused by a transiting/eclipsing system. As an example, the candidate most similar to the size and temperature of the Earth is KOI~7711.01 (KIC~004940203), with four transits that all cleanly pass the individual transit metrics. It orbits a 5734\,K star, has an insolation flux slightly less than that of Earth, and is about 30\% larger according to its DR25 catalog properties.  Plots showing visualizations of the transit data and its quality are available at the Exoplanet Archive for this object\footnote{\url{https://exoplanetarchive.ipac.caltech.edu/data/KeplerData/\allowbreak 004/004940/004940203/tcert/kplr004940203\_q1\_q17\_dr25\_obs\_tcert.pdf}} and for all of the \opstce{s}, \injtce{s}, \scrtce{s}, and \invtce{s}.

Several confirmed planets fall in our eta-Earth sample.  Kepler-186f (KOI~571.05), Kepler-439b (KOI~4005.01) and Kepler-1593b (KOI~4356.01) move into the habitable zone according to the confirmed planet properties. They are included in Table~\ref{t:hz} with a footnote indicating they would not otherwise be listed. Kepler-296d (KOI~1422.02) and Kepler-1649b (KOI~3138.01), on the other hand, move outside the HZ according to the updated properties and are noted accordingly. Note, the default properties in the confirmed planets table at the Exoplanet Archive are selected for completeness and precision. Additional values may be available from other references that represent the best, current state of our knowledge.

Kepler-560b (KOI~463.01) is a confirmed planet that is a PC in the DR25 catalog, but failed the score cut; it is included for awareness and annotated accordingly.  The low score is caused by the Centroid Robovetter (\S\ref{s:centroidrv}) detecting a possible offset from the star's cataloged position, likely due to the star's high proper motion \citep{Mann2017}.  

Two confirmed planets dispositioned as FPs in the DR25 catalog are included in Table~\ref{t:hz}: Kepler-62f (KOI~701.04) and Kepler-283c (KOI~1298.02).  Kepler-62f has only 4 transit events in the time series.  The transit observed during Quarter 9 is on the edge of a gap and narrowly fails Rubble.  The transit observed during Quarter 12 is flagged by the Skye metric.  Taken together, this leaves fewer than three unequivocal transits, the minimum required for the PC disposition. 

Kepler-283c (KOI~1298.02) fails the shape metric.  Its phase-folded transit appears v-shaped when TTVs are not included in the modeling.  We note that vetting metrics employed by the DR25 Robovetter were computed without consideration of transit timing variations, whereas the transit fits used in the KOI table, described in \S\ref{s:mcmc}, includes the timing variations as measured by \citet{Rowe2015cat}. 

\subsection{Caveats}
When selecting candidates from the KOI catalog for further study, as we did for the eta-Earth sample (\S\ref{s:hz}), it is important to remember a few caveats. First, even with a high cut on disposition score, the reliability against false alarms is not 100\%. Some candidates may still be caused by false alarms, especially those around the larger, hotter stars. Also, this reliability number does not include the astrophysical reliability. Many of our tools to detect astrophysical false positives do not work for long-period, low MES candidates. For example, it is nearly impossible to detect the centroid offset created from a background eclipsing binary and secondary eclipses are not deep enough to detect for these stars. 

Second, the measured radius and semi-major axis of each planet depends on the stellar catalog.   As discussed in \S\ref{s:stars} and \citet{Mathur2017ApJS}, the stellar radii and masses are only known to a certain precision and the quality of the data used to derive these stellar properties varies between targets. These unknowns are reflected in the 1-sigma error bars shown in Figure~\ref{f:hzPlot} and listed in the KOI table. The uncertainty in the stellar information limits our knowledge of these planets.  As an example, for Kepler-452 (KIC~8311864), the DR25 stellar catalog lists a temperature of 5579$\pm150$\,K and stellar radius of 0.798$^{+0.150}_{-0.075}$\,\rsun, while the values in the confirmation paper \citep{Jenkins2015} after extensive follow-up are 5757$\pm85$\,K for the effective temperature and 1.11$^{+0.15}_{-0.09}$ for the stellar radius.  As a result, the planet Kepler-452b is given as 1.6$\pm0.2$\,\re\ in \citet{Jenkins2015} and 1.09$^{+0.2}_{-0.1}$\,\re\ in the DR25 catalog. The radii and stellar temperature differ by less than 2-sigma, but those differences change the interpretation of the planet from a super-Earth in the middle of the habitable zone of an early G dwarf host to an Earth-size planet receiving about half the amount of flux from a late K star.  As follow-up observations of each candidate star is obtained and errors on the stellar parameters decrease, we expect this population to change in significant ways.  

Third, high-resolution imaging has proven crucial for identifying light from background and bound stars which add flux to the \Kepler{} photometric time series \citep{Furlan2017}. When this occurs, unaccounted for extra light dilutes the transit, causing the radii to be significantly underestimated \citep{Ciardi2015, Furlan2017densities}.  As a result, we fully expect that once follow-up observations are obtained for these stars, several of the PCs in this catalog, including those listed in the eta-Earth sample, will be found to have radii larger than reported in this catalog. 


\section{Using the DR25 Catalog for Occurrence Rate Calculations}
\label{s:occurates}
The DR25 candidate catalog was designed with the goal of providing a well characterized sample of planetary candidates for use in occurrence rate calculations.  For those smallest planets at the longest periods, our vetting is especially prone to miss transits and confuse other signals as transits, and this must be accounted for when doing occurrence rates.  However, the completeness and reliability presented in this paper are simply the last two pieces of a much larger puzzle that must be assembled in order to perform occurrence rates with this catalog.  In this section we endeavor to make users aware of other issues and biases, as well as all the products available to help interpret this KOI catalog, all of which are hosted at the NASA exoplanet archive.

\subsection{Pipeline Detection Efficiency}

\begin{figure*}[ht]
\centering
\hspace{-2.5em}\begin{tabular}{cc}
\includegraphics[width=0.5\linewidth]{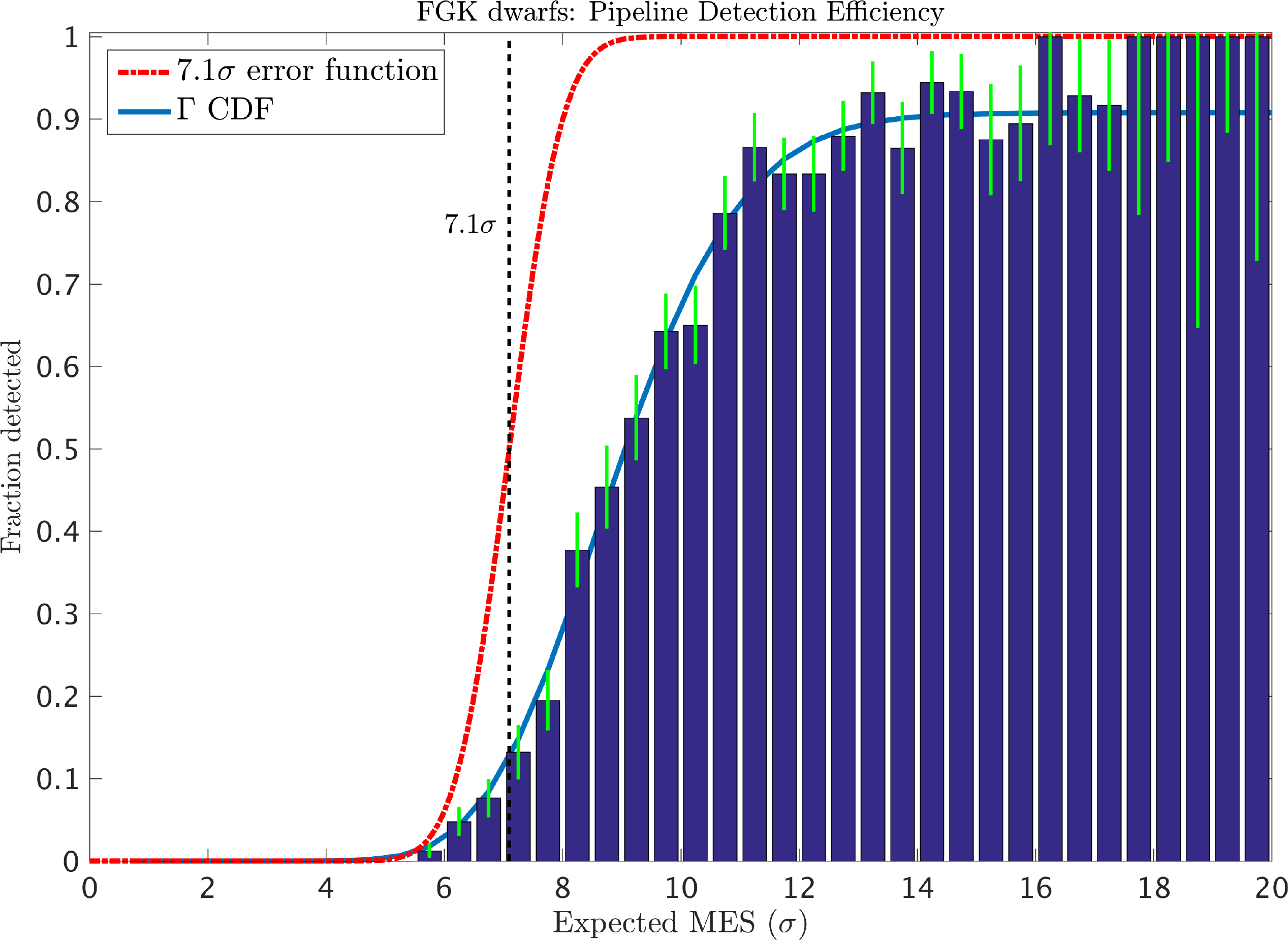} 
\includegraphics[width=0.5\linewidth]{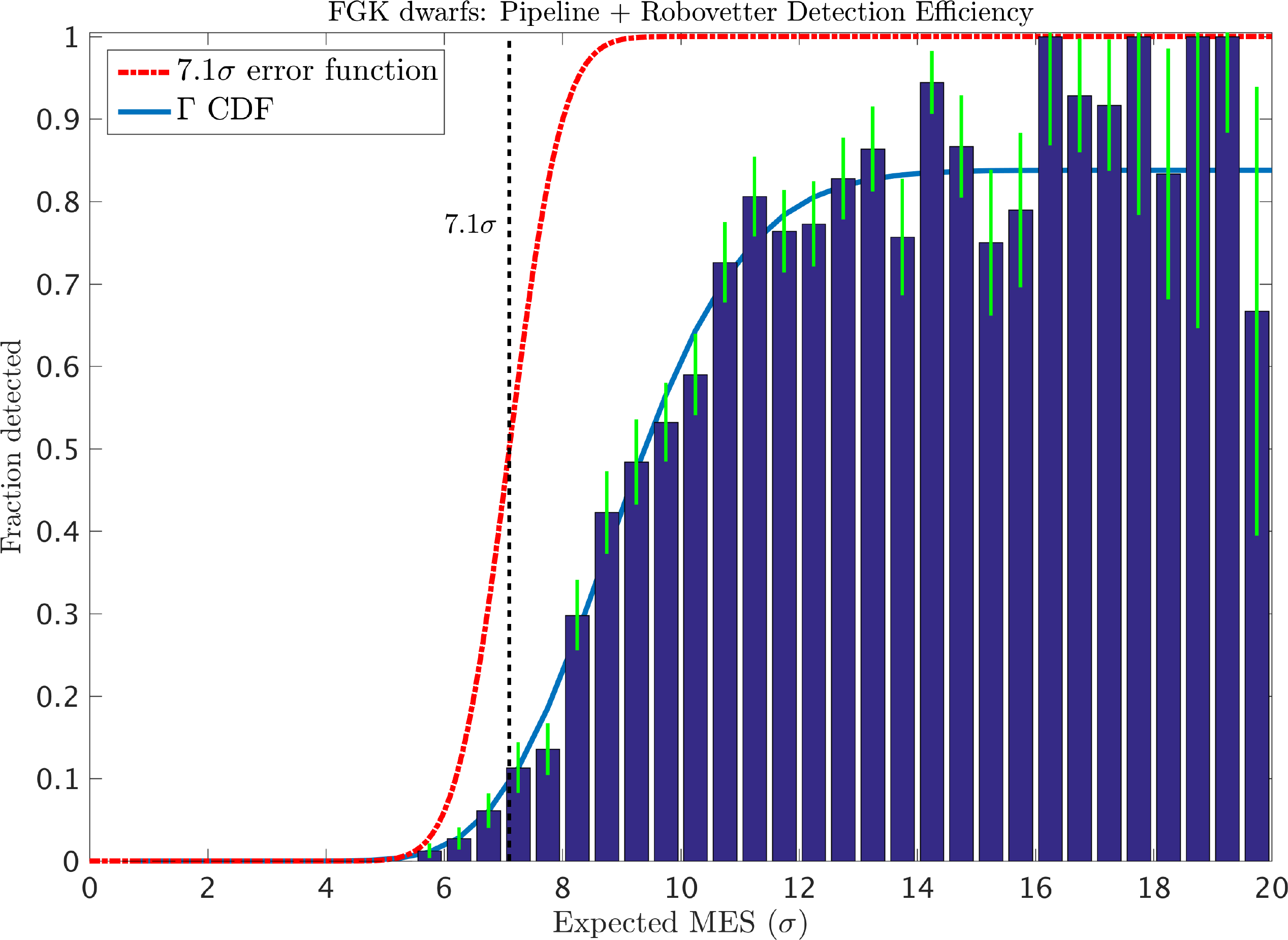}
\end{tabular}
\caption{Left: The average detection efficiency of the \Kepler{} Pipeline for a sample of FGK stars, as measured by the pixel-level transit injection experiment and described by \citet{Christiansen2017}. The solid blue line is a best-fit $\Gamma$ cumulative distribution function \citep[see Equation 1 of ][]{Christiansen2016}; the red dashed line shows the hypothetical performance for a perfect detector in TPS. Right: The average detection efficiency of the \Kepler{} Pipeline and the Robovetter, where the injections successfully recovered by the Pipeline are then subsequently evaluated as PCs by the Robovetter.}

\label{f:fulldetectionefficiency}

\end{figure*}

Any measure of the catalog completeness must include the completeness of the Robovetter and the \Kepler{} Pipeline. The Pipeline's detection efficiency has been explored in two ways: using pixel-level transit injection and using flux-level transit injection. In the former, a simulated transiting planet signal is injected into the calibrated pixels of each Kepler target, which are then processed through the pipeline. This experiment provides an estimate of the average detection efficiency over all the stars that were searched. A full description of the signals that were injected and recovered can be found in \citet{Christiansen2017}. The pixel-level measurements have the advantage of following transit signals through all the processing steps of the \Kepler{} Pipeline, and the recovered signals can be further classified with the Robovetter, as demonstrated in \S\ref{s:candr}. Figure \ref{f:fulldetectionefficiency} shows the average pipeline detection efficiency for a sample of FGK stars: the left panel shows the pipeline detection efficiency, and the right panel shows the combined Pipeline and Robovetter detection efficiency, calculated by taking the injections that were successfully recovered by the pipeline and processing them through the Robovetter. A gamma cumulative distribution function is fit to both \citep[see equation 1 of ][]{Christiansen2016}.  Notice that the detection efficiency decreases by 5--10~percentage points (of the entire set that were injected) for all MES, as expected given the results shown in Figure~\ref{f:1dcompare}.

Since the pixel-level transit injection includes only one injection per target, it does not examine potential variations in the pipeline completeness for individual targets due to differences in stellar properties or astrophysical variability. To probe these variations, a small number of individual stars had a large number of transiting signals (either several thousand or several hundred thousand, depending on the analysis) injected into the detrended photometry, which was processed only through the transit-search portion of the TPS module. The flux-level injections revealed that there are significant target-to-target variations in the detection efficiency. The flux-level injections and the resulting detection efficiency is available for the sample of stars that were part of this study. For more information on the flux-level injection study see \citet{Burke2017c}. All products associated with the flux-level and pixel-level injections can be found at the NASA Exoplanet Archive.\footnote{\url{https://exoplanetarchive.ipac.caltech.edu/docs/KeplerSimulated.html}}

\subsection{Astrophysical Reliability}
We have described the reliability of the DR25 candidates with regard to the possibility that the observed events are actually caused by stellar or instrumental noise. See \S\ref{s:candr} for how this reliability varies with various measured parameters.  However, even if the observed signal is not noise, other astrophysical events can mimic a transit.  Some of these other astrophysical events are removed by carefully vetting the KOI with \Kepler{} data alone.  Specifically, the Robovetter looks for significant secondary eclipses to rule out eclipsing binaries, and for a significant offset in the location of the in- and out-of-transit centroids to rule-out background eclipsing binaries. \citet[][]{Morton2016} developed the \textit{vespa} tool which considers the likelihood that a transit event is caused by various astrophysical events, including a planet.  The False Positive Probabilities (FPP) table\footnote{\url{https://exoplanetarchive.ipac.caltech.edu/cgi-bin/TblView/nph-tblView?app=ExoTbls&config=koifpp}} provides the results of applying this tool to the KOIs in the DR25 catalog. It provides a probability that the observed signal is one of the known types of astrophysical false positives.  The FPP table results are only reliable for high signal-to-noise (MES$\gtrapprox$10) candidates with no evidence that the transit occurs on a background source.  For more information on this table see the associated documentation at the NASA Exoplanet Archive.

To robustly determine whether a KOI's signal originates from the target star, see the Astrophysical Positional Probabilities Table\footnote{\url{https://exoplanetarchive.ipac.caltech.edu/cgi-bin/TblView/nph-tblView?app=ExoTbls&config=koiapp}}.  Using a more complete catalog of stars than the original Kepler Input Catalog \citep{Brown2011}, \citet{Bryson2017a} calculates the probability that the observed transit-like signal originates from the target star. Note, these positional probabilities are computed independent of the results from the Centroid Robovetter, and are not used by the Robovetter. 

To help understand the astrophysical reliability of the DR25 KOIs as a population, we have provided data to measure how well the Robovetter removes certain types of FPs.  As part of the pixel-level transit injection efforts, we injected signals that mimic eclipsing binaries and background eclipsing binaries. Those that were recovered by the \Kepler{} Pipeline can be used to measure the effectiveness of the Robovetter at removing this type of FP. A full description of these injections and an analysis of the Robovetter's effectiveness in detecting these signals can be found in \citet{Coughlin2017a}.


\subsection{Imperfect Stellar Information}
For those doing occurrence rates, another issue to consider is whether the measured size of the planet is correct. As discussed in \S\ref{s:stars}, the stellar catalog (i.e., radii and temperatures) provided by \citet{Mathur2017ApJS} typically has errors of 27 percent for the stellar radii. Results from \textit{Gaia} \citep{gaia2,gaia1} are expected to fix many of the shortcomings of this catalog. Also, the dilution from an unaccounted for bound or line-of-sight binary \citep{Ciardi2015, Furlan2017}, can cause planet radii to be larger than what is reported in the DR25 catalog. For occurrence rate calculations this dilution also has implications for the stars that have no observed planets because it means the search did not extend to planet radii that are as small as the stellar catalog indicates.  For this reason, any correction to the occurrence rates that might be applied needs to consider the effect on all searched stars, not just the planet hosts.




\section{Conclusions}
\label{s:conclusions}

The DR25 KOI catalog has been characterized so that it can serve as the basis for occurrence rate studies of exoplanets with periods as long as 500 days. The detection efficiency of the entire search \citep{Burke2017b,Christiansen2017} and of the Robovetter vetting process \citep{Coughlin2017a} has been calculated by injecting planetary transits into the data and determining which types of planets are found and which are missed. For this DR25 KOI catalog, the vetting completeness has been balanced against the catalog reliability, i.e., how often false alarms are mistakenly classified as PCs. This is the first \Kepler{} exoplanet catalog to be characterized in this way, enabling occurrence rate measurements at the detection limit of the mission.  As a result, accurate measurements of the frequency of terrestrial-size planets at orbital periods of hundreds of days is possible.

The measurement of the reliability using the inverted and scrambled light curves is new to this KOI catalog. We measure how often noise is labeled as a planet candidate and combine that information with the number of false alarms coming from the \Kepler{} Pipeline. Some pure noise signals so closely mimic transiting signals that it is nearly impossible to remove them all. Because of this, it is absolutely imperative that those using this candidate catalog for occurrence rates consider this source of noise. For periods longer than $\approx$200 days and radii less than $\approx$4\,\Rearth, these noise events are often labelled as PC and thus the reliability of the catalog is near 50\%.  Astrophysical reliability is another concern that must be accounted for independently.  However, even once it is shown that another astrophysical scenario is unlikely \citep[as was done for the DR24 KOIs in][]{Morton2016}, the PCs in this catalog cannot be validated without first showing that the candidates have a sufficiently high false alarm reliability. 

We have shown several ways to identify high reliability or high completeness samples. Reliability is a strong function of the MES and the number of observed transits. Also, the FGK dwarf stars are known to be quieter than giant stars and in general the true transits can be more easily separated from the false alarms. We also provide the disposition score, a measure of how robustly a candidate has passed the Robovetter; this can be used to easily find the most reliable candidates. Those doing follow-up observations of KOIs may also use this disposition score to identify the candidates that will optimize ground-based follow-up observations.  

This search of the \Kepler{} data yielded 219 new PCs. Among those new candidates are two new candidates in multi-planet systems (KOI-82.06 and KOI-2926.05).  Also, the catalog contains ten new high-reliability, super-Earth size, habitable zone candidates.  Some of the most scrutinized signals in the DR25 KOI catalog will likely be those fifty small, temperate PCs in the eta-Earth sample defined in \S\ref{s:hz}.  These signals, along with their well characterized completeness and reliability, can be used to make an almost direct measurement on the occurrence rate of planets with the size and insolation flux as Earth, especially around GK dwarf stars.  While this catalog is an important step forward in measuring this number, it is important to remember a few potential biases inherent to this catalog. Namely, errors in the stellar parameters result in significant errors on the planetary sizes and orbital distances, and unaccounted for background stars make planet radii appear smaller than reality and impact the detection limit of the search for all stars.  Also, the Robovetter is not perfect --- completeness of the vetting procedures and the reliability of these signals (both astrophysical and false alarm) must be considered in any calculation.

Ultimately, characterizing this catalog was made possible because of the Robovetter (\S\ref{s:robovetter}) and the innovative metrics it uses to vet each TCE. It has improved the uniformity and accuracy of the vetting process and has allowed the entire process to be tested with known transits and known false positives. As a result, the Robovetter could be run many times, each time improving the vetting by changing thresholds or introducing new metrics. We adapted our vetting process as we learned about the data set, ensuring the highest reliability and completeness achievable in the time allowed.  The Robovetter metrics and logic may prove useful for future transit missions that will find an unprecedented abundance of signals that will require rapid candidate identification for ground-based follow-up, e.g., K2 \citep{Howell2014}, TESS \citep{Ricker2015}, and PLATO \citep{Rauer2016}.

\acknowledgments
The authors would like to thank the anonymous referee for providing comments that improved the clarity and accuracy of this manuscript. 
This paper includes data collected by the \Kepler{} mission. The Kepler Mission was a PI-led Discovery Class Mission funded by the NASA Science Mission directorate. The authors acknowledge the efforts of the \Kepler{} Mission team for generating the many data products used to create the KOI catalog. These products were generated by the \Kepler{} Mission science pipeline through the efforts of the Kepler Science Operations Center and Science Office. The Kepler Mission is led by the project office at NASA Ames Research Center. Ball Aerospace built the Kepler photometer and spacecraft which is operated by the mission operations center at LASP.  
We acknowledge the Kepler Education and Outreach team for their efforts in making the results of this paper accessible to the public. We thank the many scientists who contributed to the \Kepler\ Mission over the years, including R. Gilliland, E. Furlan, J. Orosz and K. Col\'{o}n. We thank the managers and engineers who worked on \Kepler{} over the years, without whom we would not have had a successful \Kepler{} Mission. This research has made use of NASA's Astrophysics Data System. We thank GNU parallel for enabling rapid running of the Robovetter input metrics \citep{Tange2011a}. We thank P.P. Mullally for inspiring the names of certain algorithms. Thank you to \citet{Turbo-King2017} for a spirited discussion.
Some of the data products used in this paper are archived at the NASA Exoplanet Archive, which is operated by the California Institute of Technology, under contract with the National Aeronautics and Space Administration under the Exoplanet Exploration Program. Some of the data presented in this paper were obtained from the Mikulski Archive for Space Telescopes (MAST). STScI is operated by the Association of Universities for Research in Astronomy, Inc., under NASA contract NAS5-26555. Support for MAST for non-HST data is provided by the NASA Office of Space Science via grant NNX09AF08G and by other grants and contracts.  J.F.R acknowledges support from NASA grant NNX14AB82G issued through the Kepler Participating Scientist Program. This research was undertaken, in part, thanks to funding from the Canada Research Chairs program. This research was enabled, in part, by support provided by Calcul Qu\'{e}bec (www.calculquebec.ca) and Compute Canada (www.computecanada.ca). 
D.H. and S.M. acknowledge support by the National Aeronautics and Space Administration under Grant NNX14AB92G issued through the Kepler Participating Scientist Program. J.L.C. is supported by NASA under award No. GRNASM99G000001. J.S. is supported by the NASA \Kepler\ Participating Scientist Program NNX16AK32G.  W.F.W. gratefully acknowledges support from NASA via the \Kepler{} Participating Scientist Program grant NNX14AB91G. V.S.A. acknowledges support from VILLUM FONDEN (research grant 10118).
Funding for the Stellar Astrophysics Centre is provided by The Danish National Research Foundation (Grant DNRF106). The research was supported by the ASTERISK project (ASTERoseismic Investigations with SONG and Kepler) funded by the European Research Council (Grant agreement no.: 267864).

\added{
\software{George \citep{Ambikasaran14}, Kepler Science Data Processing Pipeline (\url{https://github.com/nasa/kepler-pipeline}), Robovetter (\url{https://github.com/nasa/kepler-robovetter}), Marshall (\url{https://sourceforge.net/projects/marshall/}), Centroid Robovetter \citep{Mullally2017}, LPP Metric \citep{Thompson2015a}, Model-Shift Uniqueness Test \citep{Rowe2015cat}, Scipy package (\url{https://www.scipy.org}), Ephemeris Match (\url{https://github.com/JeffLCoughlin/EphemMatch}), Kepler: Kepler Transit Model Codebase Release, \url{https://doi.org/10.5281/zenodo.60297}}
}

\bibliographystyle{aasjournal}
\bibliography{References.bib,KSCI-Refs.bib}

\begin{thebibliography}{}
\expandafter\ifx\csname natexlab\endcsname\relax\def\natexlab#1{#1}\fi

\bibitem[{{Aigrain} {et~al.}(2015){Aigrain}, {Llama}, {Ceillier}, {Chagas},
  {Davenport}, {Garc{\'{\i}}a}, {Hay}, {Lanza}, {McQuillan}, {Mazeh}, {de
  Medeiros}, {Nielsen}, \& {Reinhold}}]{Aigrain2015}
{Aigrain}, S., {Llama}, J., {Ceillier}, T., {et~al.} 2015, \mnras, 450, 3211

\bibitem[{{Akeson} {et~al.}(2013){Akeson}, {Chen}, {Ciardi}, {Crane}, {Good},
  {Harbut}, {Jackson}, {Kane}, {Laity}, {Leifer}, {Lynn}, {McElroy}, {Papin},
  {Plavchan}, {Ram{\'{\i}}rez}, {Rey}, {von Braun}, {Wittman}, {Abajian},
  {Ali}, {Beichman}, {Beekley}, {Berriman}, {Berukoff}, {Bryden}, {Chan},
  {Groom}, {Lau}, {Payne}, {Regelson}, {Saucedo}, {Schmitz}, {Stauffer},
  {Wyatt}, \& {Zhang}}]{Akeson2013}
{Akeson}, R.~L., {Chen}, X., {Ciardi}, D., {et~al.} 2013, \pasp, 125, 989

\bibitem[{{Ambikasaran} {et~al.}(2014){Ambikasaran}, {Foreman-Mackey},
  {Greengard}, {Hogg}, \& {O'Neil}}]{Ambikasaran14}
{Ambikasaran}, S., {Foreman-Mackey}, D., {Greengard}, L., {Hogg}, D.~W., \&
  {O'Neil}, M. 2014, ArXiv e-prints, arXiv:1403.6015

\bibitem[{{Baranec} {et~al.}(2016){Baranec}, {Ziegler}, {Law}, {Morton},
  {Riddle}, {Atkinson}, {Schonhut}, \& {Crepp}}]{Barnec2016}
{Baranec}, C., {Ziegler}, C., {Law}, N.~M., {et~al.} 2016, \aj, 152, 18

\bibitem[{{Barclay} {et~al.}(2013){Barclay}, {Rowe}, {Lissauer}, {Huber},
  {Fressin}, {Howell}, {Bryson}, {Chaplin}, {D{\'e}sert}, {Lopez}, {Marcy},
  {Mullally}, {Ragozzine}, {Torres}, {Adams}, {Agol}, {Barrado}, {Basu},
  {Bedding}, {Buchhave}, {Charbonneau}, {Christiansen},
  {Christensen-Dalsgaard}, {Ciardi}, {Cochran}, {Dupree}, {Elsworth},
  {Everett}, {Fischer}, {Ford}, {Fortney}, {Geary}, {Haas}, {Handberg},
  {Hekker}, {Henze}, {Horch}, {Howard}, {Hunter}, {Isaacson}, {Jenkins},
  {Karoff}, {Kawaler}, {Kjeldsen}, {Klaus}, {Latham}, {Li}, {Lillo-Box},
  {Lund}, {Lundkvist}, {Metcalfe}, {Miglio}, {Morris}, {Quintana}, {Stello},
  {Smith}, {Still}, \& {Thompson}}]{Barclay2013}
{Barclay}, T., {Rowe}, J.~F., {Lissauer}, J.~J., {et~al.} 2013, \nat, 494, 452

\bibitem[{{Barge} {et~al.}(2008){Barge}, {Baglin}, {Auvergne}, {Rauer},
  {L{\'e}ger}, {Schneider}, {Pont}, {Aigrain}, {Almenara}, {Alonso},
  {Barbieri}, {Bord{\'e}}, {Bouchy}, {Deeg}, {La Reza}, {Deleuil}, {Dvorak},
  {Erikson}, {Fridlund}, {Gillon}, {Gondoin}, {Guillot}, {Hatzes}, {Hebrard},
  {Jorda}, {Kabath}, {Lammer}, {Llebaria}, {Loeillet}, {Magain}, {Mazeh},
  {Moutou}, {Ollivier}, {P{\"a}tzold}, {Queloz}, {Rouan}, {Shporer}, \&
  {Wuchterl}}]{Barge2008}
{Barge}, P., {Baglin}, A., {Auvergne}, M., {et~al.} 2008, \aap, 482, L17

\bibitem[{{Batalha} {et~al.}(2011){Batalha}, {Borucki}, {Bryson}, {Buchhave},
  {Caldwell}, {Christensen-Dalsgaard}, {Ciardi}, {Dunham}, {Fressin},
  {Gautier}, {Gilliland}, {Haas}, {Howell}, {Jenkins}, {Kjeldsen}, {Koch},
  {Latham}, {Lissauer}, {Marcy}, {Rowe}, {Sasselov}, {Seager}, {Steffen},
  {Torres}, {Basri}, {Brown}, {Charbonneau}, {Christiansen}, {Clarke},
  {Cochran}, {Dupree}, {Fabrycky}, {Fischer}, {Ford}, {Fortney}, {Girouard},
  {Holman}, {Johnson}, {Isaacson}, {Klaus}, {Machalek}, {Moorehead},
  {Morehead}, {Ragozzine}, {Tenenbaum}, {Twicken}, {Quinn}, {VanCleve},
  {Walkowicz}, {Welsh}, {Devore}, \& {Gould}}]{Batalha2011Kepler10}
{Batalha}, N.~M., {Borucki}, W.~J., {Bryson}, S.~T., {et~al.} 2011, \apj, 729,
  27

\bibitem[{{Batalha} {et~al.}(2013){Batalha}, {Rowe}, {Bryson}, {Barclay},
  {Burke}, {Caldwell}, {Christiansen}, {Mullally}, {Thompson}, {Brown},
  {Dupree}, {Fabrycky}, {Ford}, {Fortney}, {Gilliland}, {Isaacson}, {Latham},
  {Marcy}, {Quinn}, {Ragozzine}, {Shporer}, {Borucki}, {Ciardi}, {Gautier},
  {Haas}, {Jenkins}, {Koch}, {Lissauer}, {Rapin}, {Basri}, {Boss}, {Buchhave},
  {Carter}, {Charbonneau}, {Christensen-Dalsgaard}, {Clarke}, {Cochran},
  {Demory}, {Desert}, {Devore}, {Doyle}, {Esquerdo}, {Everett}, {Fressin},
  {Geary}, {Girouard}, {Gould}, {Hall}, {Holman}, {Howard}, {Howell},
  {Ibrahim}, {Kinemuchi}, {Kjeldsen}, {Klaus}, {Li}, {Lucas}, {Meibom},
  {Morris}, {Pr\v{s}a}, {Quintana}, {Sanderfer}, {Sasselov}, {Seader}, {Smith},
  {Steffen}, {Still}, {Stumpe}, {Tarter}, {Tenenbaum}, {Torres}, {Twicken},
  {Uddin}, {Van Cleve}, {Walkowicz}, \& {Welsh}}]{Batalha2013}
{Batalha}, N.~M., {Rowe}, J.~F., {Bryson}, S.~T., {et~al.} 2013, \apjs, 204, 24

\bibitem[{{Borucki}(2016)}]{Borucki2016}
{Borucki}, W.~J. 2016, Reports on Progress in Physics, 79, 036901

\bibitem[{{Borucki} {et~al.}(2009){Borucki}, {Koch}, {Jenkins}, {Sasselov},
  {Gilliland}, {Batalha}, {Latham}, {Caldwell}, {Basri}, {Brown},
  {Christensen-Dalsgaard}, {Cochran}, {DeVore}, {Dunham}, {Dupree}, {Gautier},
  {Geary}, {Gould}, {Howell}, {Kjeldsen}, {Lissauer}, {Marcy}, {Meibom},
  {Morrison}, \& {Tarter}}]{Borucki2009}
{Borucki}, W.~J., {Koch}, D., {Jenkins}, J., {et~al.} 2009, Science, 325, 709

\bibitem[{{Brown} {et~al.}(2011){Brown}, {Latham}, {Everett}, \&
  {Esquerdo}}]{Brown2011}
{Brown}, T.~M., {Latham}, D.~W., {Everett}, M.~E., \& {Esquerdo}, G.~A. 2011,
  \aj, 142, 112

\bibitem[{{Bryson} \& {Morton}(2017)}]{Bryson2017a}
{Bryson}, S.~T., \& {Morton}, T.~D. 2017, Planet Reliability Metrics:
  Astrophysical Positional Probabilities for Data Release 25 (KSCI-19108-001)

\bibitem[{{Bryson} {et~al.}(2010){Bryson}, {Jenkins}, {Klaus}, {Cote},
  {Quintana}, {Hall}, {Ibrahim}, {Chandrasekaran}, {Caldwell}, {van Cleve}, \&
  {Haas}}]{Bryson2010b}
{Bryson}, S.~T., {Jenkins}, J.~M., {Klaus}, T.~C., {et~al.} 2010, in {Society
  of Photo-Optical Instrumentation Engineers (SPIE) Conference Series}, Vol.
  7740, {Society of Photo-Optical Instrumentation Engineers (SPIE) Conference
  Series}

\bibitem[{{Bryson} {et~al.}(2013){Bryson}, {Jenkins}, {Gilliland}, {Twicken},
  {Clarke}, {Rowe}, {Caldwell}, {Batalha}, {Mullally}, {Haas}, \&
  {Tenenbaum}}]{Bryson2013}
{Bryson}, S.~T., {Jenkins}, J.~M., {Gilliland}, R.~L., {et~al.} 2013, \pasp,
  125, 889

\bibitem[{{Bryson} {et~al.}(2017){Bryson}, {Abdul-Masih}, {Batalha}, {Burke},
  {Caldwell}, {Colon}, {Coughlin}, G., {Haas}, {Henze}, {Huber}, {Latham},
  {Morton}, {Romine}, {Rowe}, {Thompson}, \& {Wolfgang}}]{Bryson2017c}
{Bryson}, S.~T., {Abdul-Masih}, M., {Batalha}, N., {et~al.} 2017, The Kepler
  Certified False Positive Table (KSCI-19093-003)

\bibitem[{{Burke} \& {Catanzarite}(2017{\natexlab{a}})}]{Burke2017b}
{Burke}, C.~J., \& {Catanzarite}, J. 2017{\natexlab{a}}, Planet Detection
  Metrics: Per-Target Detection Contours for Data Release 25 (KSCI-19111-002)

\bibitem[{{Burke} \& {Catanzarite}(2017{\natexlab{b}})}]{Burke2017a}
---. 2017{\natexlab{b}}, Planet Detection Metrics: Per-Target Flux-Level
  Transit Injection Tests of TPS for Data Release 25 (KSCI-19109-002)

\bibitem[{{Burke} \& {Catanzarite}(2017{\natexlab{c}})}]{Burke2017c}
---. 2017{\natexlab{c}}, Planet Detection Metrics: Window and One-Sigma Depth
  Functions for Data Release 25 (KSCI-19101-002)

\bibitem[{{Burke} \& {Seader}(2016)}]{Burke2016}
{Burke}, C.~J., \& {Seader}, S.~E. 2016, Window and One-Sigma Depth Functions
  for Data Release 24 (KSCI-19085-002)

\bibitem[{{Burke} {et~al.}(2015){Burke}, {Christiansen}, {Mullally}, {Seader},
  {Huber}, {Rowe}, {Coughlin}, {Thompson}, {Catanzarite}, {Clarke}, {Morton},
  {Caldwell}, {Bryson}, {Haas}, {Batalha}, {Jenkins}, {Tenenbaum}, {Twicken},
  {Li}, {Quintana}, {Barclay}, {Henze}, {Borucki}, {Howell}, \&
  {Still}}]{Burke2015}
{Burke}, C.~J., {Christiansen}, J.~L., {Mullally}, F., {et~al.} 2015, \apj,
  809, 8

\bibitem[{{Byrd} {et~al.}(1995){Byrd}, {Peihuang}, {Nocedal}, \&
  {Zhu}}]{Byrd95}
{Byrd}, R.~H., {Peihuang}, L., {Nocedal}, J., \& {Zhu}, C. 1995, SIAM J. Sci.
  Comput., 16

\bibitem[{{Cacciari}(2009)}]{Cacciari2009}
{Cacciari}, C. 2009, \memsai, 80, 97

\bibitem[{{Chiavassa} {et~al.}(2017){Chiavassa}, {Caldas}, {Selsis}, {Leconte},
  {Von Paris}, {Bord{\'e}}, {Magic}, {Collet}, \& {Asplund}}]{Chiavassa2017}
{Chiavassa}, A., {Caldas}, A., {Selsis}, F., {et~al.} 2017, \aap, 597, A94

\bibitem[{{Christiansen}(2015)}]{Christiansen2015}
{Christiansen}, J.~L. 2015, Planet Detection Metrics: Pipeline Detection
  Efficiency (KSCI-19094-001)

\bibitem[{{Christiansen}(2017)}]{Christiansen2017}
---. 2017, Planet Detection Metrics: Pixel-Level Transit Injection Tests of
  Pipeline Detection Efficiency for Data Release 25 (KSCI-19110-001)

\bibitem[{{Christiansen} {et~al.}(2010){Christiansen}, {Ballard},
  {Charbonneau}, {Madhusudhan}, {Seager}, {Holman}, {Wellnitz}, {Deming},
  {A'Hearn}, \& {the EPOXI Team}}]{Christiansen2010}
{Christiansen}, J.~L., {Ballard}, S., {Charbonneau}, D., {et~al.} 2010, \apj,
  710, 97

\bibitem[{{Christiansen} {et~al.}(2012){Christiansen}, {Jenkins}, {Caldwell},
  {Burke}, {Tenenbaum}, {Seader}, {Thompson}, {Barclay}, {Clarke}, {Li},
  {Smith}, {Stumpe}, {Twicken}, \& {Van Cleve}}]{Christiansen2012}
{Christiansen}, J.~L., {Jenkins}, J.~M., {Caldwell}, D.~A., {et~al.} 2012,
  \pasp, 124, 1279

\bibitem[{{Christiansen} {et~al.}(2013{\natexlab{a}}){Christiansen}, {Jenkins},
  {Caldwell}, {Barclay}, {Bryson}, {Burke}, {Campbell}, {Catanzarite},
  {Clarke}, {Coughlin}, {Girouard}, {Haas}, {Ibrahim}, {Klaus},
  {Kolodziejczak}, {Li}, {McCauliff}, {Morris}, {Mullally}, {Quintana}, {Rowe},
  {Sabale}, {Seader}, {Smith}, {Still}, {Tenenbaum}, {Thompson}, {Twicken}, \&
  {Uddin}}]{Christiansen2013a}
---. 2013{\natexlab{a}}, {Kepler Data Characteristics Handbook
  (KSCI-19040-004)}, \url{http://archive.stsci.edu/kepler/manuals/
  Data\_Characteristics.pdf}

\bibitem[{{Christiansen} {et~al.}(2013{\natexlab{b}}){Christiansen}, {Clarke},
  {Burke}, {Jenkins}, {Barclay}, {Ford}, {Haas}, {Sabale}, {Seader}, {Claiborne
  Smith}, {Tenenbaum}, {Twicken}, {Kamal Uddin}, \&
  {Thompson}}]{Christiansen2013b}
{Christiansen}, J.~L., {Clarke}, B.~D., {Burke}, C.~J., {et~al.}
  2013{\natexlab{b}}, \apjs, 207, 35

\bibitem[{{Christiansen} {et~al.}(2015){Christiansen}, {Clarke}, {Burke},
  {Seader}, {Jenkins}, {Twicken}, {Catanzarite}, {Smith}, {Batalha}, {Haas},
  {Thompson}, {Campbell}, {Sabale}, \& {Kamal Uddin}}]{Christiansen2015b}
---. 2015, \apj, 810, 95

\bibitem[{{Christiansen} {et~al.}(2016){Christiansen}, {Clarke}, {Burke},
  {Jenkins}, {Bryson}, {Coughlin}, {Mullally}, {Thompson}, {Twicken},
  {Batalha}, {Haas}, {Catanzarite}, {Campbell}, {Kamal Uddin}, {Zamudio},
  {Smith}, \& {Henze}}]{Christiansen2016}
---. 2016, \apj, 828, 99

\bibitem[{{Ciardi} {et~al.}(2015){Ciardi}, {Beichman}, {Horch}, \&
  {Howell}}]{Ciardi2015}
{Ciardi}, D.~R., {Beichman}, C.~A., {Horch}, E.~P., \& {Howell}, S.~B. 2015,
  \apj, 805, 16

\bibitem[{{Claret}(2000)}]{Claret2000}
{Claret}, A. 2000, \aap, 359, 289

\bibitem[{{Claret} \& {Bloemen}(2011)}]{Claret2011}
{Claret}, A., \& {Bloemen}, S. 2011, \aap, 529, A75

\bibitem[{{Coughlin}(2014)}]{Coughlin2014}
{Coughlin}, J.~L. 2014, Description of the TCERT Vetting Products for the
  Q1-Q16 Catalog Using SOC 9.1 (KSCI-19103-001)

\bibitem[{{Coughlin}(2017{\natexlab{a}})}]{Coughlin2017b}
---. 2017{\natexlab{a}}, Description of the TCERT Vetting Reports for Data
  Release 25 (KSCI-19105-001)

\bibitem[{{Coughlin}(2017{\natexlab{b}})}]{Coughlin2017a}
---. 2017{\natexlab{b}}, Planet Detection Metrics: Robovetter Completeness and
  Effectiveness for Data Release 25 (KSCI-19114-001)

\bibitem[{{Coughlin} \& {L{\'o}pez-Morales}(2012)}]{Coughlin2012}
{Coughlin}, J.~L., \& {L{\'o}pez-Morales}, M. 2012, \aj, 143, 39

\bibitem[{{Coughlin} {et~al.}(2014){Coughlin}, {Thompson}, {Bryson}, {Burke},
  {Caldwell}, {Christiansen}, {Haas}, {Howell}, {Jenkins}, {Kolodziejczak},
  {Mullally}, \& {Rowe}}]{Coughlin2014a}
{Coughlin}, J.~L., {Thompson}, S.~E., {Bryson}, S.~T., {et~al.} 2014, \aj, 147,
  119

\bibitem[{{Coughlin} {et~al.}(2016){Coughlin}, {Mullally}, {Thompson}, {Rowe},
  {Burke}, {Latham}, {Batalha}, {Ofir}, {Quarles}, {Henze}, {Wolfgang},
  {Caldwell}, {Bryson}, {Shporer}, {Catanzarite}, {Akeson}, {Barclay},
  {Borucki}, {Boyajian}, {Campbell}, {Christiansen}, {Girouard}, {Haas},
  {Howell}, {Huber}, {Jenkins}, {Li}, {Patil-Sabale}, {Quintana}, {Ramirez},
  {Seader}, {Smith}, {Tenenbaum}, {Twicken}, \& {Zamudio}}]{Coughlin2016}
{Coughlin}, J.~L., {Mullally}, F., {Thompson}, S.~E., {et~al.} 2016, \apjs,
  224, 12

\bibitem[{{Devor} {et~al.}(2008){Devor}, {Charbonneau}, {O'Donovan},
  {Mandushev}, \& {Torres}}]{Devor2008a}
{Devor}, J., {Charbonneau}, D., {O'Donovan}, F.~T., {Mandushev}, G., \&
  {Torres}, G. 2008, \aj, 135, 850

\bibitem[{{Dotter} {et~al.}(2008){Dotter}, {Chaboyer}, {Jevremovi{\'c}},
  {Kostov}, {Baron}, \& {Ferguson}}]{Dotter2008}
{Dotter}, A., {Chaboyer}, B., {Jevremovi{\'c}}, D., {et~al.} 2008, \apjs, 178,
  89

\bibitem[{{Doyle} {et~al.}(2011){Doyle}, {Carter}, {Fabrycky}, {Slawson},
  {Howell}, {Winn}, {Orosz}, {Pr\v{s}a}, {Welsh}, {Quinn}, {Latham}, {Torres},
  {Buchhave}, {Marcy}, {Fortney}, {Shporer}, {Ford}, {Lissauer}, {Ragozzine},
  {Rucker}, {Batalha}, {Jenkins}, {Borucki}, {Koch}, {Middour}, {Hall},
  {McCauliff}, {Fanelli}, {Quintana}, {Holman}, {Caldwell}, {Still},
  {Stefanik}, {Brown}, {Esquerdo}, {Tang}, {Furesz}, {Geary}, {Berlind},
  {Calkins}, {Short}, {Steffen}, {Sasselov}, {Dunham}, {Cochran}, {Boss},
  {Haas}, {Buzasi}, \& {Fischer}}]{Doyle2011}
{Doyle}, L.~R., {Carter}, J.~A., {Fabrycky}, D.~C., {et~al.} 2011, Science,
  333, 1602

\bibitem[{{Dressing} \& {Charbonneau}(2013)}]{Dressing2013}
{Dressing}, C.~D., \& {Charbonneau}, D. 2013, \apj, 767, 95

\bibitem[{{Dressing} \& {Charbonneau}(2015)}]{Dressing2015}
---. 2015, \apj, 807, 45

\bibitem[{{Fabrycky} {et~al.}(2014){Fabrycky}, {Lissauer}, {Ragozzine}, {Rowe},
  {Steffen}, {Agol}, {Barclay}, {Batalha}, {Borucki}, {Ciardi}, {Ford},
  {Gautier}, {Geary}, {Holman}, {Jenkins}, {Li}, {Morehead}, {Morris},
  {Shporer}, {Smith}, {Still}, \& {Van Cleve}}]{Fabrycky2014}
{Fabrycky}, D.~C., {Lissauer}, J.~J., {Ragozzine}, D., {et~al.} 2014, \apj,
  790, 146

\bibitem[{{Fogtmann-Schulz} {et~al.}(2014){Fogtmann-Schulz}, {Hinrup}, {Van
  Eylen}, {Christensen-Dalsgaard}, {Kjeldsen}, {Silva Aguirre}, \&
  {Tingley}}]{Fogtmann2014Kepler10}
{Fogtmann-Schulz}, A., {Hinrup}, B., {Van Eylen}, V., {et~al.} 2014, \apj, 781,
  67

\bibitem[{{Ford}(2005)}]{Ford2005AJ}
{Ford}, E.~B. 2005, \aj, 129, 1706

\bibitem[{{Foreman-Mackey} {et~al.}(2016){Foreman-Mackey}, {Morton}, {Hogg},
  {Agol}, \& {Sch{\"o}lkopf}}]{ForemanMackey16}
{Foreman-Mackey}, D., {Morton}, T.~D., {Hogg}, D.~W., {Agol}, E., \&
  {Sch{\"o}lkopf}, B. 2016, \aj, 152, 206

\bibitem[{{Fuller} {et~al.}(2017){Fuller}, {Hambleton}, {Shporer}, {Isaacson},
  \& {Thompson}}]{Fuller2017}
{Fuller}, J., {Hambleton}, K., {Shporer}, A., {Isaacson}, H., \& {Thompson}, S.
  2017, ArXiv e-prints, arXiv:1706.05053

\bibitem[{{Fulton} {et~al.}(2017){Fulton}, {Petigura}, {Howard}, {Isaacson},
  {Marcy}, {Cargile}, {Hebb}, {Weiss}, {Johnson}, {Morton}, {Sinukoff},
  {Crossfield}, \& {Hirsch}}]{Fulton2017}
{Fulton}, B.~J., {Petigura}, E.~A., {Howard}, A.~W., {et~al.} 2017, ArXiv
  e-prints, arXiv:1703.10375

\bibitem[{Furlan \& Howell(2017)}]{Furlan2017densities}
Furlan, E., \& Howell, S.~B. 2017, The Astronomical Journal, 154, 66

\bibitem[{{Furlan} {et~al.}(2017){Furlan}, {Ciardi}, {Everett}, {Saylors},
  {Teske}, {Horch}, {Howell}, {van Belle}, {Hirsch}, {Gautier}, {Adams},
  {Barrado}, {Cartier}, {Dressing}, {Dupree}, {Gilliland}, {Lillo-Box},
  {Lucas}, \& {Wang}}]{Furlan2017}
{Furlan}, E., {Ciardi}, D.~R., {Everett}, M.~E., {et~al.} 2017, \aj, 153, 71

\bibitem[{{Gaia Collaboration} {et~al.}(2016{\natexlab{a}}){Gaia
  Collaboration}, {Brown}, {Vallenari}, {Prusti}, {de Bruijne}, {Mignard},
  {Drimmel}, {Babusiaux}, {Bailer-Jones}, {Bastian}, \& et~al.}]{gaia2}
{Gaia Collaboration}, {Brown}, A.~G.~A., {Vallenari}, A., {et~al.}
  2016{\natexlab{a}}, \aap, 595, A2

\bibitem[{{Gaia Collaboration} {et~al.}(2016{\natexlab{b}}){Gaia
  Collaboration}, {Prusti}, {de Bruijne}, {Brown}, {Vallenari}, {Babusiaux},
  {Bailer-Jones}, {Bastian}, {Biermann}, {Evans}, \& et~al.}]{gaia1}
{Gaia Collaboration}, {Prusti}, T., {de Bruijne}, J.~H.~J., {et~al.}
  2016{\natexlab{b}}, \aap, 595, A1

\bibitem[{Garcia(2010)}]{Garcia2010}
Garcia, D. 2010, Computational Statistics \& Data Analysis, 54, 1167

\bibitem[{{Garc{\'{\i}}a} {et~al.}(2014){Garc{\'{\i}}a}, {Ceillier},
  {Salabert}, {Mathur}, {van Saders}, {Pinsonneault}, {Ballot}, {Beck},
  {Bloemen}, {Campante}, {Davies}, {do Nascimento}, {Mathis}, {Metcalfe},
  {Nielsen}, {Su{\'a}rez}, {Chaplin}, {Jim{\'e}nez}, \& {Karoff}}]{Garcia2014}
{Garc{\'{\i}}a}, R.~A., {Ceillier}, T., {Salabert}, D., {et~al.} 2014, \aap,
  572, A34

\bibitem[{{Gilliland} {et~al.}(2015){Gilliland}, {Cartier}, {Adams}, {Ciardi},
  {Kalas}, \& {Wright}}]{Gilliland2015}
{Gilliland}, R.~L., {Cartier}, K.~M.~S., {Adams}, E.~R., {et~al.} 2015, \aj,
  149, 24

\bibitem[{{Gilliland} {et~al.}(2011){Gilliland}, {Chaplin}, {Dunham},
  {Argabright}, {Borucki}, {Basri}, {Bryson}, {Buzasi}, {Caldwell}, {Elsworth},
  {Jenkins}, {Koch}, {Kolodziejczak}, {Miglio}, {van Cleve}, {Walkowicz}, \&
  {Welsh}}]{Gilliland2011}
{Gilliland}, R.~L., {Chaplin}, W.~J., {Dunham}, E.~W., {et~al.} 2011, \apjs,
  197, 6

\bibitem[{{Hambleton} {et~al.}(2017){Hambleton}, {Fuller}, {Thompson}, {Pr{\v
  s}a}, {Kurtz}, {Shporer}, {Isaacson}, {Howard}, {Endl}, {Cochran}, \&
  {Murphy}}]{Hambleton2017}
{Hambleton}, K., {Fuller}, J., {Thompson}, S., {et~al.} 2017, ArXiv e-prints,
  arXiv:1706.05051

\bibitem[{Hampel(1974)}]{Hampel1974}
Hampel, F.~R. 1974, Journal of the American Statistical Association, 69, 383

\bibitem[{{He} \& {Niyogi}(2004)}]{He2004}
{He}, X., \& {Niyogi}, P. 2004, {Advances in Neural Inoformation Processing
  Systems}, 16, 37

\bibitem[{{Hoffman} \& {Rowe}(2017)}]{Hoffman2017}
{Hoffman}, K.~L., \& {Rowe}, J.~F. 2017, Uniform Modeling of KOIs: MCMC Notes
  for Data Release 25 (KSCI-19113-001)

\bibitem[{{Howard} {et~al.}(2012){Howard}, {Marcy}, {Bryson}, {Jenkins},
  {Rowe}, {Batalha}, {Borucki}, {Koch}, {Dunham}, {Gautier}, {Van Cleve},
  {Cochran}, {Latham}, {Lissauer}, {Torres}, {Brown}, {Gilliland}, {Buchhave},
  {Caldwell}, {Christensen-Dalsgaard}, {Ciardi}, {Fressin}, {Haas}, {Howell},
  {Kjeldsen}, {Seager}, {Rogers}, {Sasselov}, {Steffen}, {Basri},
  {Charbonneau}, {Christiansen}, {Clarke}, {Dupree}, {Fabrycky}, {Fischer},
  {Ford}, {Fortney}, {Tarter}, {Girouard}, {Holman}, {Johnson}, {Klaus},
  {Machalek}, {Moorhead}, {Morehead}, {Ragozzine}, {Tenenbaum}, {Twicken},
  {Quinn}, {Isaacson}, {Shporer}, {Lucas}, {Walkowicz}, {Welsh}, {Boss},
  {Devore}, {Gould}, {Smith}, {Morris}, {Pr\v{s}a}, {Morton}, {Still},
  {Thompson}, {Mullally}, {Endl}, \& {MacQueen}}]{Howard2012}
{Howard}, A.~W., {Marcy}, G.~W., {Bryson}, S.~T., {et~al.} 2012, \apjs, 201, 15

\bibitem[{{Howell} {et~al.}(2016){Howell}, {Ciardi}, {Giampapa}, {Everett},
  {Silva}, \& {Szkody}}]{Howell2016}
{Howell}, S.~B., {Ciardi}, D.~R., {Giampapa}, M.~S., {et~al.} 2016, \aj, 151,
  43

\bibitem[{{Howell} {et~al.}(2014){Howell}, {Sobeck}, {Haas}, {Still},
  {Barclay}, {Mullally}, {Troeltzsch}, {Aigrain}, {Bryson}, {Caldwell},
  {Chaplin}, {Cochran}, {Huber}, {Marcy}, {Miglio}, {Najita}, {Smith},
  {Twicken}, \& {Fortney}}]{Howell2014}
{Howell}, S.~B., {Sobeck}, C., {Haas}, M., {et~al.} 2014, \pasp, 126, 398

\bibitem[{{Huber} {et~al.}(2014){Huber}, {Silva Aguirre}, {Matthews},
  {Pinsonneault}, {Gaidos}, {Garc{\'i}a}, {Hekker}, {Mathur}, {Mosser},
  {Torres}, {Bastien}, {Basu}, {Bedding}, {Chaplin}, {Demory}, {Fleming},
  {Guo}, {Mann}, {Rowe}, {Serenelli}, {Smith}, \& {Stello}}]{Huber2014a}
{Huber}, D., {Silva Aguirre}, V., {Matthews}, J.~M., {et~al.} 2014, \apjs, 211,
  2

\bibitem[{{Jenkins}(2002)}]{Jenkins2002b}
{Jenkins}, J.~M. 2002, \apj, 575, 493

\bibitem[{{Jenkins}(2017{\natexlab{a}})}]{Jenkins2017}
---. 2017{\natexlab{a}}, Kepler Data Processing Handbook (KSCI-19081-002)

\bibitem[{{Jenkins}(2017{\natexlab{b}})}]{JenkinsKDPH}
---. 2017{\natexlab{b}}, Kepler Data Processing Handbook (KSCI-19081-002)

\bibitem[{{Jenkins} {et~al.}(2002){Jenkins}, {Caldwell}, \&
  {Borucki}}]{Jenkins2002a}
{Jenkins}, J.~M., {Caldwell}, D.~A., \& {Borucki}, W.~J. 2002, \apj, 564, 495

\bibitem[{{Jenkins} {et~al.}(2015){Jenkins}, {Twicken}, {Batalha}, {Caldwell},
  {Cochran}, {Endl}, {Latham}, {Esquerdo}, {Seader}, {Bieryla}, {Petigura},
  {Ciardi}, {Marcy}, {Isaacson}, {Huber}, {Rowe}, {Torres}, {Bryson},
  {Buchhave}, {Ramirez}, {Wolfgang}, {Li}, {Campbell}, {Tenenbaum},
  {Sanderfer}, {Henze}, {Catanzarite}, {Gilliland}, \& {Borucki}}]{Jenkins2015}
{Jenkins}, J.~M., {Twicken}, J.~D., {Batalha}, N.~M., {et~al.} 2015, \aj, 150,
  56

\bibitem[{{Johnson} {et~al.}(2017){Johnson}, {Petigura}, {Fulton}, {Marcy},
  {Howard}, {Isaacson}, {Hebb}, {Cargile}, {Morton}, {Weiss}, {Winn}, {Rogers},
  {Sinukoff}, \& {Hirsch}}]{CKS2}
{Johnson}, J.~A., {Petigura}, E.~A., {Fulton}, B.~J., {et~al.} 2017, ArXiv
  e-prints, arXiv:1703.10402

\bibitem[{{Kirk} {et~al.}(2016){Kirk}, {Conroy}, {Pr{\v s}a}, {Abdul-Masih},
  {Kochoska}, {Matijevi{\v c}}, {Hambleton}, {Barclay}, {Bloemen}, {Boyajian},
  {Doyle}, {Fulton}, {Hoekstra}, {Jek}, {Kane}, {Kostov}, {Latham}, {Mazeh},
  {Orosz}, {Pepper}, {Quarles}, {Ragozzine}, {Shporer}, {Southworth},
  {Stassun}, {Thompson}, {Welsh}, {Agol}, {Derekas}, {Devor}, {Fischer},
  {Green}, {Gropp}, {Jacobs}, {Johnston}, {LaCourse}, {Saetre}, {Schwengeler},
  {Toczyski}, {Werner}, {Garrett}, {Gore}, {Martinez}, {Spitzer}, {Stevick},
  {Thomadis}, {Vrijmoet}, {Yenawine}, {Batalha}, \& {Borucki}}]{Kirk2016}
{Kirk}, B., {Conroy}, K., {Pr{\v s}a}, A., {et~al.} 2016, \aj, 151, 68

\bibitem[{{Koch} {et~al.}(2010){Koch}, {Borucki}, {Rowe}, {Batalha}, {Brown},
  {Caldwell}, {Caldwell}, {Cochran}, {DeVore}, {Dunham}, {Dupree}, {Gautier},
  {Geary}, {Gilliland}, {Howell}, {Jenkins}, {Latham}, {Lissauer}, {Marcy},
  {Morrison}, \& {Tarter}}]{Koch2010}
{Koch}, D.~G., {Borucki}, W.~J., {Rowe}, J.~F., {et~al.} 2010, \apjl, 713, L131

\bibitem[{{Kopparapu} {et~al.}(2013){Kopparapu}, {Ramirez}, {Kasting}, {Eymet},
  {Robinson}, {Mahadevan}, {Terrien}, {Domagal-Goldman}, {Meadows}, \&
  {Deshpande}}]{Kopparapu2013}
{Kopparapu}, R.~K., {Ramirez}, R., {Kasting}, J.~F., {et~al.} 2013, \apj, 765,
  131

\bibitem[{{Kreiner}(2004)}]{Kreiner2004}
{Kreiner}, J.~M. 2004, \actaa, 54, 207

\bibitem[{{Kruse} \& {Agol}(2014)}]{Kruse2014}
{Kruse}, E., \& {Agol}, E. 2014, Science, 344, 275

\bibitem[{{Law} {et~al.}(2014){Law}, {Morton}, {Baranec}, {Riddle},
  {Ravichandran}, {Ziegler}, {Johnson}, {Tendulkar}, {Bui}, {Burse}, {Das},
  {Dekany}, {Kulkarni}, {Punnadi}, \& {Ramaprakash}}]{Law2014}
{Law}, N.~M., {Morton}, T., {Baranec}, C., {et~al.} 2014, \apj, 791, 35

\bibitem[{{Lissauer} {et~al.}(2011{\natexlab{a}}){Lissauer}, {Fabrycky},
  {Ford}, {Borucki}, {Fressin}, {Marcy}, {Orosz}, {Rowe}, {Torres}, {Welsh},
  {Batalha}, {Bryson}, {Buchhave}, {Caldwell}, {Carter}, {Charbonneau},
  {Christiansen}, {Cochran}, {Desert}, {Dunham}, {Fanelli}, {Fortney},
  {Gautier}, {Geary}, {Gilliland}, {Haas}, {Hall}, {Holman}, {Koch}, {Latham},
  {Lopez}, {McCauliff}, {Miller}, {Morehead}, {Quintana}, {Ragozzine},
  {Sasselov}, {Short}, \& {Steffen}}]{Lissauer2011}
{Lissauer}, J.~J., {Fabrycky}, D.~C., {Ford}, E.~B., {et~al.}
  2011{\natexlab{a}}, \nat, 470, 53

\bibitem[{{Lissauer} {et~al.}(2011{\natexlab{b}}){Lissauer}, {Ragozzine},
  {Fabrycky}, {Steffen}, {Ford}, {Jenkins}, {Shporer}, {Holman}, {Rowe},
  {Quintana}, {Batalha}, {Borucki}, {Bryson}, {Caldwell}, {Carter}, {Ciardi},
  {Dunham}, {Fortney}, {Gautier}, {Howell}, {Koch}, {Latham}, {Marcy},
  {Morehead}, \& {Sasselov}}]{Lissauer2011b}
{Lissauer}, J.~J., {Ragozzine}, D., {Fabrycky}, D.~C., {et~al.}
  2011{\natexlab{b}}, \apjs, 197, 8

\bibitem[{{Lissauer} {et~al.}(2014){Lissauer}, {Marcy}, {Bryson}, {Rowe},
  {Jontof-Hutter}, {Agol}, {Borucki}, {Carter}, {Ford}, {Gilliland}, {Kolbl},
  {Star}, {Steffen}, \& {Torres}}]{Lissauer2014}
{Lissauer}, J.~J., {Marcy}, G.~W., {Bryson}, S.~T., {et~al.} 2014, \apj, 784,
  44

\bibitem[{{Lopez} \& {Fortney}(2013)}]{Lopez2013}
{Lopez}, E.~D., \& {Fortney}, J.~J. 2013, \apj, 776, 2

\bibitem[{{Lundkvist} {et~al.}(2016){Lundkvist}, {Kjeldsen}, {Albrecht},
  {Davies}, {Basu}, {Huber}, {Justesen}, {Karoff}, {Silva Aguirre}, {van
  Eylen}, {Vang}, {Arentoft}, {Barclay}, {Bedding}, {Campante}, {Chaplin},
  {Christensen-Dalsgaard}, {Elsworth}, {Gilliland}, {Handberg}, {Hekker},
  {Kawaler}, {Lund}, {Metcalfe}, {Miglio}, {Rowe}, {Stello}, {Tingley}, \&
  {White}}]{Lundkvist2016}
{Lundkvist}, M.~S., {Kjeldsen}, H., {Albrecht}, S., {et~al.} 2016, Nature
  Communications, 7, 11201

\bibitem[{{Mandel} \& {Agol}(2002)}]{Mandel2002}
{Mandel}, K., \& {Agol}, E. 2002, \apjl, 580, L171

\bibitem[{{Mann} {et~al.}(2017){Mann}, {Dupuy}, {Muirhead}, {Johnson}, {Liu},
  {Ansdell}, {Dalba}, {Swift}, \& {Hadden}}]{Mann2017}
{Mann}, A.~W., {Dupuy}, T., {Muirhead}, P.~S., {et~al.} 2017, \aj, 153, 267

\bibitem[{{Marcy} {et~al.}(2014){Marcy}, {Isaacson}, {Howard}, {Rowe},
  {Jenkins}, {Bryson}, {Latham}, {Howell}, {Gautier}, {Batalha}, {Rogers},
  {Ciardi}, {Fischer}, {Gilliland}, {Kjeldsen}, {Christensen-Dalsgaard},
  {Huber}, {Chaplin}, {Basu}, {Buchhave}, {Quinn}, {Borucki}, {Koch}, {Hunter},
  {Caldwell}, {Van Cleve}, {Kolbl}, {Weiss}, {Petigura}, {Seager}, {Morton},
  {Johnson}, {Ballard}, {Burke}, {Cochran}, {Endl}, {MacQueen}, {Everett},
  {Lissauer}, {Ford}, {Torres}, {Fressin}, {Brown}, {Steffen}, {Charbonneau},
  {Basri}, {Sasselov}, {Winn}, {Sanchis-Ojeda}, {Christiansen}, {Adams},
  {Henze}, {Dupree}, {Fabrycky}, {Fortney}, {Tarter}, {Holman}, {Tenenbaum},
  {Shporer}, {Lucas}, {Welsh}, {Orosz}, {Bedding}, {Campante}, {Davies},
  {Elsworth}, {Handberg}, {Hekker}, {Karoff}, {Kawaler}, {Lund}, {Lundkvist},
  {Metcalfe}, {Miglio}, {Silva Aguirre}, {Stello}, {White}, {Boss}, {Devore},
  {Gould}, {Prsa}, {Agol}, {Barclay}, {Coughlin}, {Brugamyer}, {Mullally},
  {Quintana}, {Still}, {Thompson}, {Morrison}, {Twicken}, {D{\'e}sert},
  {Carter}, {Crepp}, {H{\'e}brard}, {Santerne}, {Moutou}, {Sobeck}, {Hudgins},
  {Haas}, {Robertson}, {Lillo-Box}, \& {Barrado}}]{Marcy2014}
{Marcy}, G.~W., {Isaacson}, H., {Howard}, A.~W., {et~al.} 2014, \apjs, 210, 20

\bibitem[{{Mathur} {et~al.}(2017){Mathur}, {Huber}, {Batalha}, {Ciardi},
  {Bastien}, {Bieryla}, {Buchhave}, {Cochran}, {Endl}, {Esquerdo}, {Furlan},
  {Howard}, {Howell}, {Isaacson}, {Latham}, {MacQueen}, \&
  {Silva}}]{Mathur2017ApJS}
{Mathur}, S., {Huber}, D., {Batalha}, N.~M., {et~al.} 2017, \apjs, 229, 30

\bibitem[{{Mayor} \& {Queloz}(1995)}]{Mayor1995}
{Mayor}, M., \& {Queloz}, D. 1995, \nat, 378, 355

\bibitem[{{Mazeh} {et~al.}(2012){Mazeh}, {Nachmani}, {Sokol}, {Faigler}, \&
  {Zucker}}]{Mazeh2012}
{Mazeh}, T., {Nachmani}, G., {Sokol}, G., {Faigler}, S., \& {Zucker}, S. 2012,
  \aap, 541, A56

\bibitem[{{McQuillan} {et~al.}(2014){McQuillan}, {Mazeh}, \&
  {Aigrain}}]{McQuillan2014}
{McQuillan}, A., {Mazeh}, T., \& {Aigrain}, S. 2014, \apjs, 211, 24

\bibitem[{{Meibom} {et~al.}(2011){Meibom}, {Mathieu}, {Stassun}, {Liebesny}, \&
  {Saar}}]{Meibom2011}
{Meibom}, S., {Mathieu}, R.~D., {Stassun}, K.~G., {Liebesny}, P., \& {Saar},
  S.~H. 2011, \apj, 733, 115

\bibitem[{{Mignard}(2005)}]{Mignard2005}
{Mignard}, F. 2005, in {Astronomical Society of the Pacific Conference Series},
  Vol. 338, {Astrometry in the Age of the Next Generation of Large Telescopes},
  ed. {P.~K.~Seidelmann \& A.~K.~B.~Monet}, 15--+

\bibitem[{{More} {et~al.}(1980){More}, {Garbow}, \& {Hillstrom}}]{More1980}
{More}, J., {Garbow}, B., \& {Hillstrom}, K. 1980, Argoone National Laboratory
  Report ANL-80-74

\bibitem[{{Morton} {et~al.}(2016){Morton}, {Bryson}, {Coughlin}, {Rowe},
  {Ravichandran}, {Petigura}, {Haas}, \& {Batalha}}]{Morton2016}
{Morton}, T.~D., {Bryson}, S.~T., {Coughlin}, J.~L., {et~al.} 2016, \apj, 822,
  86

\bibitem[{{Mullally}(2017)}]{Mullally2017}
{Mullally}, F. 2017, Planet Detection Metrics: Automatic Detection of
  Background Objects Using the Centroid Robovetter (KSCI-19115-001)

\bibitem[{{Mullally} {et~al.}(2016){Mullally}, {Coughlin}, {Thompson},
  {Christiansen}, {Burke}, {Clarke}, \& {Haas}}]{Mullally2016}
{Mullally}, F., {Coughlin}, J.~L., {Thompson}, S.~E., {et~al.} 2016, \pasp,
  128, 074502

\bibitem[{{Mullally} {et~al.}(2015){Mullally}, {Coughlin}, {Thompson}, {Rowe},
  {Burke}, {Latham}, {Batalha}, {Bryson}, {Christiansen}, {Henze}, {Ofir},
  {Quarles}, {Shporer}, {Van Eylen}, {Van Laerhoven}, {Shah}, {Wolfgang},
  {Chaplin}, {Xie}, {Akeson}, {Argabright}, {Bachtell}, {Barclay}, {Borucki},
  {Caldwell}, {Campbell}, {Catanzarite}, {Cochran}, {Duren}, {Fleming},
  {Fraquelli}, {Girouard}, {Haas}, {He{\l}miniak}, {Howell}, {Huber}, {Larson},
  {Gautier}, {Jenkins}, {Li}, {Lissauer}, {McArthur}, {Miller}, {Morris},
  {Patil-Sabale}, {Plavchan}, {Putnam}, {Quintana}, {Ramirez}, {Silva Aguirre},
  {Seader}, {Smith}, {Steffen}, {Stewart}, {Stober}, {Still}, {Tenenbaum},
  {Troeltzsch}, {Twicken}, \& {Zamudio}}]{Mullally2015cat}
---. 2015, \apjs, 217, 31

\bibitem[{{O'Donovan} {et~al.}(2006){O'Donovan}, {Charbonneau}, {Mandushev},
  {Dunham}, {Latham}, {Torres}, {Sozzetti}, {Brown}, {Trauger}, {Belmonte},
  {Rabus}, {Almenara}, {Alonso}, {Deeg}, {Esquerdo}, {Falco}, {Hillenbrand},
  {Roussanova}, {Stefanik}, \& {Winn}}]{ODonovan2006}
{O'Donovan}, F.~T., {Charbonneau}, D., {Mandushev}, G., {et~al.} 2006, \apjl,
  651, L61

\bibitem[{{Orosz} {et~al.}(2012){Orosz}, {Welsh}, {Carter}, {Fabrycky},
  {Cochran}, {Endl}, {Ford}, {Haghighipour}, {MacQueen}, {Mazeh},
  {Sanchis-Ojeda}, {Short}, {Torres}, {Agol}, {Buchhave}, {Doyle}, {Isaacson},
  {Lissauer}, {Marcy}, {Shporer}, {Windmiller}, {Barclay}, {Boss}, {Clarke},
  {Fortney}, {Geary}, {Holman}, {Huber}, {Jenkins}, {Kinemuchi}, {Kruse},
  {Ragozzine}, {Sasselov}, {Still}, {Tenenbaum}, {Uddin}, {Winn}, {Koch}, \&
  {Borucki}}]{Orosz2012}
{Orosz}, J.~A., {Welsh}, W.~F., {Carter}, J.~A., {et~al.} 2012, Science, 337,
  1511

\bibitem[{{Owen} \& {Wu}(2013)}]{Owen2013}
{Owen}, J.~E., \& {Wu}, Y. 2013, \apj, 775, 105

\bibitem[{{Petigura} {et~al.}(2013){Petigura}, {Howard}, \&
  {Marcy}}]{Petigura2013b}
{Petigura}, E.~A., {Howard}, A.~W., \& {Marcy}, G.~W. 2013, Proceedings of the
  National Academy of Science, 110, 19273

\bibitem[{{Petigura} {et~al.}(2017{\natexlab{a}}){Petigura}, {Howard}, {Marcy},
  {Johnson}, {Isaacson}, {Cargile}, {Hebb}, {Fulton}, {Weiss}, {Morton},
  {Winn}, {Rogers}, {Sinukoff}, {Hirsch}, \& {Crossfield}}]{Petigura2017}
{Petigura}, E.~A., {Howard}, A.~W., {Marcy}, G.~W., {et~al.}
  2017{\natexlab{a}}, \aj, 154, 107

\bibitem[{{Petigura} {et~al.}(2017{\natexlab{b}}){Petigura}, {Howard}, {Marcy},
  {Johnson}, {Isaacson}, {Cargile}, {Hebb}, {Fulton}, {Weiss}, {Morton},
  {Winn}, {Rogers}, {Sinukoff}, {Hirsch}, \& {Crossfield}}]{CKS1}
---. 2017{\natexlab{b}}, ArXiv e-prints, arXiv:1703.10400

\bibitem[{{Pr\v{s}a} {et~al.}(2011){Pr\v{s}a}, {Batalha}, {Slawson}, {Doyle},
  {Welsh}, {Orosz}, {Seager}, {Rucker}, {Mjaseth}, {Engle}, {Conroy},
  {Jenkins}, {Caldwell}, {Koch}, \& {Borucki}}]{Prsa2011}
{Pr\v{s}a}, A., {Batalha}, N., {Slawson}, R.~W., {et~al.} 2011, \aj, 141, 83

\bibitem[{{Quintana} {et~al.}(2014){Quintana}, {Barclay}, {Raymond}, {Rowe},
  {Bolmont}, {Caldwell}, {Howell}, {Kane}, {Huber}, {Crepp}, {Lissauer},
  {Ciardi}, {Coughlin}, {Everett}, {Henze}, {Horch}, {Isaacson}, {Ford},
  {Adams}, {Still}, {Hunter}, {Quarles}, \& {Selsis}}]{Quintana2014}
{Quintana}, E.~V., {Barclay}, T., {Raymond}, S.~N., {et~al.} 2014, Science,
  344, 277

\bibitem[{{Rappaport} {et~al.}(2017){Rappaport}, {Vanderburg}, {Jacobs},
  {LaCourse}, {Jenkins}, {Kraus}, {Rizzuto}, {Latham}, {Bieryla}, {Lazarevic},
  \& {Schmitt}}]{Rappaport2017}
{Rappaport}, S., {Vanderburg}, A., {Jacobs}, T., {et~al.} 2017, ArXiv e-prints,
  arXiv:1708.06069

\bibitem[{{Rasmussen} \& {Williams}(2006)}]{Rasmussen10}
{Rasmussen}, C.~E., \& {Williams}, C.~K.~I. 2006, Gaussian Processes for
  Machine Learning (The MIT Press)

\bibitem[{{Rauer} {et~al.}(2016){Rauer}, {Aerts}, {Cabrera}, \& {PLATO
  Team}}]{Rauer2016}
{Rauer}, H., {Aerts}, C., {Cabrera}, J., \& {PLATO Team}. 2016, Astronomische
  Nachrichten, 337, 961

\bibitem[{{Ricker} {et~al.}(2015){Ricker}, {Winn}, {Vanderspek}, {Latham},
  {Bakos}, {Bean}, {Berta-Thompson}, {Brown}, {Buchhave}, {Butler}, {Butler},
  {Chaplin}, {Charbonneau}, {Christensen-Dalsgaard}, {Clampin}, {Deming},
  {Doty}, {De Lee}, {Dressing}, {Dunham}, {Endl}, {Fressin}, {Ge}, {Henning},
  {Holman}, {Howard}, {Ida}, {Jenkins}, {Jernigan}, {Johnson}, {Kaltenegger},
  {Kawai}, {Kjeldsen}, {Laughlin}, {Levine}, {Lin}, {Lissauer}, {MacQueen},
  {Marcy}, {McCullough}, {Morton}, {Narita}, {Paegert}, {Palle}, {Pepe},
  {Pepper}, {Quirrenbach}, {Rinehart}, {Sasselov}, {Sato}, {Seager},
  {Sozzetti}, {Stassun}, {Sullivan}, {Szentgyorgyi}, {Torres}, {Udry}, \&
  {Villasenor}}]{Ricker2015}
{Ricker}, G.~R., {Winn}, J.~N., {Vanderspek}, R., {et~al.} 2015, Journal of
  Astronomical Telescopes, Instruments, and Systems, 1, 014003

\bibitem[{{Rogers}(2015)}]{Rogers2015}
{Rogers}, L.~A. 2015, \apj, 801, 41

\bibitem[{{Rowe} {et~al.}(2014){Rowe}, {Bryson}, {Marcy}, {Lissauer},
  {Jontof-Hutter}, {Mullally}, {Gilliland}, {Issacson}, {Ford}, {Howell},
  {Borucki}, {Haas}, {Huber}, {Steffen}, {Thompson}, {Quintana}, {Barclay},
  {Still}, {Fortney}, {Gautier}, {Hunter}, {Caldwell}, {Ciardi}, {Devore},
  {Cochran}, {Jenkins}, {Agol}, {Carter}, \& {Geary}}]{Rowe2014}
{Rowe}, J.~F., {Bryson}, S.~T., {Marcy}, G.~W., {et~al.} 2014, \apj, 784, 45

\bibitem[{{Rowe} {et~al.}(2015{\natexlab{a}}){Rowe}, {Coughlin}, {Antoci},
  {Barclay}, {Batalha}, {Borucki}, {Burke}, {Bryson}, {Caldwell}, {Campbell},
  {Catanzarite}, {Christiansen}, {Cochran}, {Gilliland}, {Girouard}, {Haas},
  {He{\l}miniak}, {Henze}, {Hoffman}, {Howell}, {Huber}, {Hunter},
  {Jang-Condell}, {Jenkins}, {Klaus}, {Latham}, {Li}, {Lissauer}, {McCauliff},
  {Morris}, {Mullally}, {Ofir}, {Quarles}, {Quintana}, {Sabale}, {Seader},
  {Shporer}, {Smith}, {Steffen}, {Still}, {Tenenbaum}, {Thompson}, {Twicken},
  {Van Laerhoven}, {Wolfgang}, \& {Zamudio}}]{Rowe2015cat}
{Rowe}, J.~F., {Coughlin}, J.~L., {Antoci}, V., {et~al.} 2015{\natexlab{a}},
  \apjs, 217, 16

\bibitem[{{Rowe} {et~al.}(2015{\natexlab{b}}){Rowe}, {Coughlin}, {Antoci},
  {Barclay}, {Batalha}, {Borucki}, {Burke}, {Bryson}, {Caldwell}, {Campbell},
  {Catanzarite}, {Christiansen}, {Cochran}, {Gilliland}, {Girouard}, {Haas},
  {He{\l}miniak}, {Henze}, {Hoffman}, {Howell}, {Huber}, {Hunter},
  {Jang-Condell}, {Jenkins}, {Klaus}, {Latham}, {Li}, {Lissauer}, {McCauliff},
  {Morris}, {Mullally}, {Ofir}, {Quarles}, {Quintana}, {Sabale}, {Seader},
  {Shporer}, {Smith}, {Steffen}, {Still}, {Tenenbaum}, {Thompson}, {Twicken},
  {Van Laerhoven}, {Wolfgang}, \& {Zamudio}}]{Rowe2015a}
---. 2015{\natexlab{b}}, \apjs, 217, 16

\bibitem[{Ruppert(2010)}]{Ruppert2010}
Ruppert, D. 2010, Statistics and Data Analysis for Financial Engineering
  (Springer Texts in Statistics), 1st edn. (Springer, Berlin)

\bibitem[{{Samus} {et~al.}(2009){Samus}, {Durlevich}, \& {et al.}}]{Samus2015}
{Samus}, N.~N., {Durlevich}, O.~V., \& {et al.} 2009, VizieR Online Data
  Catalog, 1, 2025

\bibitem[{{Santerne} {et~al.}(2012){Santerne}, {D{\'i}az}, {Moutou}, {Bouchy},
  {H{\'e}brard}, {Almenara}, {Bonomo}, {Deleuil}, \& {Santos}}]{Santerne2012}
{Santerne}, A., {D{\'i}az}, R.~F., {Moutou}, C., {et~al.} 2012, \aap, 545, A76

\bibitem[{{Santerne} {et~al.}(2016){Santerne}, {Moutou}, {Tsantaki}, {Bouchy},
  {H{\'e}brard}, {Adibekyan}, {Almenara}, {Amard}, {Barros}, {Boisse},
  {Bonomo}, {Bruno}, {Courcol}, {Deleuil}, {Demangeon}, {D{\'{\i}}az},
  {Guillot}, {Havel}, {Montagnier}, {Rajpurohit}, {Rey}, \&
  {Santos}}]{Santerne2016}
{Santerne}, A., {Moutou}, C., {Tsantaki}, M., {et~al.} 2016, \aap, 587, A64

\bibitem[{{Seader} {et~al.}(2015){Seader}, {Jenkins}, {Tenenbaum}, {Twicken},
  {Smith}, {Morris}, {Catanzarite}, {Clarke}, {Li}, {Cote}, {Burke},
  {McCauliff}, {Girouard}, {Campbell}, {Kamal Uddin}, {Zamudio}, {Sabale},
  {Henze}, {Thompson}, \& {Klaus}}]{Seader2015}
{Seader}, S., {Jenkins}, J.~M., {Tenenbaum}, P., {et~al.} 2015, \apjs, 217, 18

\bibitem[{{Shporer}(2017)}]{Shporer2017}
{Shporer}, A. 2017, \pasp, 129, 072001

\bibitem[{{Shporer} {et~al.}(2011){Shporer}, {Jenkins}, {Rowe}, {Sanderfer},
  {Seader}, {Smith}, {Still}, {Thompson}, {Twicken}, \& {Welsh}}]{Shporer2011}
{Shporer}, A., {Jenkins}, J.~M., {Rowe}, J.~F., {et~al.} 2011, \aj, 142, 195

\bibitem[{{Shporer} {et~al.}(2016){Shporer}, {Fuller}, {Isaacson}, {Hambleton},
  {Thompson}, {Pr{\v s}a}, {Kurtz}, {Howard}, \& {O'Leary}}]{Shporer2016hb}
{Shporer}, A., {Fuller}, J., {Isaacson}, H., {et~al.} 2016, \apj, 829, 34

\bibitem[{{Slawson} {et~al.}(2011){Slawson}, {Pr\v{s}a}, {Welsh}, {Orosz},
  {Rucker}, {Batalha}, {Doyle}, {Engle}, {Conroy}, {Coughlin}, {Gregg},
  {Fetherolf}, {Short}, {Windmiller}, {Fabrycky}, {Howell}, {Jenkins}, {Uddin},
  {Mullally}, {Seader}, {Thompson}, {Sanderfer}, {Borucki}, \&
  {Koch}}]{Slawson2011}
{Slawson}, R.~W., {Pr\v{s}a}, A., {Welsh}, W.~F., {et~al.} 2011, \aj, 142, 160

\bibitem[{{Stark} {et~al.}(2015){Stark}, {Roberge}, {Mandell}, {Clampin},
  {Domagal-Goldman}, {McElwain}, \& {Stapelfeldt}}]{stark2015}
{Stark}, C.~C., {Roberge}, A., {Mandell}, A., {et~al.} 2015, \apj, 808, 149

\bibitem[{{Stumpe} {et~al.}(2014){Stumpe}, {Smith}, {Catanzarite}, {Van Cleve},
  {Jenkins}, {Twicken}, \& {Girouard}}]{Stumpe2014}
{Stumpe}, M.~C., {Smith}, J.~C., {Catanzarite}, J.~H., {et~al.} 2014, \pasp,
  126, 100

\bibitem[{Tange(2011)}]{Tange2011a}
Tange, O. 2011, ;login: The USENIX Magazine, 36, 42

\bibitem[{{Tenenbaum} {et~al.}(2012){Tenenbaum}, {Christiansen}, {Jenkins},
  {Rowe}, {Seader}, {Caldwell}, {Clarke}, {Li}, {Quintana}, {Smith}, {Stumpe},
  {Thompson}, {Twicken}, {Van Cleve}, {Borucki}, {Cote}, {Haas}, {Sanderfer},
  {Girouard}, {Klaus}, {Middour}, {Wohler}, {Batalha}, {Barclay}, \&
  {Nickerson}}]{Tenenbaum2012}
{Tenenbaum}, P., {Christiansen}, J.~L., {Jenkins}, J.~M., {et~al.} 2012, \apjs,
  199, 24

\bibitem[{{Thompson} {et~al.}(2016{\natexlab{a}}){Thompson}, {Fraquelli}, {van
  Cleve}, \& {Caldwell}}]{Thompson2016KAM}
{Thompson}, S.~E., {Fraquelli}, D., {van Cleve}, J.~E., \& {Caldwell}, D.~A.
  2016{\natexlab{a}}, Kepler Archive Manual (KDMC-10008-006)

\bibitem[{{Thompson} {et~al.}(2015{\natexlab{a}}){Thompson}, {Mullally},
  {Coughlin}, {Christiansen}, {Henze}, {Haas}, \& {Burke}}]{Thompson2015b}
{Thompson}, S.~E., {Mullally}, F., {Coughlin}, J.~L., {et~al.}
  2015{\natexlab{a}}, \apj, 812, 46

\bibitem[{{Thompson} {et~al.}(2012){Thompson}, {Everett}, {Mullally},
  {Barclay}, {Howell}, {Still}, {Rowe}, {Christiansen}, {Kurtz}, {Hambleton},
  {Twicken}, {Ibrahim}, \& {Clarke}}]{Thompson2012}
{Thompson}, S.~E., {Everett}, M., {Mullally}, F., {et~al.} 2012, \apj, 753, 86

\bibitem[{{Thompson} {et~al.}(2015{\natexlab{b}}){Thompson}, {Jenkins},
  {Caldwell}, {Barclay}, {Bryson}, {Burke}, {Campbell}, {Catanzarite},
  {Christiansen}, {Clarke}, {Coughlin}, {Girouard}, {Haas}, {Ibrahim}, {Li},
  {McCauliff}, {Morris}, {Mullally}, {Rowe}, {Sabale}, {Seader}, {Smith},
  {Tenenbaum}, {Twicken}, \& {Uddin}}]{Thompson2015a}
{Thompson}, S.~E., {Jenkins}, J.~M., {Caldwell}, D.~A., {et~al.}
  2015{\natexlab{b}}, {Kepler Data Release 24 Notes (KSCI-19064-002)},
  \url{http://archive.stsci.edu/kepler/release\_notes/release\_notes24/KSCI-19064-002DRN24.pdf}

\bibitem[{{Thompson} {et~al.}(2016{\natexlab{b}}){Thompson}, {Caldwell},
  {Jenkins}, {Barclay}, {Barentsen}, {Bryson}, {Burke}, {Campbell},
  {Catanzarite}, {Christiansen}, {Clarke}, {Col\'on}, {Coughlin}, {Girouard},
  {Haas}, {Harrison}, {Ibrahim}, {Klaus}, {Li}, {McCauliff}, {Morris},
  {Mullally}, {Rowe}, {Sabale}, {Seader}, {Smith}, {Tenenbaum}, {Twicken},
  {Uddin}, \& {Van Cleve}}]{DRN25}
{Thompson}, S.~E., {Caldwell}, D.~A., {Jenkins}, J.~M., {et~al.}
  2016{\natexlab{b}}, Kepler Data Release 25 Notes (KSCI-19065-002),
  \url{http://archive.stsci.edu/kepler/release\_notes/release\_notes25/KSCI-19065-002DRN25.pdf}

\bibitem[{{Torres} {et~al.}(2015){Torres}, {Kipping}, {Fressin}, {Caldwell},
  {Twicken}, {Ballard}, {Batalha}, {Bryson}, {Ciardi}, {Henze}, {Howell},
  {Isaacson}, {Jenkins}, {Muirhead}, {Newton}, {Petigura}, {Barclay},
  {Borucki}, {Crepp}, {Everett}, {Horch}, {Howard}, {Kolbl}, {Marcy},
  {McCauliff}, \& {Quintana}}]{Torres2015}
{Torres}, G., {Kipping}, D.~M., {Fressin}, F., {et~al.} 2015, \apj, 800, 99

\bibitem[{{Turbo-King} {et~al.}(2017){Turbo-King}, {Tang}, {Habeertable},
  {Chouffe}, {Exquisit}, \& {Keg-beer}}]{Turbo-King2017}
{Turbo-King}, M., {Tang}, B.~R., {Habeertable}, Z., {et~al.} 2017, ArXiv
  e-prints, arXiv:1703.10803

\bibitem[{{Twicken} {et~al.}(2016){Twicken}, {Jenkins}, {Seader}, {Tenenbaum},
  {Smith}, {Brownston}, {Burke}, {Catanzarite}, {Clarke}, {Cote}, {Girouard},
  {Klaus}, {Li}, {McCauliff}, {Morris}, {Wohler}, {Campbell}, {Kamal Uddin},
  {Zamudio}, {Sabale}, {Bryson}, {Caldwell}, {Christiansen}, {Coughlin},
  {Haas}, {Henze}, {Sanderfer}, \& {Thompson}}]{Twicken2016}
{Twicken}, J.~D., {Jenkins}, J.~M., {Seader}, S.~E., {et~al.} 2016, \aj, 152,
  158

\bibitem[{{Van Cleve} \& {Caldwell}(2009)}]{VanCleve2009}
{Van Cleve}, J.~E., \& {Caldwell}, D.~A. 2009, {Kepler Instrument Handbook
  (KSCI-19033-001)},
  \url{http://archive.stsci.edu/kepler/manuals/KSCI-19033-001.pdf}

\bibitem[{{Van Cleve} \& {Caldwell}(2016)}]{KIH}
---. 2016, {Kepler Instrument Handbook (KSCI-19033-0012},
  \url{http://archive.stsci.edu/kepler/manuals/KSCI-19033-002.pdf}

\bibitem[{{Van Cleve} {et~al.}(2016{\natexlab{a}}){Van Cleve}, {Christiansen},
  {Jenkins}, \& {Caldwell}}]{KDCH}
{Van Cleve}, J.~E., {Christiansen}, J.~L., {Jenkins}, J., \& {Caldwell}, D.~A.
  2016{\natexlab{a}}, {Kepler Data Characteristics Handbook (KSCI-19040-005)},
  \url{http://archive.stsci.edu/kepler/manuals/Data\_Characteristics.pdf}

\bibitem[{{Van Cleve} {et~al.}(2016{\natexlab{b}}){Van Cleve}, {Howell},
  {Smith}, {Clarke}, {Thompson}, {Bryson}, {Lund}, {Handberg}, \&
  {Chaplin}}]{VanCleve2016K2}
{Van Cleve}, J.~E., {Howell}, S.~B., {Smith}, J.~C., {et~al.}
  2016{\natexlab{b}}, \pasp, 128, 075002

\bibitem[{{Welsh} {et~al.}(2011){Welsh}, {Orosz}, {Aerts}, {Brown},
  {Brugamyer}, {Cochran}, {Gilliland}, {Guzik}, {Kurtz}, {Latham}, {Marcy},
  {Quinn}, {Zima}, {Allen}, {Batalha}, {Bryson}, {Buchhave}, {Caldwell},
  {Gautier}, {Howell}, {Kinemuchi}, {Ibrahim}, {Isaacson}, {Jenkins}, {Prsa},
  {Still}, {Street}, {Wohler}, {Koch}, \& {Borucki}}]{Welsh2011}
{Welsh}, W.~F., {Orosz}, J.~A., {Aerts}, C., {et~al.} 2011, \apjs, 197, 4

\bibitem[{{Wu} {et~al.}(2010){Wu}, {Twicken}, {Tenenbaum}, {Clarke}, {Li},
  {Quintana}, {Allen}, {Chandrasekaran}, {Jenkins}, {Caldwell}, {Wohler},
  {Girouard}, {McCauliff}, {Cote}, \& {Klaus}}]{Wu2010}
{Wu}, H., {Twicken}, J.~D., {Tenenbaum}, P., {et~al.} 2010, in {Society of
  Photo-Optical Instrumentation Engineers (SPIE) Conference Series}, Vol. 7740,
  {Society of Photo-Optical Instrumentation Engineers (SPIE) Conference Series}

\bibitem[{{Youdin}(2011)}]{Youdin2011}
{Youdin}, A.~N. 2011, \apj, 742, 38

\bibitem[{{Ziegler} {et~al.}(2017){Ziegler}, {Law}, {Morton}, {Baranec},
  {Riddle}, {Atkinson}, {Baker}, {Roberts}, \& {Ciardi}}]{Ziegler2017}
{Ziegler}, C., {Law}, N.~M., {Morton}, T., {et~al.} 2017, \aj, 153, 66

\bibitem[{{Zimmerman} {et~al.}(2017){Zimmerman}, {Thompson}, {Mullally},
  {Fuller}, {Shporer}, \& {Hambleton}}]{Zimmerman2017}
{Zimmerman}, M., {Thompson}, S., {Mullally}, F., {et~al.} 2017, ArXiv e-prints,
  arXiv:1706.05434

\end{thebibliography}


\appendix
\twocolumngrid
\section{Robovetter Metric Details}
\label{s:metrics}
In this appendix we describe, in detail, each of the Robovetter tests in the order in which they are performed by the Robovetter. See \S\ref{s:robovetter} for an overview of the logic used by the Robovetter.

\subsection{Two Robovetter Detrendings}
As mentioned in \S\ref{abbrev}, for all of the Robovetter tests that require a phased light curve and model fit, we utilize two different detrendings and model fits (named ALT and DV). Both were also used by the DR24 Robovetter. Every test that is applied to the DV phased light curves is also applied to the ALT detrending, albeit with different thresholds for failure. Failing a test using either detrending results in the TCE being classified as an FP.

In the \kepler{} Pipeline, the DV module produces a harmonic-removed, median-detrended, phased flux light curve, along with a transit model fit \citep{JenkinsKDPH,Wu2010}. However, the harmonic removal software is known to suppress or distort short-period ($\lesssim$ 3 days) signals causing short-period eclipsing binaries with visible secondaries to appear as transiting planets with no visible secondaries \citep{Christiansen2013b}. It can also make variable stars with semi-coherent variability, such as star spots or pulsations, appear as transit-like signals. As an alternative, we implement the ALT detrending method that utilizes the pre-search data conditioned (PDC) time-series light curves and the non-parametric penalized least-squares detrending method of \citet{Garcia2010} which includes only the out-of-transit points when computing the filter. This ALT detrending technique is effective at accurately detrending short-period eclipsing binaries and variable stars, i.e., preserving their astrophysical signal.  These ALT detrended light curves are phased and fit with a simple trapezoidal transit model.

\subsection{The TCE is the Secondary of an Eclipsing Binary}
\label{s:issecond}
If a TCE under examination is not the first one in a system, the Robovetter checks if there exists a previous TCE with a similar period that was designated as an FP due to a stellar eclipse (see~\S\ref{sigsecsec}). (Note, TCEs for a given system are ordered from highest MES to lowest MES, and the Robovetter runs on them in this order.) To compute whether two TCEs have the same period within a given statistical threshold, we employ the period matching criteria of \citet[][see equations 1-3]{Coughlin2014a}, $\sigma_{P}$, where higher values of $\sigma_{P}$ indicate more significant period matches. We re-state the equations here as:

\begin{equation}
\label{peq1}
\Delta P = \frac{P_{A}-P_{B}}{P_{A}}\\
\end{equation}

\begin{equation}
\label{peq2}
\Delta P^{\prime} = \textrm{abs}(\Delta P - \textrm{rint}(\Delta P))\\
\end{equation}

\begin{equation}
\label{peq3}
\sigma_{P} = \sqrt{2}\cdot\textrm{erfcinv}(\Delta P^{\prime})\\
\end{equation}

\noindent where $P_{A}$ is the period of the shorter-period TCE, $P_{B}$ is the period of the longer-period TCE, $\mathrm{rint()}$ rounds a number to the nearest integer, $\mathrm{abs()}$ yields the absolute value, and $\mathrm{erfcinv()}$ is the inverse complementary error function. We consider any value of $\sigma_{P}$ $>$ 3.5 to indicate significantly similar periods.

If the current TCE is (1) in a system that has a previous TCE dispositioned as an FP due to a stellar eclipse, (2) matches the previous TCE's period with $\sigma_{P}$ $>$ 3.5, and (3) is separated in phase from the previous TCE by at least 2.5 times the transit duration, then the current TCE is considered to be a secondary eclipse. In this case, it is designated as an FP and is classified into both the not transit-like and stellar eclipse FP categories --- a unique combination that can be used to identify secondary eclipses while still ensuring they are not assigned \kepler{} Object of Interest numbers (see \S\ref{koisec}). Note that since the \kepler{} Pipeline generally identifies TCEs in order of their signal-to-noise, from high to low, sometimes a TCE identified as a secondary can have a deeper depth than the primary, depending on their relative durations and shapes. Also note that it is possible that the periods of two TCEs will meet the period matching criteria, but be different enough to have their relative phases shift significantly over the $\approx$4 year mission duration. Thus, the potential secondary TCE is actually required to be separated in phase by at least 2.5 times the previous TCE's transit duration over the entire mission time frame in order to be labeled as a secondary. Also, the \kepler{} Pipeline will occasionally detect the secondary eclipse of an eclipsing binary at a half, third, or some smaller integer fraction of the orbital period of the system, such that the epoch of the detected secondary coincides with that of the primary. Thus, when a non-1:1 period ratio is detected, we do not impose criteria (3), the phase separation requirement. Note, equations~\ref{peq1}-\ref{peq3} allow for integer period ratios.

\subsection{Not Transit-Like}
\label{nottransitlikesec}

\begin{figure*}[ht]
\centering
\includegraphics[width=\linewidth]{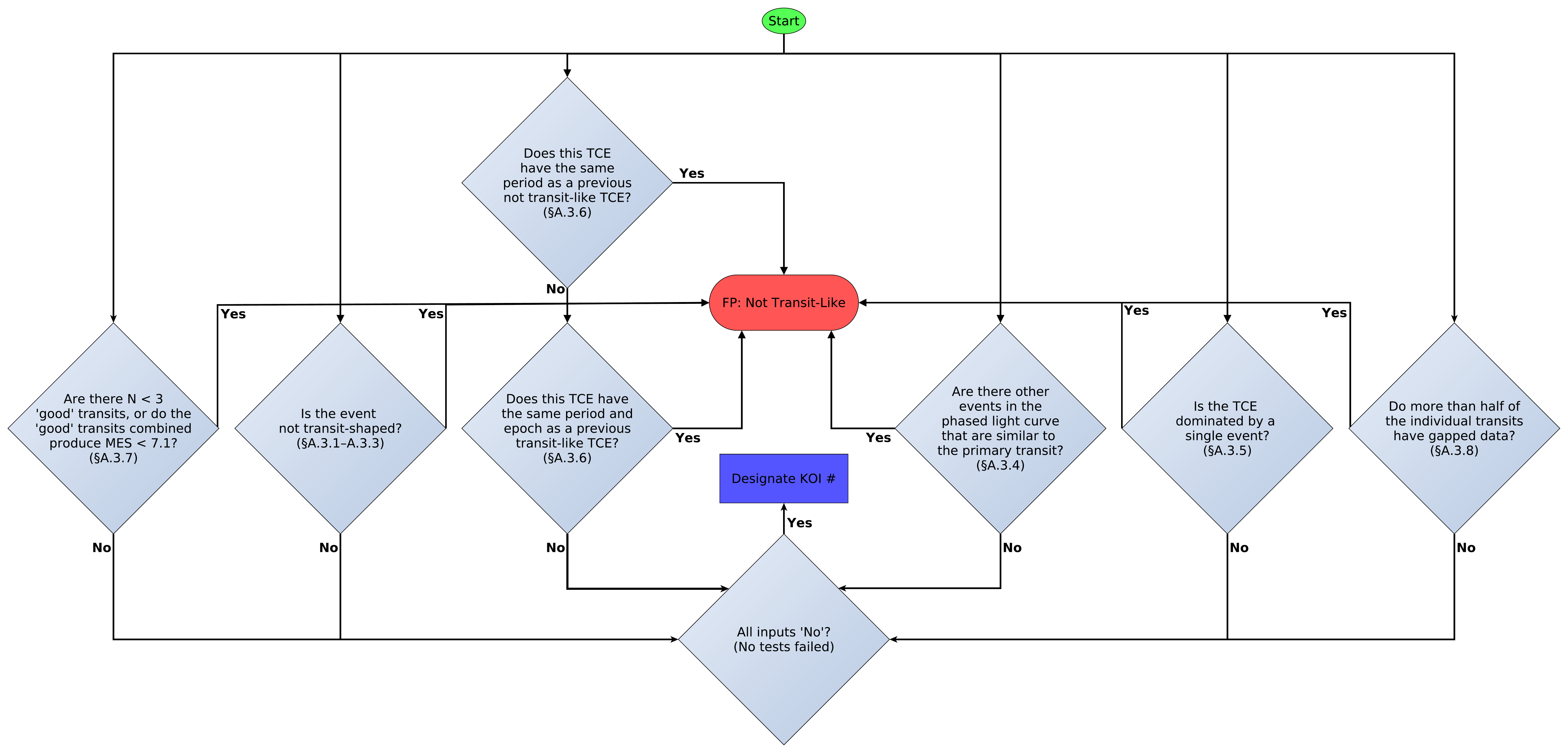}
\caption{The not transit-like flowchart of the Robovetter. Diamonds represent ``yes'' or ``no'' decisions that are made with quantitative metrics. If a TCE fails any test (via a ``yes'' response to any decision) then it is dispositioned as a not transit-like FP. If a TCE passes all tests (via a ``no'' response to all decisions), then it is given a KOI number and passed to the stellar eclipse module (see \S\ref{sigsecsec} and Figure~\ref{robovetter-sigsec-fig}). The section numbers on each decision diamond correspond to the sections in this paper where these tests are discussed.}
\label{robovetter-transitlike-fig}
\end{figure*}

A very large fraction of false positive TCEs have light curves that do not resemble a detached transiting or eclipsing object. These include quasi-sinusoidal light curves from pulsating stars, star spots, and contact binaries, as well as more sporadic light curves due to instrumental artifacts. The first step in the catalog process is to determine whether each TCE is not transit-like. All transit-like \opstce{s} are given \kepler{} Object of Interest (KOI) numbers, which are used to keep track of transit-like systems over multiple \kepler{} Pipeline runs. We employ a series of algorithmic tests to reliably identify these not transit-like FP TCEs, as shown by the flowchart in Figure~\ref{robovetter-transitlike-fig}.

\subsubsection{The LPP Metric}
\label{s:lpp}

Many short-period FPs are due to variable stars that exhibit a quasi-sinusoidal phased light curve. We implement the LPP transit-like metric described by \citet{Thompson2015b} to separate those TCEs that show a transit shape from those that do not. This technique bins the TCE's folded light curve and then applies a dimensionality reduction algorithm called Locality Preserving Projections \citep[LPP, ][]{He2004}.  It then measures the average Euclidean distance in these reduced dimensions to the nearest known transit-like TCEs to yield a single number that represents the similarity of a TCE's shape to that of known transits. 

For the DR25 KOI catalog, we deviated slightly from the method described by \citet{Thompson2015b}\footnote{The code is available here https://sourceforge.net/p/lpptransitlikemetric/}.  The DR24 LPP metric algorithm, when applied to DR25, produced LPP values that were systematically higher for short-period, low-MES TCEs. The transit duration of short period TCEs can be a significant fraction of the orbital period, so when folded and binned these transits have a noticeably different shape. And since we use \injtce{s} as our training set, which has very few short-period examples, there are very few known transits for the algorithm to match to, causing large measured distances for these transit event. The trend with MES is rooted in the fact that when the binned light curve has a lower signal-to-noise, it is less likely for two folded light curves to be similar to each other, creating more scatter in the reduced dimensions, and thus increasing the measured distance to known transits in those dimensions.  

We reduced these dependencies by altering how we calculate the LPP metric for the DR25 KOI catalog. For our set of known transit-like TCEs, we now use the union of the set of recovered \injtce{s} and the set of PCs from the DR24 KOI catalog \citep{Coughlin2016} that were re-found as \opstce{s} in DR25. Including these PCs provides more examples at short period. We also changed how the folded light curve was binned. TCEs with lower MES are given wider bins for those cadences near the transit center, while keeping the total number of bins fixed (99 bins total including 41 for the in-transit portion). Finally, we divide these raw LPP values by the 75$^{th}$ percentile of the raw LPP values for the 100 TCEs that are closest in period.  In this way we reduce the period dependence in the LPP metric.  Generally, the resulting LPP metric values lie near to a value of one, and values greater than $\approx$ 2 appear to be not-transit shaped.  To create the DR25 catalog the Robovetter adopted a threshold of 2.2 for the DV detrending and 3.2 for the ALT detrending.

\subsubsection{Sine Wave Event Evaluation Test}
\label{s:sweetntl}

On occasion, a variable star's variability will have been mostly removed by both the DV and ALT detrendings and will thus appear transit-like. To identify these cases we developed the Sine Wave Event Evaluation Test (SWEET) to examine the PDC data and look for a strong sinusoidal signal at the TCE's period. 

SWEET begins with the PDC data and normalizes each quarter by dividing the time series by the median flux value and subtracting 1.0. Outliers are robustly removed by utilizing a criterion based on the median absolute deviation (MAD) --- specifically, outliers are identified as any point that lies more than $\sqrt{2}\cdot\mathrm{erfcinv}(1/N_{\rm dat})\sigma$ from the median, where $N_{\rm dat}$ is the number of data points, erfcinv is the inverse complementary error function, and 1$\sigma$=1.4826$\cdot$MAD \citep[see][]{Hampel1974,Ruppert2010}. Three different sine curves are fitted to the resulting data, with their periods fixed to half, exactly, and twice the TCE period, with their phase, amplitude, and offset allowed to vary. Of the three fits, the one with the highest signal-to-noise ratio, defined as the amplitude divided by its error, is chosen as the strongest fit. If a TCE has a SWEET signal-to-noise ratio greater than 50, an amplitude greater than the TCE transit depth in both the DV and ALT detrendings, and has a period less than 5.0 days, it fails as not transit-like.


\subsubsection{TCE Chases}
\label{s:tcechases}

In \S\ref{s:chases} we describe an individual transit metric called Chases that assesses the detection strength of individual transit events relative to other signals nearby in time. TCE Chases takes the median value of these individual transit measurements.  When the median value is less than 0.8 the TCE fails as not-transit-like.  As with the individual Chases metric, TCE Chases is only calculated when the TCE has five or fewer transit events contributing to the signal.  With more than five transit events, the individual transit events are not expected to be statistically significant, and the assumptions of the Chases metric no longer apply.

\subsubsection{The Model-Shift Uniqueness Test}
\label{s:ms}

If a TCE under investigation is truly a PC, there should not be any other transit-like events in the folded light curve with a depth, duration, and period similar to the primary signal, in either the positive or negative flux directions, i.e., the transit event should be unique in the phased light curve. Many FPs are due to noisy, quasi-periodic signals (see \S\ref{tcesec}) and thus are not unique in the phased light curve. In order to identify these cases, we developed a ``model-shift uniqueness test'' and used it extensively for identifying false positives in the Q1--Q12 \citep{Rowe2015a}, Q1--Q16 \citep{Mullally2015cat}, and DR24 \citep{Coughlin2016} planet candidate catalogs.

See \S3.2.2 of \citet{Rowe2015a} and page 23 of \citet{Coughlin2017b} for figures and a detailed explanation of the ``model-shift uniqueness test'', \added{as well as the publicly available code\footnote{\url{https://github.com/JeffLCoughlin/Model-Shift}}.}. Briefly, after removing outliers, the best-fit model of the primary transit is used as a template to measure the best-fit depth at all other phases. The deepest event aside from the primary (pri) transit event is labeled as the secondary (sec) event, the next-deepest event is labeled as the tertiary (ter) event, and the most positive (pos) flux event (i.e., shows a flux brightening) is labeled as the positive event. The significances of these events ($\sigma_{\rm Pri}$, $\sigma_{\rm Sec}$, $\sigma_{\rm Ter}$, and $\sigma_{\rm Pos}$) are computed assuming white noise as determined by the standard deviation of the light curve residuals. Also, the ratio of the red noise (at the timescale of the transit duration) to the white noise ($F_{\rm Red}$) is computed by examining the standard deviation of the best-fit depths at phases outside of the primary and secondary events.  

When examining all events among all TCEs, assuming Gaussian noise, the minimum threshold for an event to be considered statistically significant is given by

\begin{equation}
    FA_{1} = \sqrt{2}\cdot \textrm{erfcinv}\left(\frac{T_{\rm dur}}{P \cdot N_{\rm TCEs}}\right)
\end{equation}

\noindent where $T_{\rm dur}$ is the transit duration, $P$ is the period, and N$_{\rm TCEs}$ is the number of TCEs examined. (The quantity $P$/$T_{\rm dur}$ represents the number of independent statistical tests for a single target.) When comparing two events from the same TCE, the minimum difference in their significances in order to be considered distinctly different is given by

\begin{equation}
    FA_{2} = \sqrt{2}\cdot \textrm{erfcinv}\left(\frac{T_{\rm dur}}{P}\right)
\end{equation}

\noindent We compute the following quantities to use as decision metrics:

\begin{equation}
    MS_{1} = FA_{1} - \sigma_{\rm Pri}/F_{\rm Red}
\end{equation}

\begin{equation}
    MS_{2} = FA_{2} - (\sigma_{\rm Pri} - \sigma_{\rm Ter})
\end{equation}

\begin{equation}
    MS_{3} = FA_{2} - (\sigma_{\rm Pri} - \sigma_{\rm Pos})
\end{equation}

In the Robovetter, we disposition a TCE as a not transit-like FP if either $MS_{1}$~$>$~1.0, $MS_{2}$~$>$~2.0, or $MS_{3}$~$>$~4.0 in the DV detrending, or if either $MS_{1}$~$>$~-3.0, $MS_{2}$~$>$~1.0, or $MS_{3}$~$>$~1.0 in the ALT detrending. These criteria ensure that the primary event is statistically significant when compared to the systematic noise level of the light curve, the tertiary event, and the positive event, respectively. We also fail TCEs as not transit-like if $\sigma_{\rm Pri}$ exactly equals zero in both the DV and ALT detrendings. A value of zero indicates that the fit failed for both detrendings, and suggests that something is fundamentally flawed with the TCE.

\subsubsection{Dominated by Single Event}
\label{s:sesmes}

The depths of individual transits of planet candidates should be equal to each other, and thus assuming constant noise levels, the SNR of individual transits should be nearly equivalent as well. In contrast, most of the long-period FPs that result from three or more equidistant systematic events are dominated in SNR by one event. The \kepler{} Pipeline measures detection significance via the Multiple Event Statistic (MES), which is calculated by combining the Single Event Statistic (SES) of all the individual events that comprise the TCE --- both the MES and SES are measures of SNR. Assuming all individual events have equal SES values,

\begin{equation}
{\rm MES} = \sqrt{N_{\rm Trans}} \cdot {\rm SES}
\end{equation}

\noindent where $N_{\rm Trans}$ is the number of transit events that comprise the TCE. Thus, SES/MES = 0.577 for a TCE with three transits, and less for a greater number of transits. If the largest SES value of a TCE's transit events, ${\rm SES}_{\rm Max}$, divided by the MES is much larger than 0.577 (regardless of the number of transits), this indicates that one of the individual events dominates when calculating the SNR.

In the Robovetter, for TCEs with periods greater than 90 days, if ${{\rm SES}_{\rm Max} / {\rm MES} > 0.8}$ it is dispositioned as a not transit-like FP. The period cutoff of 90 days is applied because short-period TCEs can have a large number of individual transit events, which dramatically increases the chance of one event coinciding with a large systematic feature, thus producing a large ${{\rm SES}_{\rm Max} / {\rm MES}}$ value despite being a valid planetary signal.

\subsubsection{Previous TCE With Same Period}
\label{s:sameperiod}
Most quasi-sinusoidal FPs produce multiple TCEs at the same period, or at integer ratios of each other. If a TCE in a system has been declared as not transit-like due to another test, it is logical that all subsequent TCEs in that system at the same period, or ratios thereof, should also be dispositioned not transit-like. Thus, we match the period of a given TCE to all previous not transit-like FPs via equations~\ref{peq1}-\ref{peq3}. If the current TCE has a period match with $\sigma_{P}$ $>$ 3.25 to a prior not transit-like FP, it is also dispositioned as a not transit-like FP.

Similarly, some TCEs are produced that correspond to the edge of a previously identified transit-like TCE in the system. This often results when the previous TCE corresponding to a transit or eclipse is not completely removed prior to searching the light curve for another TCE. Thus, we match the period of a given TCE to all previous transit-like TCEs via equations~\ref{peq1}-\ref{peq3}.  If the current TCE has a period match with $\sigma_{P}$ $>$ 3.25 to a prior transit-like FP, and the two epochs are separated in phase by less than 2.5 transit durations, the current TCE is dispositioned as a not transit-like FP. For clarity, we note that it is sometimes possible that the periods of two TCEs will meet the period matching criteria, but be different enough to have their epochs shift significantly in phase over the $\sim$4 year mission duration. Thus, if they are separated in phase by less than 2.5 transit durations at any point in the mission time frame, the current TCE is dispositioned as a not transit-like FP.

\subsubsection{Individual Transit Metrics}
\label{s:indivtrans}
A new approach implemented in DR25 is to examine individual transit events for each TCE and determine if they are transit-like. After rejecting these ``bad'' transit events, we check if either

\begin{itemize}
\item There are less than 3 ``good'' events left
\item The re-computed MES using only `good' events is $<$ 7.1
\end{itemize}

\noindent If either of these conditions are met, then the TCE is failed as not transit-like. This is in line with the \kepler{} mission requirement of at least three valid transit events with a MES~$\ge$~7.1 in order to generate a TCE. In the following subsections we list the various tests we apply to each individual transit event.

\paragraph{Rubble -- Missing Data}
\label{s:rubble}
A number of TCEs from the \kepler{} Pipeline are based on transit events that are missing a significant amount of data either in-transit or just before and/or after. These tend to be false positives that are triggering on edges of gaps, or cases were a large amount of data has been removed and a TCE is being created from the residuals of previous TCEs in the system. We thus devised the Rubble metric to clean-up these fragments from the TCE list. The Rubble value for each individual transit is computed by dividing the number of \Kepler\ cadences that are available in the DV time series by the number of cadences expected across two transit durations given \Kepler's regular 29.42\,min cadence and the transit duration provided by the DV fit. If the Rubble value for the transit falls below threshold, then that transit is not counted as a valid transit. We adopted a threshold value of 0.75 to generate the DR25 KOI Catalog.

\paragraph{Marshall -- Transit Shape}
\label{s:marshall}
In the DR24 KOI Catalog, \citet{Coughlin2016} used the Marshall algorithm \citep{Mullally2016} to identify and reject false alarm TCEs caused by short period transients in the data. Marshall fits the proposed transit with models of various transients and uses a Bayesian Information Criterion (BIC) to decide which model is the best explanation for the data. Simulations in \citet{Mullally2016} showed that Marshall was 95\% complete for TCEs with periods $>150$\,days and correctly rejected 66\% of simulated artifact events. The limit on Marshall's effectiveness at eliminating false alarms was that it used a parabola to describe the out-of-transit flux, which failed to capture much of the real observed stellar variability. To ensure high completeness, Marshall was tuned to prevent a variable continuum from causing true transits to be rejected, at the cost of a lower effectiveness.

For the DR25 KOI catalog, we use a Gaussian Process approach \citep[GP,][]{Rasmussen10} to provide an improved continuum model and increase our effectiveness, while maintaining our high completeness. Briefly, our approach aims to model the covariance in the light curve to better fit the trends in our data.
A similar approach was used by \citet{ForemanMackey16} to model single transits due to very long period planets ($P > 1000$\,days).

Our procedure is as follows. For each individual proposed transit event, we select a snippet of PDC data 30 times the reported transit duration centered on the event. Where the event happens near the start (or end) of a quarter, we take a snippet of similar length anchored at the start (or end) of the quarter. We use the \emph{George} package \citep{Ambikasaran14} to fit the covariance of the out-of-transit flux with an exponential squared function, $ {\mathrm{Cov(\delta t})} = A \exp{ (\delta t/\ell)^2}$, where $A$ and $\ell$ are tunable parameters. 

We next fit four models to the entire snippet.

\begin{equation}
\left.\begin{aligned}
G(t | A, \ell) + y_0 \\
G(t | A, \ell) + y_0 + S(t)\\
G(t | A, \ell) + y_0 + S(t)(1 - \exp{\beta t})\\
G(t | A, \ell) + y_0 + S(t - \tau/2) - S(t + \tau/2) 
\end{aligned}\right.
\label{e:marshall}
\end{equation}

\noindent
where $G$ is the Gaussian Process model with the tunable parameters held fixed to those found earlier, and $y_0$ is a constant offset. $S(t)$ is given by

\begin{equation}
S(t) = \frac{d}{1 + e^{-\gamma (t-t_0)} }
\end{equation}

\noindent
where $d$ and $t_0$ are tunable parameters and $\gamma$ is a positive constant. This function, known as a sigmoid (or logistic) function, has asymptotes of 0 for $t<<t_0$, and d for $t>>t_0$. The function transitions quickly, but smoothly, between the two states near $t=t_0$, where it takes on a value of d/2. 

By using a sigmoid and avoiding the discontinuities present in the models used by the original Marshall algorithm \citep{Mullally2016} we can use the L-BFGS-B algorithm \citep{Byrd95} available in the Scipy package \footnote{\url{www.scipy.org}} instead of the less robust Nelder-Mead.

The second function in equation \ref{e:marshall} models a discrete jump in the data. We fit this model seeded with a negative-going dip at the predicted time of ingress, and also with a positive-going spike at the predicted egress, as we see both types of features in \Kepler\ data. The third model fits a Sudden Pixel Sensitivity Drop (SPSD) event, probably caused by a cosmic ray hit on the detector. The last model approximates a box transit. By varying the parameter $\gamma$ we could in principle model transit ingress and egress, but find that extra degree of freedom is not necessary to fit the low signal-to-noise events of most concern.

For each transit the Marshall method returns the BIC score, the preferred model, and the difference between the BIC scores of the preferred model and the sigmoid box fit.  A transit is considered sufficiently bad when this difference (also known as the Marshall score) exceeds a particular threshold, as with the original Marshall algorithm.  However, in a few cases the Gaussian process fails and yields extremely large, unbelievable BIC values. In these cases the transit is set to always pass.  Also, for low MES transits, the expected SES of a transit is sufficiently low that Marshall will be unable to distinguish between the ``no transit'' model and a low signal-to-noise transit.  Because of this the Robovetter declares a specific transit is not valid if all of the following criteria are met:

\begin{itemize}
\item The BIC score of the best-fitting non-transit model is at least 10 lower than the BIC of the transit-model
\item The BIC score of the best-fitting non-transit model is less than 1.0E6
\item Either MES/$\sqrt{\textrm{N}_{\textrm{RealTrans}}}$~$>$~4.0 or the lowest BIC model is for the constant offset model, 
\end{itemize}

Note, N$_{\textrm{RealTrans}}$ is the total number of observed transit events for the TCE. The Marshall code used for the DR25 KOI catalog is available on sourceforge\footnote{ \url{https://sourceforge.net/projects/marshall/}}.

\paragraph{Chases -- SES artifacts}
\label{s:chases}

The Chases metric was developed to chase-down non-transit like events on long period, low MES TCEs. Qualitatively, the metric mimics the human vetting preference to classify a TCE as a PC when individual transit events ``stand-out'' as a unique, transit-like signal from a visual inspection of the \kepler\ flux time-series data.  In order to quantify this human vetting preference, we developed the the Chases algorithm. Chases uses the SES time series generated by the TPS module of the \kepler\ Pipeline \citep{JenkinsKDPH}.  The SES time series measures the significance of a transit signal centered on every cadence.  Details of calculating the SES time series is given in \citet{Jenkins2002a} and illustrative examples are given in \citet{Tenenbaum2012}. A transit produces a peak in the SES time series (as do systematic signals). TPS searches the SES time series for equally spaced peaks indicative of a series of transits. The series of individual peaks in the SES time series are combined to form the MES employed as the primary threshold for detecting a transit signal \citep{Jenkins2002a,Twicken2016,JenkinsKDPH}.

\begin{figure}[htb]
\includegraphics[width=0.5\textwidth]{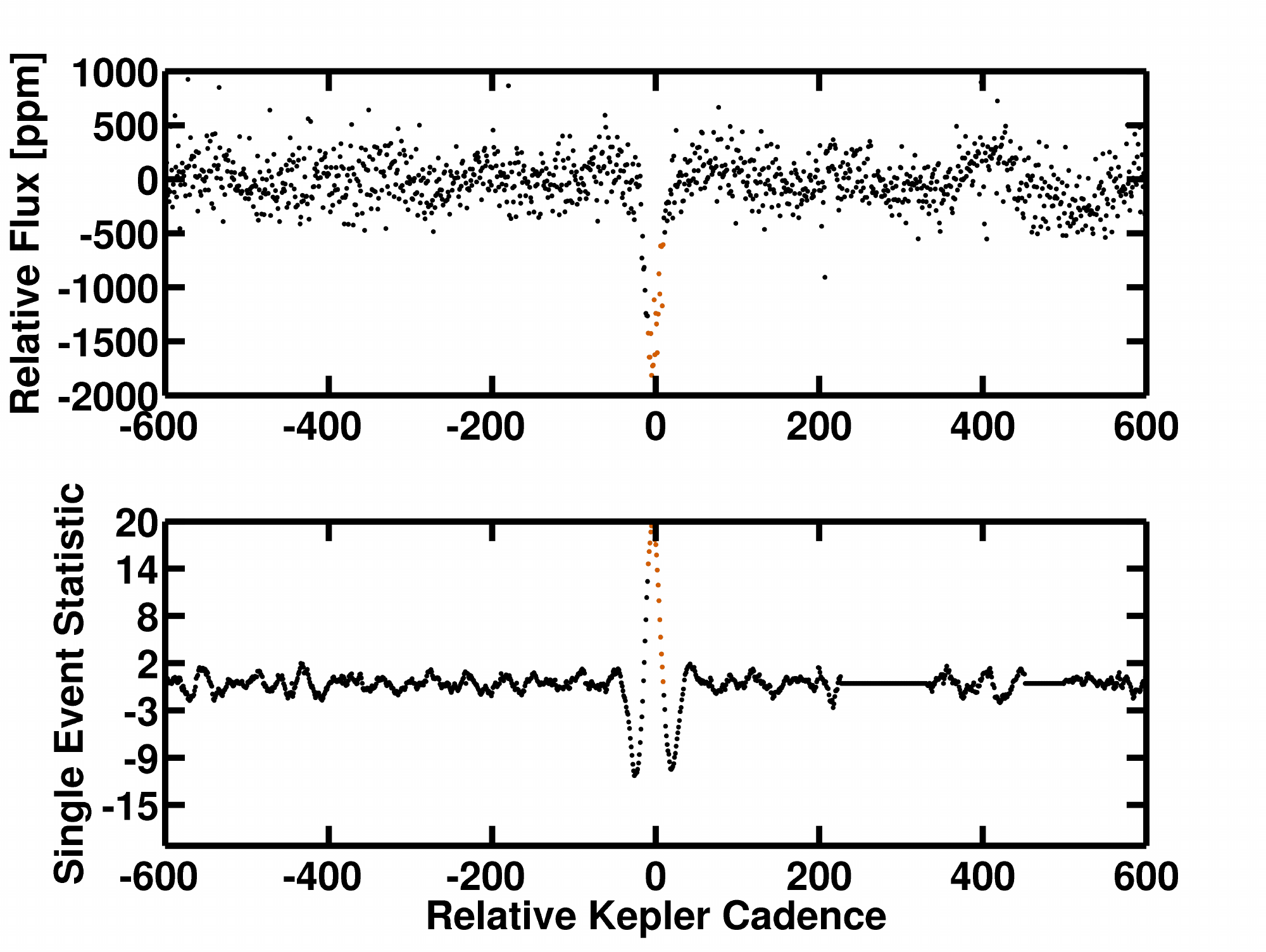}
\caption{Upper panel: flux time series for a single transit event contributing to the TCE for KOI 3900.01 on target KIC~11911580 (black points).  The cadences in transit (orange points) show a significant flux decrement relative to the baseline flux level.  Lower panel: SES time series of the transit event show in the upper panel, representing the archetypal shape of a transit signal displaying a strong central peak with two low-amplitude, symmetric side troughs. There are no other events as strong as the transit nearby in time so this signal has an individual transit event Chases metric, $Ch_{i}=1$.}
\label{fig:chases1}
\end{figure}

The Chases metric quantifies how well the SES peaks contributing to a TCE approximate the expected shape and significance (relative to neighboring data) of a bona fide transit signal.  Figure~\ref{fig:chases1} shows the detrended flux time series (upper panel) and the corresponding SES time series (lower panel) for a clear single transit event contributing to the TCE detection of K03900.01 on target KIC~11911580.  The flux time series, with a very clear decrement during in-transit cadences (orange points), has the archetypal SES time series of a strong central peak with two low-amplitude, symmetric side troughs (caused by the way TPS uses wavelets to modify the model transits when calculating the SES, see \citealt{JenkinsKDPH}). 

The Chases metric for an individual transit event is formulated by identifying the maximum SES value for cadences in transit, $SES_{\rm max}$ (in Figure~\ref{fig:chases1}, $SES_{\rm max}\approx20$).  Next, excluding cadences within 1.5$\tau_{\rm dur}$ of mid transit (to avoid the symmetric side troughs), where $\tau_{\rm dur}$ is the detected transit duration, the SES time series is searched for $\Delta_{t}$, the temporally closest feature to mid transit in the absolute value of the SES time series, $\lvert SES \rvert$. A feature is defined as when  $\lvert SES \rvert>f\, SES_{\rm max}$, where $f$ represents a tunable fraction of the peak in the SES time series.  Finally, we define a maximum window $\Delta_{tmax}=P_{orb}/10$ with which to search for a comparable peak in $\lvert SES \rvert$, and form the final Chases metric for an individual transit event as $C_{i} = {\rm min}(\Delta_{t},\Delta_{tmax})/\Delta_{tmax}$.  

\begin{figure}[htb]
\includegraphics[width=0.5\textwidth]{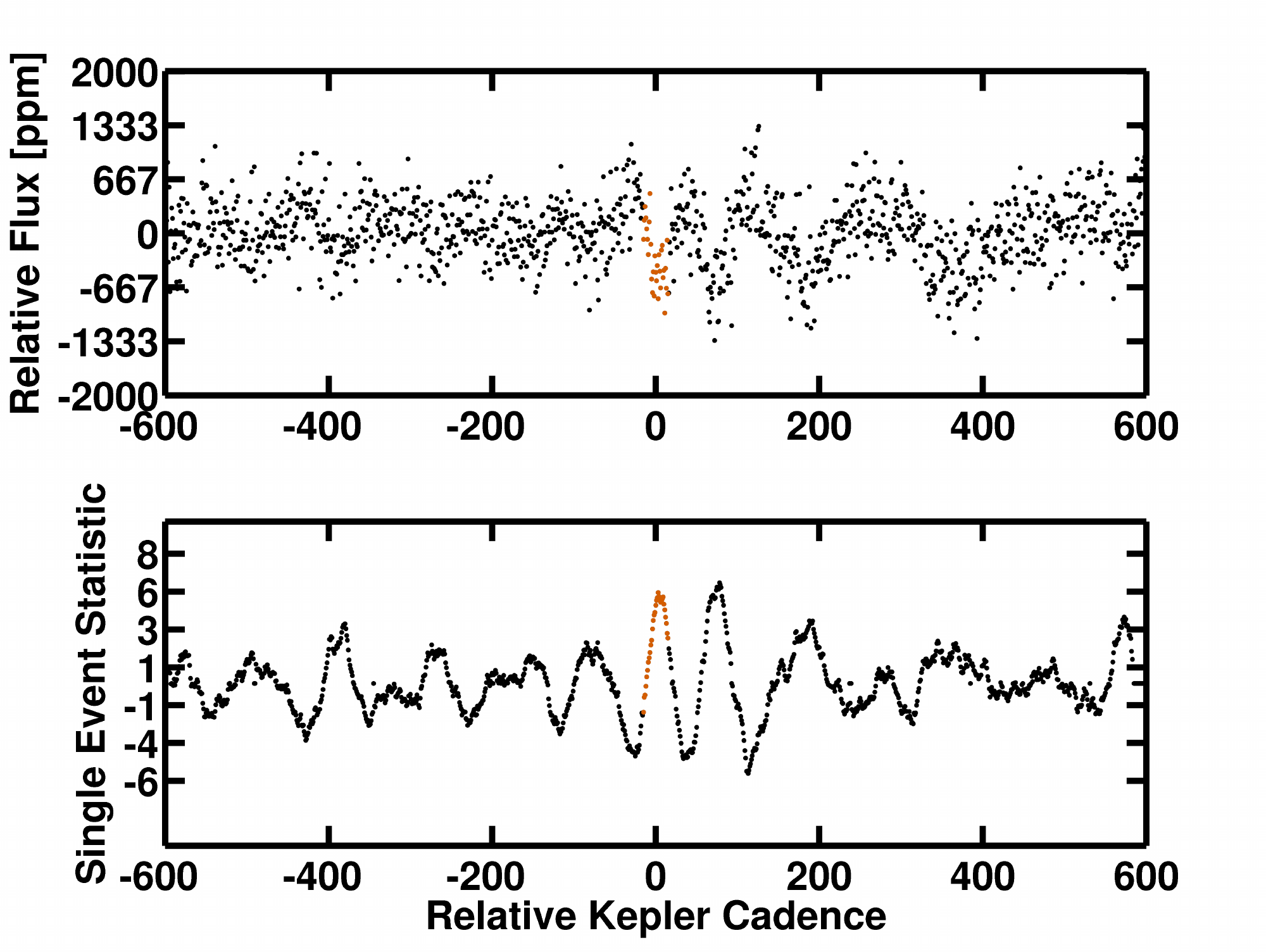}
\caption{Upper panel: flux time series for a single transit event contributing to the TCE on target KIC~11449918 (black points).  The cadences in transit (orange points) show a flux decrement, but there are numerous other flux decrements of similar depth and shape.  The instrumental ``rolling band'' pattern noise contributes systematics to the flux time series of target KIC 11449918 causing numerous signal detections.  Lower panel: SES time series of the transit event shown in the upper panel, representing the non-unique nature of the SES peak relative to surrounding data. The neighboring peak of comparable strength in the SES time series would result in $Ch_{i}=.016$ and the transit would be considered "bad" by Chases. }
\label{fig:chases2}
\end{figure}

\begin{figure}[htb]
\includegraphics[width=0.5\textwidth]{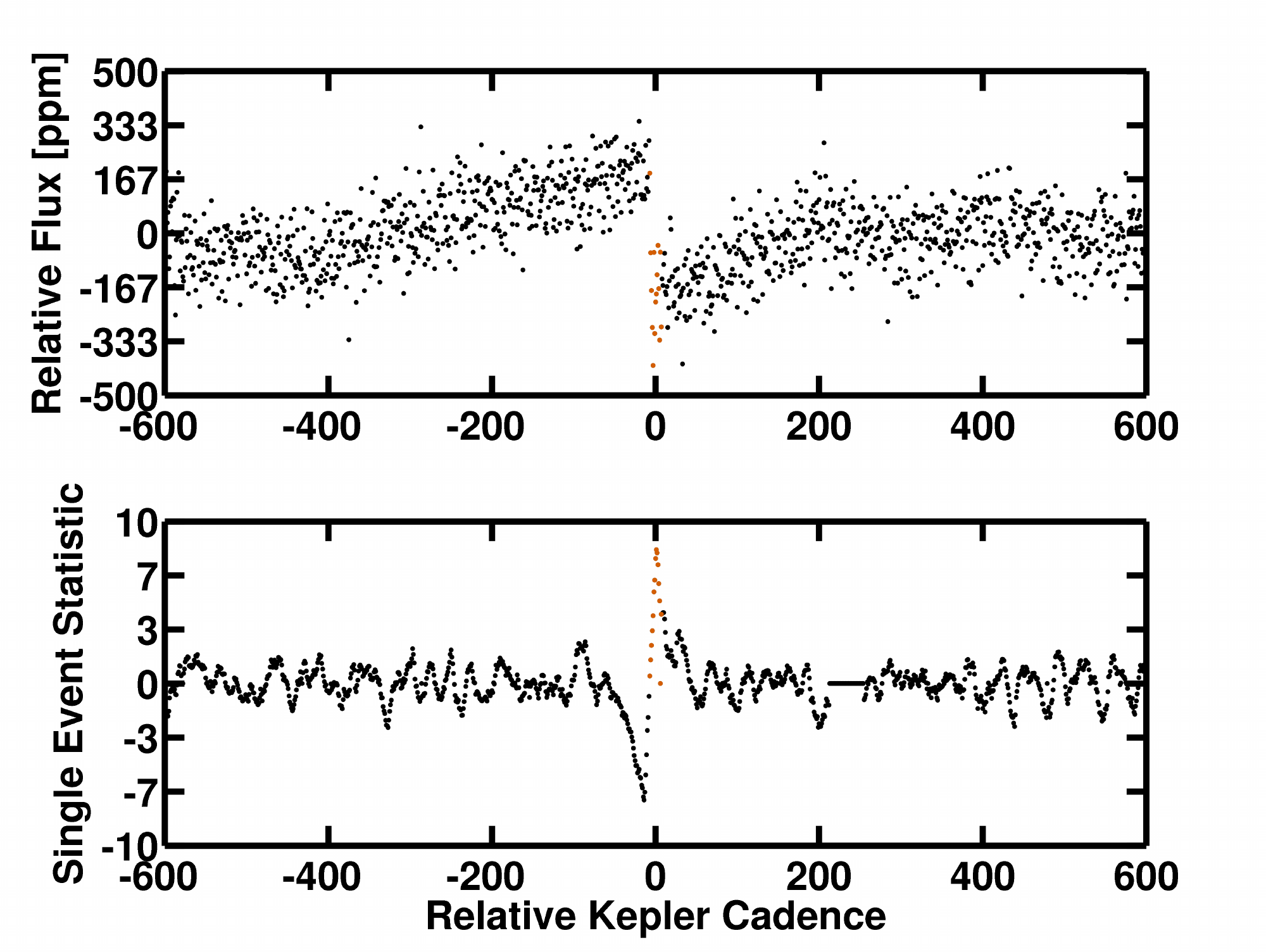}
\caption{Upper panel: flux time series for a single transit event contributing to the TCE on target KIC 12357074 (black points).  The cadences in transit (orange points) show a flux decrement, but the sudden drop in flux followed by the gradual return to the baseline is archetypal of the SPSD instrumental signature.  Lower panel: SES time series for the transit event shown in the upper panel, illustrating the strongly asymmetric SES peak having a comparable amplitude negative SES trough preceding the SES peak. The neighboring trough of comparable absolute strength to the transit's peak would result in $Ch_{i}=.005$ and the transit would be considered "bad" by Chases.}
\label{fig:chases3}
\end{figure}

A value of $C_{i}=1$ indicates that there is no comparable peak/trough in the SES time series within $f$ of $SES_{\rm max}$ over the interval $\Delta_{tmax}$ of the transit signal.  Thus, $Ch_{i}=1$ is consistent with a unique, transit-like signal.  A value of $Ch_{i}\approx0$ indicates that a comparable strength feature is present in the SES time series temporally close to the transit event, and is consistent with the human vetting tendency to dismiss such signals as spurious.  Figure~\ref{fig:chases2} shows an example of a spurious TCE detection on the target KIC 11449918.  The target is on a detector suffering from elevated levels of the ``rolling-band'' image artifacts as described in \S\ref{s:skye}.  The neighboring peak of comparable strength in the SES time series would result in $Ch_{i}\approx0$ for this individual transit event.  The Chases metric is also sensitive to the shape of the transit signal as illustrated in Figure~\ref{fig:chases3}. The SPSD shown in Figure~\ref{fig:chases3} is a spurious instrumental signal with an asymmetric shape. Because Chases uses the absolute value of the SES, $Ch_{i}\approx0$ for these types of events.


For each TCE with five or fewer transit events contributing to the signal, $Ch_{i}$ is calculated for every transit event.  With more than five transit events, the individual transit events are not expected to be statistically significant, and the assumptions of the Chases metric no longer apply. The individual transit event $Ch_{i}$ values were used to recalculate the MES (see \S\ref{s:indivtrans}). Transit events with $Ch_{i}<0.01$ were excluded from the Robovetter's MES calculation.

\paragraph{Skye -- Image Artifacts Clustered by Skygroup}
\label{s:skye}

As discussed in \ref{s:tces}, there are a number of TCEs caused by rolling-band image artifacts. These artifacts are caused by a spatial pattern in the CCD bias level that moves across the chip in response to changes in the temperature of the chip \citep[for more detail see ][]{VanCleve2009}. If a number of individual transit events from TCEs on different targets, but the same skygroup (region of the sky that falls on the same CCD each quarter), occur at the same time, they are very likely systematic in origin. The metric called Skye looks for an excess in the number of individual events occurring at the same time in the same skygroup. If an excess is identified we consider these events to be caused by artifacts. 

\begin{figure}[ht]
\centering
\begin{tabular}{c}
\includegraphics[width=\linewidth]{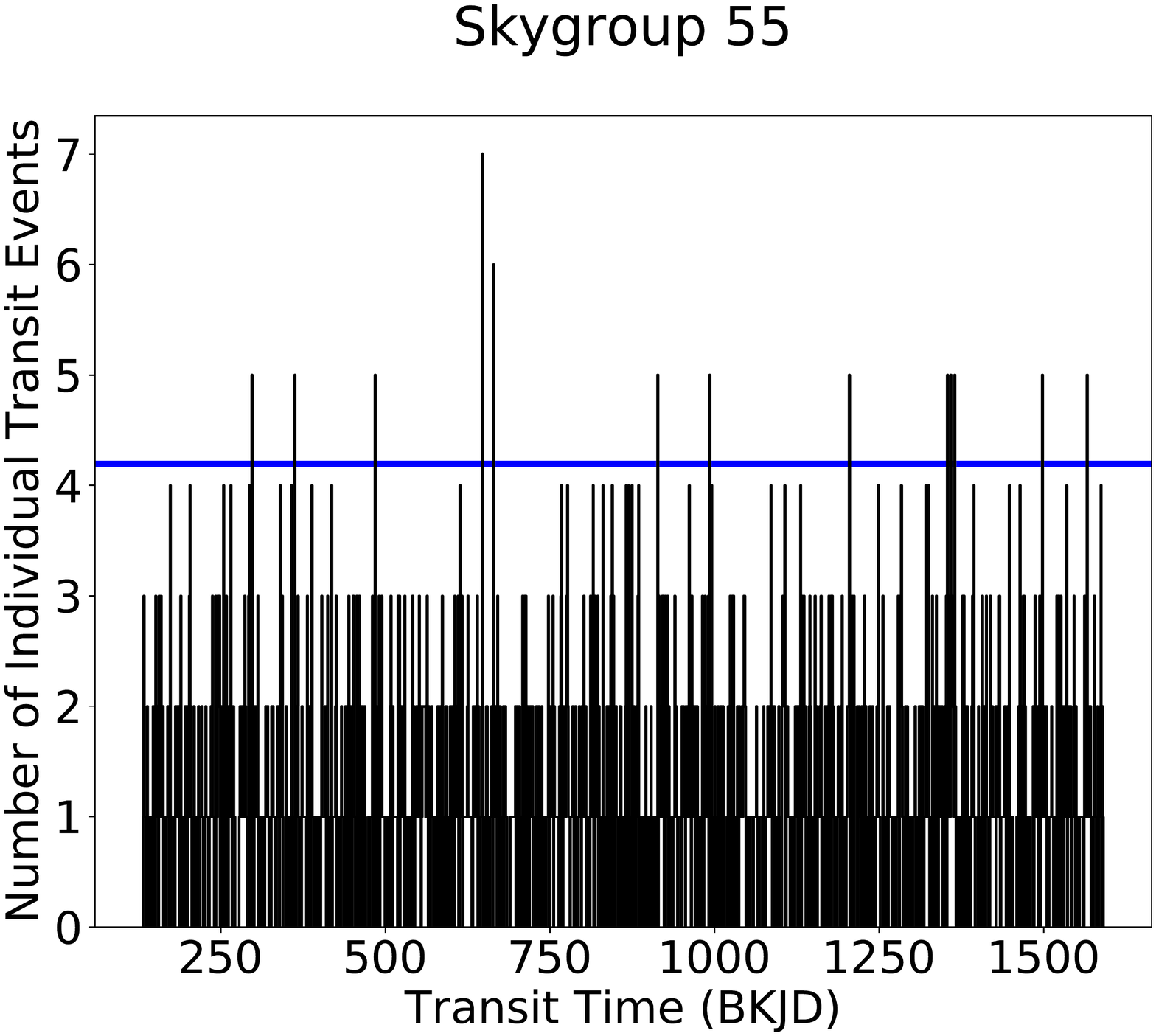}\\
\includegraphics[width=\linewidth]{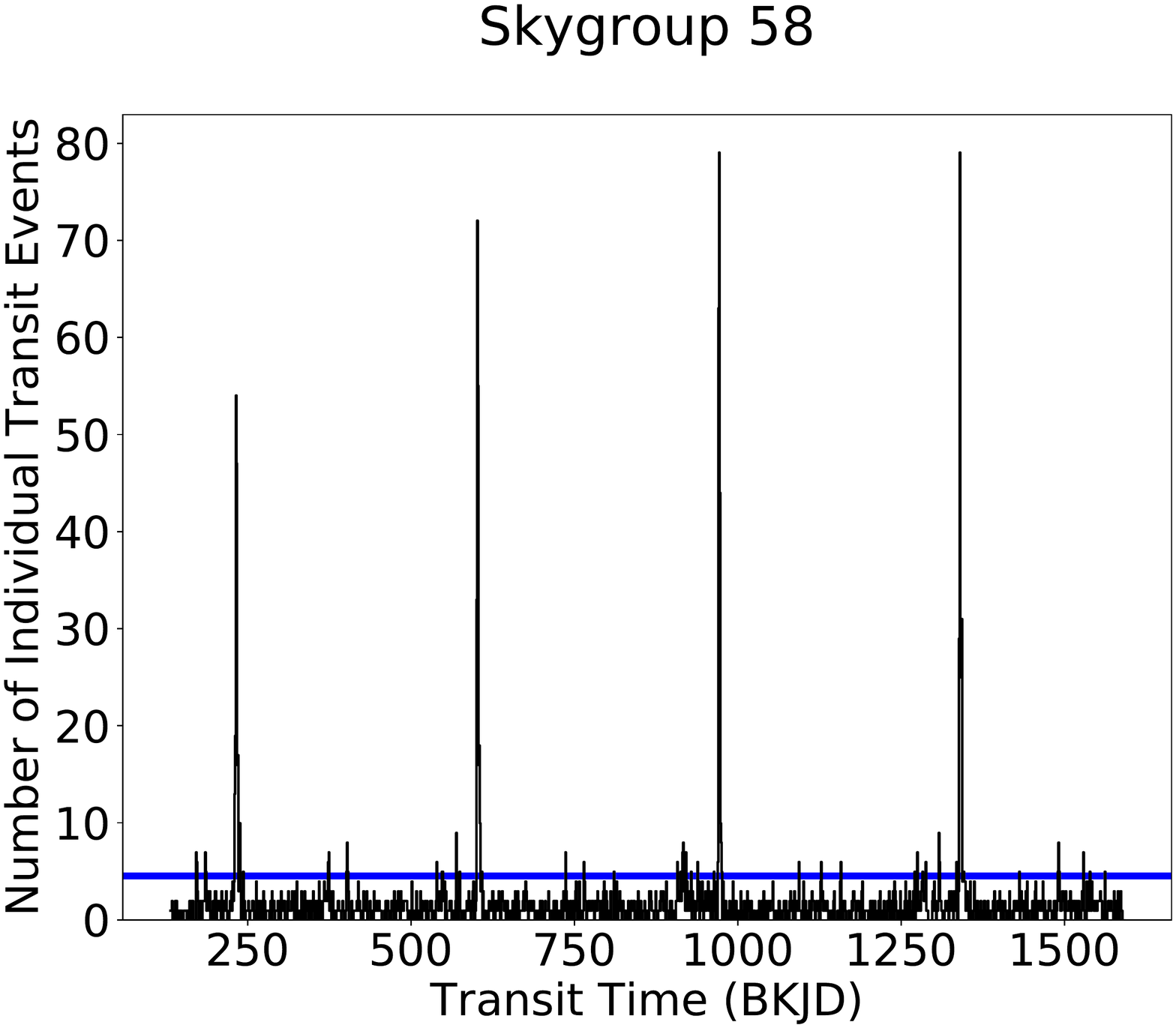}
\end{tabular}
\caption{An example of how the Skye metric flags individual transit events. The plots show the number of individual transit events (from TCEs with periods greater than 45 days) that occur in one-day time bins throughout the mission duration. Two of the 84 skygroups were chosen to be shown as examples, with skygroup 55 plotted on top, and skygroup 58 plotted on bottom. Skygroup 58 (lower panel) has a strong clustering of transit events at times that correspond to the $\sim$372~day orbital period of the spacecraft, as the stars belonging to skygroup 58 fall on CCD channels with strong rolling-band signal. In contrast, skygroup 55 is nearly uniform. Individual transits that occur in a one-day time bin with a number of transit events above the threshold (shown by the blue horizontal line; see Equation~\ref{eq:skye}) are flagged as bad transits due to the Skye metric.}
\label{skyefig}
\end{figure}

More specifically, for each skygroup we bin the individual events into 1.0\,d bins. We only use those \opstce\ with periods greater than 45\,d ($\sim$half a \Kepler\ quarter) for each skygroup. The reason for the period cut is that the long-period \opstce{s} are likely to be affected by rolling-band systematics, but the short-period ones are not.  Including shorter period TCEs would dramatically increase the number of individual transits and would reduce the significance of the anomalous peaks.  See Figure~\ref{skyefig} for an example of the anomalous peaks seen in some skygroups when the data is binned in this way.

To determine which events are anomalous, for each skygroup, we compute the average rate ($R$) of transits, by dividing the overall number of individual transit events in the skygroup by the number of 1.0\,d bins. Assuming the majority of transits are randomly distributed in time, and utilizing Poisson counting statistics, any peaks greater than:

\begin{equation}
\label{eq:skye}
\mathrm{threshold} = R + N\cdot\sqrt{R}
\end{equation}

\noindent are statistically significant and indicative of temporal clustering, given a chosen value for $N$. We choose a value of $N = 3.0$, and robustly determine the rate for each skygroup by first computing the threshold using all the bins, then iteratively rejecting all bins with a height greater than threshold and re-computing threshold until it converges and does not change with further iterations.

For each skygroup and its threshold, we identify the individual times-of-transit for TCEs belonging to the skygroup that fall in bins that are above the threshold. We assign Skye a value of 1.0 to these individual transits to indicate they are bad transits. The Skye value for all other transit times are set to zero. \added{The Skye code is publicly available on github\footnote{\url{https://github.com/JeffLCoughlin/skye}}.}

\paragraph{Zuma -- Negative Significance}
\label{s:zuma}

A valid transit-like TCE should be comprised of individual events that correspond to flux decrements. If any event instead shows an increase of flux then that event is suspect. We thus designate any individual transit event with SES~$<$~0 as ``bad''.

\paragraph{Tracker -- Ephemeris Slip}
\label{s:tracker}
After the TPS module of the \kepler{} Pipeline detects a TCE, it is sent to DV to be fit with a full transit model. DV allows the period and epoch to vary when fitting in order to provide as accurate a fit as possible. Sometimes the TPS ephemeris and DV ephemeris can end up significantly different. When this occurs it indicates that the underlying data is not transit-like and the TCE is likely due to quasi-sinusoidal systematics, which cause the ephemeris to wander when fitting.

Tracker measures (i.e., keeps track of) the time difference between the TPS and DV linear ephemerides in units of the TCE's duration for each transit. When Tracker is greater than  0.5 $T_{\mathrm{dur}}$ for any transit we designate the transit as bad.

\subsubsection{Fraction of Gapped Events}
\label{s:rocky}

Due to the method of data gapping employed in TPS, sometimes the \kepler{} Pipeline can create a TCE that has a majority of its individual events occur where there is no actual in-transit data. This tends to happen particularly in multi-TCE systems, because once the \kepler{} Pipeline detects a TCE in a given system, it removes the data corresponding to the in-transit cadences of that TCE, and re-searches the light curve. 

We thus measure the number of individual transit events that actually contain data. Specifically, we compute the fraction of individual events with either SES~$\ne$~0 or Rubble~$>$~0.75, which indicate there is sufficient in-cadence data present. If the fraction of transits meeting these criteria is~$\le$~0.5, we fail the TCE as not transit-like and give it the flag TRANS\_GAPPED.

\subsubsection{No Data Available}
\label{s:nofits}

In a very small number of cases, neither the DV nor the ALT detrending produces a light curve and model fit for a TCE. This happens when the TCE is extremely not transit-like, usually due to a combination of severe systematics and a lack of substantial in-transit data. As a result, if no data from either detrending is available, the Robovetter fails a TCE as not transit-like.

\subsection{Stellar Eclipse}
\label{sigsecsec}

\begin{figure*}[ht]
\centering
\includegraphics[width=\linewidth]{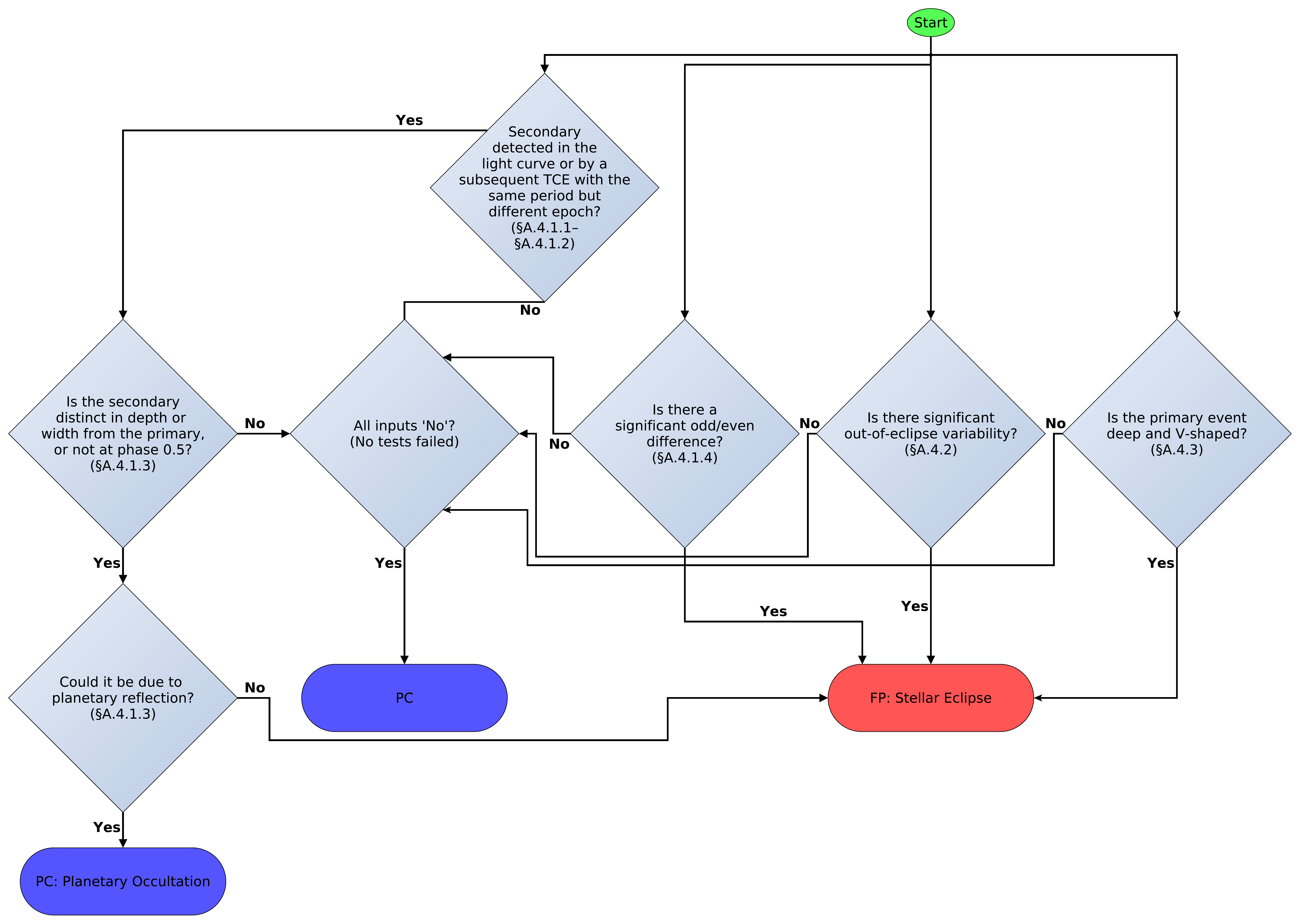}
\caption{Flowchart describing the stellar eclipse tests of the Robovetter. Diamonds represent ``yes'' or ``no'' decisions that are made with quantitative metrics. The multiple arrows originating from ``Start'' represent decisions that are made in parallel.}
\label{robovetter-sigsec-fig}
\end{figure*}

If a TCE is deemed transit-like by passing all of the tests presented in \S\ref{nottransitlikesec} on both detrendings, it is given a KOI number (see flowchart in Figure~\ref{robovetter-transitlike-fig}).
However, many of these KOIs are FPs due to eclipsing binaries and contamination from nearby variable stars. We employ a series of robotic tests to detect systems that are due to stellar companions, as shown by the flowchart in Figure~\ref{robovetter-sigsec-fig}.



\subsubsection{Secondary Eclipse}

One of the most common methods to detect a stellar system is the presence of a significant secondary in the light curve. With the exception of some hot Jupiter type planets \citep[e.g., HAT-P-7,][]{Borucki2009}, the visibility of a secondary eclipse in \kepler{} data is a telltale sign of a stellar eclipsing binary.

\paragraph{Subsequent TCE With Same Period}
\label{s:secondTce}
Once the \kepler{} Pipeline detects a TCE in a given system, it removes the data corresponding to this event and re-searches the light curve. It is thus able to detect the secondary eclipse of an eclipsing binary as a subsequent TCE, which will have the same period, but different epoch, as the primary TCE. Thus, using equations~\ref{peq1}-\ref{peq3}, the Robovetter dispositions a TCE as a stellar system FP if its period matches a subsequent TCE within the specified tolerance ($\sigma_{P}$ $>$ 3.25) and they are separated in phase by at least 2.5 times the transit duration. For clarity, we note again that it is sometimes possible that the periods of two TCEs will meet the period matching criteria, but be different enough to have their epochs shift significantly in phase over the $\sim$4 year mission duration. The phase separation requirement must be upheld over the entire mission duration in order to disposition the TCE as an FP due to a stellar eclipse.  

Occasionally the \kepler{} Pipeline will detect the secondary eclipse of an eclipsing binary at half, third, or some smaller integer fraction of the orbital period of the system. In these cases, the epoch of the TCE corresponding to the secondary will overlap with that of the primary. These cases are accounted for by not requiring a phase separation of at least 2.5 transit durations when a period ratio other than unity is detected. (Note that equations~\ref{peq1}-\ref{peq3} allow for integer period ratios.) While this approach will likely classify any multi-planet system in an exact 2:1 orbital resonance as an FP due to a stellar eclipse, in practice this is non-existent. Exact 2:1 orbital resonances, where ``exact'' means the period ratio is close enough to 2.0 over the $\sim$4 year mission duration to avoid any drift in relative epoch, appear to be extremely rare \citep{Fabrycky2014}. Also, they might produce strong transit timing variations, which would likely preclude their detection. The \kepler{} Pipeline employs a strictly linear ephemeris when searching for TCEs, and thus while planets with mild transit timing variations (TTVs), e.g., deviations from a linear ephemeris less than the transit duration, are often detected, planets with strong TTVs, e.g., deviations from a linear ephemeris greater than the transit duration, are often not detected.

\paragraph{Secondary Detected in Light Curve}
\label{secdetectsec}
\label{s:second}

There are many cases when a secondary eclipse does not produce its own TCE, most often when its MES is below the \kepler{} Pipeline detection threshold of 7.1. The model-shift uniqueness test, discussed in \S\ref{s:ms}, is well-suited to automatically detect secondary eclipses in the phased light curve, as it searches for the next two deepest events aside from the primary event. It is thus able to detect the best-candidate secondary eclipse in the light curve and assess its significance. We compute the following quantities to use as secondary detection metrics

\begin{equation}
    MS_{4} = \sigma_{\rm Sec}/F_{\rm Red} - FA_{1}
\end{equation}

\begin{equation}
    MS_{5} = (\sigma_{\rm Sec} - \sigma_{\rm Ter}) - FA_{2}
\end{equation}

\begin{equation}
    MS_{6} = (\sigma_{\rm Sec} - \sigma_{\rm Pos}) - FA_{2}
\end{equation}

Recall that $\sigma$ indicates a significance and was defined in \S\ref{s:ms}. If $MS_{4}$~$>$1, $MS_{5}$~$>$0, and $MS_{6}$~$>$0, in either the DV or alternate detrendings, the Robovetter dispositions the TCE as a stellar system FP. These criteria ensure that the secondary event is statistically significant when compared to the systematic noise level of the light curve, the tertiary event, and the positive event, respectively.

\paragraph{Candidates with Stellar Eclipses}
\label{s:sscand}
There are two exceptions when the above-mentioned conditions are met, but the Robovetter does not designate the TCE as an FP. First, if the primary and secondary widths and depths are statistically indistinguishable, and the secondary is located at phase 0.5, then it is possible that the TCE is a PC that has been detected at twice the true orbital period. Thus, the Robovetter labels a TCE with a stellar eclipse as a PC when ${\sigma_{\rm Pri} - \sigma_{\rm Sec} < FA_{2}}$ and the phase of the secondary is within 1/4 of the primary transit's duration of phase 0.5. Second, hot Jupiter PCs can have detectable secondary eclipses due to planetary occultations via reflected light and thermal emission \citep{Coughlin2012,Christiansen2010}. Thus, a TCE with a detected stellar eclipse is labeled as a PC with the stellar eclipse flag (in order to facilitate the identification of hot Jupiter occultations) when the geometric albedo required to produce the observed secondary eclipse is less than 1.0, the planetary radius is less than 30~\re{}, the depth of the secondary is less than 10\% of the primary, and the impact parameter is less than 0.95. The additional criteria beyond the albedo criterion are needed to ensure that this test is only applied to potentially valid planets and not grazing eclipsing binaries. We calculate the geometric albedo by using the stellar mass, radius, and effective temperature from the DR25 stellar catalog \citep{Mathur2017ApJS}, and the values of the period and radius ratio from the original DV fits.

\paragraph{Odd/Even Depth Difference}

\label{s:oddeven}
If the primary and secondary eclipses of eclipsing binaries are similar in depth, and the secondary is located near phase 0.5, the \kepler{} Pipeline may detect them as a single TCE at half the true orbital period of the eclipsing binary. In these cases, if the primary and secondary depths are dissimilar enough, it is possible to detect it as an FP by comparing the depths of the odd- and even-numbered transit events and their associated uncertainties, via the following statistic,

\begin{equation}
\sigma_{\rm OE} = \frac{abs\left(d_{\rm odd} - d_{\rm even}\right)}{\sqrt{\sigma_{odd}^{2} + \sigma_{even}^{2}}} 
\end{equation}

\noindent where $d_{\rm odd}$ is the measured depth using the odd-numbered transits, with associated uncertainty $\sigma_{odd}$, $d_{\rm even}$ is the measured depth using the even-numbered transits, with associated uncertainty $\sigma_{even}$, and abs() returns the absolute value.

We use two different methods to compute $d_{\rm odd}$, $\sigma_{odd}$, $d_{\rm even}$, $\sigma_{even}$, and thus $\sigma_{\rm OE}$, for both for the DV and ALT detrending. For the first method, the depths are computed by taking the median of all the points near the center of all transits, and the uncertainty is the standard deviation of those points, both using only the odd- or even-numbered transits. For the ALT detrending with a trapezoidal fit, we use all points that lie within $\pm$30 minutes of the central time of transit, as well as any other points within the in-transit flat portion of the trapezoidal fit. For the DV detrending, we use all points within $\pm$30 minutes of the central time of transit. (This threshold corresponds to the long-cadence integration time of the \kepler{} spacecraft. Including points farther away from the central time of transit degrades the accuracy and precision of the test.) If $\sigma_{\rm OE}$ $>$ 1.1 for either the DV or ALT detrending then the TCE is labeled as an FP due to a secondary eclipse and given the DEPTH\_ODDEVEN\_DV and/or DEPTH\_ODDEVEN\_ALT flag(s). The value of 1.1 was empirically derived using manual checks and transit injection. This method is very robust to outliers and systematics, but not extremely sensitive as it does not take into account the full transit shape to measure the depth.

The second method measures the depths and uncertainties by running the model-shift test separately on the portions of the light curve within half a phase of the odd- and even-numbered transits. Model-shift measures the depths and associated uncertainties using the entire transit model and taking into account the measured noise level of the entire light curve. This method is more sensitive to small odd/even differences, but also more sensitive to outliers and light curve systematics compared to the above method. If $\sigma_{\rm OE}$ $>$ 11.2 for the DV detrending, or $>$ 19.8 for the ALT detrending, then the TCE is labeled as an FP due to a stellar eclipse and given the MOD\_ODDEVEN\_DV and/or MOD\_ODDEVEN\_ALT flag(s). The thresholds of 11.2 and 19.8 were empirically derived using manual checks and transit injection. This method is susceptible to outliers and systematics (and why the thresholds are set fairly high), but can also detect small, yet significant odd/even differences that the other method listed above cannot.

\subsubsection{Out of Eclipse Variability}
\label{s:sweeteb}
Short-period eclipsing binaries will often show out-of-eclipse variability due to tidal forces that deform the star from a perfect spheroid. The variability manifests as quasi-sinusoidal variations at either the period, or half the period, of the binary.

We use the information from SWEET (see~\S\ref{s:sweetntl}) to detect these cases. If a transit-like TCE has a SWEET SNR greater than 50, an amplitude less than the TCE transit depth in either the DV and ALT detrendings, an amplitude greater than 5,000~ppm, and a period less than 10 days, we fail it as a stellar system.

\subsubsection{V-Shape Metric}
\label{s:shapemetric}
There are cases of eclipsing binaries that do not show a secondary eclipse, either due to the secondary star being too low luminosity for the eclipse to be detectable, or the binary has significant eccentricity and a longitude of periastron such that geometrically no eclipse occurs. Also, most detached eclipsing binaries will not exhibit detectable out-of-eclipse variability. In these cases, the only remaining way to infer that the signal is due to a stellar system and not a planet is to utilize the shape and depth of the transit.

In previous catalogs \citep{Rowe2015cat,Mullally2015cat,Coughlin2016} TCEs were not failed based on their inferred radii alone. This was deliberate as the catalogs attempted to be as agnostic to stellar parameters as possible, such that dispositions would remain applicable if and when better stellar parameters were obtained, e.g., by GAIA \citep{Cacciari2009,Mignard2005}. This resulted in some PC KOIs with large depths that were known to very likely be eclipsing binaries, and in fact were later confirmed as such by follow-up observations \citep{Santerne2016}.

In this catalog, we attempt to strike a balance between identifying these binary systems, while still remaining agnostic to stellar parameters. We adapted a simple shape parameter, originally proposed in \citet{Batalha2013}, and express it as the sum of the modeled radius ratio and the impact parameter. This metric reliably identifies eclipsing binaries both due to being too deep (large $R_{p}$/$R_{\star}$) and due to grazing eclipses (large impact parameter, $b$). Specifically we fail a transit-like TCE as a stellar system if $R_{p}/R_{\star} + b >1.04$.

\subsection{Centroid Offset}
\label{s:centroids}
\subsubsection{Centroid Robovetter}
\label{s:centroidrv}
The Robovetter relies on a piece of code called the Centroid Robovetter\footnote{\url{https://sourceforge.net/projects/keplercentroidrobovetter/}} \citep{Mullally2017} to detect when a transit signal originates from a background or nearby star instead of from the target star. The Centroid Robovetter has not changed since its implementation for the DR24 KOI catalog; we summarize it below for completeness. 

Given that \kepler{'s} pixels are 3.98\arcsec{} square \citep{Koch2010}, and the typical photometric aperture has a radius of 4--7 pixels \citep{Bryson2010b}, it is quite common for a given target star to be contaminated by light from another star. If that other star is variable, then that variability will be visible in the target aperture at a reduced amplitude. If the variability due to contamination results in a TCE, then it is a false positive, whether the contaminator is an eclipsing binary, planet, or other type of variable star \citep{Bryson2013}. For example, if a transit or an eclipse occurs on a bright star, a shallower event may be observed on a nearby, fainter star. Similarly, a star can be mistakenly identified as experiencing a shallow transit if a deep eclipse occurs on a fainter, nearby source.

The DV module of the \kepler{} Pipeline produces difference images for each quarter, which are made by subtracting the average flux in each pixel during each transit from the flux in each pixel just before, and after, each transit \citep{Bryson2013}. If the resulting difference image shows significant flux at a location (centroid) other than the target, then the TCE is likely an FP due to a centroid offset.

In our robotic procedure to detect FPs due to centroid offsets, we first check that the difference image for each quarter contains a discernible stellar image and is not dominated by background noise. This is done by searching for at least 3 pixels that are adjacent to each other and brighter than a given threshold, which is set by the noise properties of the image. We use an iterative sigma clipping approach to eliminate bright pixels when calculating the background noise, as the star often dominates the flux budget of a substantial number of pixels in the aperture.

For the difference images that are determined to contain a discernible stellar image, we first search for evidence of contamination from sources that are resolved from the target. Since resolved sources near the edge of the image may not be fully captured, attempts to fit models of the stellar profile often fail to converge. Instead, we check if the location of the brightest pixel in the difference image is more than 1.5 pixels from the location of the target star. If at least two-thirds of the quarterly difference images show evidence of an offset by this criterion, we disposition the TCE as an FP due to a centroid offset. 

If no centroid offset is identified by the previous method, we then look for contamination from sources that are unresolved from the target. We fit a model of the pixel response function (PRF) to the difference images and search for statistically significant shifts in the centroid with respect to the PRF centroid of the out-of-transit images, or the catalog position of the source. Following \citet{Bryson2013}, a TCE is marked as an FP due to a centroid offset if there are at least three difference images with a discernible stellar image, and a 3$\sigma$ significant offset larger than 2$\arcsec$, or a 4$\sigma$ offset larger than 1$\arcsec$ is measured.  

The Centroid Robovetter gives the \kepler\ Robovetter several flags to indicate whether a centroid offset was detected and whether that detection can be trusted. The names of those flags have been changed for DR25 to be consistent with our minor flag naming scheme. A list of the minor flags are available in Appendix~\ref{s:minorflags}.

\subsubsection{Ghost Diagnostic}
\label{s:ghost}
The last method we use to detect a centroid offset is the ghost diagnostic, which was added to the DR25 \kepler{} Pipeline \citep[see \S\,11.3.7 of][]{JenkinsKDPH}. It determines whether a transit signal is likely contamination from a ghost image of a star located away from the target star in the focal plane. Ghost reflections occur when light from a bright star is reflected off the CCD and again from the field flattener plate and back onto the CCD. It appears as a diffuse, out-of-focus image of the pupil of the telescope. A similar type of false positive results from direct PRF (Pixel Response Function) contamination, when flux from the broad wings of a bright star near the target star on the CCD overlaps the target star's PRF.  If a ghost reflection (or the PRF of a nearby star) containing a transit-like signature (e.g. an eclipsing binary signal) overlaps the PRF of the target star, then the contaminating transit signal will be equally strong in the periphery and the core of the target. 

To detect this type of false alarm, the ghost diagnostic essentially measures the strength of the TCE signal in two separate light curves --- one created using the average of the pixels inside the target's optimal aperture minus the average of the pixels in an annulus surrounding the target aperture (core aperture correlation statistic), and the other using the average of the pixels in the annulus surrounding the target aperture (halo aperture correlation statistic). If the ratio of the halo aperture to core aperture statistic is greater than 4.0, the TCE is marked as an FP with the major flag set to Centroid Offset. This ghost diagnostic is not available to vet the \scrtce{s} and thus the reliability measured with that set of TCEs will be too small by an insignificant amount.

\subsection{Ephemeris Matching}
\label{ephemmatchsec}
\label{s:ephemmatch}

Another method for detecting FPs due to contamination is to compare the ephemerides (periods and epochs) of TCEs to each other, as well as other known variable sources in the \kepler{} field. If two targets have the same ephemeris within a specified tolerance, then at least one of them is an FP due to contamination. \citet{Coughlin2014a} used Q1--Q12 data to compare the ephemerides of KOIs to each other and eclipsing binaries known from both \kepler{}- and ground-based observations. They identified over 600 FPs via ephemeris matching, of which over 100 were not known as FPs via other methods. They also identified four main mechanisms of contamination. The results of \citet{Coughlin2014a} were incorporated in \citet[][see \S3.3]{Rowe2015a}, and with some small modifications to \citet[][see \S5.3]{Mullally2015cat} and \citet{Coughlin2016}. 

\begin{figure*}[ht]
\centering
\begin{tabular}{cc}
\includegraphics[width=0.5\linewidth]{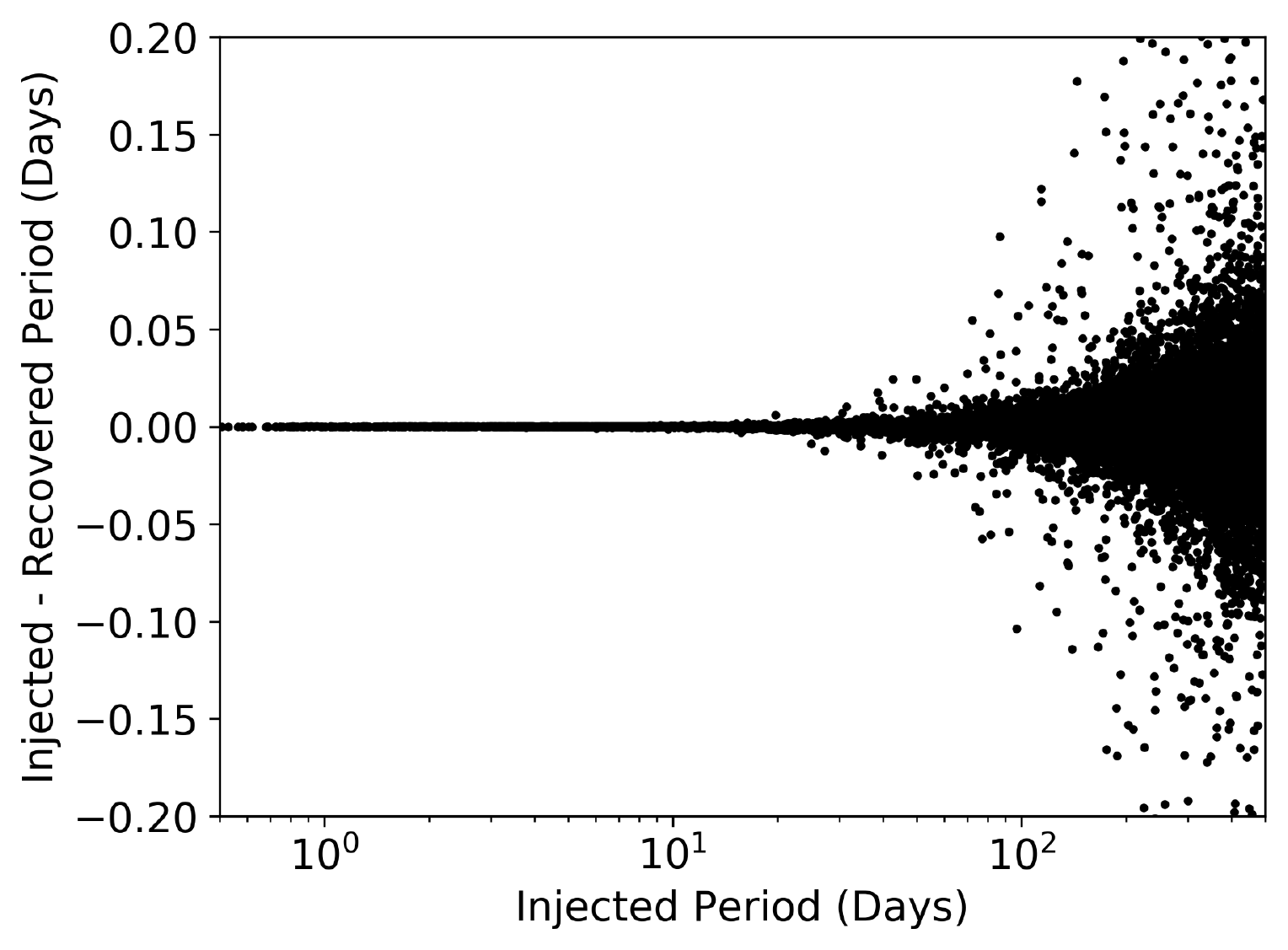} &
\includegraphics[width=0.5\linewidth]{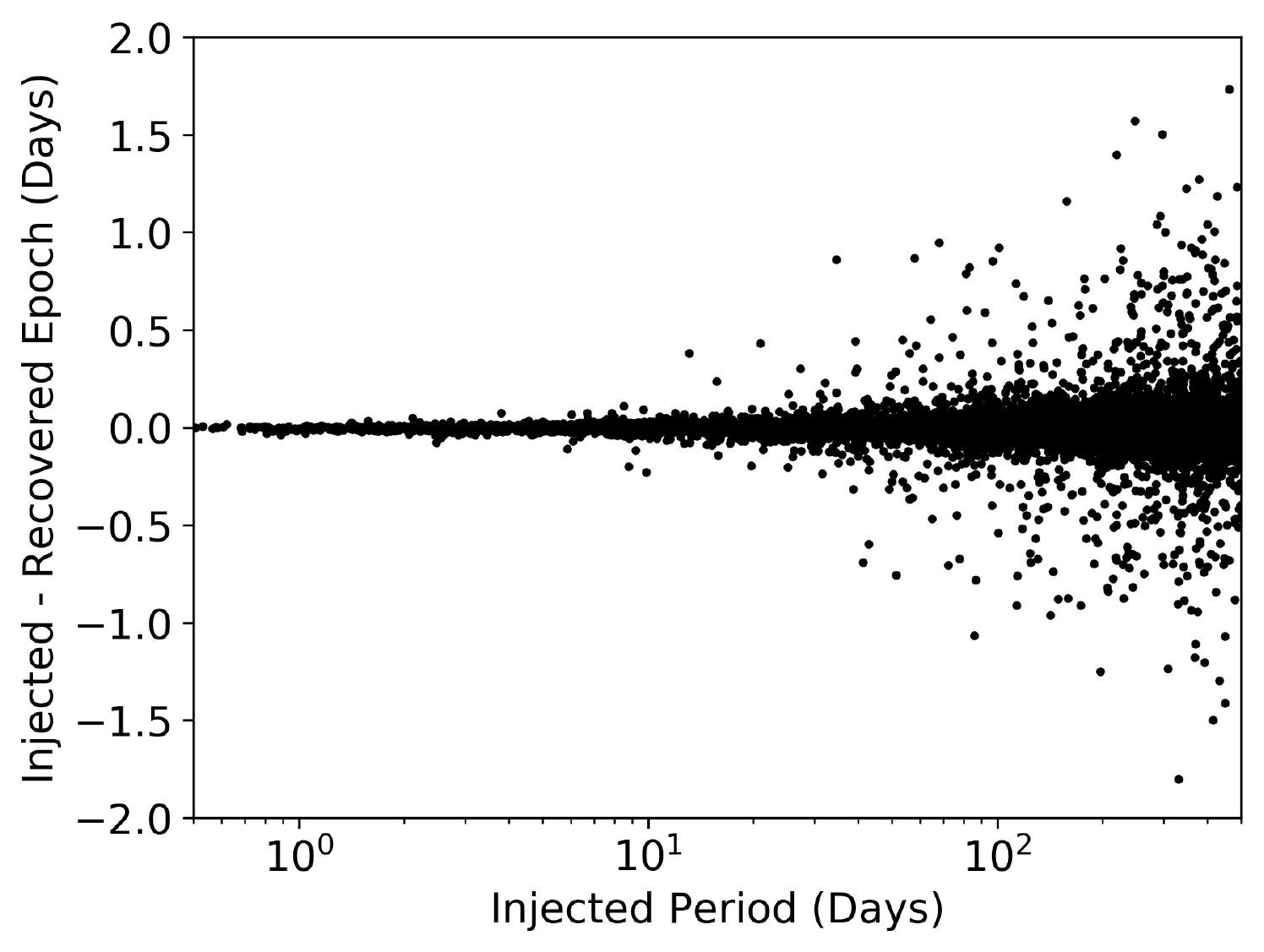} \\
\includegraphics[width=0.5\linewidth]{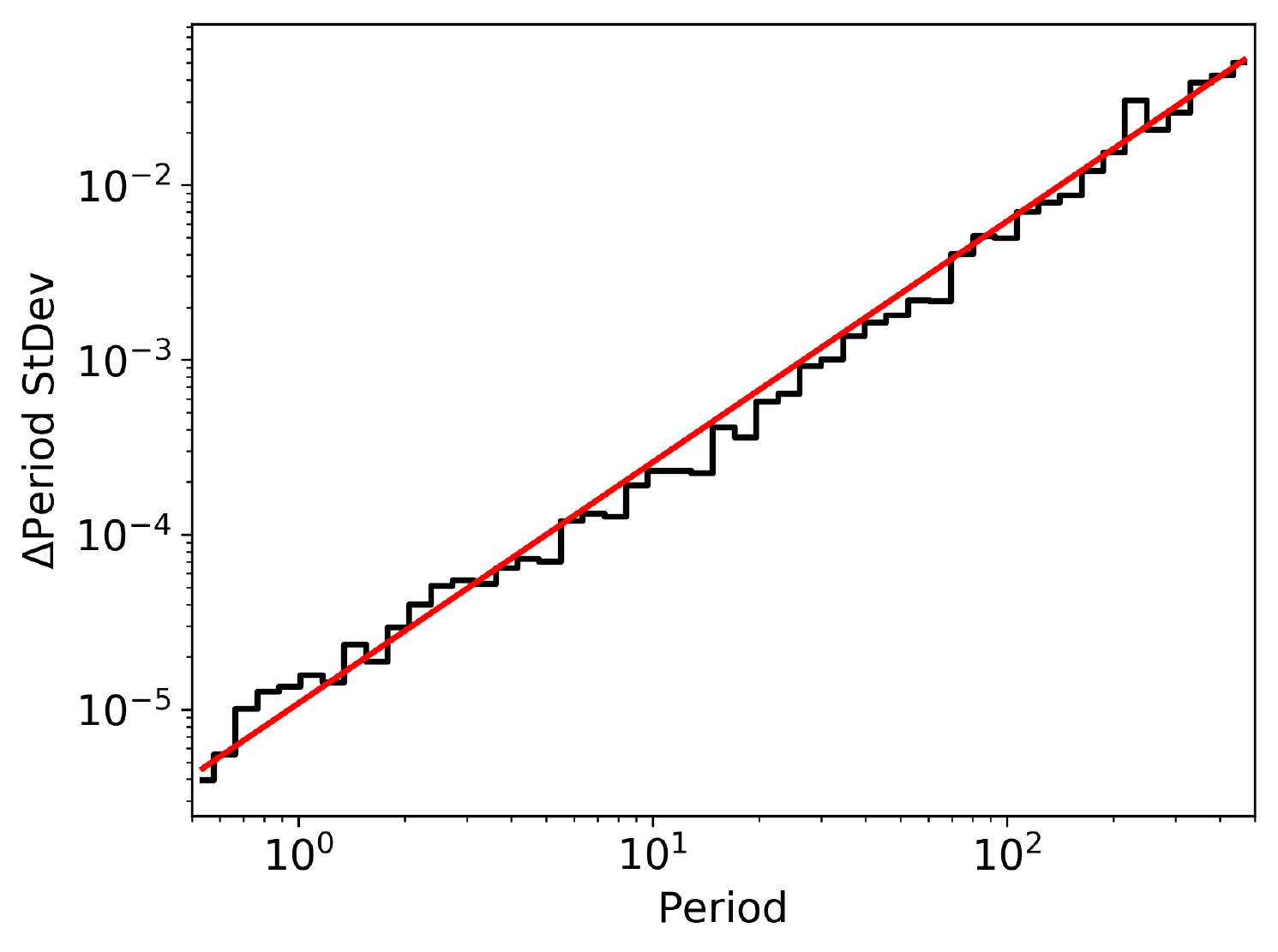} &
\includegraphics[width=0.5\linewidth]{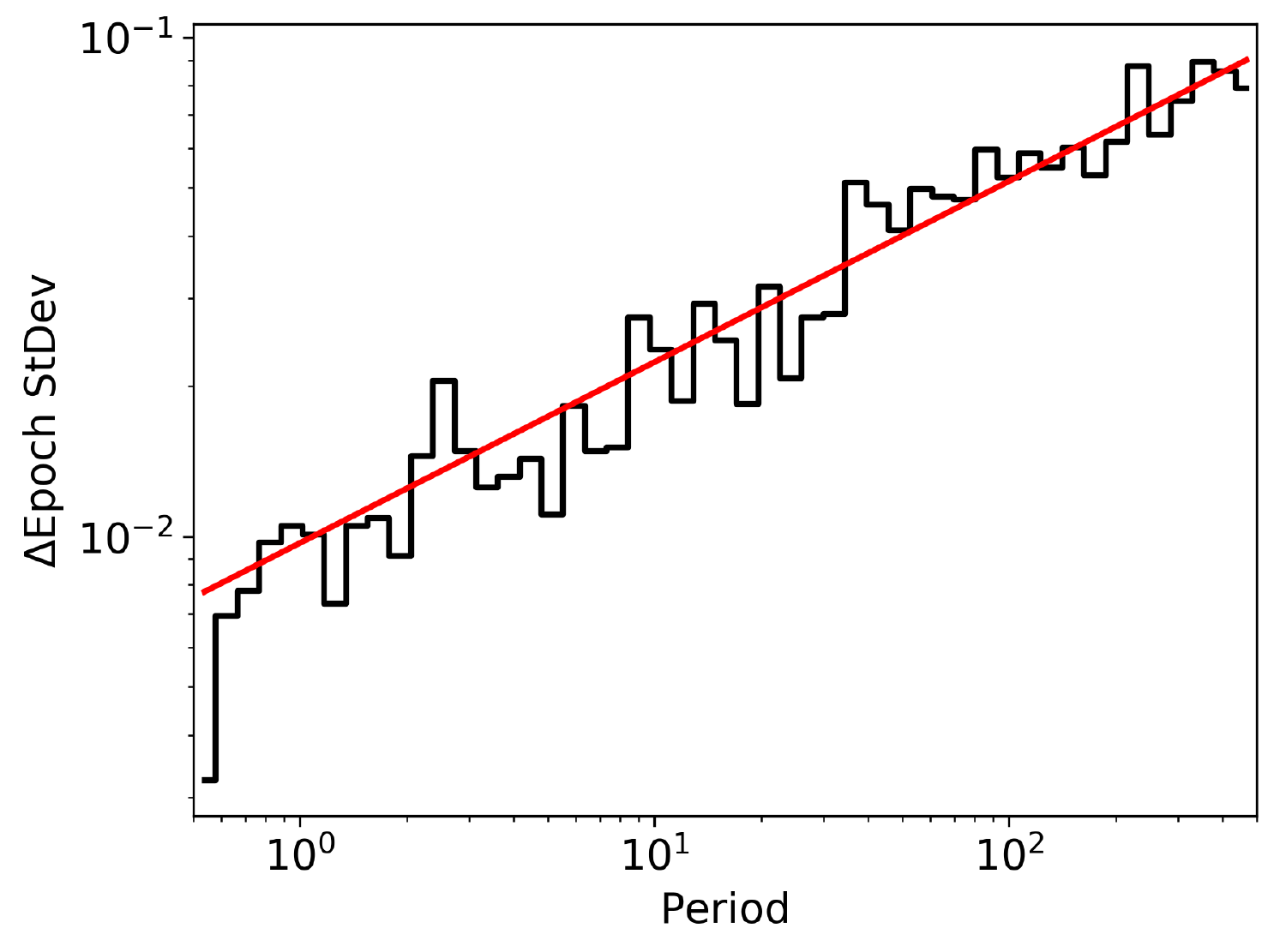}
\end{tabular}
\caption{A plot of injected versus recovered periods and epochs of injected on-target planets. The top plots shows the difference between the injected and recovered periods (top) and epochs (right) as a function of period. The bottom plots show the measured standard deviation of the differences in period (left) and epoch (right) in logarithmic space. The red line shows the best-fit power-law in each case.}
\label{injephemfig}
\end{figure*}

We modified the matching criteria used in previous catalogs to improve performance. We use the results of the transit injection run (\S\ref{s:simulated}) to measure the ability of the original DV fits by the \kepler{} Pipeline to recover period and epoch as a function of period. (Note that while the DV fits do produce an error on the measured period, it is not a robustly measured error, and thus not sufficient for our purposes.) In Figure~\ref{injephemfig} we show, in the top two panels, the difference in the injected and recovered period and epoch, as a function of the injected period. The bottom panels show the measured standard deviation of the difference as a function of period, in linear and logarithmic space respectively. The red line is the result of a best-fit power law.

When comparing two objects, A and B, where A is defined to have the shorter period, the new matching metrics we use, $S_{P}$ and $S_{T}$ for period and epoch respectively, are:

\begin{equation}
    S_{P} = \frac{\left|P_{r} \cdot P_{A} - P_{B}\right|}{\sqrt{2}\cdot\sigma_{P}(P_{A})}
\end{equation}

\begin{equation}
    S_{T} = \frac{\left| T_{A} - T_{B} - T_{r} \cdot P_{A}\right|}{\sqrt{2}\cdot\sigma_{T}(P_{A})}
\end{equation}

\noindent where $P_{A}$ and $P_{B}$ are the periods of objects A and B, $T_{A}$ and $T_{B}$ are similarly the epochs of objects A and B, $\sigma_{P}(P_{A})$ and $\sigma_{T}(P_{A})$ are the errors in period and epoch, given period $P_{A}$, derived from the best-fit power law to the standard deviation of the injected versus recovered periods and epochs, respectively. The period ratio, $P_{r}$, and epoch ratio, $T_{r}$, are defined by:

\begin{equation}
P_{r} = \textrm{rint}\left(\frac{P_{B}}{P_{A}}\right)
\end{equation}

\begin{equation}
T_{r} = \textrm{rint}\left(\frac{T_{A} - T_{B}}{P_{A}}\right)
\end{equation}

\noindent where rint() rounds a number to the nearest integer. Thus, a perfect match has $S_{P}$~=~0 and  $S_{T}$~=~0, with worse matches having increasingly larger values of $S_{P}$ and $S_{T}$. 

We consider matches with $S_{P}$~$<$~5 and $S_{T}$~$<$~5, with period ratios of 50 or less ($P_{r}$~$<$~50), to be statistically significant enough to constitute a match. We also require:

\begin{enumerate}

\item The two objects do not have the same KIC ID,

\item The two objects satisfy at least one of the following conditions: 

    \begin{enumerate}
    
    \item A separation distance less than $d_{\rm max}$ arcseconds, where
    \begin{equation}
    \label{disteq}
    d_{\rm max}(\arcsec) = 55\cdot\sqrt{10^{6}\cdot 10^{-0.4 \cdot m_{\rm kep}}+1}
    \end{equation}

\noindent with the \kepler{} magnitude of the brighter source being used for $m_{\rm kep}$,  

    \item Located on opposite sides of the field-of-view center, but equidistant from the center to within a 100$\arcsec$ (25 pixel) tolerance.
    
    \item Located on the same CCD module and within 5 pixels of the same column value in any of the 4 quarters.
    
    \item Located on the same CCD module and within 5 pixels of the same row and column value in any of the 4 quarters.

   \end{enumerate}

\end{enumerate}

\noindent Criterion 1 ensures that no star is ever matched to itself. Criterion 2a is a semi-empirically determined formula derived to account for direct PRF contamination and reflection off the field flattener lens, assuming the average wings of a \emph{Kepler} PSF can be approximated by a Lorentzian distribution. The formula allows for any two stars to match within a generous 55$\arcsec$ range, but allows for bright stars to match to larger distances, e.g., a 10$^{\rm th}$ mag star could match up to 550$\arcsec$ away, and a 5th mag star could match up to 5500$\arcsec$ away. Criterion 2b accounts for antipodal reflection off the Schmidt Corrector. Criterion 2c accounts for the column anomaly \citep[see \S3.5 of][]{Coughlin2016}, and criterion 2d accounts for video crosstalk.

In this Q1--Q17~DR25 catalog, we match the ephemerides of all Q1--Q17~DR25 TCEs \citep{Twicken2016}, including rogue TCEs, to the following sources:

\begin{itemize}
 \item Themselves.
 \item The list of \npredrtwentyfivekois{} KOIs from the NASA Exoplanet Archive cumulative KOI table after the closure of the Q1--Q17~DR24 table and publication of the last catalog \citep{Coughlin2016}.
 \item The \kepler{} Eclipsing Binary Working Group list of \nkebs{} ``true'' eclipsing binaries found with \kepler{} data as of 2016 October 13 \citep{Prsa2011,Slawson2011,Kirk2016}.
 \item J.M. Kreiner's up-to-date database of ephemerides of ground-based eclipsing binaries as of 2016 October 13 \citep{Kreiner2004}.
 \item Ground-based eclipsing binaries found via the TrES survey \citep{Devor2008a}.
 \item The General Catalog of Variable Stars \citep[GCVS][]{Samus2015} list of all known ground-based variable stars, published 2016 October 05.
\end{itemize}

\added{The ephemeris matching code used for the DR25 catalog is publicly available on github\footnote{\url{https://github.com/JeffLCoughlin/EphemMatch}}.}

\begin{figure*}[ht]
\centering
\includegraphics[width=\linewidth]{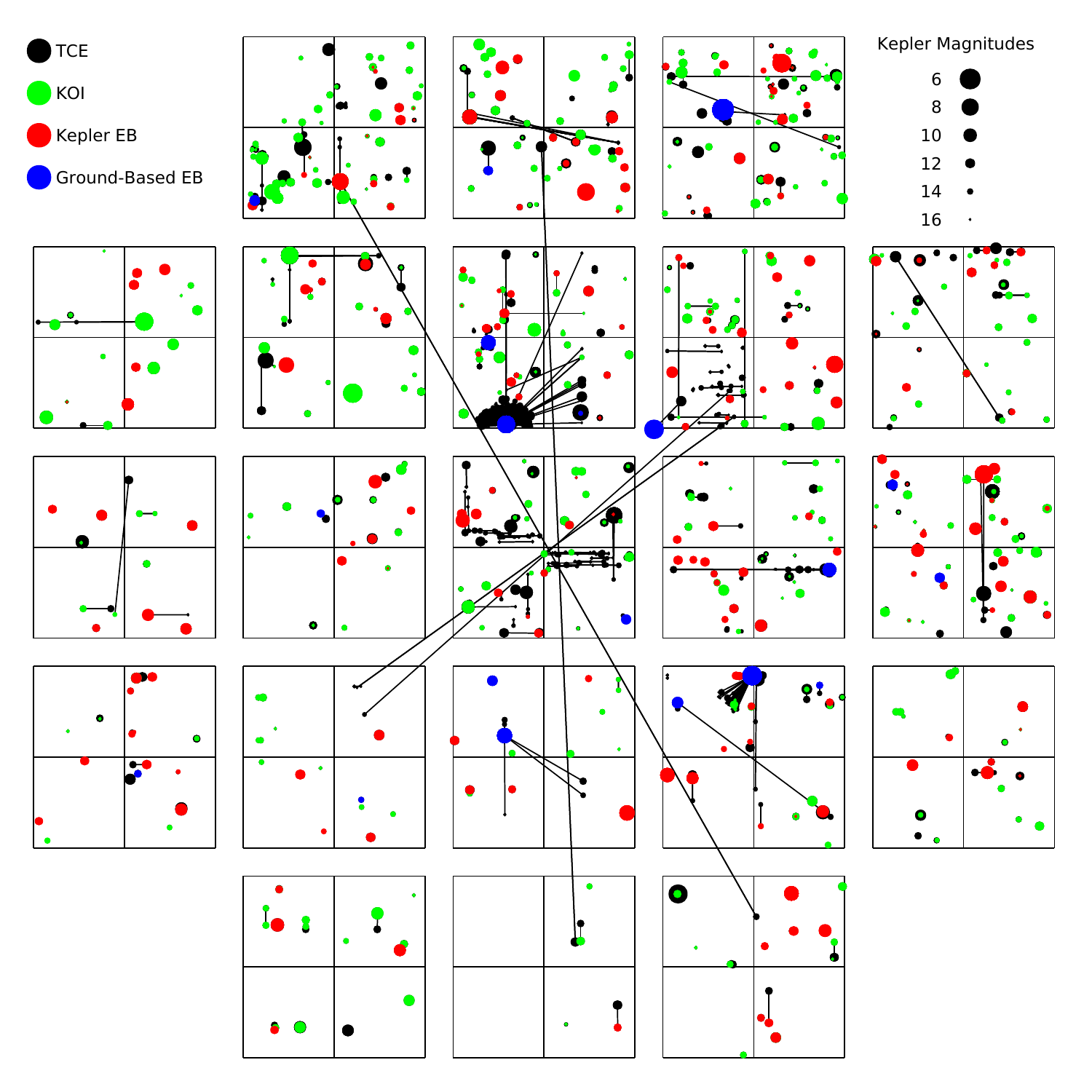}
\caption{Distribution of ephemeris matches on the focal plane. Symbol size scales with magnitude, while color represents the catalog in which the contaminating source was found. Blue indicates that the true transit is from a variable star only known as a result of ground-based observations. Red circles are stars listed in the \kepler{} EBWG catalog \citep[][\url{http://keplerebs.villanova.edu/}]{Kirk2016}, green are KOIs, and black are TCEs. Black lines connect false positive matches with the most likely contaminating parent. In most cases parent and child are so close that the connecting line is invisible.}
\label{ephemmatchfig}
\end{figure*}

Via ephemeris matching, we identify \nephemmatch{} Q1--Q17~DR25 TCEs as FPs. Of these, \nonlyephemmatch{} were identified as FPs only due to ephemeris matching. We list all \nephemmatch{} TCEs in Table~\ref{ephemmatchtab}, as this information is valuable for studying contamination in the \kepler{} field. In this table each TCE is identified by its KIC ID and planet number, separated by a dash. We also list in Table~\ref{ephemmatchtab} each TCE's most likely parent, the period ratio between child and parent (P$_{\rm rat}$), the distance between the child and parent in arcseconds, the offset in row and column between the child and parent in pixels ($\Delta$Row and $\Delta$Col), the magnitude of the parent (m$_{\rm Kep}$), the difference in magnitude between the child and parent ($\Delta$Mag), the depth ratio of the child and parent (D$_{\rm rat}$), the mechanism of contamination, and a flag to designate unique situations. In Figure~\ref{ephemmatchfig} we plot the location of each FP TCE and its most likely parent, connected by a solid line. TCEs are represented by solid black points, KOIs are represented by solid green points, eclipsing binaries found by \kepler{} are represented by solid red points, eclipsing binaries discovered from the ground are represented by solid blue points, and TCEs due to a common systematic are represented by open black points. The \kepler{} magnitude of each star is shown via a scaled point size. Most parent-child pairs are so close together that the line connecting them is not easily visible on the scale of the plot.

\begin{deluxetable*}{ccccccccccc}
\tablecolumns{11}
\tabletypesize{\scriptsize}
\tablewidth{\linewidth}
\tablecaption{The \nephemmatch{} Q1--Q17~DR25 TCEs Identified as FPs due to Ephemeris Matches}
\tablehead{\colhead{TCE} & \colhead{Parent} & \colhead{P$_{\rm rat}$} & \colhead{Distance} & \colhead{$\Delta$Row} & \colhead{$\Delta$Col} & \colhead{m$_{\rm Kep}$} & \colhead{$\Delta$Mag} & \colhead{D$_{\rm rat}$} & \colhead{Mechanism} & \colhead{Flag} \\ & & & (\arcsec) & (Pixels) & (Pixels) & & & & & }
\startdata
001433962-01 & 3924.01 & 1:1 & 13.5 & 3 & -2 & 14.91 & 0.56 & 4.7434E+02 & Direct-PRF & 0\\
001724961-01 & 001724968-01 & 1:1 & 4.7 & 1 & -1 & 13.39 & -2.96 & 2.1190E+00 & Direct-PRF & 0\\
002166206-01 & 3735.01 & 1:1 & 8.3 & -1 & -2 & 17.64 & -4.34 & 5.6706E+02 & Direct-PRF & 0\\
002309585-01 & 5982.01 & 1:1 & 11.7 & -2 & 1 & 13.93 & 1.45 & 2.0011E+02 & Direct-PRF & 0\\
002437112-01 & 3598.01 & 1:1 & 19.7 & -5 & 1 & 17.63 & -1.48 & 1.0525E+03 & Direct-PRF & 0\\
002437112-02 & 002437149-02 & 2:1 & 19.7 & -5 & 1 & 17.63 & -1.48 & 6.9253E+02 & Direct-PRF & 0\\
002437488-01 & 6268.01 & 1:1 & 10.6 & 0 & 3 & 16.98 & -2.02 & 2.5330E+02 & Direct-PRF & 0\\
002437804-01 & 002437783-01 & 1:1 & 14.4 & 4 & -1 & 17.30 & -3.14 & 1.4225E+02 & Direct-PRF & 0\\
\nodata & \nodata & \nodata & \nodata & \nodata & \nodata & \nodata & \nodata & \nodata & \nodata & \nodata\\
\enddata
\tablecomments{A suffix of ``pri'' in the parent name indicates the object is an eclipsing binary known from the ground, and the child TCE matches to its primary. Similarly a suffix of ``sec'' indicates the child TCE matches the secondary of a ground-based EB. Parent names are listed, in priority order when available, by (1) their Bayer designation (e.g., RR-Lyr-pri), (2) their EBWG (Eclipsing Binary Working Group; \citealt{Kirk2016}) designation (e.g., 002449084-pri), (3) their KOI number (e.g., 3924.01), and (4) their TCE number (e.g., 001724968-01). A flag of 1 indicates that the TCE is a bastard, which are cases where two or more TCEs match each other via the Direct-PRF contamination mechanism, but neither can physically be the parent of the other via their magnitudes, depths, and distances, and thus the true parent has not been identified. A flag of 2 indicates cases of column anomalies that occur on different outputs of the same module. These cases likely involve cross-talk to carry the signal from one output to another. TCEs due to the common systematic do not have information listed for a parent source, as they are not caused by a single parent. Note that  Table~\ref{ephemmatchtab} is published in its entirety in the electronic edition of the Astrophysical Journal. A portion is shown here for guidance regarding its form and content.}
\label{ephemmatchtab}
\end{deluxetable*}

Since \kepler{} does not observe every star in its field of view, it can often be the case that a match is found between two objects, but given their relative magnitude, distance, and depths it is clear that neither is the parent of the other, so these are classified as ``bastards'' \citep{Coughlin2014a}. To identify the bastards due to direct PRF contamination, we performed a robust fit of the Kepler PRF model described by equations~9~and~10 of \citet{Coughlin2014a} to the depth ratio, magnitude difference, and distance between each object identified as due to direct PRF contamination and its most likely parent. After iteratively rejecting outliers greater than 4.0 times the standard deviation, the fit converged with values of $\alpha$ = 6.93\arcsec and $\gamma$ = 0.358\arcsec. Outliers greater than 4.0 times the standard deviation of the final iteration, with these resulting fit parameters, were labeled as bastards. For the mechanism of column anomaly and reflection, if the depth ratio of the two objects is between 0.01 and 100, then it is labeled as a bastard, as these mechanisms should produce depth ratios of at least 1E-3 or 1E3. All bastards are identified with a flag of 1 in Table~\ref{ephemmatchtab}. Additionally, it can sometimes be the case that objects are matched via the column anomaly, but are on different outputs of the same module --- these cases likely involve the column anomaly working in conjunction with cross-talk, and thus are complicated, and given a flag of 2 in Table~\ref{ephemmatchtab}. Finally, a flag of 3 indicates a combination of flags 1 and 2.

\subsection{Informational Only Tests}
\label{s:infoonly}

There are a couple tests that the Robovetter performs that do not influence the disposition of a TCE. While failing one of these tests indicates a likely FP, it is not reliable enough to declare a TCE an FP. Instead, TCEs that fail these tests are given information only flags (see \S\ref{s:minorflags}) as a way to notify users that a manual inspection of the TCE and the Robovetter results is likely warranted.

\subsubsection{Planet In Star}
\label{s:planetinstar}

In some cases, the DV fit returns a semi-major axis of the planetary orbit that is smaller than the radius of the host star. Such a fit is unphysical, as the planet would be orbiting inside the star; this is usually indicative of an FP. However, since many of the stellar parameters have large errors and their accuracy can vary, this situation does not guarantee the TCE is an FP. Thus, if a TCE is dispositioned as transit-like (the NT flag is not set), and if the semi-major axis from the DV fit is less than the stellar radius from the DR25 stellar properties catalog \citep{Mathur2017ApJS}, the TCE is flagged as PLANET\_IN\_STAR.

\subsubsection{Seasonal Depth Differences}
\label{s:seasonaldiff}

Due to the \kepler{} spacecraft's rotation every $\approx$90~days, each target and the surrounding stars will fall on a new CCD every quarter, and return to the same CCD once every four quarters. All of the quarters that correspond to the same CCD are labeled as being in a given season (e.g., Q2, Q6, Q10, and Q14 belong to Season 0, Q3, Q7, Q11, and Q15 belong to Season 1, etc., \citealt{Thompson2016KAM}). The shape and size of the optimal aperture for a given star is seasonally dependant and can change significantly season-to-season. As a result, a target will have differing amounts of third light in its optimal aperture from nearby stars. If the source of the signal that triggers a TCE is not from the target star, but rather from another source (as just discussed in \S\ref{s:centroids} and \S\ref{s:ephemmatch}), the level of contamination, and thus observed depth of the TCE, will have significant seasonal variation. Observation of seasonal depth differences is usually a good indication that the target is contaminated and a centroid offset is likely. However, depth differences can also arise when the signal is truly coming from the target, but significant third light exists in the aperture and the seasonal variations are not sufficiently corrected.

In order to automatically detect seasonal depth differences, if a TCE has been dispositioned as transit-like (the NT flag is not set), we measure the depth and associated error of the primary event in each season utilizing the first method described in the second paragraph of \S\ref{s:oddeven}, i.e., we compute the median and standard deviation of all the points within $\pm$15 minutes of the center of transit. We then obtain an average depth over all seasons, $D_{a}$, by computing the mean of the depths of all four seasons. 

The significance of the seasonal depth differences, $S_{\rm Diff}$, is then computed via,

\begin{equation}
    S_{\rm Diff} = \frac{\sum_{n=0}^{3} \left|D_{n} - D_{a}\right|}{\sqrt{\sum_{n=0}^3 \sigma_{n}^{2} + N\cdot\sigma_{a}^{2}}}
\end{equation}

\noindent where $n$ denotes a particular season (0, 1, 2, or 3), N is the total number of seasons with a measured depth and uncertainty, $D_{n}$ is the measured depth in a given season, $\sigma_{n}$ is the measured error on the depth in a given season, $D_{a}$ is the measured averaged depth, and $\sigma_{a}$ is the measured error of the average depth, given by,

\begin{equation}
    \sigma_{a} = \frac{\sqrt{\sum_{n=0}^{3}\sigma_{n}^2}}{N}
\end{equation}

\noindent For either the DV or ALT detrending, if $S_{\rm Diff}$~$>$~3.6 then the TCE is flagged as having significant seasonal depth differences via the flag SEASONAL\_DEPTH\_(ALT|DV).

\subsubsection{Period Aliasing}
\label{s:periodalias}

In some cases, the \kepler{} Pipeline detects a signal (and produces a TCE) that is at an integer multiple of the signal's true period. In most cases, this is due to the presence of seasonal depth differences, as the Pipeline ends up only locking onto events in the quarters with the strongest (deepest) signal. While this usually indicates an FP due to a centroid offset, as discussed in \ref{s:seasonaldiff}, it is not a definitive measure. Also, the Pipeline will detect real planets with significant TTVs at longer (near integer multiple) periods.

In order to detect a period alias, we utilize the Model-shift results --- if the TCE's period is an integer multiple of the signal's true period, then several, equally spaced events should be visible in the phased light curve. If the TCE has been dispositioned as transit-like (the NT flag is not set), the Robovetter first checks if Model-shift detected significant secondary and tertiary events, by ensuring that $\sigma_{\rm Sec}/F_{\rm Red} > FA_{1}$ and $\sigma_{\rm Ter}/F_{\rm Red} > FA_{1}$. If so, the phases of the secondary and tertiary events, $\phi_{\rm Sec}$ and $\phi_{\rm Ter}$, are then expressed as the absolute value of the their distance in phase from the primary event, i.e., constrained to be between 0.0 and 0.5. (For example, if secondary and tertiary events were initially detected at phases of 0.1 and 0.7, then $\phi_{\rm Sec} = 0.1$ and $\phi_{\rm Ter} = 0.3$.) If period aliasing is present, then $\phi_{\rm Sec}$ and $\phi_{\rm Ter}$ should be $\approx$~$n / N$, where $N$ is the integer multiple of the true signal that the Pipeline detected it at, and $n$ is an integer between 1 and $N-1$ that is different for the secondary and tertiary events. (E.g., in the case of $\phi_{\rm Sec} = 0.1$ and $\phi_{\rm Ter} = 0.3$, this implies $N = 10$, $n = 1$ for $\phi_{\rm Sec}$, and $n = 3$ for $\phi_{\rm Ter}$). 

We derive metrics to measure how close $\phi_{\rm Sec}$ and $\phi_{\rm Ter}$ each are to an exact integer period alias, called $S_{\rm Sec}$ and $S_{\rm Ter}$. Specifically,

\begin{equation}
\begin{aligned}
S_{\rm Sec} = \sqrt{2}\cdot\textrm{erfcinv}\left(\left|\frac{1}{\phi_{\rm Sec}} - {\textrm{rint}\left(\frac{1}{\phi_{Sec}}\right)}\right|\right) \\[1em]
S_{\rm Ter} = \sqrt{2}\cdot\textrm{erfcinv}\left(\left|\frac{1}{\phi_{\rm Ter}} - {\textrm{rint}\left(\frac{1}{\phi_{Ter}}\right)}\right|\right)
\end{aligned}
\end{equation}

\noindent where erfcinv() is the inverse complementary error function, and rint() rounds a number to the nearest integer. The higher the values of $S_{\rm Sec}$ and $S_{\rm Ter}$, the more closely the measured phases of the significant secondary and tertiary events correspond to an integer period ratio. These computations are performed independently for the DV and ALT detrendings. If $S_{\rm Sec} > 2.0$ and $S_{\rm Ter} > 2.0$, for either detrending, the Robovetter considers a period alias detected, and the TCE is flagged as PERIOD\_ALIAS\_(ALT|DV).

\section{Minor False Positive Flag Definitions}
\label{s:minorflags}
The Robovetter produces a flag each time it gives a disposition of FP, and sometimes when it gives a disposition of PC.   Here we give a definition for each flag.  Table~ref{t:minorstats} shows the number and percentage of \opstce{s} (not including rogue and banned) that were flagged with each minor flag. These flags are available for the KOIs through the comment column in the KOI table at the Exoplanet Archive. See the Robovetter output files\footnote{The Robovetter output files have the format kplr\_dr25\_XXX\_robovetter\_output.txt (XXX represents the data set name) and can be found in the Robovetter github repository, \url{https://github.com/nasa/kepler-robovetter}} for the flags for all the \opstce{s}, \injtce{s}, \invtce{s}, \scrtce{s}. A summary of the Robovetter metrics is given in Table~\ref{t:metrics}.

\textbf{ALL\_TRANS\_CHASES}: This flag is set when the per TCE Chases metric is above threshold. This indicates that the shapes of the individual transits are generally not reliable and the TCE is dispositioned as an FP with the not transit-like major flag set. See \S\ref{s:tcechases}.

\textbf{CENT\_CROWDED}: This flag is set as a warning that more than one potential stellar image was found in the difference image, and thus a reliable centroid measurement cannot be obtained. See \S\ref{s:centroidrv}.

\textbf{CENT\_FEW\_DIFFS}: Fewer than 3 difference images of sufficiently high SNR are available, and thus very few tests in the pipeline's centroid module are applicable to the TCE. If this flag is set in conjunction with the CENT\_RESOLVED\_OFFSET flag, it serves as a warning that the source of the transit may be on a star clearly resolved from the target. See \S\ref{s:centroidrv}.

\textbf{CENT\_FEW\_MEAS}: The PRF centroid fit used by the pipeline's centroid module does not always converge, even in high SNR difference images. This flag is set as a warning if centroid offsets are recorded for fewer than 3 high SNR difference images. See \S\ref{s:centroidrv}.

\textbf{CENT\_INVERT\_DIFF}: One or more difference images were inverted, meaning the difference image claims the star got brighter during transit. This is usually due to variability of the target star and suggests the difference image should not be trusted. When this flag is set, it is a warning that the TCE requires further scrutiny, but the TCE is not marked as an FP due to a centroid offset. See \S\ref{s:centroidrv}.

\textbf{CENT\_KIC\_POS}: This measured offset distance is relative to the star's recorded position in the Kepler Input Catalog (KIC), not the out of transit centroid. Both are useful, since the KIC position is less accurate in sparse fields, but more accurate in crowded fields. If this is the only flag set, there is no reason to believe a statistically significant centroid shift is present. See \S\ref{s:centroidrv}.

\textbf{CENT\_NOFITS}: The transit was not fit by a model in DV and thus no difference images were created for use by the pipeline's centroid module, so this flag is set as a warning that the TCE cannot be evaluated. This flag is typically set for very deep transits due to eclipsing binaries. See \S\ref{s:centroidrv}.

\textbf{CENT\_RESOLVED\_OFFSET}: The TCE has a significant centroid offset because the transit occurs on a star that is spatially resolved from the target. The TCE is marked as an FP with the centroid offset flag set unless one of the other Centroid Robovetter flags is also set, casting doubt on the measurement. See \S\ref{s:centroidrv}.

\textbf{CENT\_SATURATED}: The star is saturated, so the Robovetter's centroiding assumptions break down. This flag is set as a warning, indicating that the TCE cannot be reliably evaluated.  See \S\ref{s:centroidrv}.

\textbf{CENT\_UNCERTAIN}: The significance of the centroid offset cannot be measured to high enough precision, so this flag is set as a warning that the TCE cannot be confidently dispositioned as an FP. This is typically due to having only a very small number (i.e., 3 or 4) of offset measurements, all with low SNR. See \S\ref{s:centroidrv}.

\textbf{CENT\_UNRESOLVED\_OFFSET}: There is a statistically significant shift in the centroid during transit. This indicates the is not on the target star. Thus, the TCE is dispositioned as an FP with the centroid offset major flag set, unless another Centroid Robovetter flag is also set, casting doubt on the measurement. See \S\ref{s:centroidrv}.

\textbf{DEEP\_V\_SHAPED}: The V-shape metric is above threshold. This metric uses the fitted DV radius ratio and impact parameter to determine whether the event is likely to be caused by a stellar eclipse. When the flag is set, the TCE is dispositioned as an FP with the stellar eclipse major flag set. See \S\ref{s:shapemetric}.

\textbf{DEPTH\_ODDEVEN\_(ALT|DV)}: The TCE failed the odd-even depth test using the ALT or DV detrending. This determines whether the difference in the depths of the odd and even transits is greater than the standard deviation of the measured depths. The transit-like TCE is marked as an FP with a stellar eclipse major flag set. See \S\ref{s:oddeven}.


\textbf{EPHEM\_MATCH}: The TCE has been identified as an FP due to an ephemeris match with a source that could plausibly induce the observed variability on the target. See \S\ref{ephemmatchsec} and Table~\ref{ephemmatchtab} for the contaminating source.

\textbf{HALO\_GHOST}: The ghost diagnostic value is too high. This diagnostic measures the transit strength for the out- and in-aperture pixels and determines if the transit is localized on the target star, or if it is due to contamination from a distant source. The TCE is an FP and the centroid offset major flag is set. See \S\ref{s:ghost}.

\textbf{HAS\_SEC\_TCE}: Another TCE on the same target with a higher planet number has the same period as the current transit-like TCE, but a significantly different epoch. This indicates that the current TCE is an eclipsing binary with the other TCE representing the secondary eclipse. If the PLANET\_OCCULT\_DV and PLANET\_OCCULT\_ALT flags are not set, the TCE is dispositioned as an FP with a stellar eclipse major flag set. See \S\ref{s:secondTce}.

\textbf{INCONSISTENT\_TRANS}: The ratio of the maximum SES value to the MES value is above threshold and the TCE has a period greater than 90 days. This flag indicates that the TCE has only a few transits and the MES is dominated by a single large event. Thus, the TCE is dispositioned as an FP with the not transit-like major flag set. See \S\ref{s:sesmes}.

\textbf{INDI\_TRANS\_(CHASES|MARSHALL|\\SKYE|ZUMA|TRACKER|RUBBLE)}: One or more of the individual transit metrics (Chases, Marshall, Skye, Zuma, Tracker, or Rubble) removed a transit causing the TCE's recalculated MES to drop below threshold, or the number of transits to drop below 3. The TCE is dispositioned as an FP with the not transit-like major flag set. See \S\ref{s:indivtrans}.

\textbf{IS\_SEC\_TCE}: The TCE has the same period, but a different epoch, as a previous transit-like TCE on the same target. This indicates that the current TCE corresponds to the secondary eclipse of an eclipsing binary (or a planet if the PLANET\_OCCULT\_DV or PLANET\_OCCULT\_ALT flags are set). Thus, the current TCE is dispositioned as an FP with both the not transit-like and stellar eclipse major flags set. See \S\ref{s:issecond}.

\textbf{LPP\_(ALT|DV)}: The Locality Preserving Projections (LPP) value\citet{Thompson2015b}, as computed using the ALT or DV detrending, is above threshold. This indicates that the TCE is not transit-shaped, and thus is dispositioned as an FP with the not transit-like major flag set. See \S\ref{s:lpp}.


\textbf{MOD\_NONUNIQ\_(ALT|DV)}: The Model-shift 1 test, performed with the ALT or DV detrending, is below threshold. This test calculates the significance of the primary event, taking into account red noise, and compares it to the false alarm threshold. This flag indicates the primary event is not significant compared to the amount of systematic noise in the light curve, and thus the TCE is dispositioned as an FP with the not transit-like major flag set. See \S\ref{s:ms}.


\textbf{MOD\_ODDEVEN\_(ALT|DV)}: The odd/even statistic from the Model-shift test is calculated with the ALT or DV detrending. This statistic compares the best-fit transit model to the odd and even transits separately and determines that the difference in the resulting significance values is above threshold. When set, the transit-like TCE is dispositioned as an FP with the stellar eclipse major flag set. See \S\ref{s:oddeven}.


\textbf{MOD\_POS\_(ALT|DV)}: The Model-shift 3 test, performed with the ALT or DV detrending, is below threshold. This test compares the significance of the primary and positive-going events in the phased light curve to help determine whether the primary event is unique. This flag indicates that the TCE is likely noise and thus is dispositioned as an FP with the not transit-like major flag set. See \S\ref{s:ms}.


\textbf{MOD\_SEC\_(ALT|DV)}: The Model-shift 4, 5, and 6 values, calculated using the ALT or DV detrending, are above threshold. This test calculates the significance of the secondary event divided by F$_{\mathrm{red}}$, the ratio of red noise to white noise in the light curve. The same calculation is done for the difference between the secondary and tertiary event significance values, and the difference between the secondary and positive event significance values. They indicate that there is a unique and significant secondary event in the light curve (i.e., a secondary eclipse). Thus, assuming the PLANET\_OCCUL\_(ALT|DV) flag is not set, the TCE is dispositioned as an FP with the stellar eclipse major flag set. See \S\ref{s:second}.


\textbf{MOD\_TER\_(ALT|DV)}: The Model-shift 2 test, performed with the ALT or DV detrending, is below threshold. This test calculates the difference between the primary and tertiary event significance values. This flag indicates that the primary event is not unique in the phased light curve, and thus the TCE is likely noise and dispositioned as an FP with the not transit-like major flag set. See \S\ref{s:ms}.


\textbf{NO\_FITS}: Both the trapezoidal and the original DV transit fits failed to converge. This indicates the signal is not sufficiently transit-shaped in either detrending to be fit by a transit model. The TCE is dispositioned as an FP with the not transit-like major flag set. See \S\ref{s:nofits}.

\textbf{PERIOD\_ALIAS\_(ALT|DV)}: Using the phases of the primary, secondary, and tertiary events from the Model-shift test run on the ALT or DV detrended data, a possible period alias is seen at a ratio of $N$:1, where $N$ is an integer of 3 or greater. This indicates the TCE has likely been detected at a period that is $N$ times longer than the true orbital period. This flag is currently informational only and not used to declare any TCE an FP. See \S\ref{s:periodalias}.


\textbf{PLANET\_IN\_STAR}: The original DV planet fits indicate that the fitted semi-major axis of the planet is smaller than the stellar radius. As it is possible that the stellar data is not accurate, this flag is currently informational only and not used to declare any TCE an FP. See \S\ref{s:planetinstar}.

\textbf{PLANET\_OCCULT\_(ALT|DV)}: A significant secondary eclipse was detected in the ALT or DV detrending, but it was determined to possibly be due to planetary reflection and/or thermal emission. While the stellar eclipse major flag remains set, the TCE is dispositioned as a PC. See \S\ref{s:sscand}.


\textbf{PLANET\_PERIOD\_IS\_HALF\_(ALT|DV)}: A significant secondary eclipse was detected in the ALT or DV detrending, but it was determined to be the same depth as the primary within the uncertainties. Thus, the TCE is possibly a PC that was detected at twice the true orbital period. When this flag is set, it acts as an override to other flags such that the stellar eclipse major flag is not set, and thus the TCE is dispositioned as a PC if no other major flags are set. See \S\ref{s:sscand}.


\textbf{RESIDUAL\_TCE}: The TCE has the same period and epoch as a previous transit-like TCE. This indicates the current TCE is simply a residual artifact of the previous TCE that was not completely removed from the light curve. Thus, the current TCE is dispositioned as an FP with the not transit-like major flag set. See \S\ref{s:sameperiod}.

\textbf{SAME\_NTL\_PERIOD}: The current TCE has the same period as a previous TCE that was dispositioned as an FP with the not transit-like major flag set. This indicates that the current TCE is due to the same not transit-like signal. Thus, the current TCE is dispositioned as an FP with the not transit-like major flag set. See \S\ref{s:sameperiod}.

\textbf{SEASONAL\_DEPTH\_(ALT|DV)}: There appears to be a significant difference in the computed TCE depth from different seasons using the ALT or DV detrending. This indicates significant light contamination, usually due to a bright star at the edge of the aperture, which may or may not be the origin of the transit-like event. As it is impossible to determine whether or not the TCE is on-target from this flag alone, it is currently informational only and not used to declare any TCE an FP. See \S\ref{s:seasonaldiff}.


\textbf{SWEET\_EB}: The sine wave event evaluation test (SWEET) is above threshold, the detected signal has an amplitude less than the TCE's depth, and the TCE period is less than 5 days. This flag indicates that there is a significant sinusoidal variability in the PDC data at the same period as the TCE due to out-of-eclipse EB variability. The transit-like TCE is dispositioned as an FP with the stellar eclipse major flag set. See \S\ref{s:sweeteb}.

\textbf{SWEET\_NTL}: The sine wave event evaluation test (SWEET) is above threshold, the detected signal has an amplitude greater than the TCE's depth, and the TCE period is less than 5 days. This flag indicates that there is a significant sinusoidal variability in the PDC data at the same period as the TCE, and the detected event is due to stellar variability and not a transit. The TCE is dispositioned as an FP with the not transit-like major flag set. See \S\ref{s:sweetntl}.

\textbf{TRANS\_GAPPED}: The fraction of gapped transit events is above threshold. This flag indicates that a large number of observable transits had insufficient in-cadence data. The TCE is dispositioned as an FP with the not transit-like major flag set. See \S\ref{s:rocky}.

\begin{deluxetable}{lrl}
\tablecolumns{4}
\tablewidth{\linewidth}
\tabletypesize{\scriptsize}
\tablecaption{\opstces\ Minor Flag Statistics}
\tablehead{
\colhead{Minor Flag} & \colhead{Num. Flagged} & \colhead{\% Flagged}
}
\startdata
ALL\_TRANS\_CHASES & 8176 & 25.145 \\
CENT\_CROWDED & 42 &  0.129 \\
CENT\_FEW\_DIFFS & 8957 & 27.547 \\
CENT\_FEW\_MEAS & 589 &  1.811 \\
CENT\_KIC\_POS & 1635 &  5.028 \\
CENT\_NOFITS & 1952 &  6.003 \\
CENT\_RESOLVED\_OFFSET & 1956 &  6.016 \\
CENT\_SATURATED & 3820 & 11.748 \\
CENT\_UNCERTAIN & 89 &  0.274 \\
CENT\_UNRESOLVED\_OFFSET & 743 &  2.285 \\
DEEP\_V\_SHAPED & 895 &  2.753 \\
DEPTH\_ODDEVEN\_ALT & 220 &  0.677 \\
DEPTH\_ODDEVEN\_DV & 177 &  0.544 \\
EPHEM\_MATCH & 1841 &  5.662 \\
HALO\_GHOST & 3150 &  9.688 \\
HAS\_SEC\_TCE & 1141 &  3.509 \\
INCONSISTENT\_TRANS & 7219 & 22.202 \\
INDIV\_TRANS\_ & 14541 & 44.721 \\
\_CHASES & 5468 & 16.817 \\
\_MARSHALL & 7614 & 23.417 \\
\_SKYE & 4790 & 14.732 \\
\_ZUMA & 2103 &  6.468 \\
\_TRACKER & 1880 &  5.782 \\
\_RUBBLE & 7137 & 21.950 \\
IS\_SEC\_TCE & 1136 &  3.494 \\
LPP\_ALT & 9948 & 30.595 \\
LPP\_DV & 19271 & 59.268 \\
MOD\_NONUNIQ\_ALT & 11376 & 34.987 \\
MOD\_NONUNIQ\_DV & 11380 & 34.999 \\
MOD\_ODDEVEN\_ALT & 487 &  1.498 \\
MOD\_ODDEVEN\_DV & 401 &  1.233 \\
MOD\_POS\_ALT & 5578 & 17.155 \\
MOD\_POS\_DV & 4672 & 14.369 \\
MOD\_SEC\_ALT & 1407 &  4.327 \\
MOD\_SEC\_DV & 1161 &  3.571 \\
MOD\_TER\_ALT & 5340 & 16.423 \\
MOD\_TER\_DV & 4970 & 15.285 \\
NO\_FITS & 113 &  0.348 \\
PERIOD\_ALIAS\_ALT & 5 &  0.015 \\
PERIOD\_ALIAS\_DV & 2 &  0.006 \\
PLANET\_IN\_STAR & 87 &  0.268 \\
PLANET\_OCCULT\_ALT & 18 &  0.055 \\
PLANET\_OCCULT\_DV & 39 &  0.120 \\
PLANET\_PERIOD\_IS\_HALF\_ALT & 18 &  0.055 \\
PLANET\_PERIOD\_IS\_HALF\_DV & 4 &  0.012 \\
RESIDUAL\_TCE & 107 &  0.329 \\
SAME\_NTL\_PERIOD & 2061 &  6.339 \\
SEASONAL\_DEPTH\_ALT & 89 &  0.274 \\
SEASONAL\_DEPTH\_DV & 83 &  0.255 \\
SWEET\_EB & 209 &  0.643 \\
SWEET\_NTL & 1377 &  4.235 \\
TRANS\_GAPPED & 5428 & 16.694 \\
\enddata
\tablecomments{For these statistics the \opstce{} set does not include the rogue or banned TCEs. Most \opstces\ fail more than one test, so the percentages are not expected to add up to 100\%.}
\label{t:minorstats}
\end{deluxetable}

\clearpage

\listofchanges

\end{document}